# *Constructive Patterns of Logical Truth*

Elnaserledinellah Mahmood Abdelwahab
makmad.org e.V., Hanover (Germany)
elnaser@makmad.org

## ABSTRACT

The simplified linguistic relation between syntax and semantics as intrinsic property of classic Arabic motivates a dedicated look at P vs. NP in light of efforts and solutions presented by ancient Arab- and Muslim scholars to facilitate logical- and mathematical deduction. In Islamic Jurisprudence (*Fikh*) it has recently been shown [Abdelwahab et al. 2014] that if a formal system expressing *Fikh* is chosen in such a way that it is both, logically complete and decidable, the question of a complete and consistent legislation is *decidable*. If this formal *Fikh*-system is additionally chosen to be at least as expressive as propositional logic, the deduction of detailed sentences is efficient while the deduction of general rules is NP-complete. Further investigation reveals that ancient scholars adopted a very efficient approach for checking the validity of assertions with regard to both, language *and* logical argument which was mainly characterized by the extensive use of *syntactical patterns* already existent in input-variables. Accordingly, the structure of input-variables in 3-SAT-problems is investigated. The discovered variable-structure divides 3-SAT-formulas into three interrelated types of which one enables efficient pattern-oriented procedures. A novel 3-SAT-solver technique is introduced which binds the resolution of 3-SAT-formulas to the construction of FBDDs using the newly proposed pattern procedures. Finally, both 3-SAT and the question of finding 2-approximation algorithms for MinFBDD (the problem of minimizing FBDDs) are addressed. Positive results are reported for both questions. A practically important consequence is the ability to construct near to minimal FBDDs with a polynomial number of nodes for all Boolean functions expressed in a compact way. Eventually, an application of this new 3-SAT-solver is shown to enable polynomial upper bounds of the number of nodes in FBDDs constructed for finite projective planes problems overhauling the currently known exponential lower bounds.

## INTRODUCTION

How fast can two numbers be multiplied?

The answer to this question is different from the rather classical question: How fast can two numbers be multiplied in a universal way? While the latter currently is $O(n\log n\ 2^{O(\log^* n)})$ for $n$-digit numbers via *Fürer*'s algorithm, where $\log^* n$ represents what is called the *iterated logarithm operation*, the former depends on the structure numbers can reveal. If, e.g., the two factors belong to the Set $\{10^i \mid i>0\}$, an optimal algorithm would simply append all zeros of one of them to those of the other performing this operation in just O($n$). Such an algorithm expects *clues* in its input-variables - *depending only on the syntactical form of the input* - to yield an efficient solution[1].

In what follows we shall call this type of algorithms *pattern-oriented (*AP*)*. In contrast, a *universal algorithm* (AU) shall be a procedure where input-variables are not assumed to possess any structural information serving to reduce complexity[2].

Can AP substitute AU?

The above example of multiplication doesn't seem to make the case since arbitrary numbers are not necessarily of a form similar to $\{10^i \mid i>0\}$.

Since algorithms are not necessarily about numbers, let us take a look at those used in other domains such as Natural Language Processing (NLP) determining the meaning of an English word. A simple AU searches in a pre-arranged lexicon which stores meanings for all admissible syntactical forms as well as other useful information. For such an AU, "admissible" simply means *existent in a pre-defined list.* Consider on the other hand the problem of determining the meaning of a word in the classical Arabic language. We could apply the former AU to solve this problem. We could also write an AP which makes use of an intrinsic property of Arabic, namely that the morphology of words is based on "roots" which represent semantic categories. In Arabic, e.g., the root of "write" has the form *k-t-b* where terms are completed by supplemental vowels and additional consonants, e.g., *kitāb* "book", *kutub* "books", *kātib* "writer", *kuttāb* "writers", *kataba* "he wrote", *yaktubu* "he writes", etc.

Sometimes, morphological rules are applied and letters are either transformed to others or deleted.

The root for "water" for example is *mauh* but it is written and pronounced *mā'*. Syntactic forms of nouns and verbs are both, *well defined* and *easily distinguishable.* According to an unproven but commonly accepted conjecture among ancient Arab linguists, this specific word structure as well as morphological rules associated with it helps avoiding ambiguous meanings and enable very compact word-lexica [Abdelwahab 1986].

Evidently, such an AP would be equivalent to AU.

Turning our attention to the problem of finding the semantics of whole sentences in English, the required AU becomes much more complex and assuming that AU uses LFG (Lexical-Functional Grammar), word-problems for LFGs are known to be NP-hard [Berwick 1982].

LFG views language as being made up by multiple dimensions of structure[3]. Each of these dimensions is represented as a distinct structure with its own rules, concepts, and form. The primary structures that have figured in LFG research are:

- Representation of grammatical functions (*f-structure*)
- Structure of syntactic constituents (*c-structure*)

---

1 An *efficient solution* always implies a complexity within P.

2 There are many other aspects of *universality* which may serve as a basis for different definitions of *universal algorithm.*

3 Lexical functional grammar. (2015, May 26). In Wikipedia, The Free Encyclopedia. Retrieved 11:33, May 11, 2015, from https://en.wikipedia.org/w/index.php?title=Lexical_functional_grammar&oldid=664164917

This dissociation of syntactic structure from predicate argument structures (essentially a rejection of *Chomsky*'s Projection Principle[4]) is crucial to the LFG framework. While *c*-structure varies somewhat across languages, the *f*-structure representation, which contains all necessary information for the semantic interpretation of an utterance, is claimed to be universal.

The lexical entry (or semantic form) includes information about the meaning of the lexical item, its arguments, and the grammatical functions (e.g., subject, object, etc.) that are associated with those arguments. Grammatical functions play an essential role in LFG, however, they have no intrinsic significance and are located at the interface between the lexicon and the syntax. LFG imposes the restriction of Direct Syntactic Encoding, which prevents any syntactic process from altering the initial assignment of a grammatical function [Neidel 1994].

A thorough look into the 3-SAT-reduction used in the proof of NP-completeness of LFG word problems given in [Berwick 1982] reveals the deep reason for the presumable intractability:

"One and the same terminal item can have two distinct lexical entries, corresponding to distinct lexical categorizations; for example, baby can be both a noun and a verb. If we had picked baby to be a verb, and hence had adopted whatever features are associated with the verb entry for baby to be propagated up the tree, then the string that was previously well-formed, "the baby is kissing John", would now be considered deviant. If a string is ill-formed under all possible derivation trees and assignments of features from possible lexical categorizations, then that string is not in the language generated by the LFG. *The ability to have multiple derivation trees and lexical categorizations for one and the same terminal item plays a crucial role in the reduction proof:* it is intended to capture the satisfiability problem of deciding whether to give an atom $X_i$ a value of T or F."[5]

Considering the same problem in classic Arabic, the above AU could be build upon LFGs with another very useful property of classic Arabic becoming apparent: All words used in sentences are annotated with signs revealing their functions in grammar and meaning[6]. Those signs are called *tashkīl*[7]. The exercise of using *tashkīl* as an interface between syntax and semantics to deduce grammatical structures and intended meanings in an Arabic sentence[8] is called *I'rāb*[9]. The main purpose of *I'rāb* is to remove semantic ambiguity resulting from syntactical similarity[10]. *I'rāb* can be an efficient procedure when it uses predetermined word- and sentence structure rules applying them directly to syntactical objects of the concerned phrase. If, e.g., we have two sentences: Naserun yashkuru Allaha (= Naser thanks Allah), Nasara Alragulu Akhahu (= *the man supported his brother*) the terminal Naser which might be - without *tashkīl* - confused between verb and noun can be identified using *I'rāb* via its *tashkīl*: "ara" to be a verb in the second phrase while it must be a noun in the first one because the "un" (also called *Tanween*) is only used for nouns[11]. Ancient Arab- and Muslim scholars devised a set of rules for the syntactical recognition of the three main categories of an Arabic sentence: Noun, verb and preposition [Ibn Malek 1274][12]. Deducing grammatical functions using *I'rāb* is similar to deducing *c&f*-structures in LFGs, the difference being that it is guided by *tashkīl*, a purely syntactical attribute of words. Thus, an LFG-based algorithm doing *I'rāb* would be AP.

[4] Under the Projection Principle, the properties of lexical items must be preserved while generating the phrase structure of a sentence.

[5] [Berwick 1982] p. 103

[6] This is consensus among of the majority of ancient Arabic Linguists of whom *Ibn Djinni* (920-1002) and *Sibaweih* (765-796) are two prominent names.

[7] Arabic diacritics. (2015, July 17). In Wikipedia, The Free Encyclopedia. Retrieved 12:02, July 29, 2015, from https://en.wikipedia.org/w/index.php?title=Arabic_diacritics&oldid=671851118

[8] An exercise considered to be in the core of any Arabic language course.

[9] ʾ*I*ʿrab. (2015, July 22). In Wikipedia, The Free Encyclopedia. Retrieved 12:03, July 29, 2015, from https://en.wikipedia.org/w/index.php?title=%CA%BEI%CA%BFrab&oldid=672637105

[10] The meaning of the word *I'rāb* is in fact: Clarify. Also: *I'rāb* has the same root as the word: Arab- and ancient linguists believe that this is attributed to the fact, that Arabs needed to distinguish themselves from people who cannot express clearly (all non-Arabs are called: *Aājem* which essentially means "not clear").

According to another ancient conjecture, there are subsets of the classic Arabic language where a word in a sentence is completely disambiguated when its *tashkīl* is known and *I'rāb* does not need contextual information to deduce grammatical functions. The proof of this conjecture would imply our LFG-AP for this subset of Arabic to be efficient[13].

In the past paragraphs we discussed examples of applications where AP presumably can (e.g., NLP) or cannot (e.g., multiplication) be substituted without loss of generality (w.l.o.g.) for AU. Do we have to go through all possible algorithms and all possible applications to be able to answer the general question of whether an AP can *always* be found and substituted for an AU?

*No.*

According to complexity theory, it is sufficient to show and investigate the existence and properties of one single AP for any known NP complete problem, the prominent candidate being the 3-SAT-problem. 3-SAT is the problem of finding a satisfying assignment for propositional formulas having max. three variables in their clauses[14]. It is hence a problem of logic.

Recognizing computability-related, intrinsic properties of classic Arabic makes it worthwhile to take a look at efforts done by ancient Arab- and Muslim scholars to facilitate logical and mathematical deduction using syntax-oriented techniques.

[11] There are of course few exceptions where *tashkīl* is not sufficient to distinguish between verbs and nouns (like in the case of the word: "Ahmadu"). Those cases have very distinguished lexical as well as grammatical categories and can be recognized via contextual information.

[12] The first five verses of the beginning of the 1000 verses long poem describing rules of Arabic. Alfiya. (2014, October 16). In Wikipedia, The Free Encyclopedia. Retrieved 16:14, July 29, 2015, from https://en.wikipedia.org/w/index.php?title=Alfiya&oldid=629819477

[13] Modern approaches to Arabic NLP (like the one described in [Attia 2008]) apply linguistic theories to modern Arabic which is written and spoken today without *tashkīl* (making therefore any classical *I'rāb* attempt fruitless). While those approaches have certainly practical validity, they fail to address the important theoretical question: How far can the classical Arabic language, written and spoken by ancient, genuine Arabic speakers be formalized and thus mechanized in a computable way? Such a study may not only shed light on intrinsic natural Arabic language properties, but also and foremost on computability related ones.

[14] Boolean satisfiability problem. (2015, July 8). In Wikipedia, The Free Encyclopedia. Retrieved 16:15, July 29, 2015, from https://en.wikipedia.org/w/index.php?title=Boolean_satisfiability_problem&oldid=670583991

It is commonly known that the term *Algorithm* is a Latin short-version of the name *Muhammad ibn Musa Al-Khwarizmi* (163-235AH / 780-850AD), who is the author of the Arabic book *Kitab Al-Jabr wa-l-Muqabala* (215AH / 830AD) [Al-Khwarizmi], i.e., *The Compendious Book on Calculation by Completion and Balancing*. This book was translated into Latin in the 12th century A.D. entitled *Liber Algebrae et Almucabola* [Chester] with *algebrae* and *Almucabola* being transliterated into Latin from the Arabic title where the term *Algebra* is derived from *Al-Jabr* in the title of *Al-Khwarizmi's* book [Abdelwahab et al.].

While there is widespread belief that *Kitab Al-Jabr wa-l-Muqabala* is a textbook for mathematics which, among others, introduced general rules to solve algebraic problems with one variable reducible to quadratic equations, it is first and foremost a textbook for *Fiqh*:

As traditionally and practically done in *Fiqh*, the first half of *Kitab Al-Jabr wa-l-Muqabala* introduced the applied methodology and term definitions.

The remaining half of his book solves legal questions on trade (commercial transactions), geometry (plane surface distributions) as well as testimonies.

Based on the newly introduced algebraic method, the by far most important part of *Kitab Al-Jabr wa-l-Muqabala* deals with Islamic heritage law.

Accordingly, *Al-Khwarizmi's* Algebra is just what its author says in the introduction: "[a] work on algebra, confining it to the fine and important parts of its calculations, such as people constantly require in cases of inheritance, legacies, partition, law-suits, and trade, and in all their dealings with one another, or where surveying, the digging of canals, geometrical computation, and other objects of various sorts and kinds are concerned." [Al-Khwarizmi]

His method of solving linear- and quadratic equations consisted of first reducing the equation to one of six standard *syntactical forms*[15].

Other Muslim scholars working on logical deduction, crucial for correct application of *Fikh*-rules, adopted *Aristotelian syllogisms*[16] as means of reaching correct rulings from right premises [AlGhazali 505H]. Syllogisms are simple deductive arguments depending largely on properties of terms used in logical assertions which can also be checked in efficient syntactical ways[17]. The completeness of various formulations of syllogistic logic has been demonstrated in [Lukasiewicz 1951]. Although syllogistic systems provide some quantification properties,

---

[15] where $b$ and $c$ are positive integers: ($ax^2=bx$), ($ax^2=c$), ($bx=c$), ($ax^2+bx=c$), ($ax^2+c=bx$), ($bx+c=ax^2$)

[16] Syllogism. (2015, July 1). In Wikipedia, The Free Encyclopedia. Retrieved 16:25, July 29, 2015, from https://en.wikipedia.org/w/index.php?title=Syllogism&oldid=669410736

[17] To be considered valid, a syllogism must follow six basic rules.

A syllogism must contain exactly three terms. The violation of this rule is called the *fallacy of four terms*.

A syllogism must have exactly three propositions.

The middle term must be distributed at least one time. Violating this rule results in the *fallacy of the undistributed middle*. (When checking for this and the next rule, it is useful to mark the distribution of every term in the syllogism.)

No term that is undistributed in the premise may be distributed in the conclusion. The violation of this rule is either the fallacy of the *illicit major* or the fallacy of the *illicit minor* depending on whether the minor or major term contains the fallacy.

A syllogism cannot have two negative premises.

If a syllogism contains a negative premise, the conclusion must be negative; conversely, if it contains a negative conclusion, it must contain a negative premise.

there is a lack of predicates which have *arity* more than one as well as the possibility to consider functions. Thus, they can be seen as subsystems of Monadic First-Order logic (MFO), which is also less expressive than full FOL. It has recently been shown [Abdelwahab et al. 2014] that if a formal system expressing *Fikh* is chosen in such a way that it is both *logically complete* and *decidable*, the question of a complete[18] and consistent[19] legislation is *decidable*[20]. If this formal *Fikh*-system is additionally chosen to be at least as expressive as propositional logic, the deduction of detailed sentences is *efficient* while the deduction of general rules is *NP-complete*.

Summarizing the above quick look into ancient Arab- and Muslim scholars efforts it appears that they adopted a highly efficient approach of finding validity of assertions in both, language and logical argumentation which was characterized by the extensive use of *syntactical patterns already existent in input variables simplifying solutions of otherwise hard problems* (i.e., essentially finding and applying AP instead of AU).

Does this approach work for 3-SAT?

An affirmative answer to this question follows these steps:

Identify structure in logical variables, similar to the case of *tashkīl*, linking syntax to semantics of 3-SAT-formulas (materialized in the combinatory space, i.e., the truth table).[21] This identification reveals a distinct *pattern property* of variables as opposed to the classic *container* property. Having identified this pattern structure, different classes of 3-SAT-formulas appear to be distinguishable. One of them can be efficiently implemented in a constructive AP.

This AP must be formalized, optimized, and its properties studied and proven. It should also be clear that the AP has general validity and can indeed be used as a substitute for known AU in any instance of the 3-SAT-problem.

Eventually, algorithmic transformations yield an application showing validity of the new results.

To transparently demonstrate the step-wise arguments, the following organization of this paper is chosen:

**Section I:** Two experiments are conducted in an informal way presenting 3-SAT-problem instances and Binary Decision Diagrams (BDDs) used to solve them with intrinsic structure unveiling self-evidently. Eventually, a precise conjecture and an objective are formulated.

**Section II:** The main formalism is presented. In the center of this formalism a defined AP is shown to possess 10 properties which are used in lemmas showing different behavior of multiple versions of it. Proofs leverage either straightforward induction or proof by contradiction. The unveiled variable structure is also shown to divide 3-SAT-formulas into three interrelated types of which only one produces efficient results with the AP. Lemmas pertaining to constructive properties and efficiency of the chosen optimized version of AP enable main theorems (Theorems 1&2) and support the correctness of the proposed conjecture and realization of the objective.

[18] Complete Legislation means that every *Fikh*-question has a ruling (verdict)

[19] Consistent Legislation means that every verdict has a reason

[20] It is still an open research question whether using syllogistic approaches for modern, automatic *Fikh*-systems is a more appropriate way to define efficient *Fikh*-algorithms or not

[21] The reader is reminded that the beginning of last century was marked by many thorough investigations related to variables used in logical statements. In fact, most fundamental ideas about arithmetic were bound to ways of viewing and using variables, e.g., *Russell*'s famous letter to *Frege* showing that *Frege*'s Basic Law V entails a contradiction. This argument has come to be known as *Russell's paradox* described more thoroughly in *Principia Mathematica* [Russell 1910] as being caused by the concept of a "true-variable" (German: *echte Veränderliche*) which is an entity defined to be ranging over a totality of entities to which it belongs itself. In his simple type-theory, the basis for all constructive approaches in modern computer science, free variables - in defining formulas - range over entities to which the collection to be defined does not belong.

**Section III**: Discusses the problem of constructing a FBDD to identify blocking sets in finite projective planes[22] where an application of the new AP is shown to overcome known exponential lower bounds.

This is achieved by showing that the assumed problem structure in the proof(s) of the exponential lower bound(s) is resolved if a 3-SAT-Solver using AP is put into action (Theorem 3). The result is a polynomial upper bound in the number of points/lines of the projective planes instance.

The appendix contains exhibits of AP-runs and extracts from final results for Projective Planes of order 2 (*Fano Planes*) and of order 3.

## I) EXPERIMENTS, CONJECTURE, AND OBJECTIVE

In what follows we are going to informally introduce a new way of visualizing variables/literals in 3-SAT-CNF Clause-Sets and explain known phenomena with its help.

Specifically, a variable is usually considered to be a *universal container* of data, i.e.:

> "[…] a storage location paired with an associated symbolic name (an identifier), which contains some known or unknown quantity of information referred to as a value. The variable name is the usual way to reference the stored value; this separation of name and content allows the name to be used independently of the exact information it represents."[23]

With reference to above container nature of variables theoretically enabling unrestricted and/or unstructured substitution of domain values contained therein, we call this classical way of considering variables the *container-view.*

In contrast, this paper shows a variable, especially a logical one, to possess an *intrinsic pattern* revealing the canonical distribution of its truth-values, hereafter referred to as *pattern-view*.

The Clause-Sets used in this section are very simple yet sufficient to show the desired properties of variables and provide with clues to more elaborate thoughts. The end of this section is marked by a **conjecture** and an **objective**. They shall both constitute the motivation behind the remainder of this paper.

**Experiment I: State-of-the-Art**

Suppose $S=\{\{a,b,c\}\{x,y,z\}\}$ is a 3-SAT-CNF Clause-Set where $a<b<c<x<y<z$ is the variable ordering used in its truth table[24] (hereafter referred to as *canonical ordering*

---

22 Blocking set. (2015, February 18). In Wikipedia, The Free Encyclopedia. Retrieved 18:14, July 29, 2015, from https://en.wikipedia.org/w/index.php?title=Blocking_set&oldid=647680966

23 Variable (computer science). (2015, July 23). In Wikipedia, The Free Encyclopedia. Retrieved 07:15, July 27, 2015, from https://en.wikipedia.org/w/index.php?title=Variable_(computer_science)&oldid=672714656

24 i.e., *a* is the leftmost, *z* the rightmost variable.

Below are ways of instantiating S according to four different orderings leading to different BDDs[25]. As can be seen, BDD4 where instantiations did not violate the canonical ordering represents a relatively small diagram:

BDD1

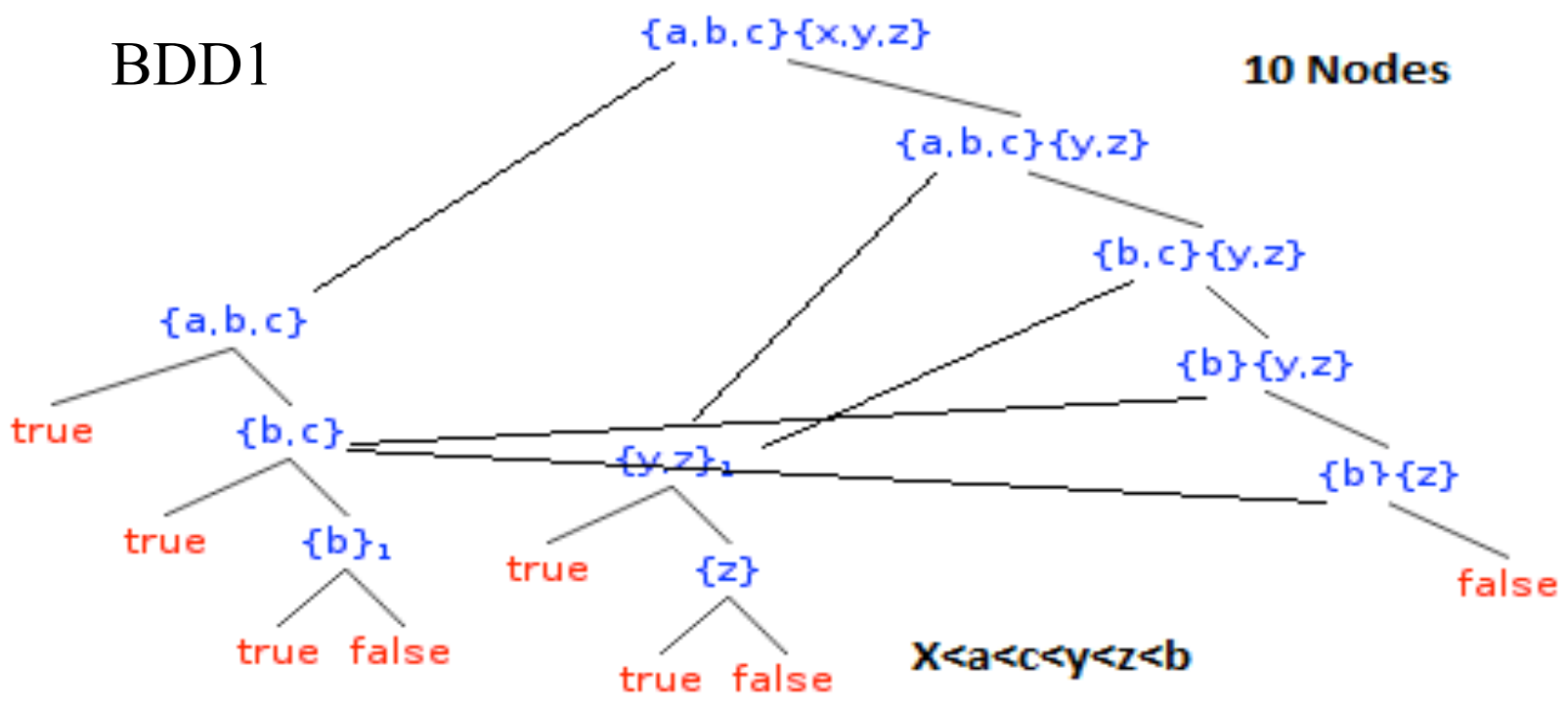

BDD2

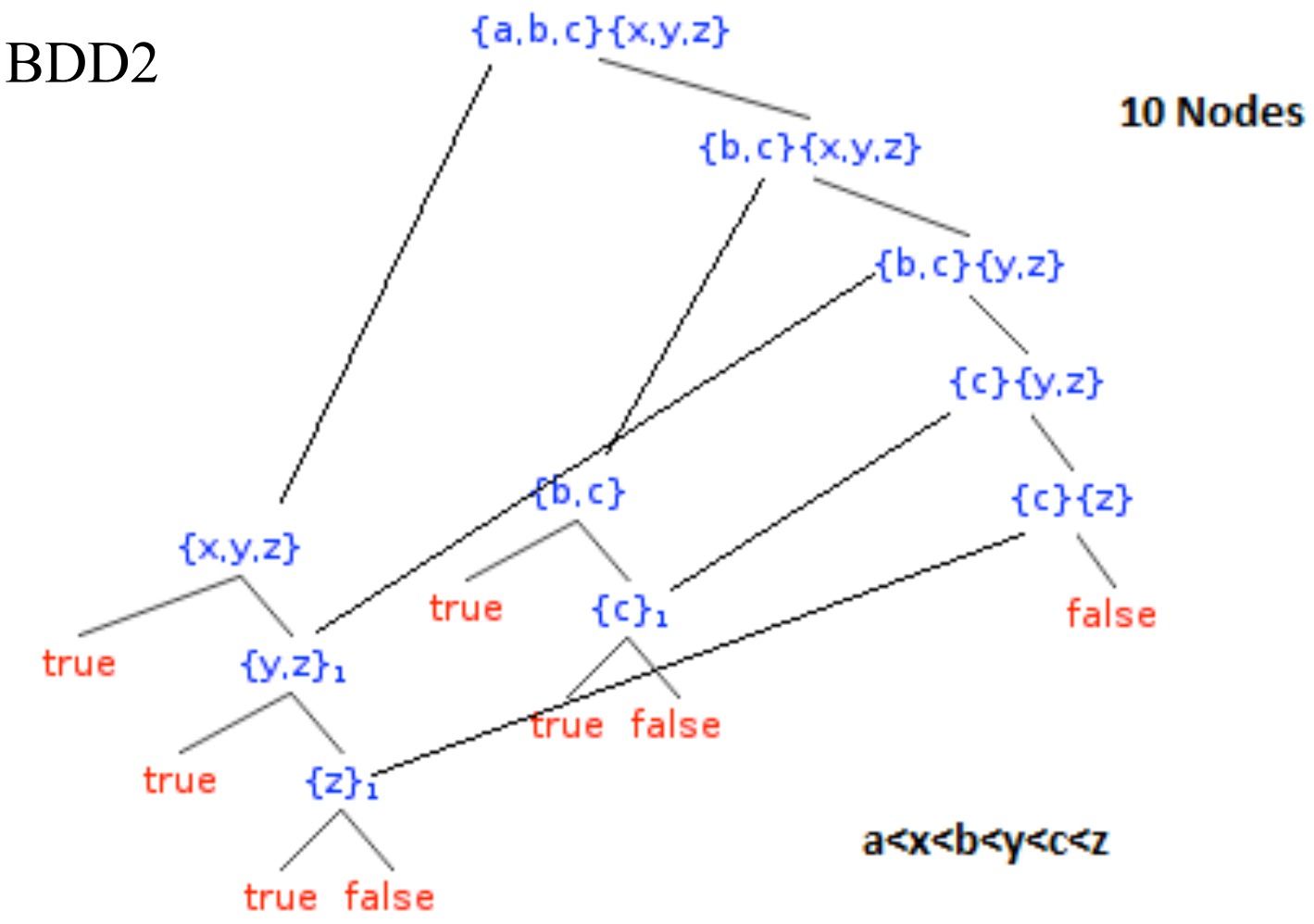

BDD3

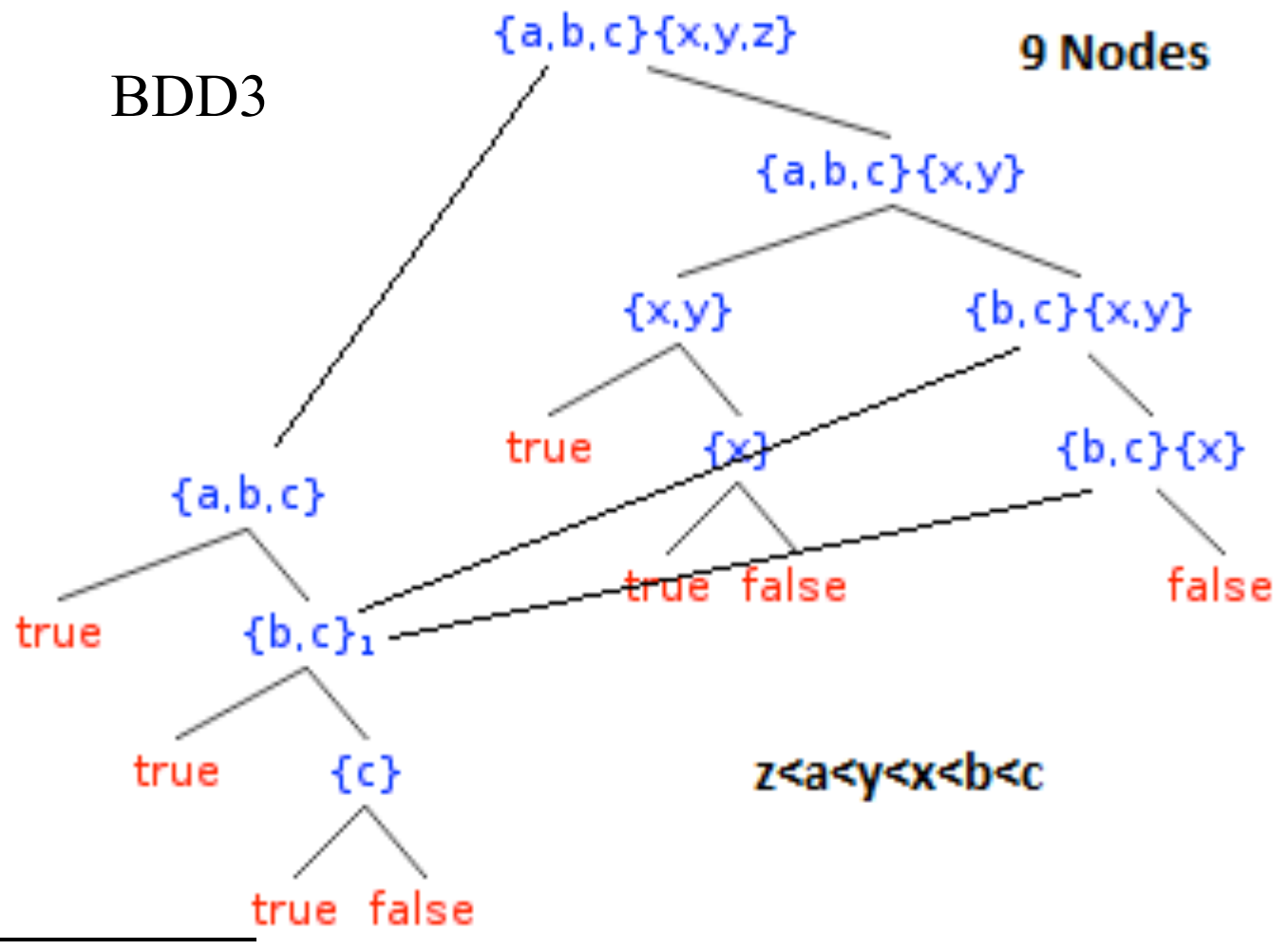

25 We use a formulation of BDDs which allows Clause-Sets instead of single literals/variables to be used in nodes. This shall be properly précised in the next sections (c.f. [Wegener 2000]).

BDD4

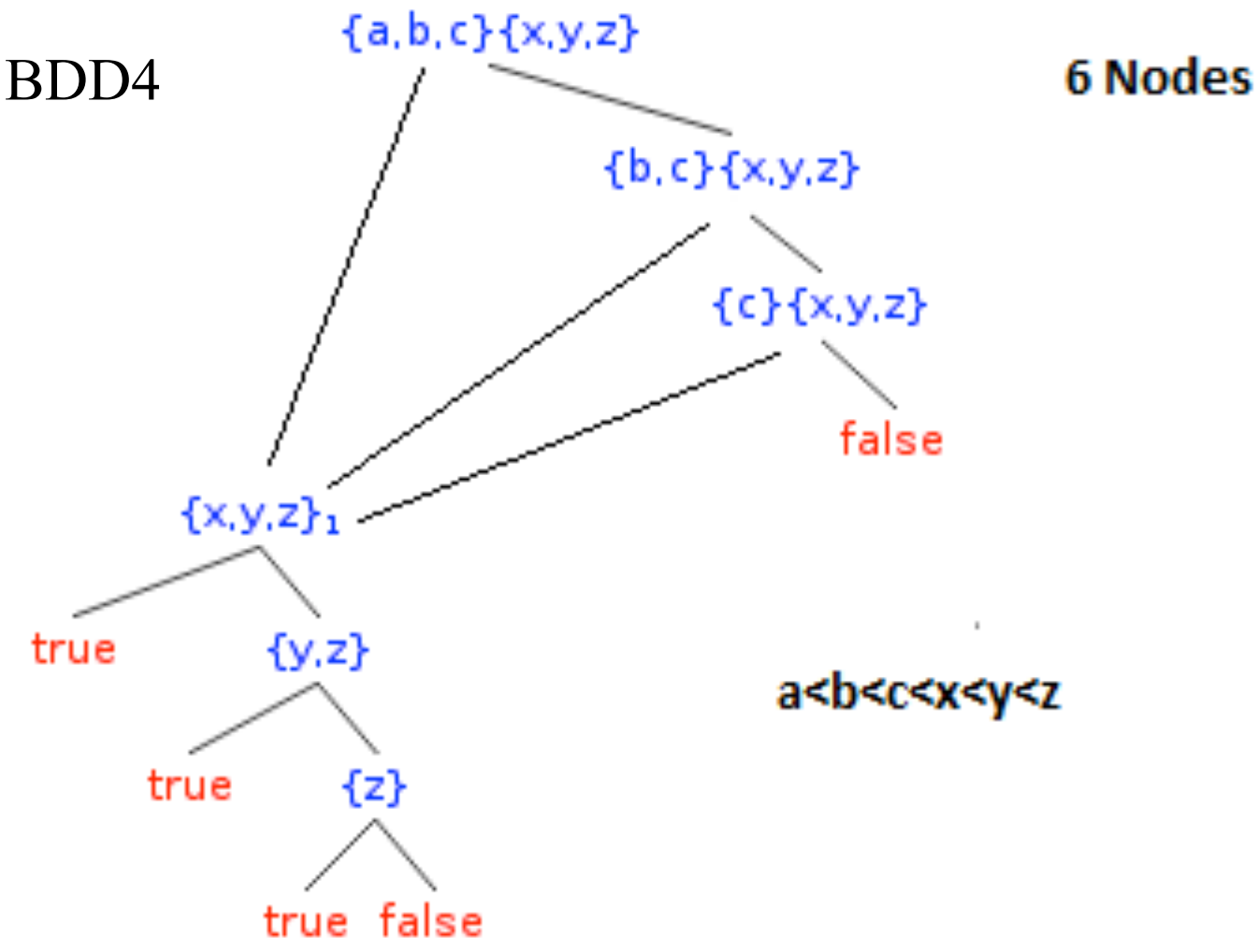

Does this mean that any instantiation procedure following the canonical ordering of an arbitrary 3-SAT-CNF Clause-Set will possess the smallest number of nodes?

***No.***

Take a look at the resolution of the extension of S:

S'={{*a*,*b*,*c*}{*x*,*y*,*z*}{*a*,*c*,*x*}}.

The following BDD5 and BDD6 show that ordering $a<c<b<x<y<z$ produces a BDD with only 7 unique-nodes compared to 8 achieved by the canonical one.

BDD5

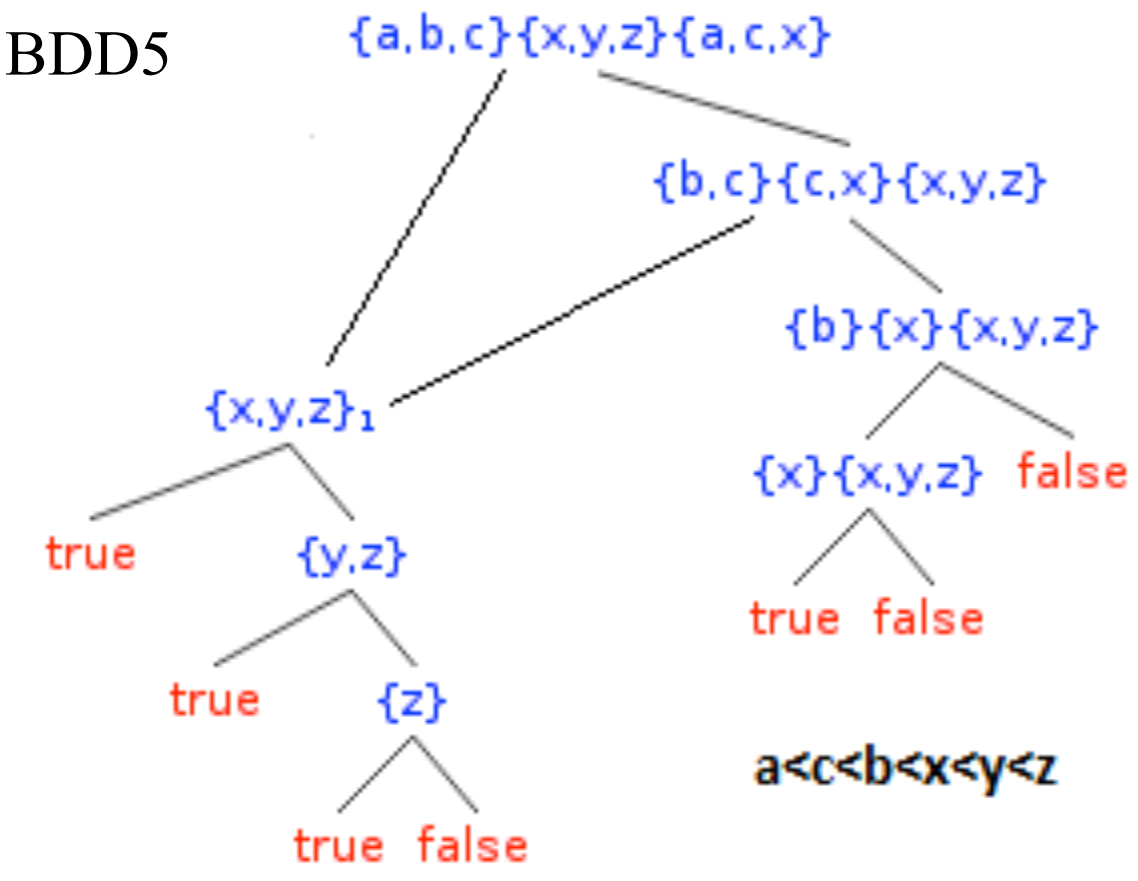

BDD6

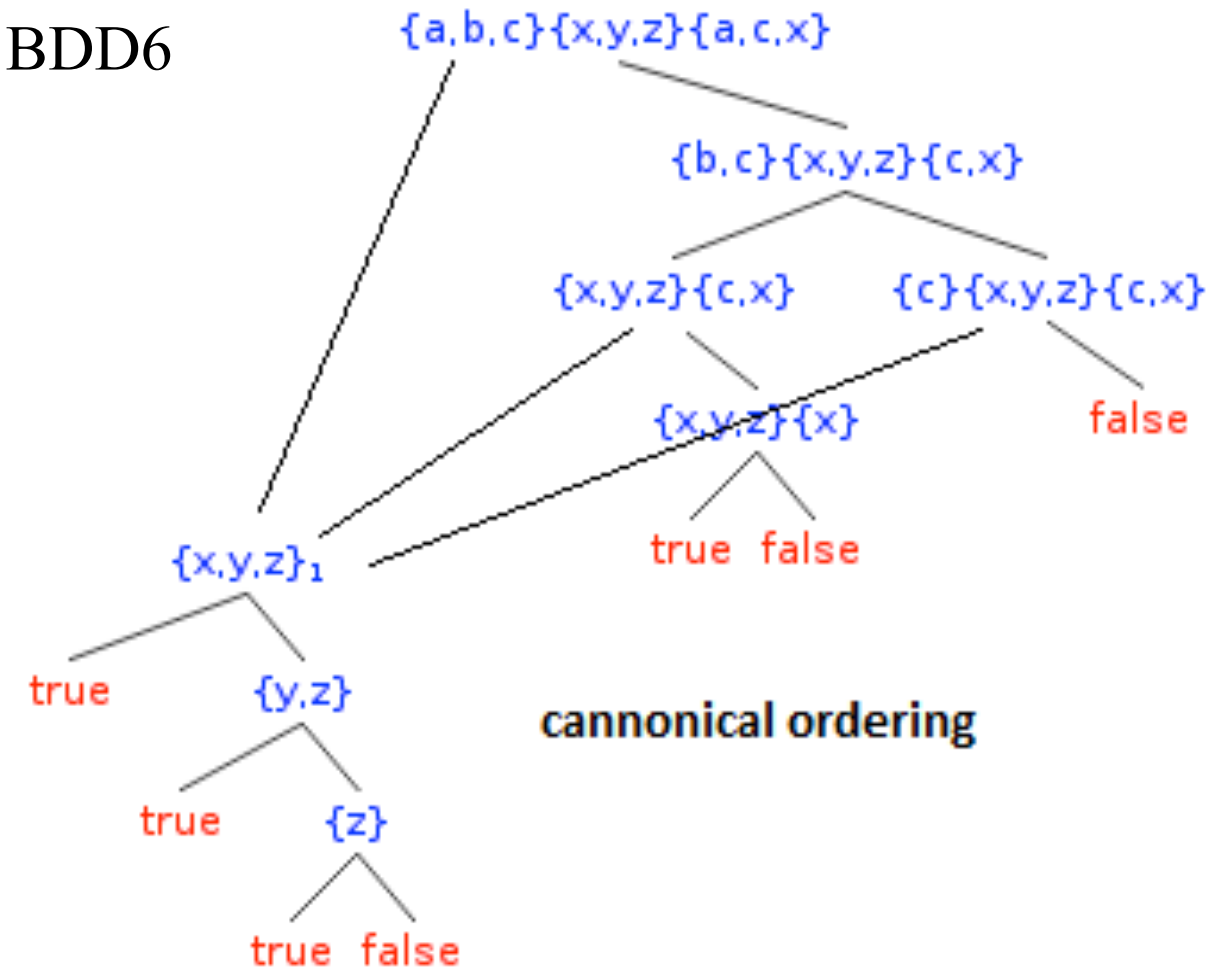

How can this be explained?

First: Using the canonical ordering in prioritizing variable instantiations for S produces better results than using arbitrary ones.

Second: This very canonical ordering fails to produce the best result in the extended case S'.

More generally: How do variable orderings affect sizes of BDDs constructed while resolving 3-SAT-CNF clauses?[26]

In diverse literature related to subject matter this central question is left unanswered. The two related problems of finding the best variable ordering for BDDs (of various types) and finding a minimal BDD are both known to be NP-complete (c.f. [Bolling et al. 1996], [Tani et al. 1993], [Guenther et al. 1999], and [Sieling 1999]).

Is there a way to acquire a deeper understanding of the nature of variables and their orderings in the context of resolving 3-SAT-CNF-clauses?

**Experiment II: Truth Patterns**

Let S={$\{x_1,x_2\}\{x_3,x_4\}\{x_0,x_5\}$} be a CNF Clause-Set. Then the truth table below (TT. 1) can be constructed:

| X0 | X1 | X2 | X3 | X4 | X5 | S |
|---|---|---|---|---|---|---|
| 0 | 0 | 0 | 0 | 0 | 0 | 0 |
| 0 | 0 | 0 | 0 | 0 | 1 | 0 |
| 0 | 0 | 0 | 0 | 1 | 0 | 0 |
| etc. | …. | … | … | … | … | … |

**Truth Table 1**

We shall define the following strings for each variable constituting what we call a *Truth Pattern* (TP) for that very variable[27]. Bits in such a string represent rows in above table. If the value of the variable in the respective row *i* is 0, the string has in bit *i* the value 0, else it has the value 1. Doing this for the six variables in S yields the following Set of strings:

$S_{X0}$ = 32(0) 32(1)[28] $S_{X1}$ = 2(16(0)16(1))[29] $S_{X2}$ = 4(8(0)8(1)) $S_{X3}$ = 8(4(0)4(1))
$S_{X4}$ = 16(2(0)2(1)) $S_{X5}$ = 32(1(0)1(1))

Forming the equivalent TPs for clauses $\{x_1,x_2\},\{x_3,x_4\},\{x_0,x_5\}$ which contain pairs of variables/literals is eventually a simple bit-OR operation which shall be called PatternOr giving the following results:

$S_{\{X1,X2\}}$=PatternOr($S_{X1}$, $S_{X2}$)=2(8(0)8(1)16(1))[30]
$S_{\{X3,X4\}}$=PatternOr($S_{X3}$, $S_{X4}$)=8(2(0)2(1)4(1))
$S_{\{X0,X5\}}$=PatternOr($S_{X1}$, $S_{X2}$)=16(1(0)1(1))32(1)[31]

Let $S_1$= $S_{\{X1,X2\}}$, $S_2$= $S_{\{X3,X4\}}$, $S_3$= $S_{\{X0,X5\}}$ and let $S_1$'=8(0)8(1), $S_2$'=2(0)2(1), $S_3$'=1(0)1(1), then the following trees represent the above strings:

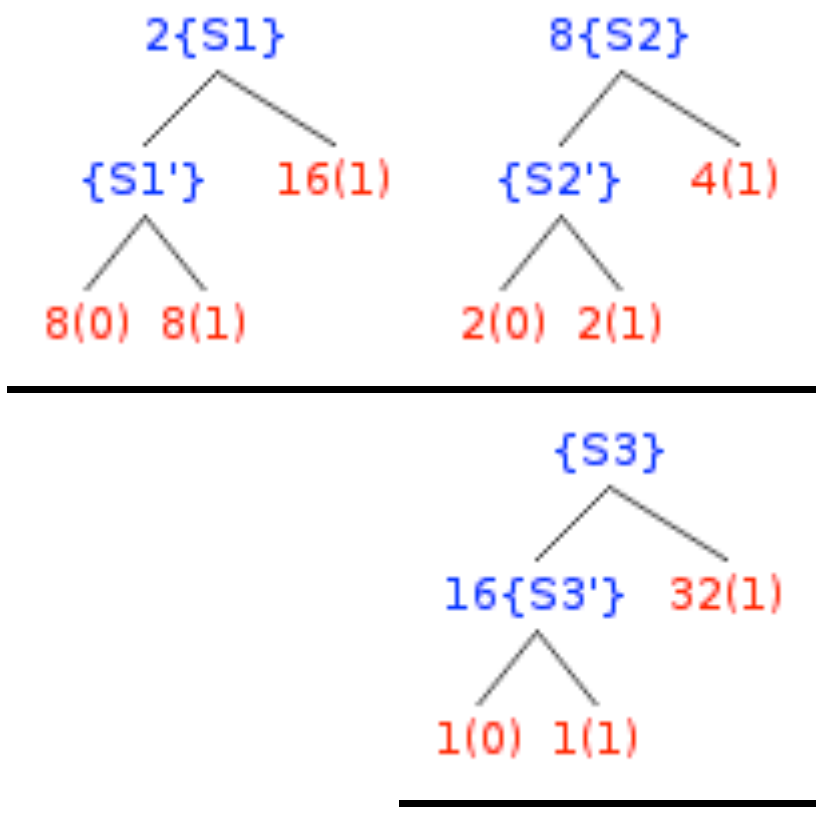

26 That variable ordering does affect sizes of BDDs is undisputed.

27 Truth Patterns represent the enumeration of all possible interpretations of the given Clause-Set S from the perspective of the given variable(s).

28 This string can be understood as follows: in the first 32 rows $X_0$ is 0 and in the second 32 rows $X_0$ is 1.

29 Meaning: the pattern 16(0)16(1) is repeated twice.

30 Meaning: a pattern containing 8(0)8(1)16(1) is repeated twice.

31 Meaning: the pattern contains 32(1) to its right and 16 times the pattern 1(0)1(1) to its left.

Resolving those strings/trees in sequence amounts to performing a bit-AND Operation (hereafter referred to as PatternAnd) which does not need to be applied on similar sub-patterns more than once, e.g.:

> PatternAnd(2(8(0)8(1)16(1)), 8(2(0)2(1)4(1))) = 2 *x* PatternAnd(8(0)8(1)16(1), 4(2(0)2(1)4(1))) etc.

Thus, step 1 of this Sequential, Pattern-oriented Resolution (SPR) shall be (TP1):

PatternAnd(

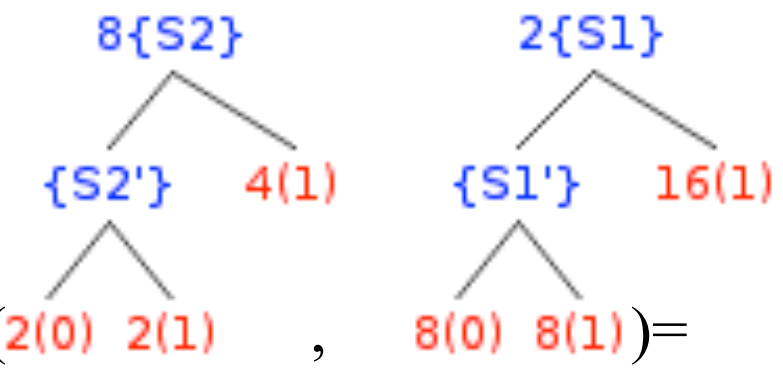

)=

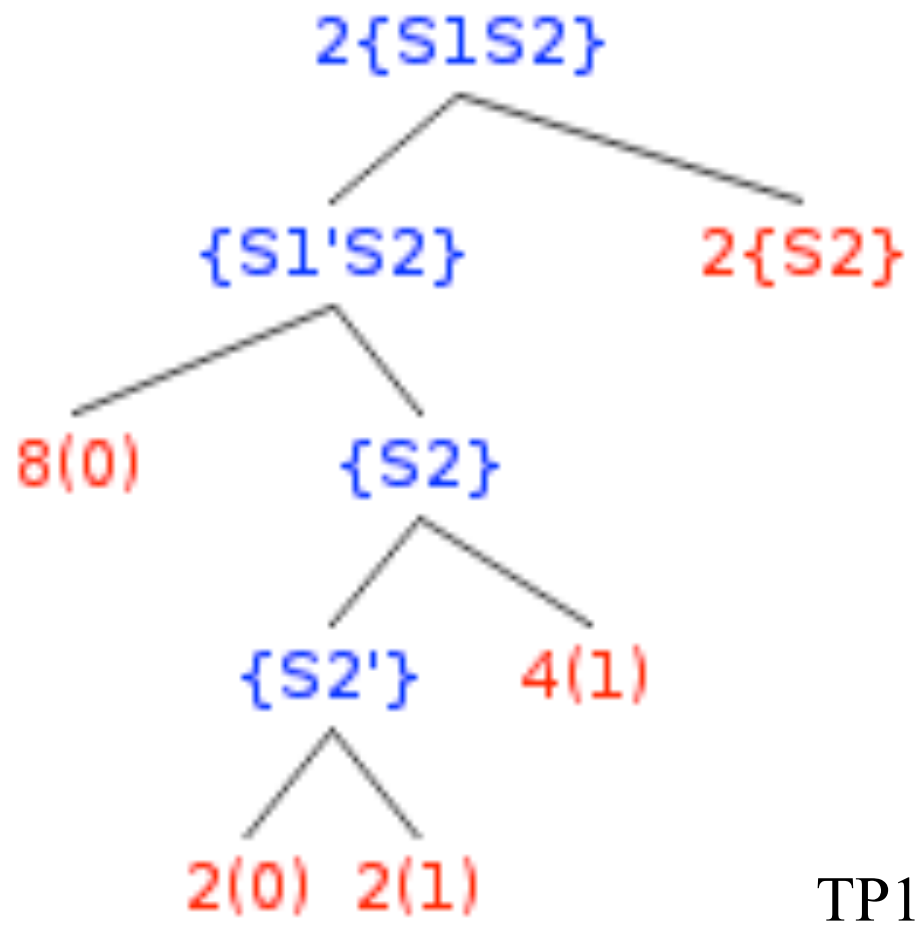

TP1

When $\{S_3\}$ is resolved in step 2 with the intermediate tree $2\{S_1S_2\}$ we get the following (TP2):

TP2

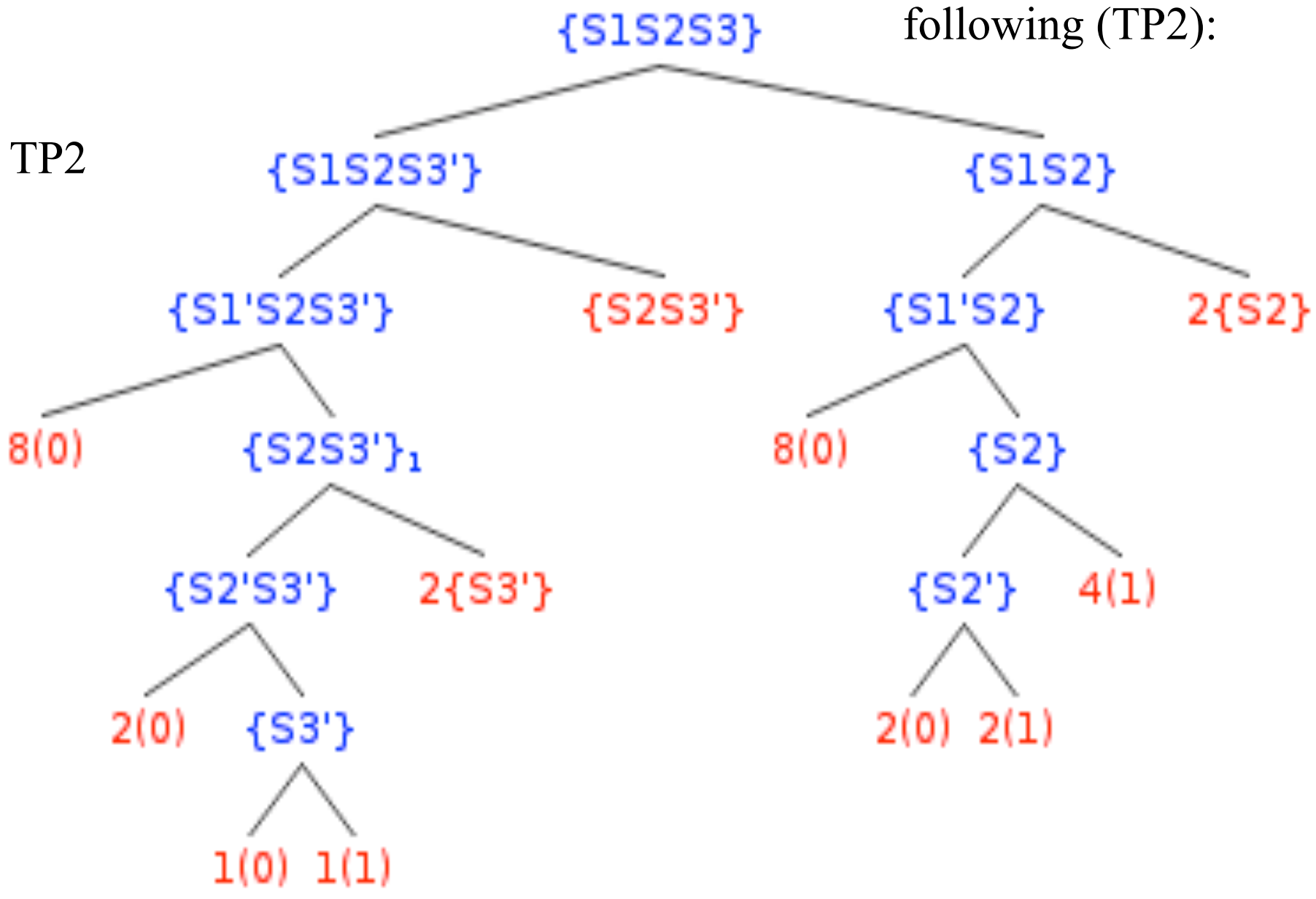

Counting the unique non-leaf-nodes (blue)[32] of this resulting tree yields 10. The BDDs corresponding to the above patterns generated in steps 1 and 2 look like BDD7 and BDD8[33]:

---

32 Such nodes represent TPs repeating in specific places within the overall pattern {S1S2S3}, e.g.: Starting with bit 16 we have sub-pattern: {S1S2}, before that it is: {S1S2S3'}.

33 BDDs in Experiment 2 differ from BDDs in Experiment 1 in the convention that *-ve* instantiations of literals/variables go through left edges of the respective nodes rather than through right ones.

BDD7

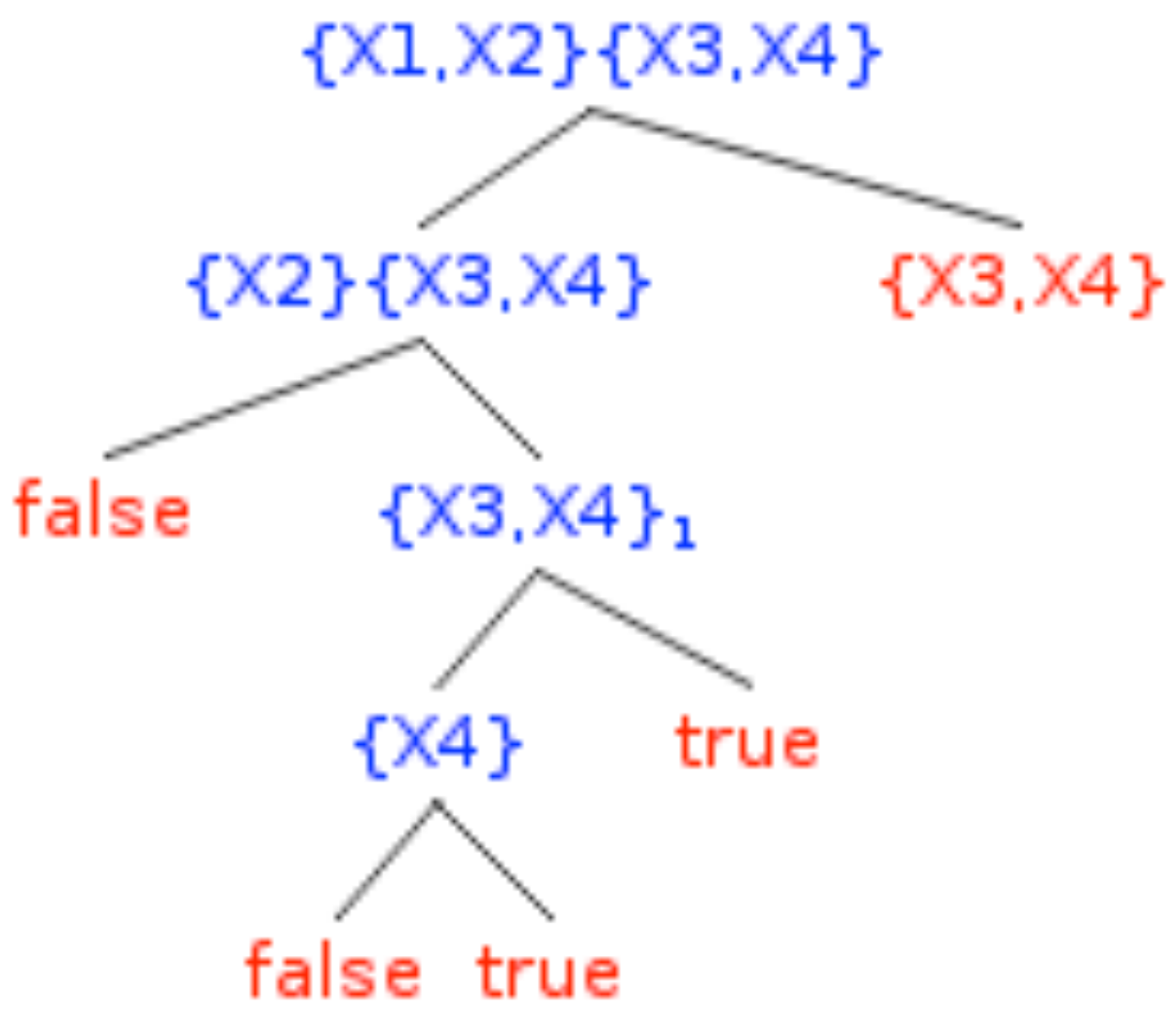

BDD8

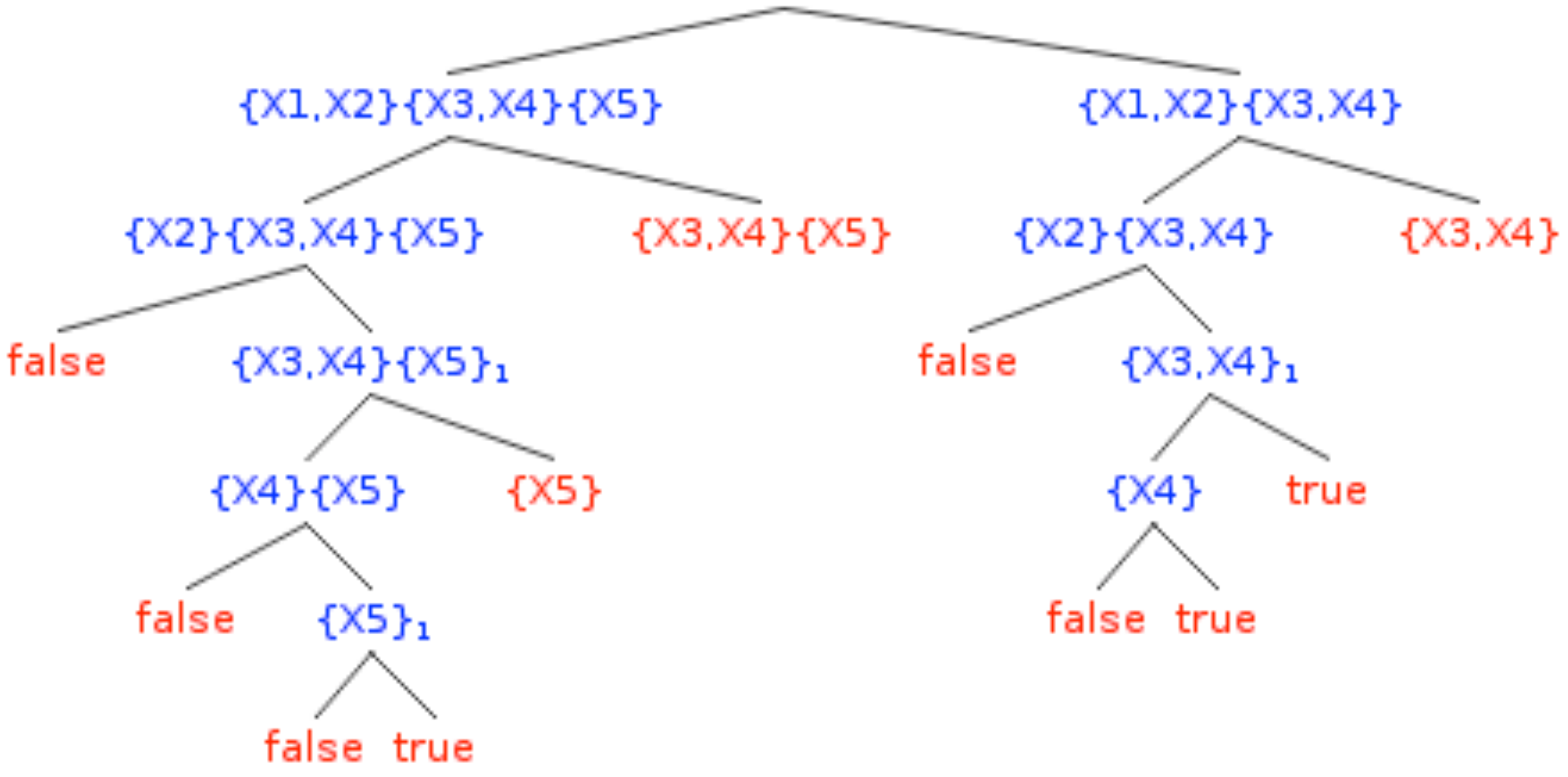

Now we repeat this exercise starting with $S_3$ instead of $S_1$ in TP3 and TP4 as follows:

TP3

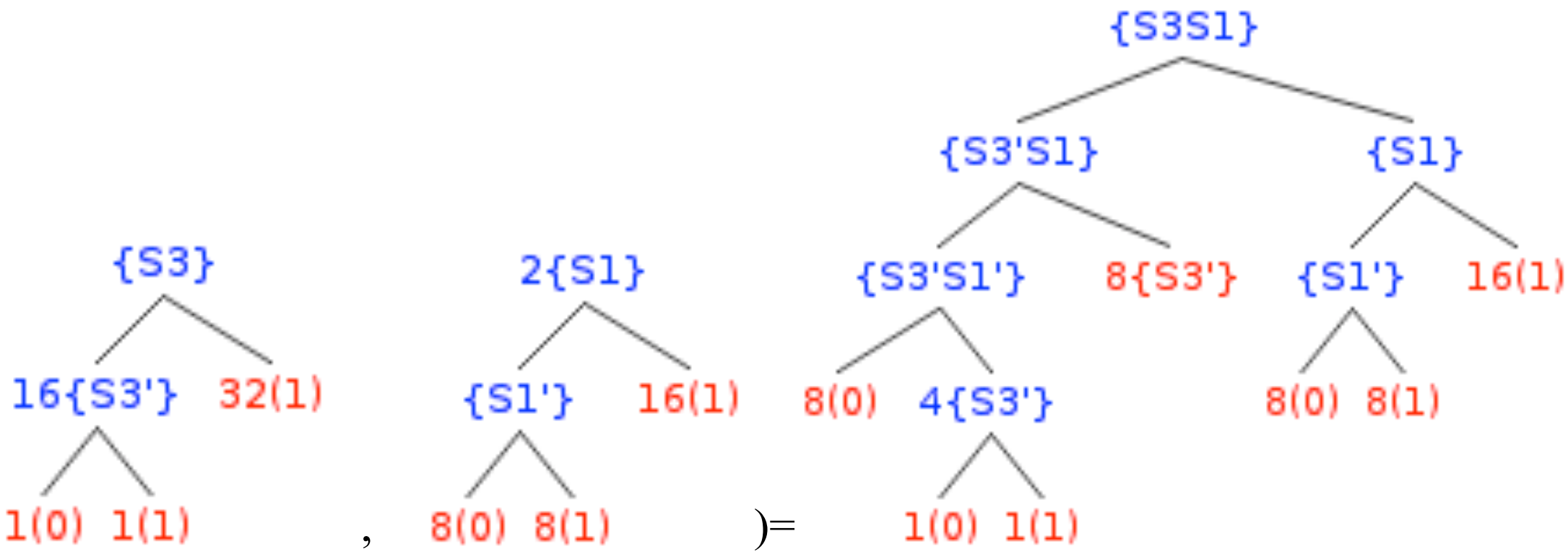

PatternAnd( , )=

TP4

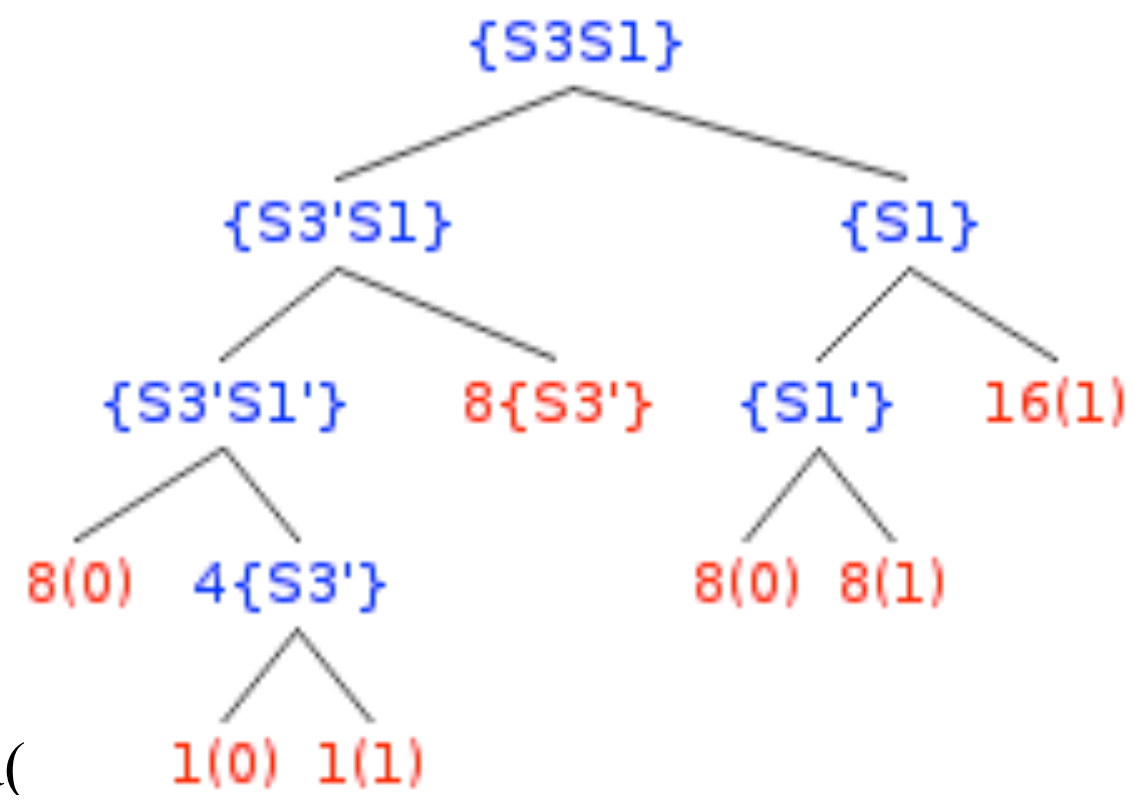

& PatternAnd( , )=

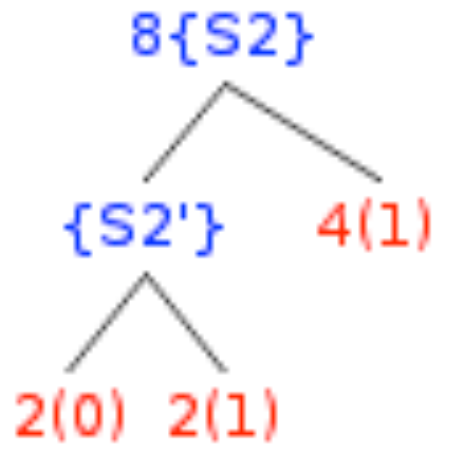

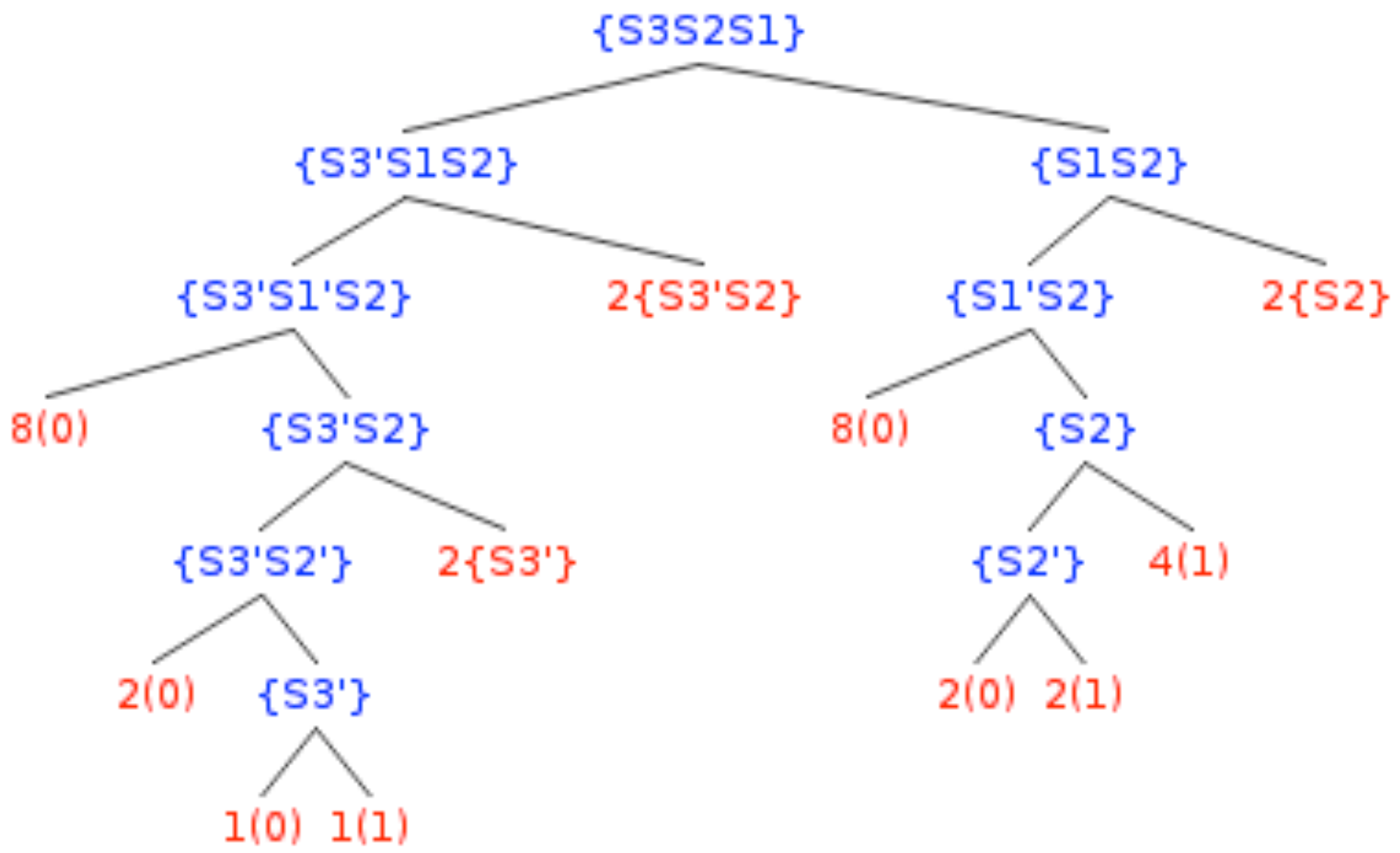

This new resulting tree TP4 again has 10 nodes.
Can we get a better result for S={{$x_1$,$x_2$}{$x_3$,$x_4$}{$x_0$,$x_5$}}? Let us try to rename variables to form: S'={{$x_0$,$x_1$}{$x_2$,$x_3$}{$x_4$,$x_5$}}. Obviously S=S'.

TPs for the new clauses and their tree representations:

$SS_{\{X0,X1\}}$=PatternOr($SS_{X0}$, $SS_{X1}$)=16(0)16(1)32(1)
$SS_{\{X2,X3\}}$=PatternOr($SS_{X2}$,$SS_{X3}$)=4(4(0)4(1)8(1))
$SS_{\{X4,X5\}}$=PatternOr($SS_{X4}$, $SS_{X5}$)=16(1(0)1(1)2(1))

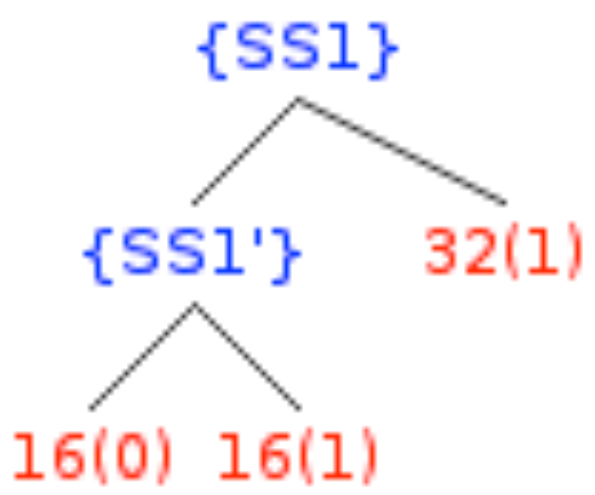

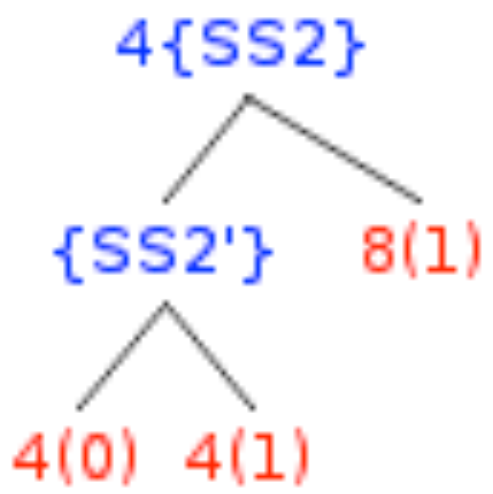

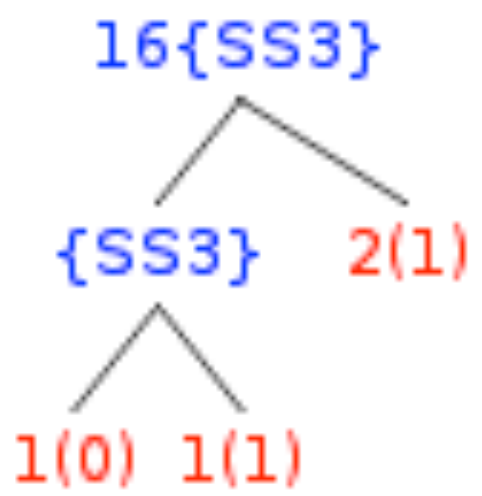

Now let us apply PatternAnd again:

TP4

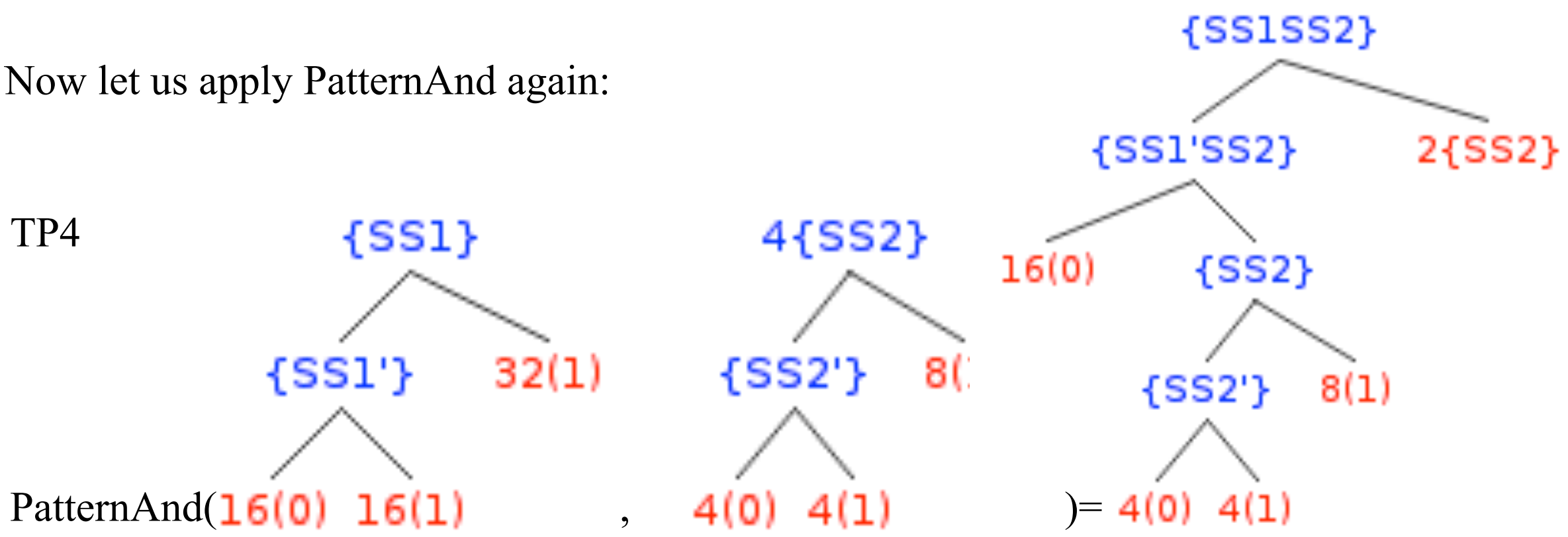

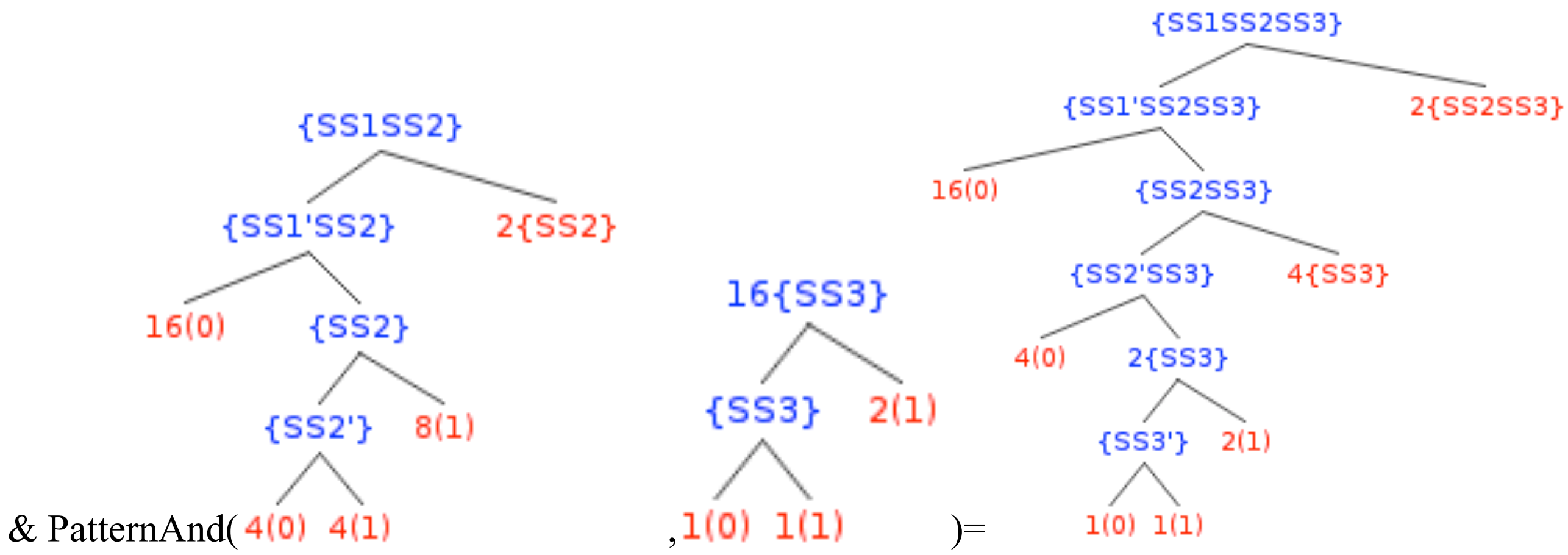

The corresponding BDDs are as follows:

BDD9

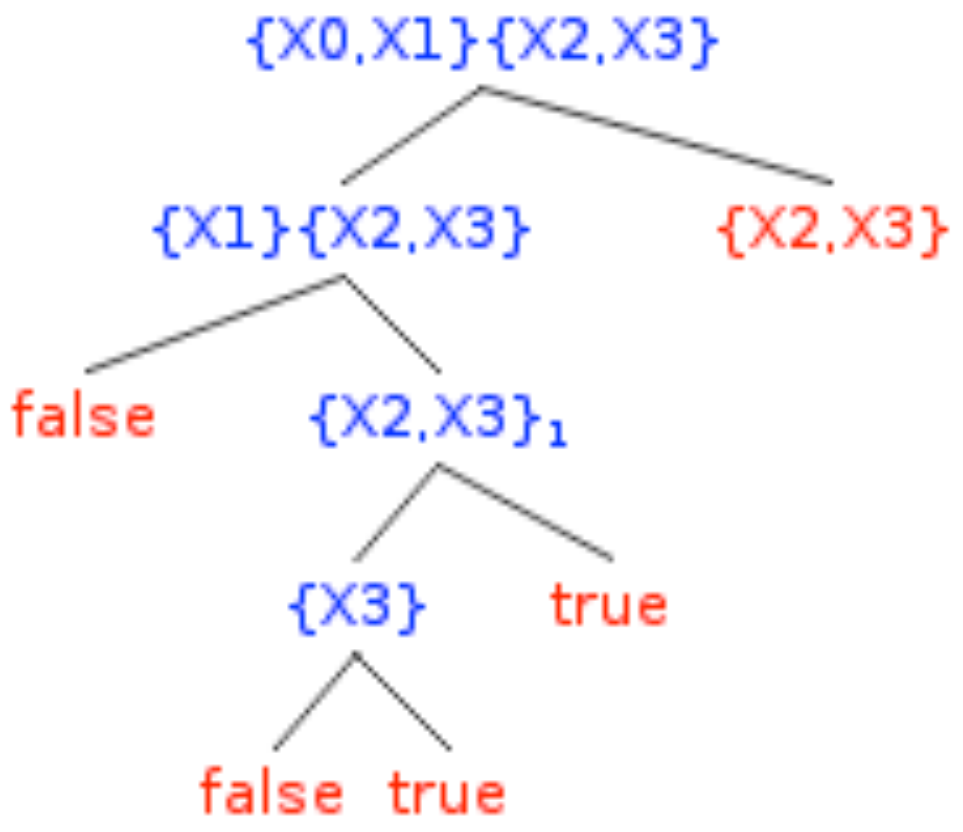

BDD10

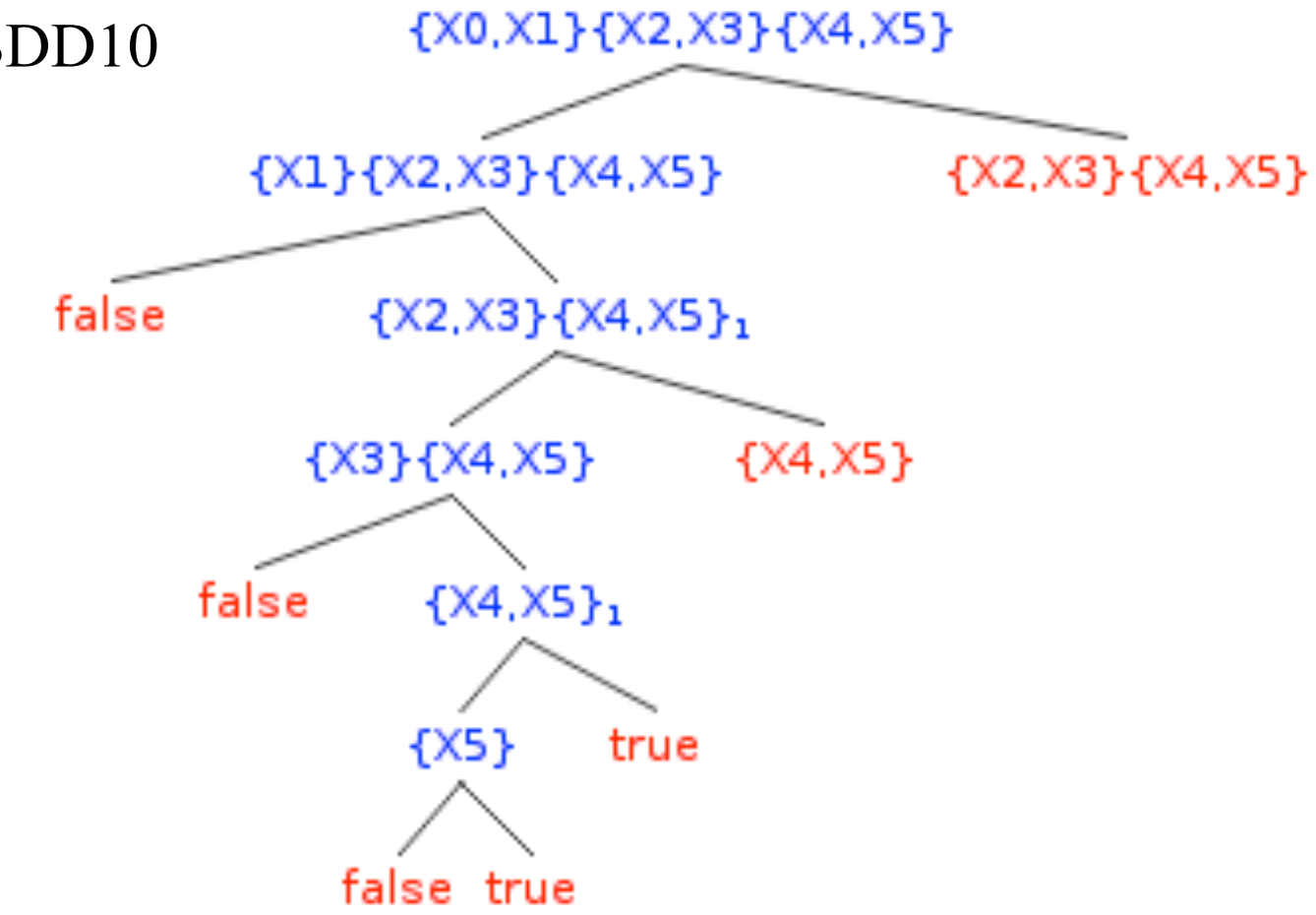

For the final result we only count six unique-nodes. How do we explain this? What did the renaming operation change?

**Conjecture and Objective**

Let's elaborate on the formation of TP2:

TP2

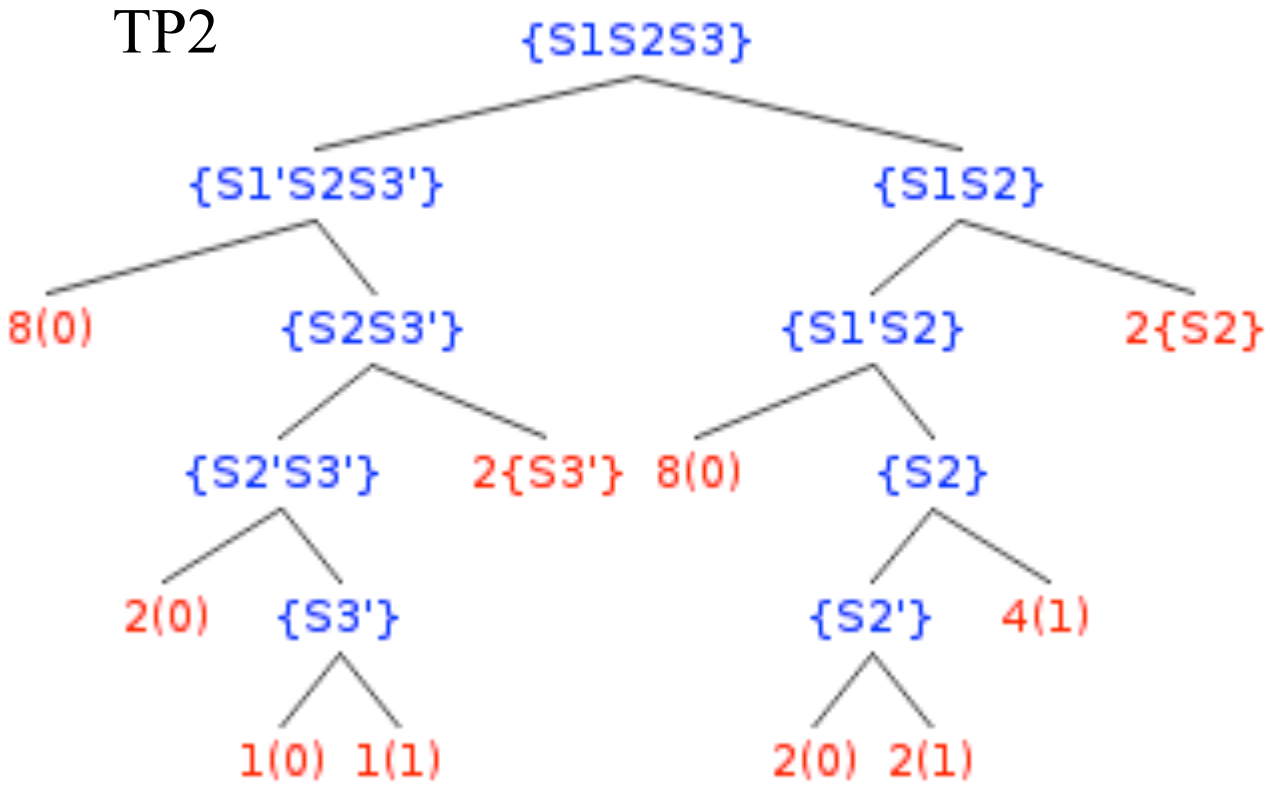

by PatternAnd from

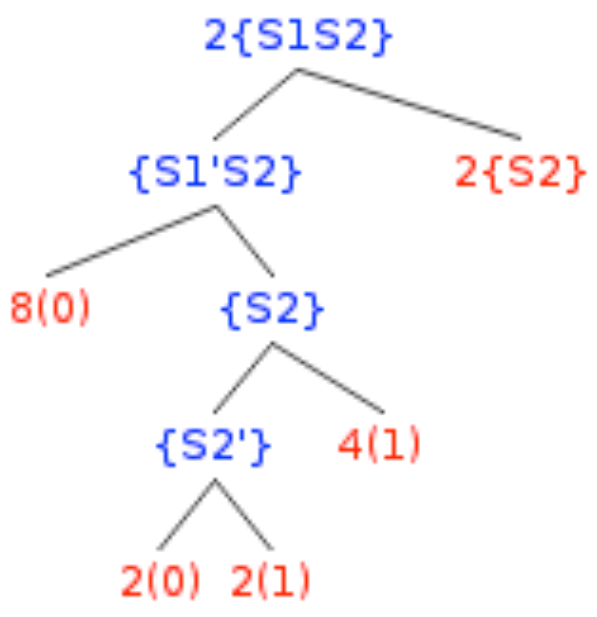

and

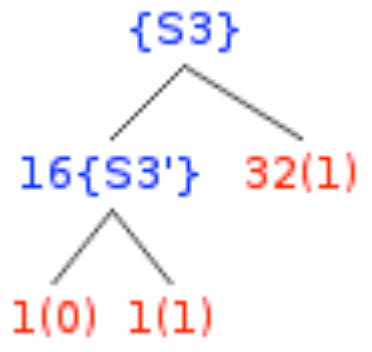

Note first that the two patterns $2\{S_1S_2\}$ and $\{S_3\}$ are 64bits long. $S_3$ has its whole right part constituted of 1-valued bits while its left part is the pattern $16\{S_3'\}$. $2\{S_1S_2\}$ is basically repeating pattern $\{S_1S_2\}$ throughout its length. $\{S_3\}$ is therefore twice as big as $\{S_1S_2\}$. We say that $\{S_3\}$ has a larger Pattern Length (PL) than $\{S_1S_2\}$[34]. This means that PatternAnd is going to resolve for the left part of the result $\{S_1S_2\}$ with $16\{S_3')$ and for the right part $\{S_1S_2\}$ with 32(1). The right part will consist of $\{S_1S_2\}$ while the resolution of the left part looks like TP5:

[34] Written: PL({S3})>PL({S1S2}) or $PL(\{X_0,X_5\})>PL(\{\{X_1,X_2\}\{X_3,X_4\}\})$.

TP5

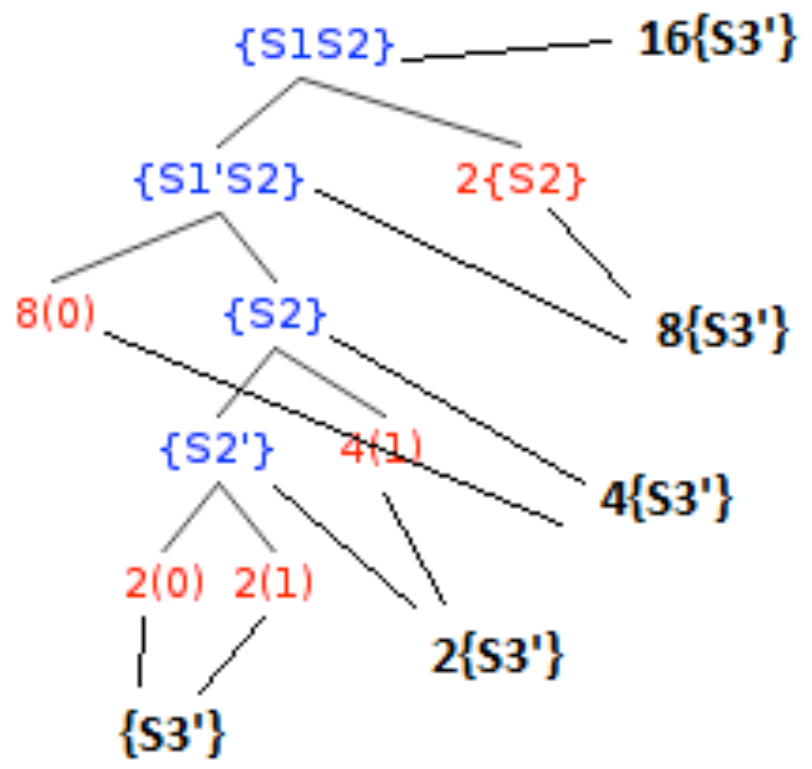

Where pattern $2^x\{S_3'\}$ is resolved with sub-nodes of $\{S_1S_2\}$ and $x$ is adjusted according to the respective PLs of the involved sub-patterns. Because $PL(\{S_3\})>PL(\{S_1S_2\})$, i.e., the new pattern to be resolved with the intermediate tree had a larger PL than the tree itself, a copy of that tree or parts of it, here: $\{S_1S_2\}$ had to be included in the final result. We say that the intermediate tree/node $2\{S_1S_2\}$ has been split and we call this type of split N-split[35]. Splits are defined more formally in next sections and play an important role in the development of main ideas of this paper. When a split occurs, it simply means that a number of new nodes is added to the resulting tree which is equal to the number of nodes of the original tree/sub-tree, thus causing a blow-up of the final result. In the above example the only N-split happening is the one caused by $\{S_3\}$, since for the rest: $PL(\{S_3'\}) < $ PL of any sub-node in $\{S_1S_2\}$.

Let us take a look at the second example of Experiment 2: When the tree $\{S_3\}$ is resolved with $2\{S_1\}$ a N-split occurs between $16\{S_3'\}$ and $\{S_1\}$ causing the formation of pattern $8\{S_3'\}$ in the first step. Then when the intermediate tree is resolved with $8\{S_2\}$, another N-split occurs when pattern $8\{S_3'\}$ is resolved with $2\{S_2\}$ causing the pattern $2\{S_3'S_2\}$ to be formed (TP6):

[35] i.e., Node-Split.

TP6

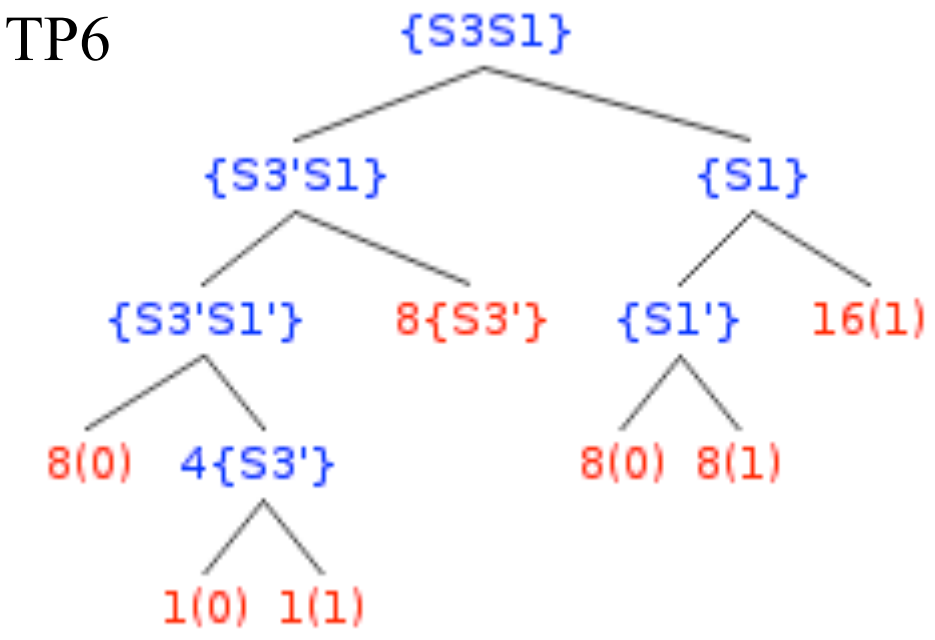

The reader may have noticed that when we renamed variables in the last example of Experiment II, all newly resolved patterns had PLs smaller than the ones already existing in intermediate trees (in the first step for example: {SS$_2$} has a PL smaller than any pattern in any node of {SS$_1$}). Therefore, no N-splits occurred during resolution and the total number of nodes remained small. Note also that all BDDs for the trees produced by the SPR-procedure shown above have one thing in common: They all use the canonical ordering of variables.

Summary of Experiment II: Splits are the cause of blow-ups in resolution-trees of SPR-procedures. N-splits occur when a newly resolved truth pattern TP has a pattern length PL (in the canonical ordering) larger than the PL of the corresponding pattern in the node of the tree/sub-tree to be resolved with TP. Expressing this in terms of BDDs: If in an SPR resolution-procedure - which uses the canonical ordering as guideline for variable instantiations - a clause C to be resolved with a node of a tree/sub-tree T has a leading literal/variable whose index/order[36] is less than the index/order of **all or any** leading literal/variable of clauses in the node of T, then an N-split occurs. We can reformulate this assertion as a safe-condition posed on Clause-Sets:

[36] Variables/Literals of smaller indices/order have longer PLs.

If ∀S (S = Clause-Set to be resolved in an SPR-procedure), ∀$C_i$, $C_j$ clauses ∈ S, $i<j$: PL($C_i$)>PL($C_j$), then no N-splits occur. This property is elaborated and précised formally in Section II.

Can this explain the trees in Experiment I?

It obviously applies to the canonical ordering case: {a,b,c}{x,y,z}{a,c,x}

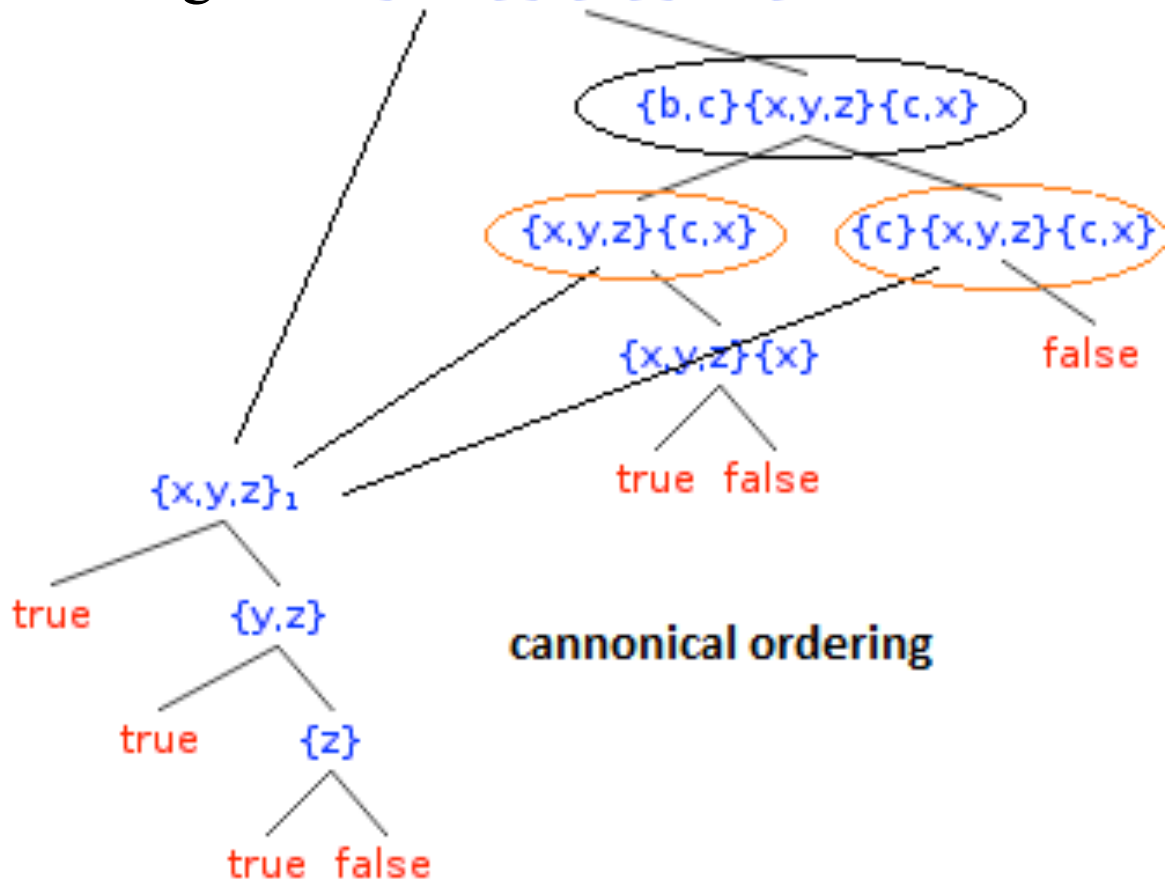

Red ellipses show the result of an N-split of sub-tree {{*x,y,z*}{*c,x*}} which is a child of the Clause-Set {{*b,c*}{*x,y,z*}{*c,x*}} where indeed PL({*c,x*})>PL({*x,y,z*}). What happens if we attempt renaming? We can rename S={{*a,b,c*}{*x,y,z*}{*a,c,x*}} to become: S'={{*a,b,c*}{*a,b,x*}{*x,y,z*}}[37].

The following BDD11 is then the final result of resolution:

BDD11

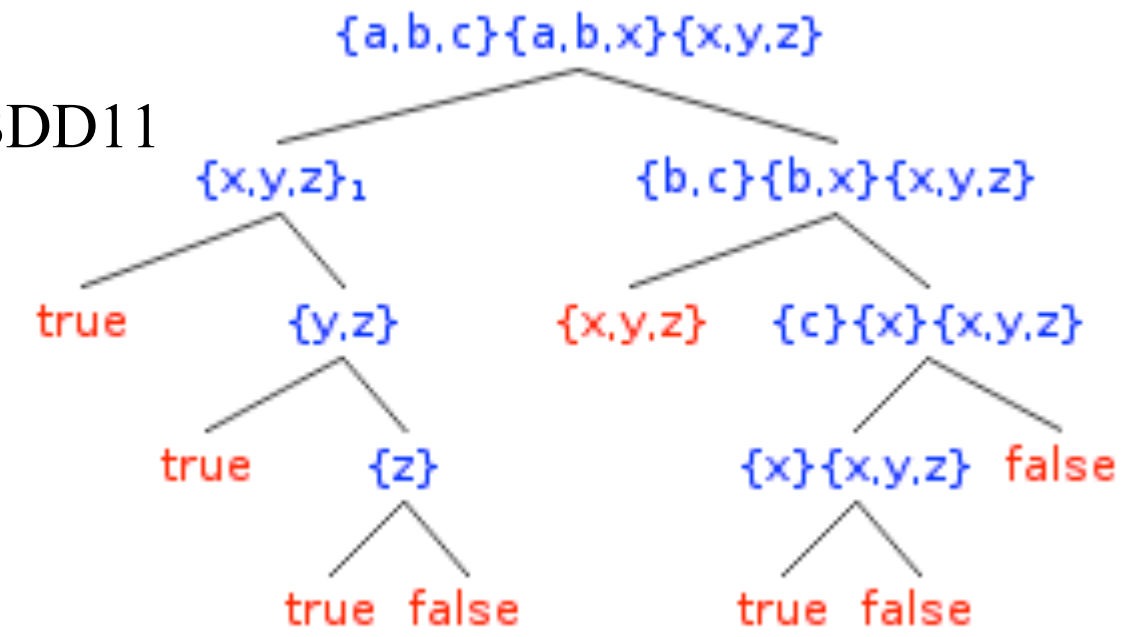

The number of nodes is reduced as expected. Note that the achieved node-count of seven seems to be the minimal as well.

[37] The reader is encouraged to verify the correctness of this renaming. The renaming procedure used here is elaborated and formalized in Section II.

We are ready to propose the following conjecture and to formulate the then following objective:

**Conjecture**
*If in an SPR-procedure solving a problem P by resolving a 3-SAT-CNF Clause-Set S, all generated sub-problems P' can be expressed using Clause-Sets S' which have safe-conditions preventing big splits, then the canonical ordering ∏ of variables in S is a near-to-optimal ordering producing a near-to-minimal BDD for P whose number of nodes is polynomial in the size of S. In case some sub-problem P' produces a Clause-Set S' which is not safe, in this particular sense, S' can be renamed and arranged to form a safe Set S'' which imposes another canonical ordering ∏' on P' resulting - in addition to ∏ - in a BDD for P with a polynomial number of nodes in the size of S and near-to-optimal as well.*

**Objective**
*Construct a 3-SAT-CNF-Solver based on SPR-procedures using canonical orderings and safe-conditions. The Solver should be able to generate BDDs whose number of* unique-nodes *is polynomial in the length of input-sets.*

Despite the conjecture and objective being described separately, they represent interdependent ways of tackling both, the NP-problem and the related BDD-minimization-problem with SPR-procedures being on center-stage. The remainder of this section is dedicated to the proof-strategy of the conjecture and achieving the objective. The indicated sequence is not exactly the one followed in Section II, but enables a consistent overview about the argumentation.

**First**, SPR-procedures are formalized and their generic properties proven:

A new AP/canonical order-based resolution-algorithm (GSPRA) is defined binding the BDD construction to Clause-Set-instantiation. The only variable/literal instantiation rule applied by this algorithm (materializing the canonical order doctrine) being the least-literal/head-clause-rule (Definition 2).

The safety condition proposed above is elaborated to divide arbitrary 3-SAT Clause-Sets into three types:

- linearly ordered
- linearly ordered, but unsorted
- almost arbitrary Sets

GSPRA behaves differently for each one of those types. For linearly ordered Sets it is shown that no 'big splits' of BDDs can be produced by GSPRA during resolution. Big splits refer in this work to ones causing exponential behavior. Properties of GSPRA or its extension $GSPRA^+$ relevant to the conjecture include:

a) The fact that any arbitrary variable ordering used in solving a 3-SAT-problem can always be converted to a canonical one using renaming so that BDD-minimization is reduced to minimization of canonically ordered BDDs (Property 9).
b) The fact that the concept of Algorithmic Equivalence[38] of nodes in BDDs constructed by SPR-procedures is essentially Syntactical Equivalence of their Clause-Sets, thus facilitating the efforts to avoid redundancies (Property 10).
c) The fact that BDDs generated by GSPRA/$GSPRA^+$ possess always a structure which guarantees a minimal number of nodes in their top-part. (Properties 8, eventually 8')

[38] i.e., BDD nodes are only equivalent when all their sub-nodes are equivalent as well.

Properties of GSPRA/GSPRA$^+$ relevant to the objective include:

a) Expansion Property 2: Stating that in constructed BDDs no nodes which were not connected in steps <*k* can be connected in steps >=*k* except in trivial cases.
b) Property 4 then 4': The only non-trivial common-nodes created by GSPRA/GSPRA$^+$ are elements of a fixed Set (called the ACS-Set).
c) Uniqueness of instantiation results (Property 5): Stating that children of a Clause-Set are equivalent whenever instantiation literals leading to them are equivalent as well.
d) Properties 8, eventually 8' (same as the foregoing).

**Second**, lemmas related to the conjecture are formulated and proven. Their interdependencies are shown in Fig. 1.

Lemma 14 and 15 show that BDDs produced by GSPRA$^+$ are minimal compared to those produced by any procedure using canonical orderings with a.a. and l.o.u Sets (respectively). Both lemmas make use of Property 8'. Lemma 16 uses Lemma 14 and 15 as well as Property 9 to show that GSPRA$^+$ produces BDDs which are minimal even for procedures using orderings other than canonical ones. It also uses Property 10 to assert that those BDDs are redundancy free. Lemma 17 provides evidence that BDDs produced by GSPRA$^+$ are always near-to-optimal.

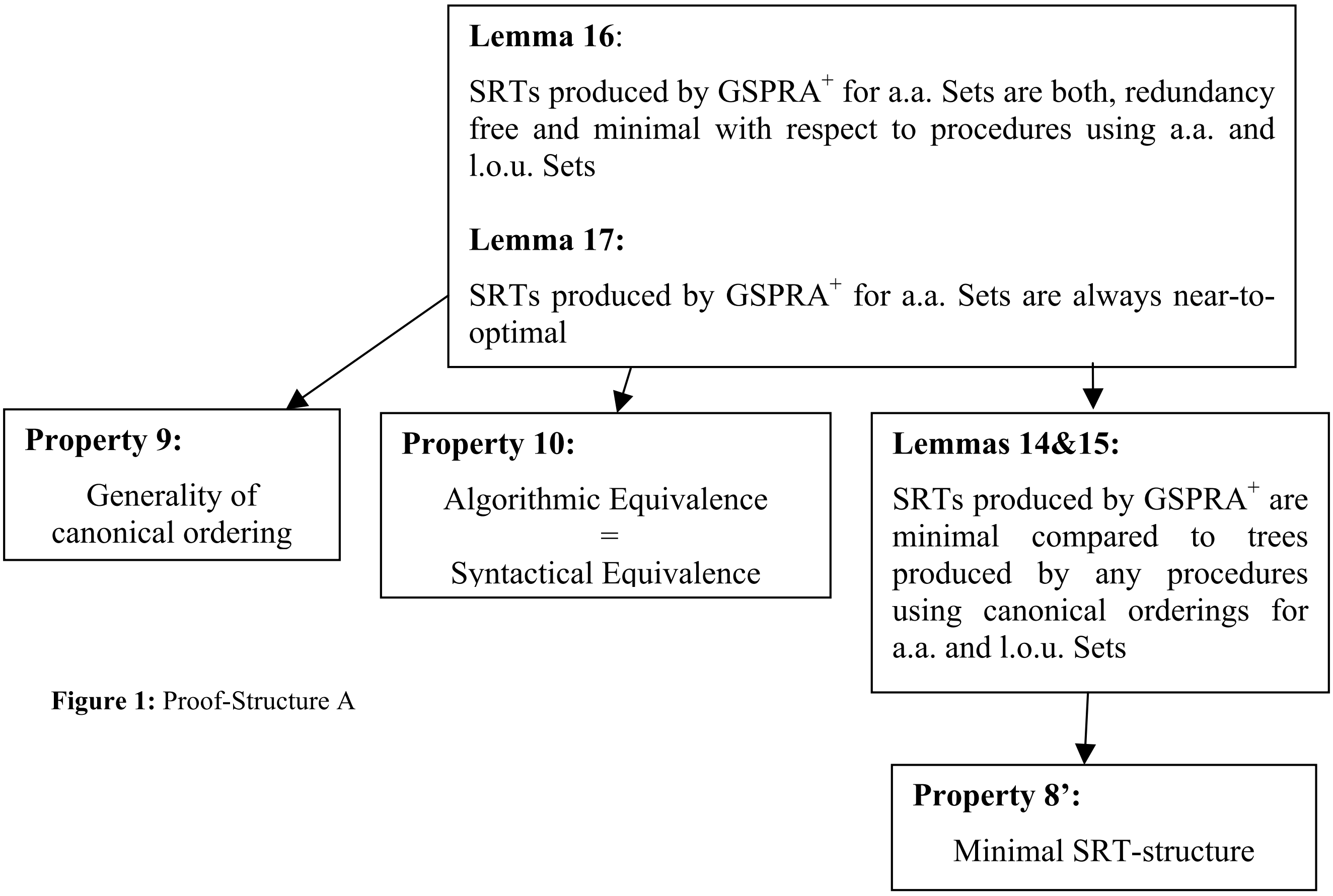

**Figure 1:** Proof-Structure A

**Third**, lemmas related to the objective are formulated and proven. Their interdependencies are shown in Fig. 2[39]:

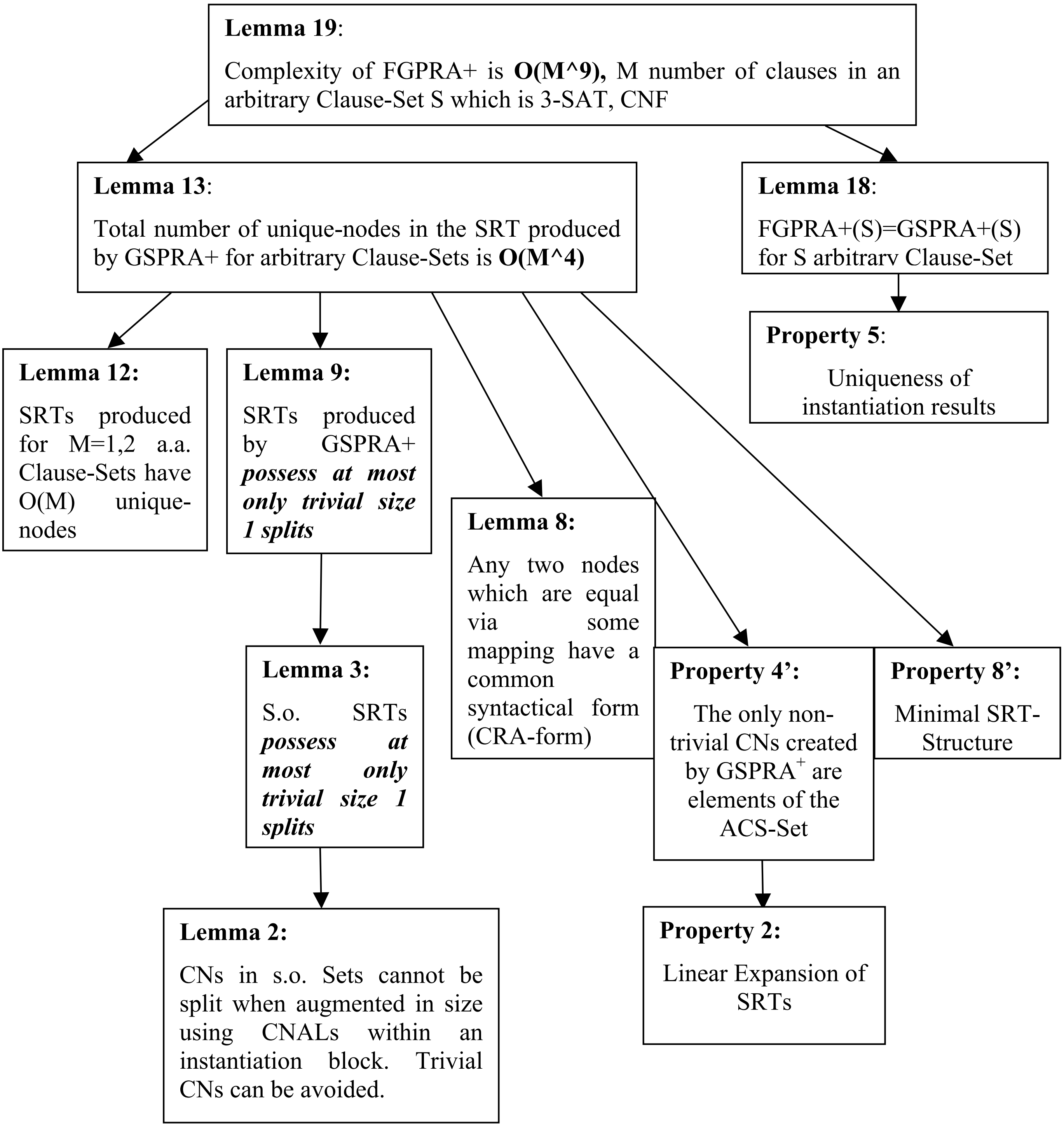

**Figure 2:** Proof-Structure B

---

[39] Proof-Structures provided here are meant to assist critical readers to find flaws in arguments and/or conclusions drawn in this paper.

Lemma 18 shows that the BDDs produced by $FGPRA^+$, the algorithm which resolves Clause-Sets in parallel, are equivalent to those produced by $GSPRA^+$. This lemma uses Property 5.

Lemma 13 is central showing that the number of unique-nodes in an SRT produced by $GSPRA^+$ for Clause-Sets of size M is in $O(M^4)$.

The main observation is that all new nodes generated in any step by $GSPRA^+$ are only non-trivial common-nodes, i.e., members of a fixed Set (Property 4'). It uses Lemma 12 in its Base-Case which asserts that for M=1,2 the number of unique-nodes is in O(M) for a.a. Clause-Sets. Lemma 9, the one related to existent splits in $GSPRA^+$ products, assures that there cannot be big splits in a final $GSPRA^+$ tree. This is a generalization of the findings in Lemma 3 which is concerned with s.o. products of GSPRA. Lemma 3 in its turn uses among other insights the fact that if CNs in s.o. Sets are guaranteed not to be augmented in size except using CNALs, then they won't split if their sized become >1 in any further steps within an instantiation block (i.e., when they are not supported). Lemma 8 is used as well. It states that any two Clause-Sets which are equivalent-via-mapping possess a common syntactical form, facilitating thus the process of forming common-nodes. Property 8' is also used by this lemma. Lemma 19 studies complexity of every operation used by $FGPRA^+$ assuming that there are always $O(M^4)$ unique-nodes in a final BDD (Lemma 13,18) to find out that the overall complexity of $FGPRA^+$ is in $O(M^9)$. The detailed plan of Section II looks like this:

II A) all concepts discussed above are formally defined:

Definitions 1&2 concern almost arbitrary (a.a.), linearly ordered (l.o.) and linearly ordered, but unsorted (l.o.u.) Clause-Sets and the GSPRA as well as tree/graph structures (called SRTs) used by it. SRTs are generalizations of BDDs binding nodes to instantiated Clause-Sets instead of single literals. GSPRA implements one instantiation rule (called least-literal/head-clause-rule).

Definitions 3&4 introduce properties as well as special types of SRTs: strongly ordered (s.o.) and loosely ordered (lo.o.).

Definition 5 formalizes the concept of a common-node (CN) in an SRT showing different types of CNs.

Definition 6 introduces the concept of a Dependency Graph (deduced from an SRT) which is equivalent to the known FBDD.

Definition 7 formalizes the notion of Algorithmic Equivalence of nodes which is central in defining minimal SRTs[40].

Section II B) introduces Lemma 1,2 and Corollary 1 which are mainly concerned with the existence and form of CNs in s.o. and lo.o. SRTs as well as the following intrinsic properties of GSPRA:

Property 1: completeness/truth table equivalence of GSPRA
Property 2: expansion of SRT
Property 3: linear derivation of clauses
Property 4: generation of non-trivial CNs
Property 5: uniqueness of instantiation results
Property 6: Syntactical Equivalence
Property 7: FBDD equivalence, branch linearity
Property 8: SRT structure
Property 9: generality of canonical orderings
Property 10: Algorithmic Equivalence = Syntactical Equivalence

[40] Binding nodes in SRTs to Clause-Sets (instead of single literals as in typical BDDs) allows minimization efforts to be reduced to finding syntactically equivalent Clause-Sets instead of fulfilling semantic criteria dependent on the nature of the Boolean function on hand as is usually the case.

II C) studies split conditions in s.o. Sets. Lemma 3 demonstrates that "big" splits cannot exist indicating already that GSPRA procedures working with such Sets are not exponential in nature.

II D) studies renaming algorithms CRA/CRA+ and their properties (Lemma 4,5,6,7) where termination and algorithmic fulfilling of l.o. conditions are shown.

II E) summarizes all ideas in an algorithm GSPRA+ which is shown to possess many interesting properties, the main one being that it produces SRTs which are "aligned"[41] with a total unique-nodes count in $O(M^4)$ where M is the number of clauses in an arbitrary 3-SAT-CNF-Set (Lemma 13). Those SRTs are shown in Section E-5 to be near-to-minimal as well.

II F) shows a parallel version of the same algorithm (FGPRA+) whose complexity is in $O(M^9)$ concluding Section II with a definition of a new Solver algorithm and Theorems 1,2 which prove that P=NP and that FGPRA+ is a polynomial 2-approximation algorithm of MinFBDD, the problem of minimizing an FBDD of a Boolean function, respectively. Theorem 2 shows that Boolean functions possess minimal FBDDs which have polynomial node-counts (in M, the number of clauses used in expressing them as 3-SAT-CNF-formulas).

## II) 3-SAT-CNF CLAUSE-SETS AND THEIR RESOLUTION

### A) DEFINITIONS

**Definition 0: nomenclature**

For a Set S of general 3-SAT-CNF-clauses of the form:

$\{\{a_1,b_{11},c_{11}\}\{a_1,b_{12},c_{12}\}..\{a_1,b_{1i},c_{1i}\}$
$\{a_2,b_{21},c_{21}\}\{a_2,b_{22},c_{22}\}..\{a_2,b_{2j},c_{2j}\}\ldots$
$\{a_m,b_{m1},c_{m1}\}\{a_m,b_{m2},c_{m2}\}....\{a_m,b_{mk},c_{mk}\}\}$

a) LIT (S): is the Set of all unique literal names/indices in S

b) LEFT($x$,C)/RIGHT($x$,C), $x$ literal name/index: Is the Set of all literal names/indices occurring in S to the left/right of literal $x$ in clause C.

c) SortOrder($x$,S), for $x$ = clause and S = Set: Is an integer number representing the sort order of $x$ with respect to other clauses of S.

d) First literals in any clause are called head- while last ones are called tail-literals

e) If C is a 3-SAT-clause, then the cardinality of the Set of all clauses which are permutations of literals in C (called short: perm(C)) is called Resolution Complexity Coefficient (RCC). It is given by the formula: $RCC_{k\text{-}SAT}={}^kP_k+{}^kP_{k-1}+{}^kP_{k-2}\ldots+{}^kP_1$ i.e., for 3-SAT $RCC_{3\text{-}SAT}={}^3P_3+{}^3P_2+{}^3P_1=15$[42]

f) Clauses created through instantiations of literals of a clause C with TRUE or FALSE are called derivations of C. They are called linear derivations if consecutive instantiations respect the linear order of literals in C[43].

g) Indices are used to stand for literal names (i.e., 1,2, etc. instead of $x_1$,$x_2$,..).

[41] A useful property amounting to making all Clause-Sets in an SRT l.o. (formally précised in Section E).

[42] Recall that ${}^nP_r=n!/(n-r)!$

[43] Examples of derivations of clause C={$x$,$y$,$z$} for any ordered indices $x$,$y$,$z$ are {$x$,$z$} and {$y$,$z$} of which only the latter is a linear derivation.

**Definition 1:** For a Set S of the above form, S is called linearly ordered (l.o.)[44] **if** the following Conditions hold:

a) $\forall a_i, b_{ij}, c_{ij} \in C_{i+j}$: $a_i < b_{ij} < c_{ij}$ literal names/indices are sorted in ascending order within clauses.
b) **S is sorted by $a_i$ & $b_{ij}$ & $c_{ij}$ in ascending order** taking into consideration negation signs[45]. **In other words:** $\forall i,j$ indices of clauses: if $i<j$ then head-literal of $C_j$ >= head-literal of $C_i$.
c) **$\forall x \in$ LIT(S), $\forall$C $\in$ S: if $x$ not $\in$ LEFT($x$,C) then $\forall y \in$ LEFT($x$,C): $x>y$** (all new names/indices of literals occurring in a clause C of S are strictly greater than all the literal names/indices to their left).
d) **Clauses appear only once in S.**

If S fulfills Conditions a), c), d), but not b) it is called linearly ordered, but unsorted (abbreviated l.o.u.)[46]. If S fulfills Conditions a), d) only it is called almost arbitrary (a.a.)[47]. Clause-Sets of the form: S={ $\{a_x, b_{x1}, c_{x1}\}\{a_x, b_{x2}, c_{x2}\}$ .. $\{a_x, b_{xi}, c_{xi}\}$ } are called blocks and are referred to by the name of the leading literal (in this case S is called $a_x$-block). Clauses having $a_x$ as leading literal are said to belong to the $a_x$-block.

[44] This formulation leads to the condition: $\forall C_i$, $C_j$ clauses $\in$ S, $i<j$: PL($C_i$)>PL($C_j$) proposed in Section I for all Clause-Sets S in a resolution-tree.

[45] i.e., {1,2,3} comes before {1,2,4} and {¬*1,2,3*} before {1,2,3}or vice versa.

[46] Corresponding to the condition: $\exists C_i$, $C_j \in$S, $i<j$: PL($C_i$)<PL($C_j$) for some Clause-Sets S in a resolution-tree.

[47] Corresponding to the condition: $\forall C_i$, $\exists C_j \in$S, $i<j$: PL($C_i$)<PL($C_j$) for some Clause-Sets S in a resolution-tree.

**Definition 2:** The Generic Sequential Patterns Resolution Algorithm (GSPRA) applied on a Set of a.a. 3-SAT-CNF-clauses S consists of the following procedure:

0. **Preliminary step:** Choose the shortest clause $C_i$ of S to be instantiated first. Sort S so that $C_0=C_i$[48].
1. Take $C_0$ and create the binary tree[49]:

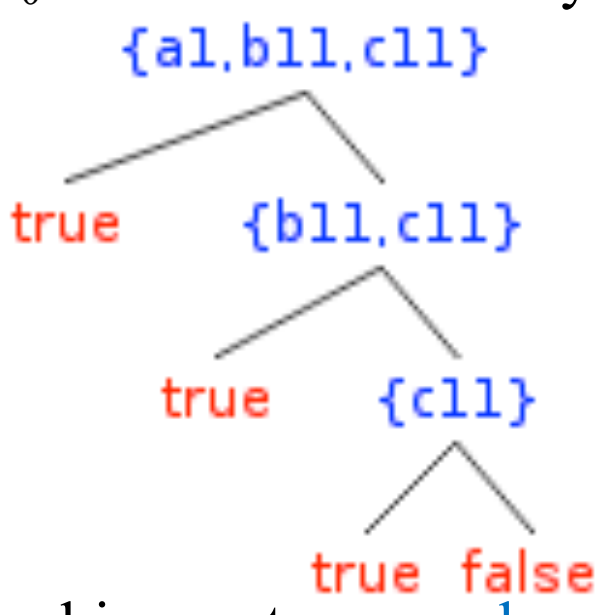

**T0**

In this binary tree, nodes contain Clause-Sets and edges represent truth assignments of single literals (called instantiations of Clause-Sets and/or literals). Branches are lists of nodes starting with base 1 until a True- or False-leaf. Note: Head-literals such as $a_1, b_{11}$, etc. have precedence in instantiation over other literals. Also note that left edges always represent *+ve* instantiations of head-literals of the current clause while right edges represent *-ve* instantiations of the same.

2. Resolve each following clause $C_i$ of S with the intermediate resolution-tree (IRT) created in the previous step, which in the beginning is equal to the tree shown in T0, as follows:

[48] Choice of $C_0$ may be crucial for GSPRA. Usually minimization of the number of nodes in the top-part of the tree is sought by choosing the shortest clause as is done here for the sake of illustration. But this doesn't necessarily lead to minimal SRTs when unique Clause-Sets are considered. In Section E a procedure is dedicated for this part of the algorithm.

[49] An example with only *+ve* literals is used (w.l.o.g.).

a) Follow left- then right edges of the current IRT-node in a depth first way substituting the literal written on the edges by TRUE and FALSE respectively if it exists in $C_i$. Set $C_i$’ and $C_i$’’ to be the resulting derivation ($C_i$’ for the left side, $C_i$’’ for the right side). If the literal does not exist in $C_i$, put $C_i$’=$C_i$’’=$C_i$.

b) Resolve resulting derivations $C_i$’, $C_i$’’ of step a. with the left- then right IRT-nodes (respectively) by calling step 2. recursively

c) If left- or right Clause-Sets of current IRT-nodes are TRUE, then substitute them with the trees for $C_i$’ or $C_i$’’ (formed like in 1.)

d) If left- or right Clause-Sets of current IRT nodes are FALSE, keep them FALSE

e) Return the final tree as a result of the resolution-procedure. It is called Sequential Resolution Tree (SRT). This tree can be simplified to form a directed graph if nodes/leafs have the same Clause-Sets/values joined together. In the remainder of this paper we shall call those graphs simply trees or SRTs as well. Thus, an SRT is (also) a directed, acyclic graph <V, E> where V is the Set of all Clause-Sets, E the Set of ordered pairs $<v_1,v_2>$ , $v_1,v_2 \in V$ representing instantiations of Clause-Sets having a parent-child relationship and produced during GSPRA(S).

3. A node in an SRT is symbolized by [$x$] **if** the lead clause in its Clause-Set is headed by a least-literal $x$. Moreover: $x$ is called the Name Literal (NL) of this Clause-Set/node.
4. As per 2 a): Edges going out of an SRT node [$x$] represent instantiations of the NL $x$ of the Clause-Set of that node (this fact is called the least-literal/head-clause-rule).
5. The Clause-Set in an SRT-root-node is called Base Clause-Set of the base-node of the SRT or simply ‘Base Clause-Set’
6. The rank of a clause is the number of literals contained in that clause. Rank of a node/Clause-Set in an SRT is an integer representing the maximum number of literals in any clause in the Clause-Set of that node.
7. The size of a node in an SRT is an integer representing the number of clauses in the Clause-Set of that node.
8. Nodes of sizes 0 or 1 (TRUE- or FALSE-leafs) are called Resolution Termination Nodes (RTNs) of the SRT.
9. A variable ordering of a problem $p$ ($\prod_p$) expressed as a 3-SAT-CNF Clause-Set S and resolved by any resolution procedure PR is a list of integers $<i,j,k,\ldots>$ representing indices of literal/variable names indicating priorities of instantiations of literals/variables of S used by PR.
   If all sub-problems of $p$ have the same ordering, subscript $p$ is omitted and we call $\prod$: BDD-ordering.
   If either $\prod_p$ or $\prod$ represent the canonical ordering of variables the following notation is used: $\prod^c{}_p$ or $\prod^c$.

**Motivation 1: Example Resolving Monotone, +*ve*, 2-SAT Clause-Sets using GSPRA[50]**

To see GSPRA for a.a. Sets in action refer to below SRTs generated in sequence for Set S={{0,3}{0,7}{1,2}{1,4}{5,6}{3,8}}:

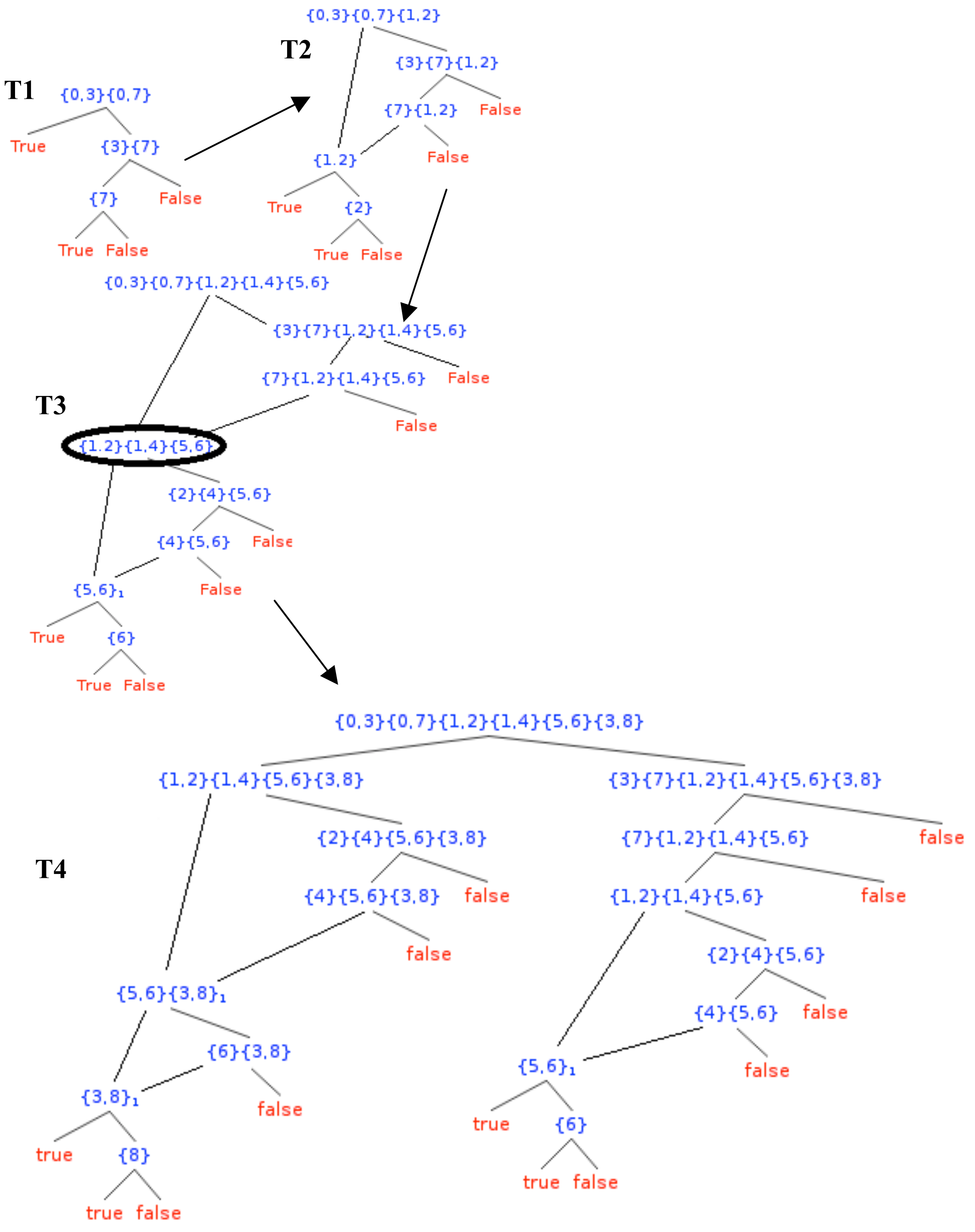

50 Observing the algorithm work on monotone, +*ve* 2-SAT does not reduce the generality of properties discussed below which occur even with this restricted version of the problem.

Note the following aspects apparent in the above GSPRA(S) sequence:

1. Number of nodes (not counting leafs) increase from T1 to T4 in the following way: 3,5,8,15, i.e., almost doubling between trees T3 & T4.
2. ({1,2}{1,4}{5,6}), the node marked with an ellipse in T3 is a node common between two branches and a subset of the original Clause-Set S.
3. This common-node as well as all its sub-nodes is copied once when T4 is formed and the copy is processed with the new clause {3,8} while retaining the original node.
   Thus, two nodes are formed in T4: ({1,2}{1,4}{5,6}) and ({1,2}{1,4}{5,6}{3,8}). This is called a CN-split of ({1,2}{1,4}{5,6}) and plays an important role in the complexity of GSPRA (c.f. Definition 8 below).
4. The common-node ({1,2} {1,4}{5,6}) is of rank 2 which is the same rank as the base-node rank. We can explain splitting of such a common-node as follows: While solving a problem having a certain order of magnitude the algorithm needs to duplicate the result of solving a sub-problem having the same order of magnitude.

   *This behavior is the cause of inherent exponential complexity.*

**Definition 3:** An SRT of a Set S of a.a. 3-SAT-CNF-clauses is called sequentially-ordered **if** every Clause-Set in any non-leaf node of the SRT has only one clause or its derivation **or** has the form: S={$C_i$, $C_j$, … $C_M$ } for some $i$<$j$<….<M, M number of clauses in S, where $C_x$'s are clauses or derivations of clauses in S.

**Definition 4:** An SRT of a Set S of 3-SAT-CNF-clauses is called strongly ordered (s.o.) **if** every Clause-Set formed during resolution is linearly ordered (l.o.). In that case the Set S is also called strongly ordered. Strongly ordered Sets are always linearly ordered whereby the inverse is not always the case, i.e., some s.o. Sets may have Clause-Sets in their IRTs which are not l.o.
If a Set S has a base Clause-Set which is l.o. while some other Clause-Sets in its generated IRTs are l.o.u., then S as well as its SRT is called loosely ordered (lo.o.), e.g.:

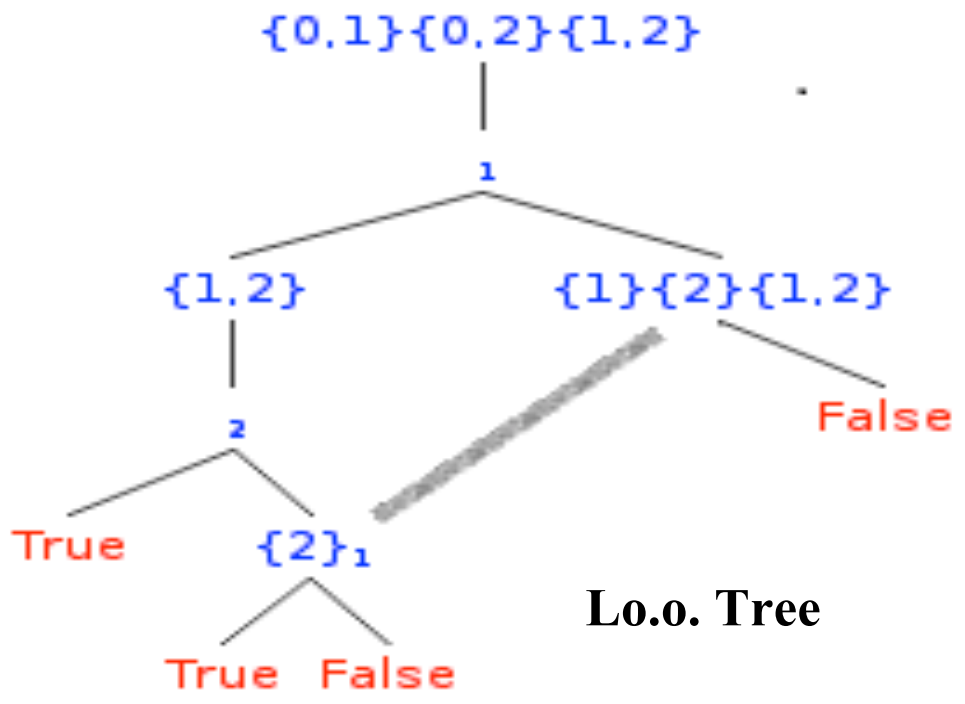

**Lo.o. Tree**

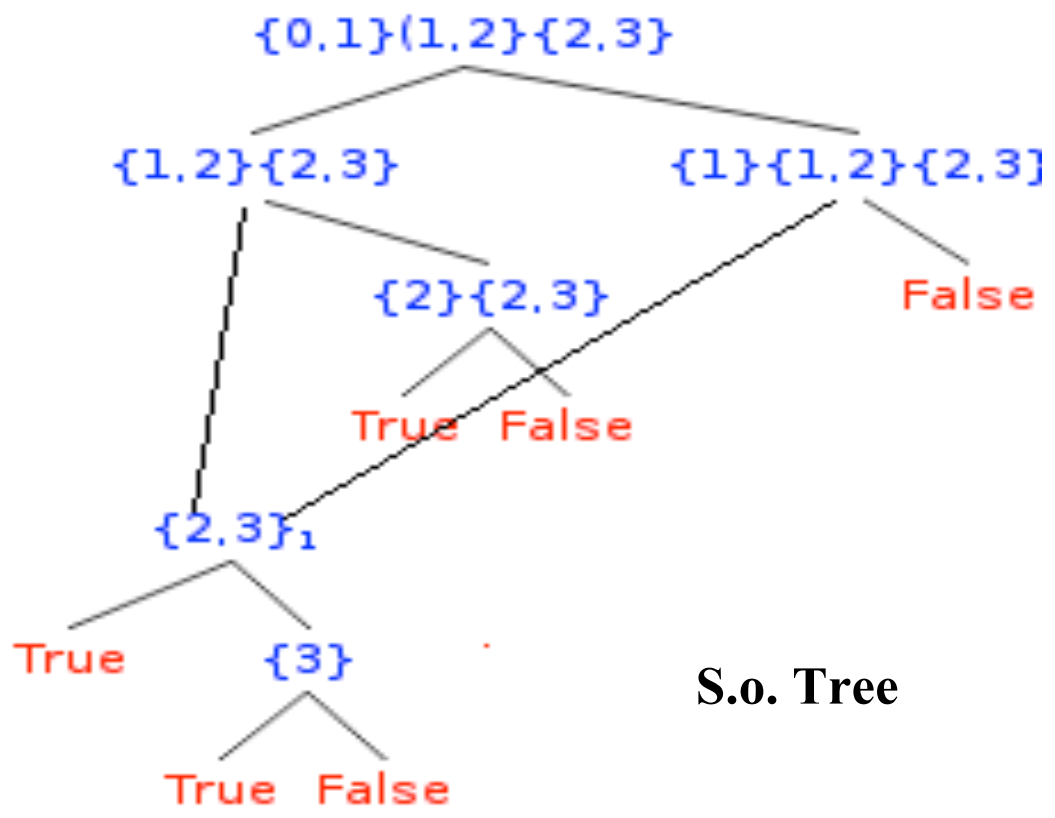

**S.o. Tree**

**Definition 5:** A node [$q$] is called common-node (CN) in an SRT of a Set of 3-SAT-CNF-clauses S **if** in step $k$ of the resolution it becomes a common child to two or more nodes ([$x$], [$y$], [$z$], … (Fig.2)). This happens when $x$,$y$,$z$,… literals are replaced by TRUE or FALSE in their respective Clause-Sets. The common-node [$q$] contains the first appearance of its name literal (NL) $q$ in all branches of the SRT containing [$x$],[$y$],[$z$],..

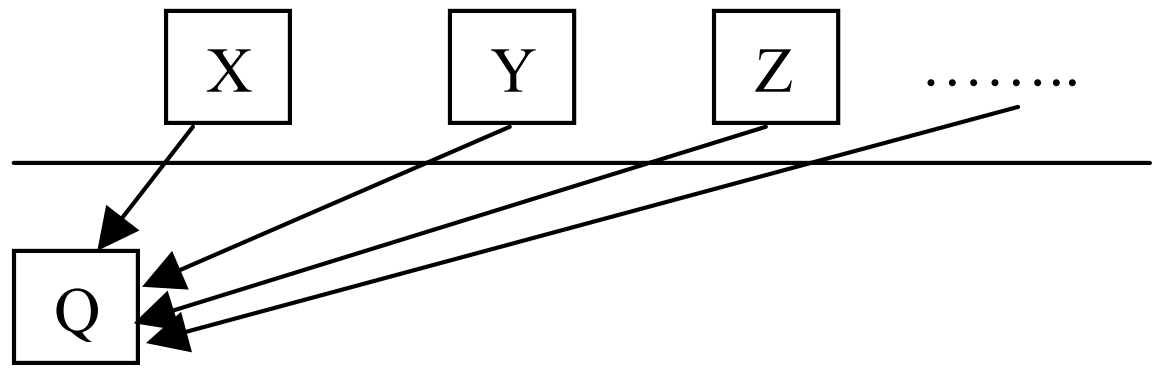

**Figure 2:** Common-node generated in <=$k$.

Types of common-nodes for 3-SAT-CNF-clauses are: Head-, Middle- and Tail Common-nodes (HCN, MCN, and TCN).

More precisely:

- A CN [$q$] is called HCN **if** its Clause-Set has a leading clause C ∈ S, NL $q$ is head of C
- A CN [$q$] is called MCN **if** its Clause-Set has a leading clause C' which is derivation of a C ∈ S, NL $q$ is middle of C
- A CN [$q$] is called TCN **if** its Clause-Set has a leading clause C' which is derivation of a C ∈ S, NL $q$ is tail of C

Examples for both HCN and TCN are provided in Lemma 2 and its respective remarks.

A CN produced in step $k$ is called "supported" in a step $>k$ if its Clause-Set gets clauses appended to its head which don't belong to any block which was instantiated in steps <=$k$ by **one or more** of its parents. A parent-set of such a CN is called "supporting". In Fig. 3 an example is shown for the CN $\{b,c\}$ which is supported by clause $\{d,e\}$ not belonging to block $B_a$:

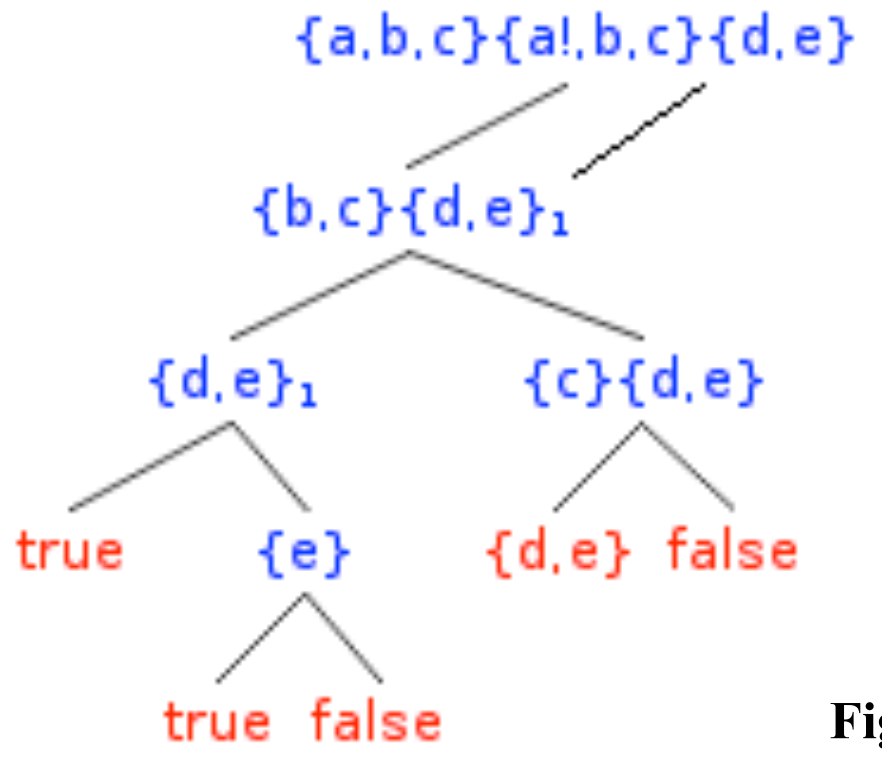

**Figure 3**

If a head-clause of a CN is also a clause of one of its parent-sets, then this parent-set is called "direct parent" of the CN. The CN itself is called "direct child" (Fig.4-a):

**Figure 4-a**

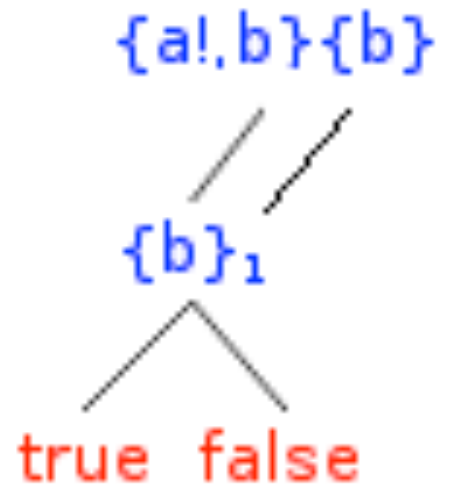

**Definition 6:** A dependency graph (DG) of a Set of 3-SAT-CNF-clauses S is a directed, acyclic graph <V,E> where V is the Set of all NLs, E the Set of ordered pairs $<v_1,v_2>$ , $v_1,v_2$ $\in V$ representing instantiations of NLs produced during GSPRA(S). DGs can be deduced from SRTs in a canonical, straightforward way[51] and used as practical alternatives for truth tables (c.f. Property 1). They are equivalent to Free Binary Decision Diagrams (FBDDs)[52] as shown in Property 7. The following two properties define a DG:

1. Each NL can appear only once in a branch.
2. Branches can have different literal/variable orderings $\prod_p$ depending on the sub-problem $p$ they belong to[53].

---

[51] By abstracting in each resolution-step for each node of the SRT the least-literal of the head-clause and building out of it a corresponding node in the DG.

[52] FBDDs are normally generated independent of SAT-Solvers or by recording - on top of resolution-procedures - variable assignment decisions encountered while trying to find a solution. The methods described here produce a canonically ordered FBDD(=DG) representing existent variable alignments in the used clauses. This FBDD is the core product of our Solver rather than a mere byproduct.

[53] In contrast to OBDDs in which one literal/variable-ordering is governing the whole graph.

A leaf of a DG is a node whose value is TRUE or FALSE. Positive leafs have the value TRUE. Fig. 4-b shows an example of a DG for the exemplary s.o. tree in Definition 4.

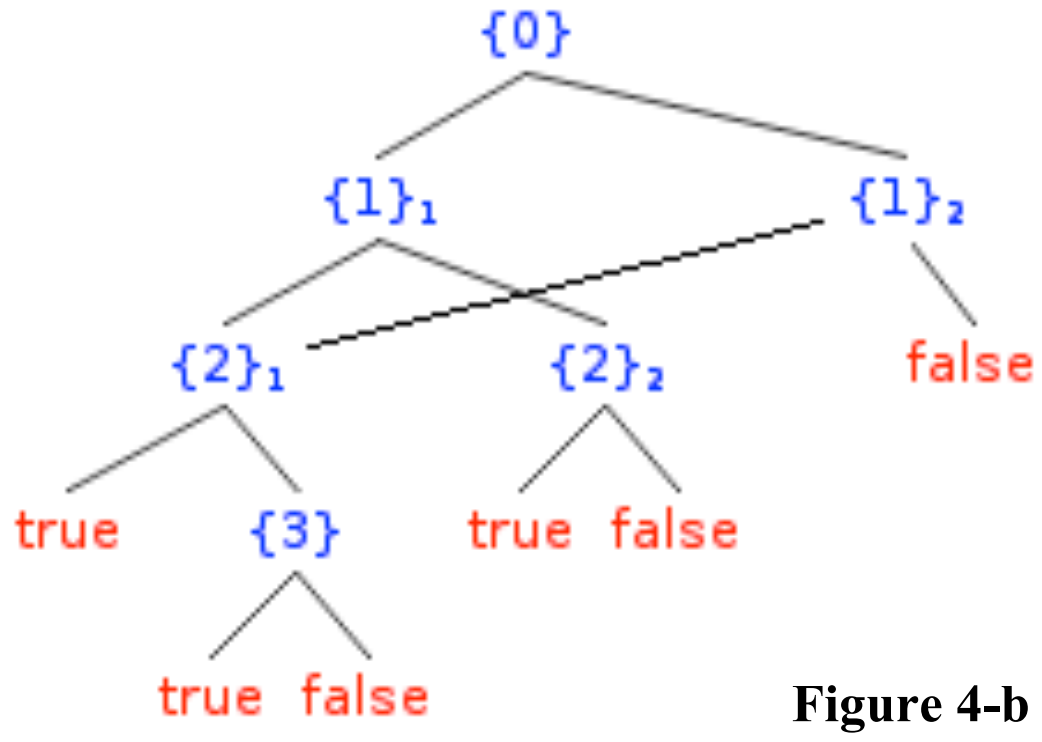

**Figure 4-b**

**Definition 7: Algorithmic Equivalence of nodes**

Two nodes of similar size >1, $n_1 \in$ SRT$_1$, $n_2$ $\in$ SRT$_2$ are said to be Algorithmically Equivalent ($n_1 \approx n_2$) **iff**

a) their Clause-Sets are (not necessarily Syntactically) Equivalent[54] and
b) their respective left- and right sub-nodes are Algorithmically Equivalent.

For size =1: $n_1 \approx n_2$ **iff** their Base-Clause-Sets form isomorphic SRTs, i.e., SRTs having the same structure of leaf/non-leaf-nodes, e.g. (Fig. 5 and 6):

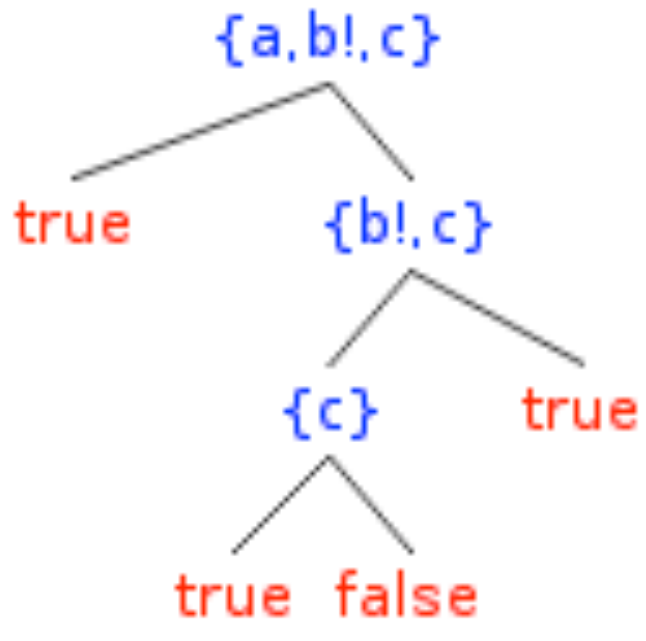

**Figure 5**

Is not isomorphic to: {a,b,c}

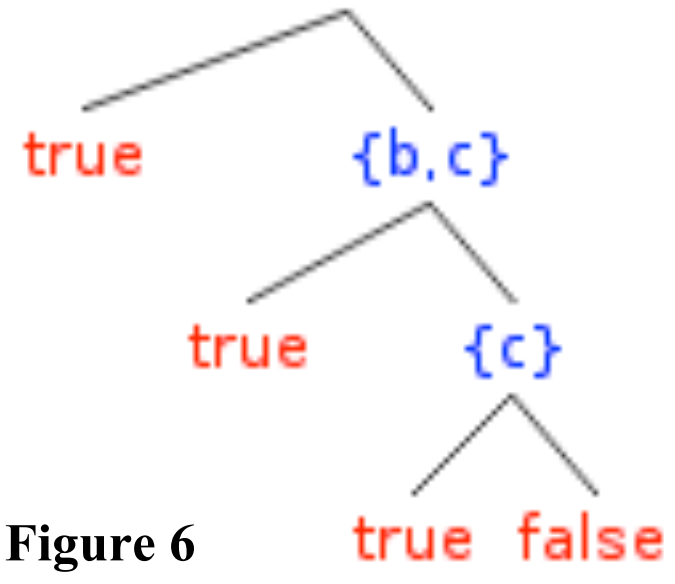

**Figure 6**

**B) PROPERTIES OF GSPRA AND DEFINED STRUCTURES**

The following ten properties of the GSPRA algorithm as well as Lemma 1, Corollary 1 and Lemma 2 are valid for s.o., lo.o. and/or a.a. 3-SAT-CNF-Sets as indicated in each respective place below. When nothing is indicated, a.a. Clause-Sets are meant. For a summary of the main results of this section refer to Fig. 41 at the very end of this section.

**Property 1 (completeness, truth table equivalence):** GSPRA is a complete, truth table equivalent algorithm, i.e., it returns TRUE **iff** there exists a variable assignment in the truth table constructed for the Set S which satisfies it and FALSE otherwise.

**Proof:** (by induction on M, the number of clauses in S)

**Base: M=1**[55] for the following tree:

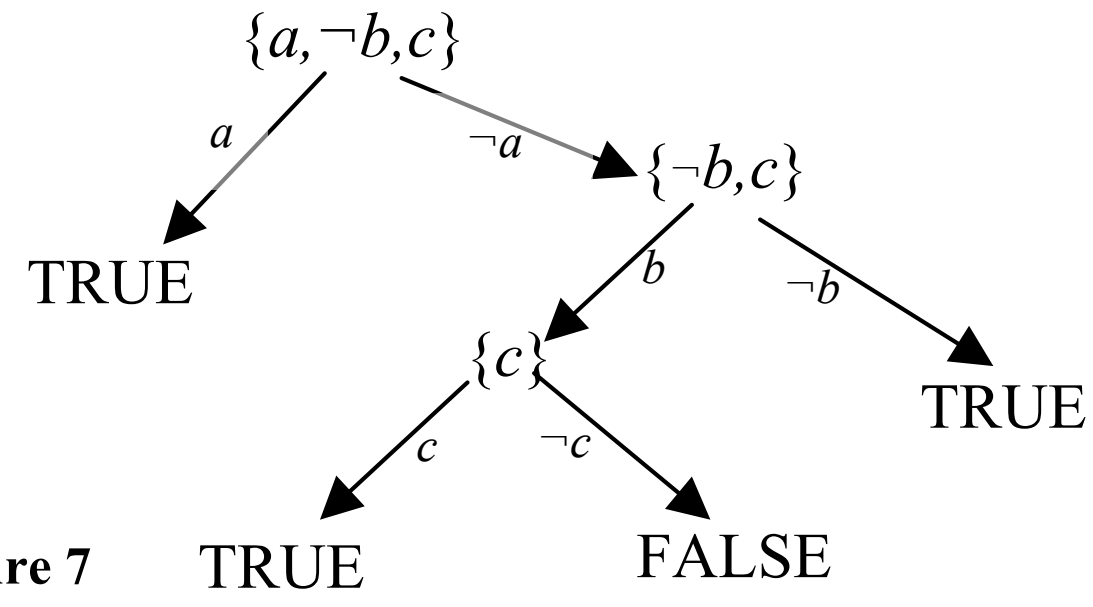

**Figure 7**

---

[54] They might be equivalent modulo renaming of variables. This loose notion of equivalence shall be hardened for GSPRA to become a syntactical one as shall be seen in the properties below.

[55] The case used here (w.l.o.g.) is not the only permutation of *+ve*/*-ve* literals *a,b,c* combined in a clause. The reader is encouraged to check other permutations and verify the validity of the property for M=1 in a similar way to the one shown here.

If we construct the truth table

| A | B | *c* | S |
|---|---|---|---|
| 0 | 0 | 0 | 1 |
| 0 | 0 | 1 | 1 |
| 0 | 1 | 0 | 0 |
| 0 | 1 | 1 | 1 |
| 1 | 0 | 0 | 1 |
| 1 | 0 | 1 | 1 |
| 1 | 1 | 0 | 1 |
| 1 | 1 | 1 | 1 |

and use the following tree propagation rule applied to any node in the tree:
"If the input value of the literal written on the edges of the node is TRUE go left, else go right. Apply this rule to all nodes in the tree until you reach a leaf". Eventually, the obtained results are equivalent to the ones found in the truth table. Let us check the two marked cases using the tree. For "010" the base-node will take us right through edge $\neg a$, then left through edge *b*, then right again through edge $\neg c$, making the overall value FALSE as stated in the truth table. For "101" we are taken by edge ***a*** directly to the value TRUE which is the value of the truth table as well. The reader is encouraged to check all the other truth table entries for validation.

**Induction Hypothesis:** SRT for M-clauses is equivalent to the truth table constructed for all variables whose literals are used in the M-clauses.

**Induction Step:** When clause $C_{M+1} = \{x,y,z\}$ is processed the following cases can be distinguished:

1. *x,y,z are new variables in S:* GSPRA will propagate $C_{M+1}$ until leafs are reached. If leafs are +*ve* then the tree representing $C_{M+1}$ will substitute them, otherwise FALSE is left (as instructed in Definition 2, step 2 c) and d). Each branch not ending with FALSE will thus have as extension a tree giving all possibilities of variable assignments for the three new variables (as seen in the Base-Case). A branch which terminates with FALSE is guaranteed by induction hypothesis to reflect the fact that the Clause-Set is not satisfiable even without taking the new clause into consideration. Thus, the newly constructed tree is logically equivalent to an extended truth table taking into account the new variables[56].
2. *x exists in S, while y,z are new:* When $C_{M+1}$ is propagated through branches of the tree, those terminating with FALSE - as seen in the previous case - are not dependent on the new clause and will keep their values and guarantee (per induction hypothesis) that the Clause-Set is not satisfiable. For all those branches which terminate with TRUE it either might be the case that this truth value is independent of the new variables and thus the truth value is kept as it is per induction hypothesis[57], **or** the branch is extended to give all possibilities of assignments of the new variable(s) as before[58]. In both cases the newly constructed tree logically corresponds to an extended truth table which contains values for two more variables in all branches where it is relevant.
3. *x,y or x,y,z are already in S:* **Either** no new nodes are added to the tree in

[56] Although syntactically the number of entries of the truth table is bigger, since the tree is discarding all unnecessary variable/value combinations (such is the case when the Clause-Set has already reached the value FALSE and adding new variables cannot change this fact).

[57] like the case of node {2}{2,3} in the above s.o. tree: Adding the clause {2,3} to the node {2} did not change the truth value of its children which were leafs.

[58] in the same s.o. tree compare the case of node {1,2} before and after adding {2,3}.

all those branches where variables already exist and where per induction hypothesis the tree is already equivalent to the right truth table portion in those branches **or** *x* and/or *y* and/or *z* are new in some branch. In that case they will be added to the +*ve* leafs accordingly and correspond to specifications of truth table values which were *don't cares* before[59]. (Q.E.D.)

**Illustration of Property 1 for the case (M=2)** [60]**:**

Suppose S={{¬*a*,*b*,¬*c*}{*b*,*c*,¬*d*}}, its SRT is as in Fig. 8:

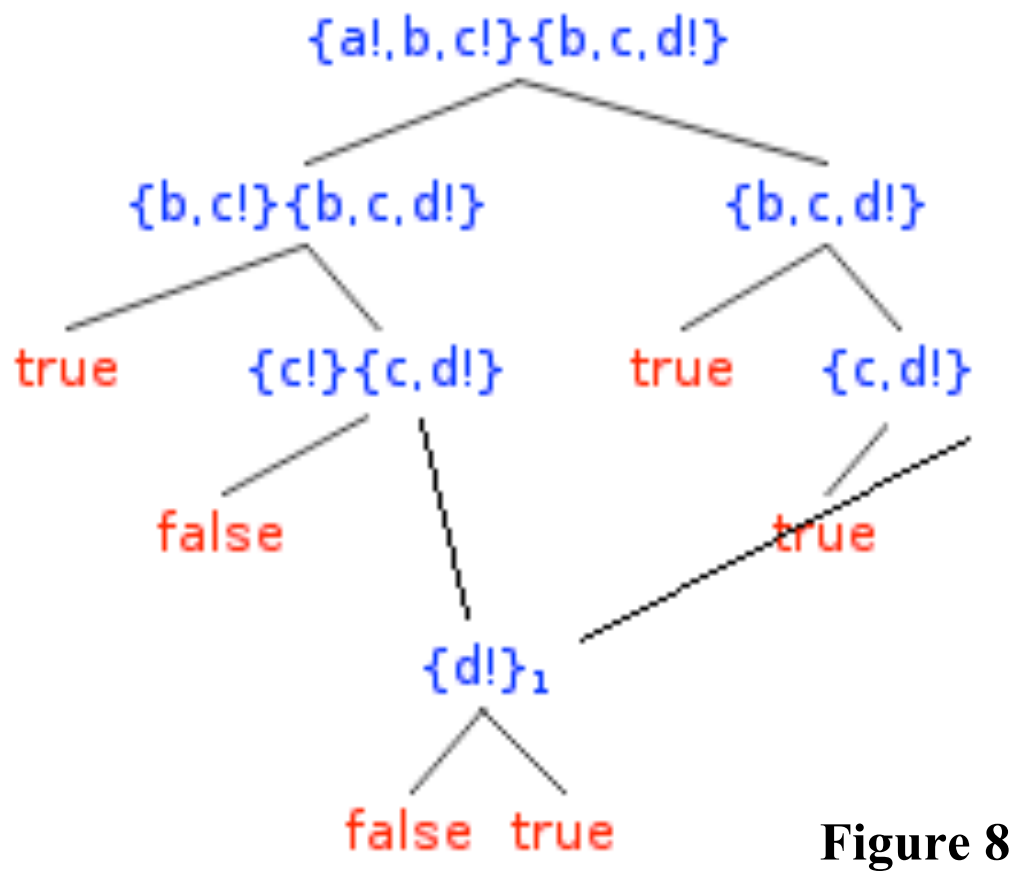

**Figure 8**

The following Truth Table 2 is the truth table for the above tree. If a variable value of the truth table does not apply to a tree node (simply because the literal does not exist in the Clause-Set), skip it.

Example: Entry *a*=1,*b*=1,*c*=0,*d*=0 in the table gives S=1. If starting at the base-node and going left (because *a*=1) through the +*ve* ***a*** edge, the node {*b*,¬*c*}{*b*,*c*,¬*d*} will lead straight to the value TRUE (since *b*=1).

**Truth Table 2**

| A | B | C | D | S |
|---|---|---|---|---|
| 0 | 0 | 0 | 0 | 1 |
| 0 | 0 | 0 | 1 | 0 |
| 0 | 0 | 1 | 0 | 1 |
| 0 | 0 | 1 | 1 | 1 |
| 0 | 1 | 0 | 0 | 1 |
| 0 | 1 | 0 | 1 | 1 |
| 0 | 1 | 1 | 0 | 1 |
| 0 | 1 | 1 | 1 | 1 |
| 1 | 0 | 0 | 0 | 1 |
| 1 | 0 | 0 | 1 | 0 |
| 1 | 0 | 1 | 0 | 0 |
| 1 | 0 | 1 | 1 | 0 |
| 1 | 1 | 0 | 0 | 1 |
| 1 | 1 | 0 | 1 | 1 |
| 1 | 1 | 1 | 0 | 1 |
| 1 | 1 | 1 | 1 | 1 |

**Property 2 (expansion of SRTs):**

∀$n_1$,$n_2$ nodes ∈ SRT, $n_1$<> $n_2$: if $n_1$,$n_2$ are not directly connected in steps <=*k*, then they cannot be directly connected in steps >*k* except in the trivial case when the new clause belongs to a block, parents of nodes were instantiating in steps <=*k* and $n_1$, $n_2$ become equivalent. Moreover: Nodes of sizes *j*<=M generated in step *k* are at most as many as nodes of sizes *j*-1 existent in step *k*-1 not counting nodes generated through splits in size-level *j* of the SRT.

**Proof:** One intrinsic property of GSPRA is that it directly connects - per definition - two nodes **iff** the Clause-Set of one of them can be instantiated - in a way respecting the least-literal/head-clause-rule - to become the parent of the other. Suppose we have at step <=*k* a situation in the SRT as seen in following Fig. 9 (left part): nodes $n_1$,$n_2$ are **not** connected. They both get instantiated through their least-literals *a,b* to different directions in the tree. Any further clause {*x,y*} in steps >*k* will keep this situation intact, since ***a*** and ***b*** remain the least-literals in their respective Clause-Sets and cannot be bypassed by clause {*x,y*} in the new tree (Fig. 9 right part) even in the worst case:[61]

[59] For illustration: Consider the case where {1,2} is added to {0,1}{0,2}. The left branch of the tree {0,1}{0,2} which is the leaf TRUE, corresponds to the fact that values of 1&2 are not relevant for the overall value of the formula {0,1}{0,2} when literal 0 is set to TRUE following this particular assignment branch, i.e., they are *don't cares*. When {1,2} is added, its tree replaces TRUE indicating for what values of 1 & 2 the same truth table gives truth values capturing satisfiability conditions of the newly added clause {1,2}.

[60] This property is known to hold for BDDs in general and thus FBDDs as well, so there is no surprise that DGs or SRTs possess it (c.f. [Friedmann 1986]).

Fig. 10 shows a trivial exception of this situation where both nodes are merged in steps >*k* (right) as the new clause {*i,a,b*} belongs to a block $B_i$ parents of both nodes were instantiating in steps <=*k*. The added clause makes N1 equivalent to N2 as seen. We call those types of CNs: Trivial Common Nodes (tCNs). They are formed in what we call: Symmetric Blocks (SBs) to be defined below and are included in the properties/lemmas dealing with the generation of CNs.

Furthermore: As resolution is sequential, then per definition the only source of new *j*-sized nodes, *j*<=M, in step *k* are *j*-1 sized ones in step *k*-1 which are resolved with the new clause. This is not counting any nodes copied in split-operations in the *j*-size-level of the SRT of course. Thus, the number of generated *j*-sized nodes in step *k* is always bounded above by the number of existent *j*-1 sized nodes in step *k*-1 if split operations are not counted.

(Q.E.D.)

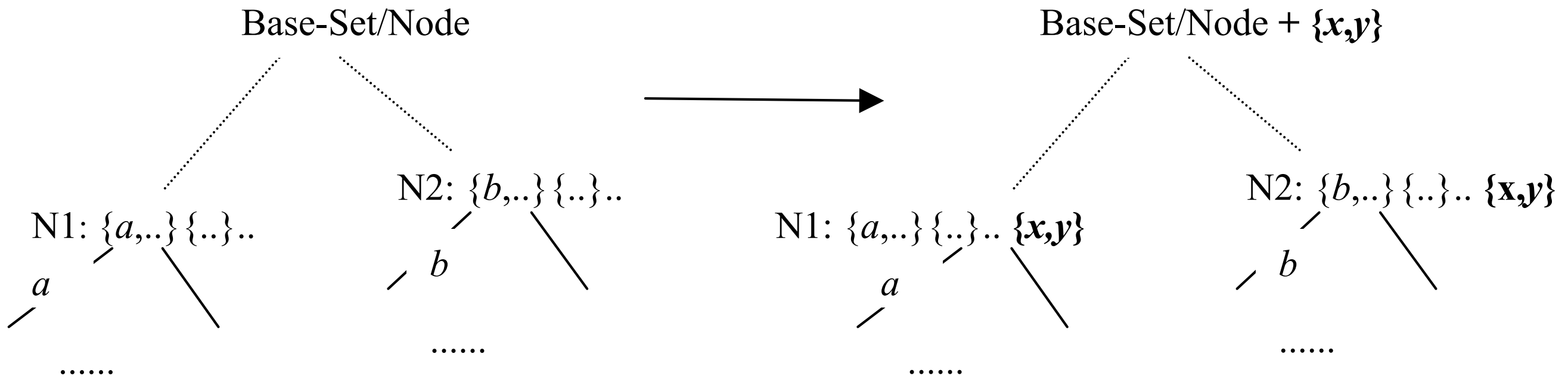

**Figure 9**

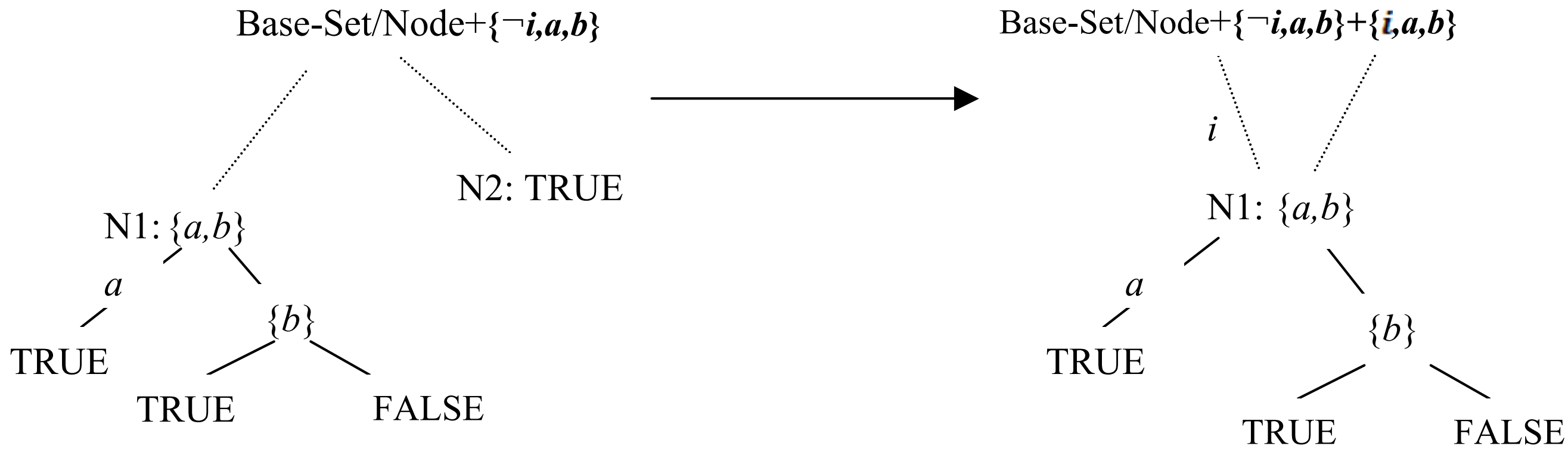

**Figure 10**

[61] Worst case here means when {*x,y*} is propagated through all the nodes as seen in the figure. Otherwise it might be that some nodes remain the same as in step *k* and are thus not connected either.

**Property 3 (linear derivation of clauses):** GSPRA produces for s.o. and lo.o. Clause-Sets only linear derivations of newly resolved clauses at any resolution-step[62].

**Proof:** This property is also caused by the least-literal/head-clause-rule (Definition 2 5.). To see this consider the case when during resolution of a new clause C={*x*,*y*,*z*} with the IRT, C is processed and gets instantiated by putting *y*=FALSE for example resulting thus is a derivation C'={*x*,*z*} which is not linear. This can only happen - as per the least-literal/head-clause-rule - if *y* is the least-literal of a head-clause in that step, i.e., there is a Clause-Set of the form: {{*y*,...}...{*x*,*y*,*z*}....} which is being instantiated. If variable *x* is bypassed during this instantiation process, then either *x*'s first occurrence is in {*x*,*y*,*z*}, i.e., it is a new variable, and in that case it will constitute a breach of the new variable names condition (Definition 1 c)), **or** *x* occurred before {*x*,*y*,*z*}. As *x* could not have occurred in {*y*,...} because of the sorting condition of literals within clauses (Definition 1 a)), it must be the case that *x* occurred in a clause between {*y*,...} and {*x*,*y*,*z*}. But then *x* is still new with respect to *y* and we still have a breach of Condition 1 c). Thus, no non-linear derivations are possible for newly processed clauses through GSPRA if SRTs are s.o. or lo.o.
(Q.E.D.)

**Property 4 (generation of non-trivial CNs):** The only non-trivial CNs generated in any step *k* by GSPRA while resolving clause C of a Set of 3-SAT-CNF-clauses which are s.o. or lo.o., are identical with either C or linear derivations of C.

**Proof:** Recall - as per Property 2 - that non-trivial CNs generated in step k are the ones which are **not** formed, because C belongs to a block, some parents were instantiating in steps <*k*. Suppose now such a non-trivial CN is neither C nor a linear derivations of it. This means that a "legacy" node constructed in steps <*k* became non-trivial CN in step *k*. For a node to become CN, at least two nodes have to be connected to it in a parent-child relation as per Definition 5. This can only mean that at least one new connection has been established in step *k* between two nodes which were previously not connected. As per Property 2 this can only happen in the trivial case when C belongs to a block, some parents were instantiating in steps<*k* and the formed CN is a tCN. Contradiction. Moreover: The linear derivation property (Property 3) tells us that only linear derivations of C can become non-trivial CNs for s.o. and lo.o. SRTs
(Q.E.D.)

**Property 5 (uniqueness of instantiation results):** Let S be an a.a. Clause-Set, $S_1$, $S_2$ any direct children of S produced - through instantiations of literals *i,j* respectively - by an instantiation procedure using the least-literal/head-clause-rule, then $S_1=S_2$ **iff** *i=j*.

**Proof:** If *i=j* then it is obvious that $S_1=S_2$=unique Clause-Set per definition of any instantiation procedure[63]. If $S_1=S_2$ then the question is: Can we instantiate two different literals in S and still get the same direct child Clause-Set (i.e., can *i<>j* imply $S_1=S_2$). In general this is possible. For example, the Set S={{*a,b,c*}{*x,y,z*}} can yield $S_1$={{*x,y,z*}}=$S_2$ if we use either

[62] Linear derivations of C being a proper subset of perm(C).

[63] Remember that instantiation of literals is done by replacing them in clauses with TRUE or FALSE depending on used signs.

*i*=*a*=TRUE or *j*=*b*=TRUE. However, when an instantiation procedure uses least-literal/head-clause-rules this cannot happen, because at any step *k,* producing a direct child, and for any Clause-Set S there is **only one least-literal** chosen for instantiation from either *a* or *b*. Thus, if in a step and for the same S literals are instantiated to give the same child-set, then they must be the same. Moreover: Once S is instantiated this way in step *k*, the same S is never instantiated again in steps >*k*.
(Q.E.D.)

**Property 6 (Syntactical Equivalence):** Let $S_1$, $S_2$ be two a.a. Clause-Sets of nodes $n_1$, $n_2$ of an SRT produced by GSPRA such that $S_1$ is instantiated in a step *k* using literal ***a*** to produce $S_1$Left, $S_1$Right, $S_2$ is instantiated in the same step *k* using literal ***i*** to produce S2Left and S2Right, then: ($S_1$Left=$S_2$left and $S_1$Right=$S_2$Right) **iff** $S_1$=$S_2$.

**Proof:** If $S_1$=$S_2$=S, then ***a***=***i*** and left- and right direct-children Clause-Sets of S are unique as per Property 5.
To see why the other direction is also valid, consider Fig. 11 where $S_1$ and $S_2$ are instantiated using ***a***, ***i*** (respectively) to produce a common left Clause-Set and using ¬***a***,¬ ***i*** to produce a common right one (rectangles). Suppose that $S_1$<>$S_2$, i.e., there exists at least a clause C ∈ $S_2$ , but not ∈ $S_1$. If C doesn't contain literals ***a***, ***i*** then it should appear in both left- and right Clause-Sets when $S_2$ is instantiated using ***i***. This cannot be the case, because $S_1$ doesn't contain C. If C contains ***i*** but not ***a***, then it shouldn't appear in one node (left or right according to the sign of ***i*** in C), but its derivation C' containing literals other than ***i*** (and ***a***) has to appear in the other side contradicting again the fact that C doesn't exist in $S_1$. If C contains ***a***, but not ***i***, then it should appear in the left- and right child nodes contradicting the fact that those nodes shouldn't contain any literal ***a***, since it has been instantiated through $S_1$. Also: $S_1$ doesn't contain C or any of its derivations in the first place. Finally, if C contains both ***a***,***i***, then some derivation C' will have to appear in a child node contradicting the fact that $S_1$ doesn't contain C. Therefore $S_1$=$S_2$.
(Q.E.D.)

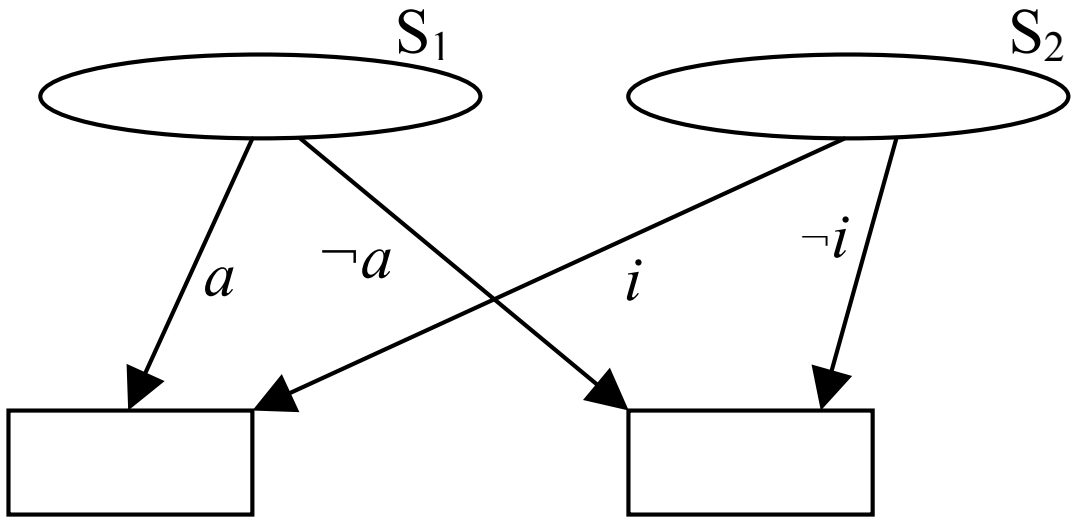

**Figure 11**

**Property 7 (FBDD-equivalence, Branch Linearity):** Let *f* be a function expressible in a.a. 3-SAT-CNF-form, then:

a) The DG produced by GSPRA is an FBDD.
b) Any FBDD produced for *f* can also be produced by a procedure which uses variable orderings, not necessarily canonical, to guide instantiation of literals in the 3-SAT-CNF-representation of *f*.

**Proof:**

1. Remember that a DG can be abstracted in a straightforward way from an SRT (c.f. Footnote 51, Definition 6). We have therefore to show that: If *b* is a branch of the DG, then the maximum size of *b* is N, where N is the number of variables in S, i.e., any variable appears only once in *b*. Moreover: Each resolution-step expands *b* by at most 3 new nodes. It is sufficient to see that when the least-literal-rule is applied to form a +*ve* edge, starting from a node *n* with NL *x*, all occurrences of *x* are instantiated with TRUE. Similarly with -*ve* edges and ¬*x*. Therefore: Clause-Sets in nodes below *n* do not have literal *x* in any of their clauses per definition. This makes it impossible to produce another edge bearing the same variable/literal name on any branch in any further resolution-step. Also, a new clause in a resolution-step may contain a maximum of 3 new literals whose edges are not already present in *b*. This makes the maximum amount of newly added nodes in *b* per step: 3.

2. Suppose *f* has an FBDD. Branches of this FBDD define variable orderings (not necessarily the same for all branches) which can be used as instruction guidelines by any procedure PR instantiating variables/literals in the 3-SAT-CNF expressing *f*. The final output of PR is then the same FBDD (with possible isomorphic sub-graphs). This PR may look like this:

**PR:**
**Inputs**: Clause-Set S representing function *f* in 3-SAT-CNF-form, FBDD for *f*
**Outputs**: Final Tree (not necessarily SRT)
**Steps**:

1. For the currentFBDDNode (initially the root node)
   i. **if** currentFBDDNode is a leaf **then** return leaf
   ii. read the variable in the node
   iii. instantiate S using this variable, left and right.
   iv. create in the resultTree two nodes representing left and right instantiations of S (call them S' and S'' respectively)
   v. call yourself recursively for left side: PR(S', leftNodeOf(currentFBDDNode))
   vi. call yourself recursively for right side: PR(S'', rightNodeOf(currentFBDDNode))
2. return resultTree

(Q.E.D.)

**Property 8 (SRT-structure):** SRTs produced by GSPRA for s.o. and lo.o. 3-SAT-CNF Clause-Sets S of size M have in their top-part, i.e., until all literals of the chosen first clause $C_0$ are instantiated[64], at most *k* M-sized unique and *k* <M-sized not necessarily unique-nodes, where *k*<=3, *k* = number of literals in $C_0$. Moreover: No SPR-unlike[65] resolution-procedure using any variable ordering on S can produce less than *k* M-sized unique-nodes and *k* <M-sized not necessarily unique-nodes in the top-part of its resolution-tree.

**Proof:** When a literal in a clause C is instantiated, it creates two types of Clause-Sets, one in which a linear derivation of C is formed and the other in which C becomes TRUE. The number of Clause-Sets resulting after the instantiation of all literals of C for each type (including the base one) being at most 2**k*, *k* = breadth of C. As prescribed in Definition 2, step 1, GSPRA builds upon the tree shown there which represents all such instantiations of clause $C_0$, which must be the shortest. As sequential resolution proceeds and clauses are propagated one by one through all branches of IRTs, the overall structure of this tree is not altered (per definition). Instead, sizes of its nodes are subject to change. All non-leaf-nodes whose total number is *k*, of this first tree containing $C_0$ or unique linear derivations of it, get new clauses appended to their Clause-Sets rendering them of size M in the final SRT. Leaf-nodes marked with TRUE in the first tree of $C_0$ are filled stepwise with Clause-Sets which can only possess <M clauses, since they miss at least $C_0$. Their number is also *k<=3*. Those <M-sized nodes might not be unique, since subsequent instantiations of S which make $C_0$ TRUE may result in equivalent Clause-Sets. Fig. 12 shows a generic picture of the top of a possible final SRT (w.l.o.g.). Note that both blue (called: Top Head Nodes, **THNs**) and red nodes (called: Top Body Nodes, **TBNs**) of the figure are considered to be in the top-part of the overall SRT. One thing has to be remembered: Signs of literals determine the position (i.e., left or right) of TBNs and THNs. In the generic case shown here, TBN-clauses are all on the left side because all literals are assumed (w.l.o.g.) to be positive.

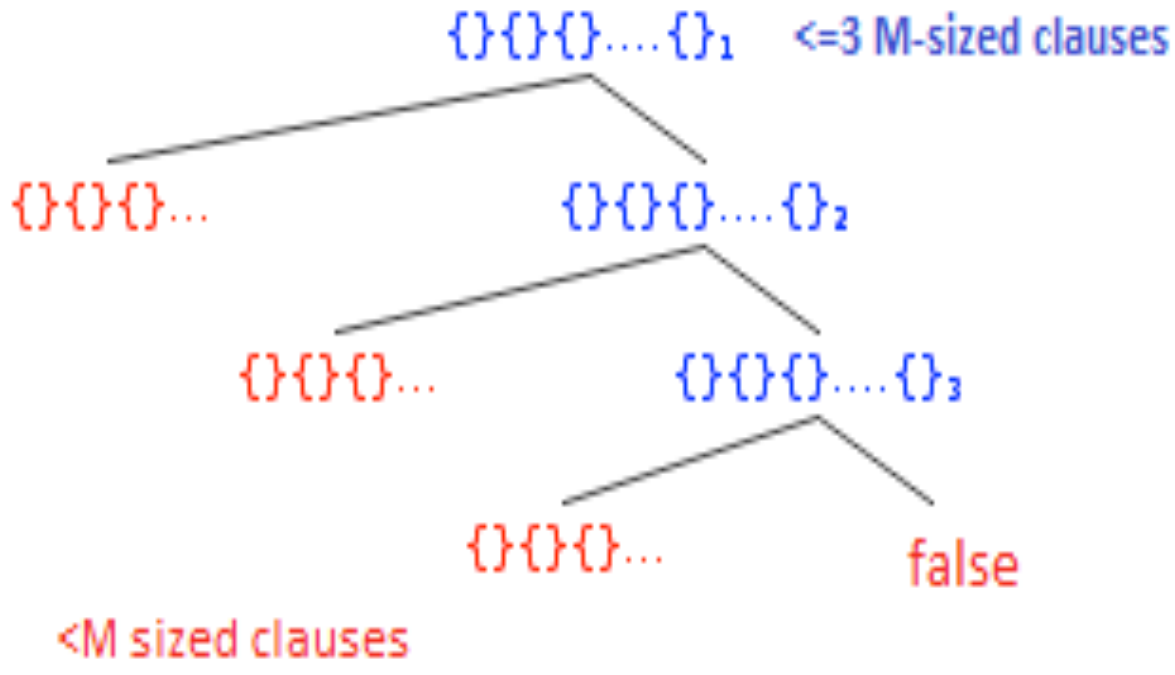

**Figure 12:** Top Head Nodes (THNs) and Top Body Nodes (TBNs)

**For the second claim**: Any *k'*-consecutive literals picked up for instantiation in the beginning of any resolution-procedure will have either to be from the same clause or from different clauses in S. If S is M-sized and *k'* literals are chosen from the same clause, then the procedure is said to be SPR-like and <M-sized Clause-Sets in both left- and right sides of instantiations, included in the boundary of the SRT-top, are reached in *k'* steps at least (where k'>=k, k size of $C_0$ )[66].

---

64 $C_0$ is chosen in Preliminary Step 0, Definition 2.

65 SPR-unlike is any procedure which does not choose only one clause ($C_0$) to fully instantiate in the top-part of its resolution-tree. It allows thus non-minimal top-parts.

66 It might be the case that an SPR-like procedure picks an initial clause different from the shortest one picked by GSPRA.

If on the other hand *k'*-consecutive literals are chosen from different clauses (SPR-unlike), then emptying at least one clause will definitely take $>k$ steps, i.e., since $C_0$ is the shortest, one needs always at least $k$ steps to empty any clause. Thus, $k$ M-sized unique-nodes and $k$ <M-sized not necessarily unique-nodes is less than or equal to any possible number of nodes of either type reached in the top-part of a tree produced by any SPR-unlike procedures.
(Q.E.D.)

**Property 9 (generality of canonical orderings):** Any ordering $\prod_p$ applied to a node solving problem $p$ - formalized using an a.a. 3-SAT-CNF Clause-Set S (in a tree generated by a procedure PR not necessarily SPR-like) can be converted to a canonical ordering $\prod^c{}_p$ used by PR with a least-literal-rule. Moreover: If an SRT produced by SPR-like procedures has a minimal number of unique-nodes with respect to all possible canonical orderings used by such a PR, then it is minimal for all non-canonical orderings used by PR as well.

**Proof:** Suppose $\prod_p = \{a,i,k,h,b,c,\ldots\}$ where $a<i<k<h<b<c,\ldots$ is an arbitrary ordering containing instantiation precedence of literals in S applied by any procedure PR, then literals/variables in S can be renamed using a bijective function[67] $f$: N => N in the following way:

$f=\{(a,a)(i,b)(k,c)(h,d)(b,e)(c,f)\ldots\}$. $\prod_p$ shall become $\prod^c{}_p=\{a,b,c,d,e,f,\ldots\}$

which is canonical. PR can obviously use $\prod^c{}_p$ with a least-literal-rule to achieve the same results as it did with $\prod_p$. Suppose now $SRT_{Min}$ constructed for S using GSPRA via a $\prod^c$ is minimal for all possible canonical orderings, i.e., the number of nodes in $SRT_{Min}$ <= number of nodes in any tree constructed for S by PR using any canonical orderings[68]. If $\prod_p$' is an ordering used by PR which is not canonical such that: Number of nodes of a tree constructed by PR using $\prod_p$' < number of nodes of $SRT_{Min}$, then S can be renamed so that $\prod_p$' becomes $\prod^c{}_p$' as seen above and produces a smaller SRT than $SRT_{Min}$ when used by PR via a least-literal-rule contradicting the minimalism assumption of $SRT_{Min}$ for all canonical orderings. This means that $SRT_{Min}$ is minimal whether canonical orderings are used by PR or not.
(Q.E.D.)

**Property 10 (Algorithmic Equivalence = Syntactical Equivalence)**

Let $n_1,n_2$ be nodes $\in$ SRT of a Set of a.a. 3-SAT-CNF-clauses, $S_1,S_2$ their respective base Clause-Sets: $n_1 \approx n_2$ **iff** $S_1=S_2$.

**Proof:** (by induction on M, the size of nodes)

**Base-Case M=1:** If $n_1 \approx n_2$ then per Definition 7 $SRT_1$ of $n_1$ is isomorphic to $SRT_2$ of $n_2$. Let $b_1 = \{\{a,b,c\}\}$ be base Clause-Set of $SRT_1$ and $b_2 = \{\{x,y,z\}\}$ be base Clause-Set of $SRT_2$. We can make $b_1=b_2$ by renaming literals in $b_2$ ($x>a,y>b,z>c$) without affecting the truth value of the Clause-Set. The other direction is trivial[69].

**Induction Hypothesis:** $\forall n_1,n_2$ nodes $\in$ SRT, $S_1,S_2$ their respective base Clause-Sets of size M: $n_1 \approx n_2$ **iff** $S_1=S_2$.

---

[67] This function is called in next sections a mapping.

[68] Different canonical orderings can be created using different renaming functions $f$.

[69] The case shown here (w.l.o.g.) has only *+ve* literals. The same property is valid for all other cases as the reader may wish to verify.

**Induction Step:** For size M+1:
If $n_1 \approx n_2$ then: As per Definition 7 left- and right sub-nodes of $n_1$, $n_2$ are equivalent as seen in Fig. 13 (dashed lines) in which rectangles represent M-sized and ellipses M+1 sized nodes. Applying Definition 7 recursively also renders other M+1, M-sized nodes equivalent (solid lines). For all rectangular nodes the induction hypothesis applies, i.e., their Clause-Sets are also equivalent. This makes it possible to apply Property 6 to the lower, purple part of Fig. 13 thus deducing that Clause-Sets of nodes $n_1$' and $n_2$' are equivalent, then applying Property 6 again to the blue part to reveal that Clause-Sets of nodes $n_1$'' and $n_2$'' are equivalent before applying Property 6 to the top-part to finally infer that Clause-Sets of nodes $n_1$ and $n_2$ are equivalent.

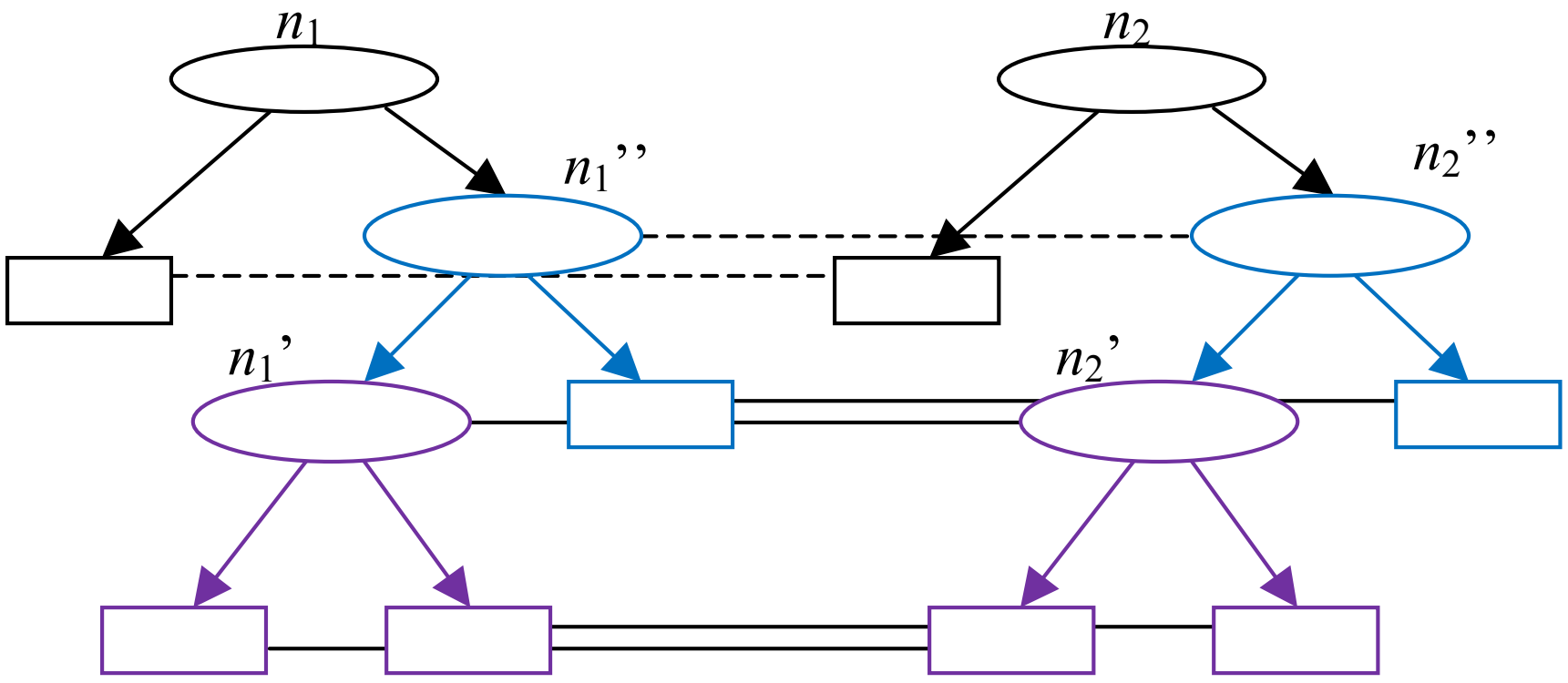

**Figure 13:** Algorithmic Equivalence = Syntactical Equivalence

Other direction: If $S_1=S_2$, then as per Definition 7 left- and right sub-nodes have to be algorithmically equivalent as well to make $n_1 \approx n_2$. We know that left- and right Clause-Sets are syntactically equivalent (Property 6). For the rectangular nodes, the induction hypothesis can be applied. For the elliptical nodes we go a level deeper applying Definition 7 and Property 6 again and so on. Thus $n_1 \approx n_2$.
(Q.E.D.)

**Lemma 1:** SRTs of 3-SAT-CNF Clause-Sets S - whether s.o. or lo.o. - are sequentially-ordered.

**Proof:** (by induction on M, the number of clauses is S)

**Base-Case M=1:** The SRT produced in Definition 2, step 1 is sequentially-ordered per definition where $C_0$', $C_0$'' are derivations of $C_0$ (Fig. 14):

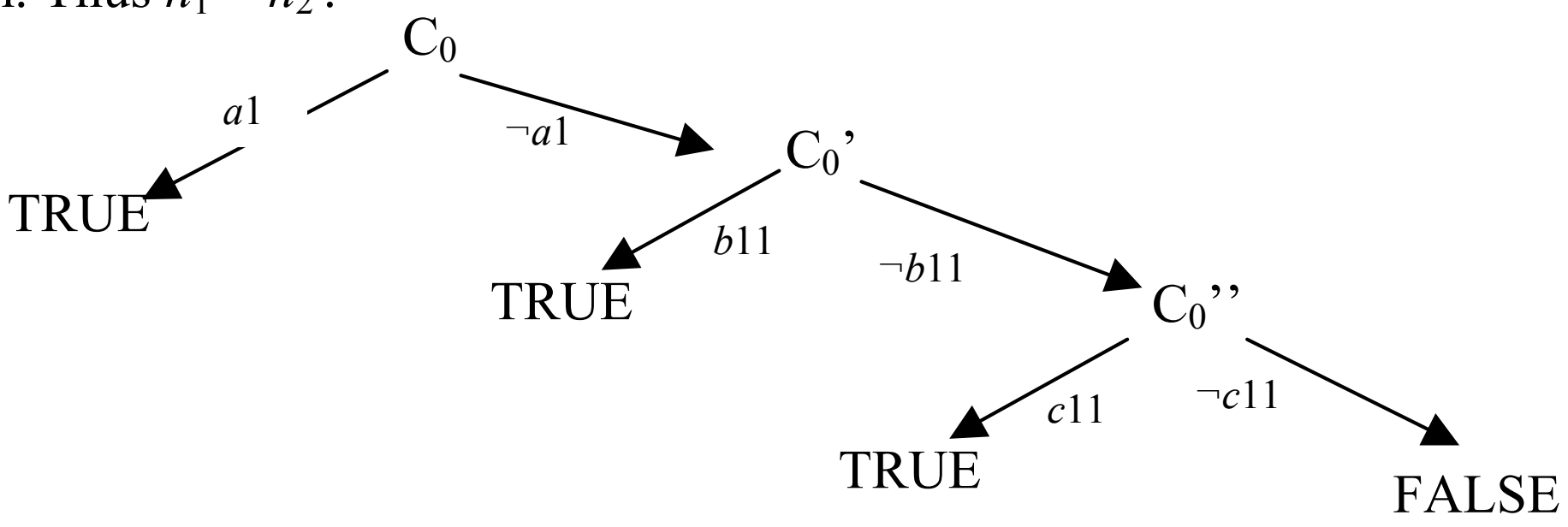

**Figure 14:** sequentially-ordered SRT

**Hypothesis: For s.o. and lo.o. Sets S of M clauses, SRT is sequentially-ordered**

**In resolution-step M+1**: Let S' = S *U* {*x*,*y*,*z*}[70]. Following cases can be distinguished:

1. *x*,*y*,*z* are new literals => Then $C_{M+1}$={*x*,*y*,*z*} is going to be added to all Clause-Sets in the resolution-tree except the leafs which may be transformed to the tree representing $C_{M+1}$ => According to the induction hypothesis all nodes are already sequentially-ordered, adding $C_{M+1}$ preserves this order for M+1
2. *x* is already in S while *y*, *z* are new => for any general node in the SRT for steps <=M of the form (Fig. 15):

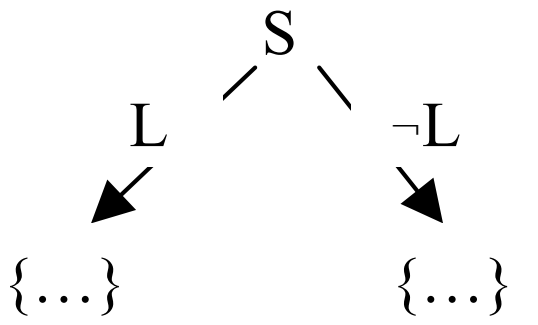

**Figure 15**

If L<>*x* => then this part of the tree (and similar ones) will look like Fig. 16 after step M+1, hence, preserving the property:

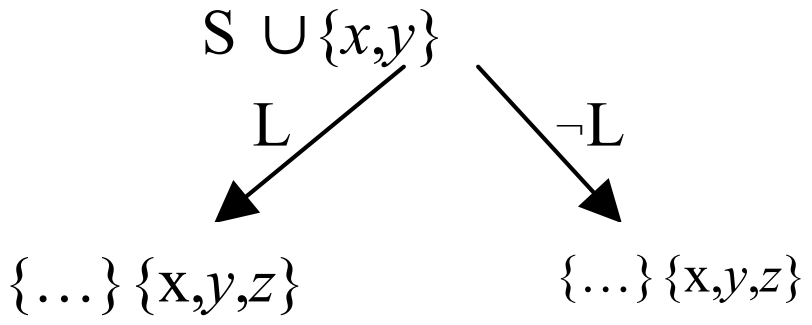

**Figure 16**

If L=*x* => the sub-tree will look like Fig. 17 also preserving the property:

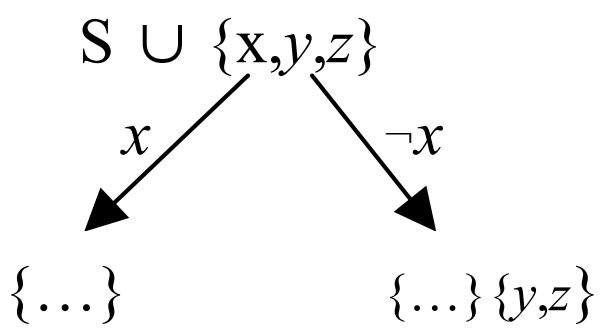

**Figure 17**

3. *x*,*y* already in S, while *z* new or *x*,*y*,*z* already in S => similar to case 2, for any node in the SRT:
   a. If *x*=L and *y*, *z*<>L a derivation of the new clause $C_{M+1}$ is not added to the left Clause-Set of the node and added to the right one (after setting *x*=FALSE)
   b. If *x*<>L and *y* or *z* =L: this case can never happen because of the least-literal-rule
   c. If *x*<>L and *y*,*z* <>L the new clause $C_{M+1}$ is added to both sides

thus preserving the sequential ordering property in all the above cases.
(Q.E.D.)

[70] An example permutation with only *+ve* literals is used here for simplification and (w.l.o.g.).

**Illustration of Lemma 1 for sample case (2-SAT, M=3, +*ve* literals only):**

Let S be {{*a,b*},{*c,d*},{*e,f*}}={$C_1$,$C_2$,$C_3$}:

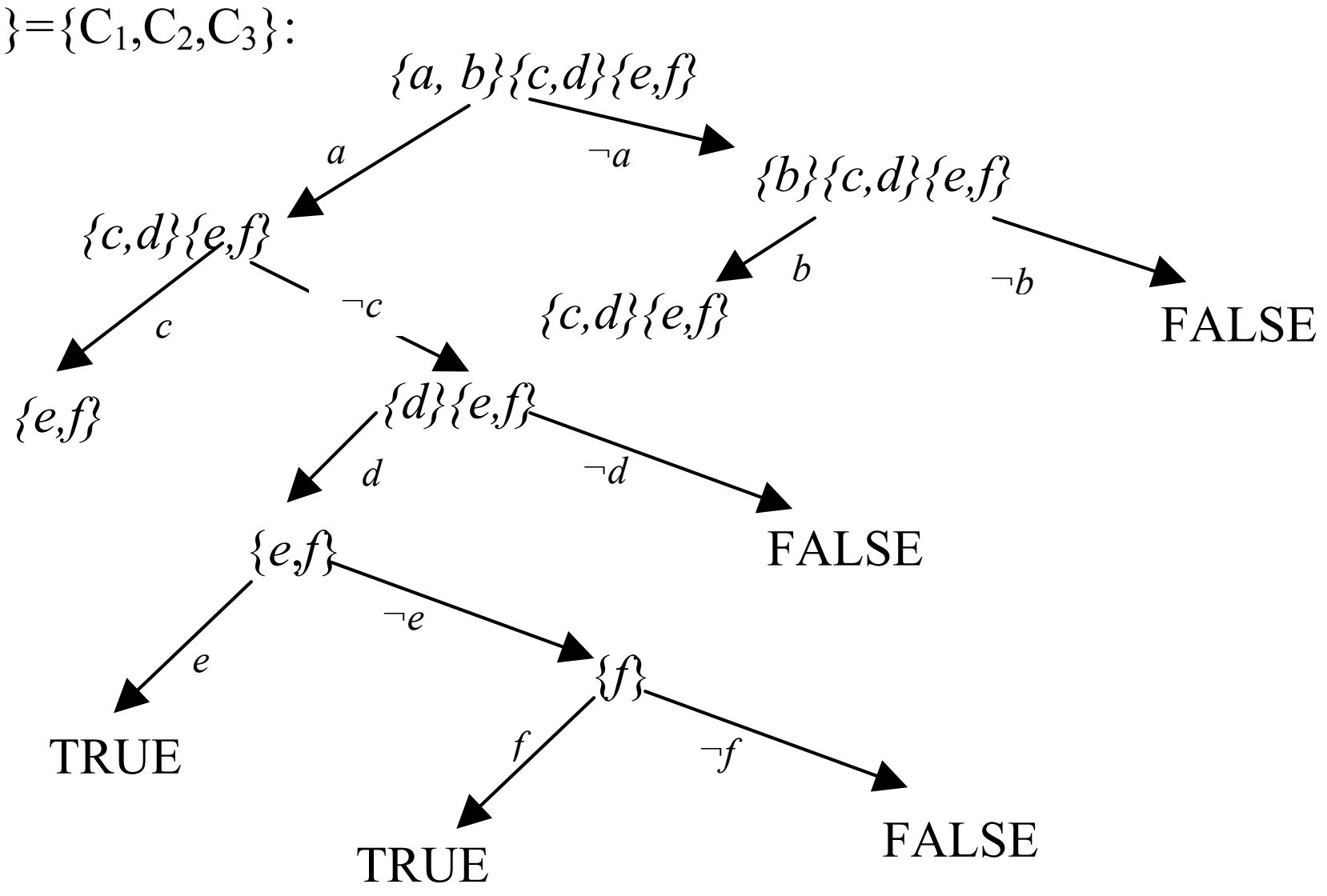

**Figure 18:** canonical SRT for S

Note that all nodes (except the leafs) are either of the form {$C_1$,$C_2$,$C_3$} or {$C_2$,$C_3$} or {$C_3$}. Trying to equalize any literals without breaching the s.o. or lo.o. property of S[71] (for example *a*=*c*=*e* or *a*=*c* only, but not *a*=*f*) will yield all the resulting tree(s) sequentially-ordered. Fig. 19 shows the above tree if *a*=*c* and *d*=*e*.

71 Recall that the main difference between s.o. and lo.o. Sets is the fact that Clause-Sets (other than the Base-Set) may be l.o.u. rather than l.o. While clauses in a l.o.u. Set may not be sorted as required by Definition 1 b), they do keep their original clause sequence, which is relevant here, intact.

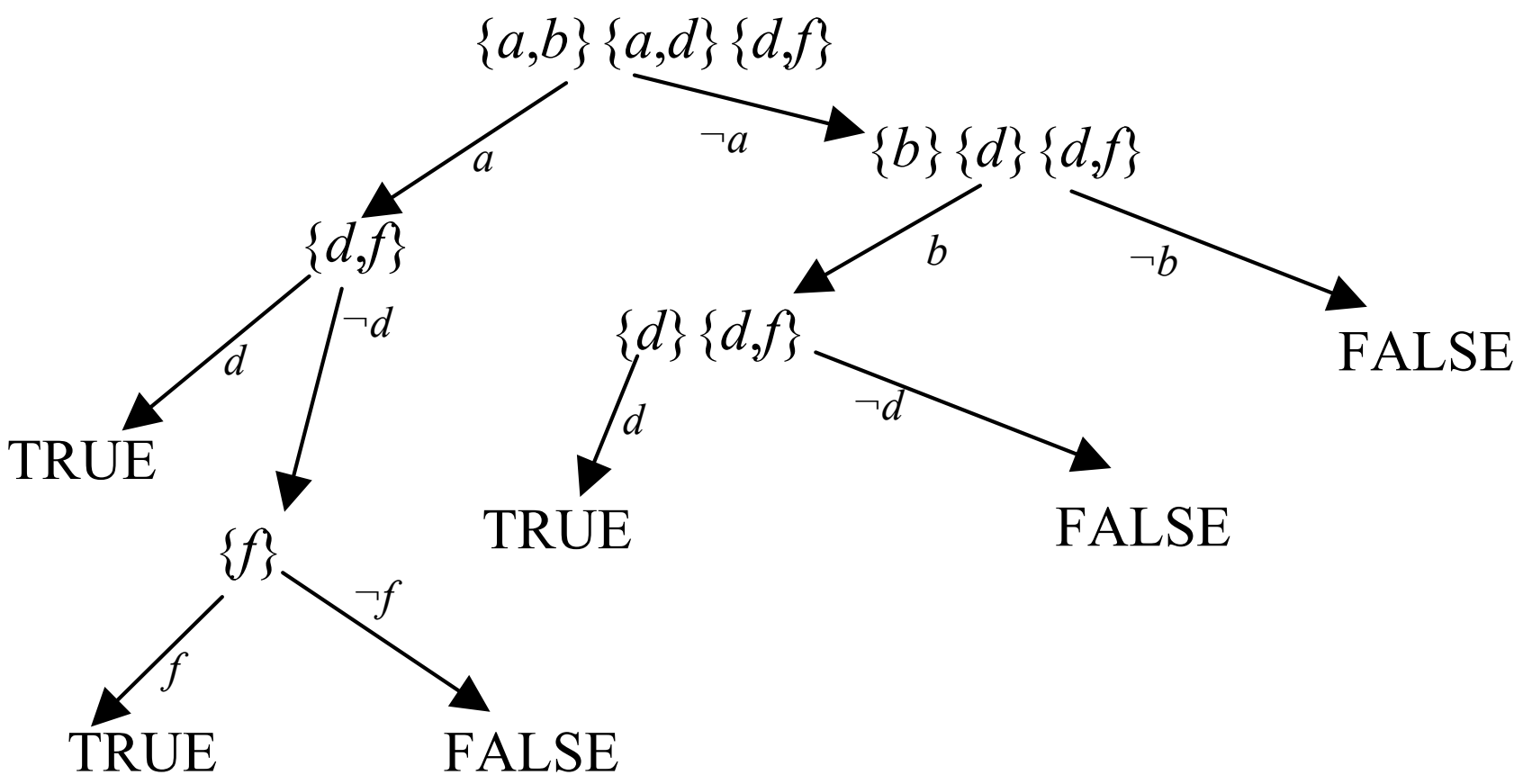

**Figure 19**

**Illustration of Lemma 1 for sample Case (2-SAT, M=3, +*ve*/-*ve* literals):**

The below tree (Fig. 20) shows for a sample Base-Case of 2-SAT that a sequential order of clauses is preserved. The least-literal/head-clause-rule is seen not to be affected neither by negation nor by breadth of clauses.

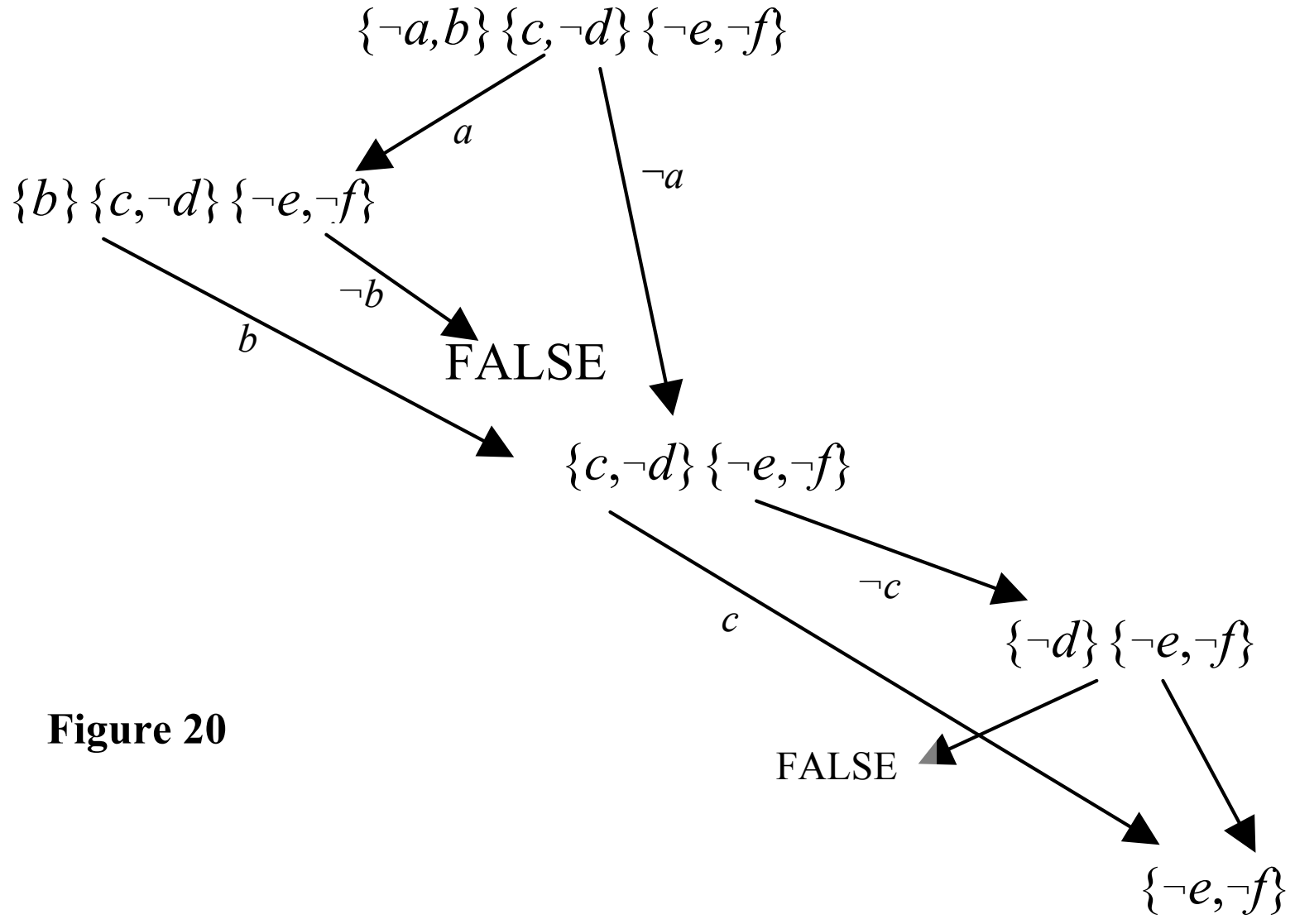

**Figure 20**

**Corollary 1:** Suppose $C_i$, $C_j$ are clauses of a s.o. Set S of a 3-SAT-CNF-problem, where $i<j$, then some literals of $C_i$ must appear in branches of the SRT containing $C_j$ and get instantiated before literals of $C_j$..

**Proof:** Two evident properties of GSPRA should be emphasized first:

1. Any clause C or its derivation C' is resolved against all branches of the IRT constructed in intermediate steps (c.f. Definition 2 of GSPRA).
2. If C or C' disappear from Clause-Sets of resolution-tree nodes, this is because some of the literals got instantiated either with TRUE or FALSE (according to signs of those literals) and made the overall truth value of C or C' =TRUE.

Knowing that, there are following possibilities for any Clause-Set S' of any node in the resolution-tree of S with respect to $C_i$, $C_j$: **Either** $C_i$, $C_j$ both appear in S' in which case (as per Lemma 1) they are already sequentially-ordered, i.e., $C_i$ literals appear before $C_j$ ones and accordingly get instantiated in the same sequence in branches, **or** only $C_j$ appears in which case (because of properties 1,2 above) the only reason for the disappearance of $C_i$ would be that some of its literals got instantiated with TRUE or FALSE before S' **or** only $C_i$ appears meaning (also because of properties 1,2 above) that a literal of $C_j$ got instantiated before any literal in $C_i$ making the overall value of $C_j$ TRUE which cannot happen, because of the least-literal-rule and the fact that all Clause-Sets are l.o.[72] **or** $C_i$, $C_j$ both don't appear in S', but in parent-nodes in the SRT. In that case the same argumentation as in the previous three cases applies.

(Q.E.D.)

**Illustrations of Corollary 1 (Fig. 21):** Base Clause-Sets of the form {… {*a*,X} ..{*b*,Y} ..} (for +*ve*, monotone 2-SAT) where *a*<*b* can have SRTs of the form:

**Figure 21**

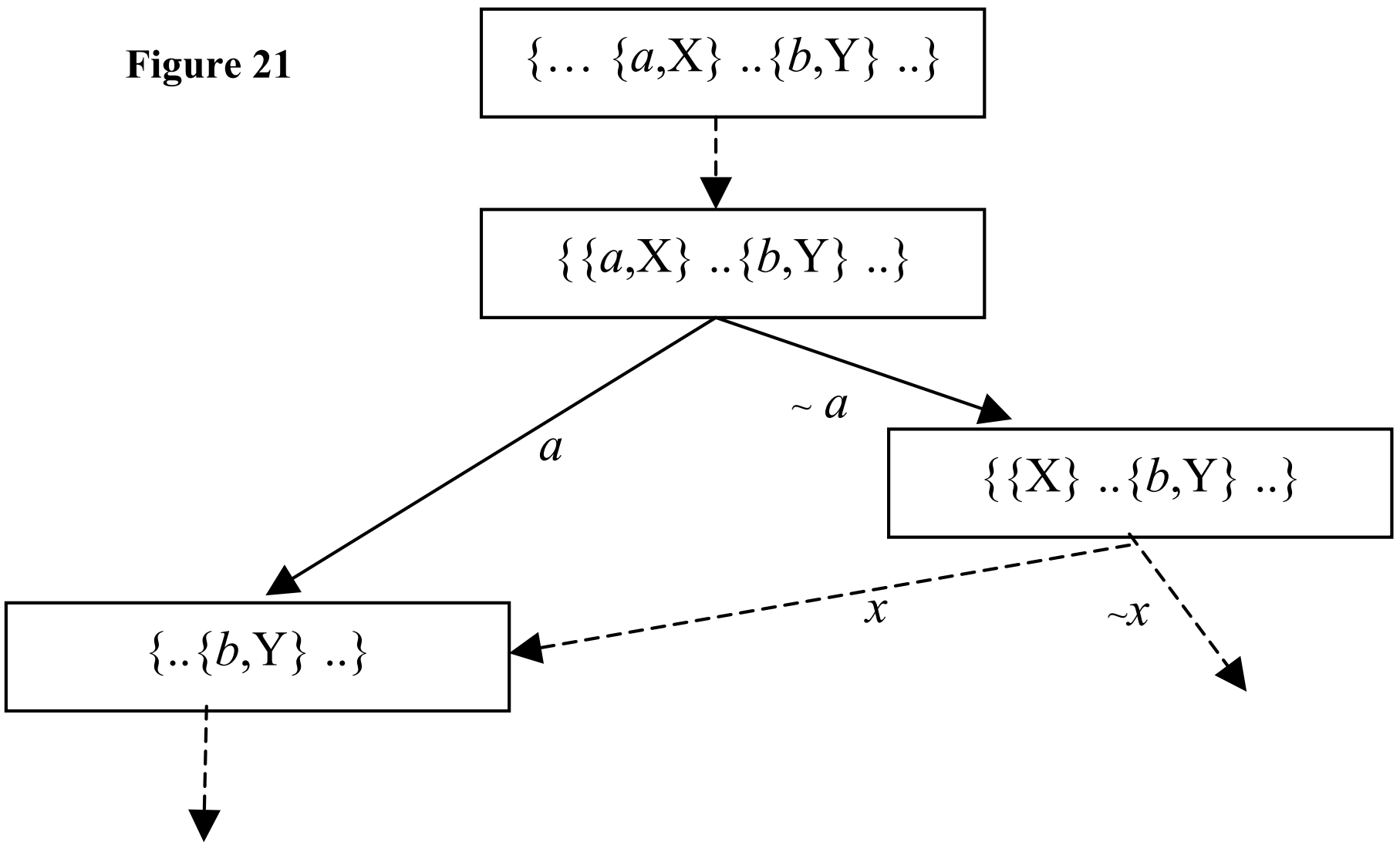

While the same tree for the ordinary 2-SAT case may look like Fig. 22 for a sample permutation:

**Figure 22**

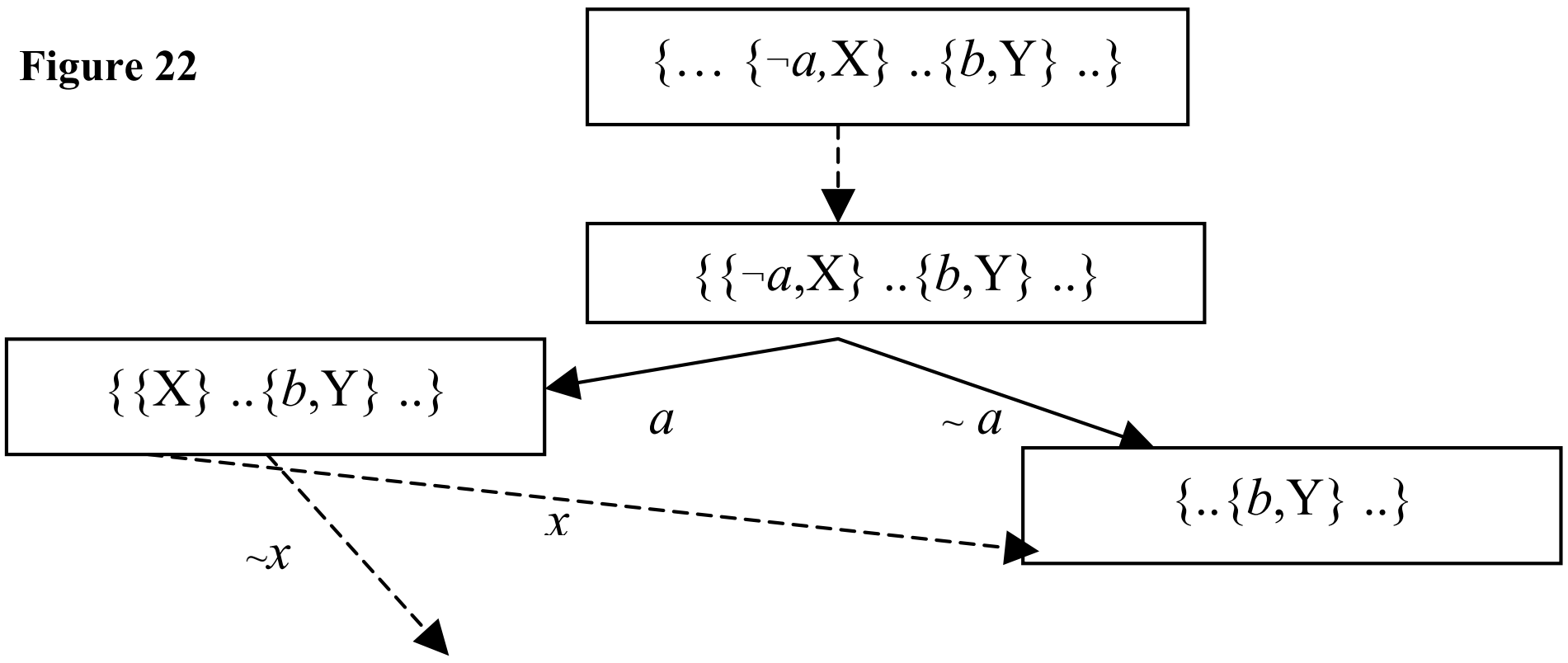

[72] Remember that as per l.o. condition for any parent-node of S': $\forall i,j$ indices of clauses: if $i<j$ then head-literal of $C_j$ >= head-literal of $C_i$ which means (because of the least-literal-rule) that no head-literal of a $C_j$ can be instantiated before a head-literal of a $C_i$.

Note that all branches in both figures (Fig. 21&22) containing {*b*,Y} also contain instantiations of ***a*** only or **X** only or both.

**Lemma 2:** Suppose S is an s.o. Set of a 3-SAT-CNF-problem. The following are properties of its SRT:

a) For any branch *b*: If *i,j* are edges corresponding to literal instantiations in *b*, $i<j$ then *i* appears before *j* in that branch.
b) For all common-nodes [*q*] and nodes [X],[Y],[Z], etc. (c.f. Fig. 2 in Definition 5), NL $q>$ NLs *x,y,z*, etc.
c) HCNs as well as TCNs exist even for 2-SAT.
d) Clauses belonging to the same block $B_x$, *x* block literal, can only be scattered between mutually exclusive branches if $B_x$ is at least partially embedded[73] in another parent block.
e) A CN [*q*] formed within a block $B_x$ through +*ve* as well as -*ve* edge- or branch-literals *x* is called: Double-Sided CN from the perspective of *x*, $DSCN_x$. Such an *x* is called distinguished literal for [*q*]. A CN [*q*] formed within a block $B_x$ through only +*ve* or only -*ve* edge- or branch-literals *x* is called: Single-Sided CN from the perspective of *x*, $SSCN_x$, *x* is called non-distinguished literal for [*q*]. If for a CN [q] there is no distinguished literal x such that the CN is $DSCN_x$, then [q] is called simply SSCN. If a non-distinguished literal *x* for a CN [*q*] formed in steps $<k$ is used to augment the size of [*q*] in step *k*, i.e., *x* is instantiated in a clause added to the clauses of [*q*] in *k*, then *x* is called: CN-Augmenting Literal (CNAL) for [*q*]. For all CNs [q] in an SRT it is true that:
   i. $DSCN_x$-nodes which are not $SSCN_y$ for any *y* can only be augmented in size when supported through parents in block $B_x$
   ii. SSCNs may split[74], but only before augmented to sizes>1.
   iii. If *x* is a CNAL for [*q*] in step *k* then it cannot be used to split it in any further step $>k$.
   iv. If the size of [*q*] is augmented using any CNAL in step *k* to become >1, then it cannot be split in steps $>k$.
f) For tCNs [q] in an SRT: [q] can be augmented to sizes >1 using a distinguished literal x within a symmetric block $B_x$. It may :
   1- Never split after being formed if $B_x$ remains symmetrical **or**
   2- Avoided altogether when relaxing the l.o. condition for symmetric or dissymmetric blocks to a special l.o.u. one called l.o.s. [75]
   3- Imposing l.o.s. or l.o. conditions on symmetric or dissymmetric blocks generates the same SRT.

[73] Defined in the proof below.

[74] We are using here the loosely defined formulation of a "split" discussed in the introduction, the exact formulation being subject of Section C.

[75] Definitions of symmetric and dissymmetric blocks as well as the l.o.s. condition is shown in the proof below.

**Proof:**

a) Suppose $i<j$: Because of the least-literal-rule the only way literal ***i*** can be instantiated after ***j*** in a branch $b$ would be either that ***j*** comes before ***i*** in the same clause thus breaching Condition a) in Definition 1 **or** clauses containing ***i*** are resolved after clauses containing ***j*** in $b$ (as per Definition 2 of GSPRA) which means ***i*** is a new variable from the perspective of clauses containing ***j***. But in that latter case ***i*** should have been **>*j*** as per Condition 1 c).

b) This follows immediately from a) (recall that node $[q]$ has two edges coming out from it marked $q$ and $\neg q$ where those edges come after $x,y,z,..$ edges).

c) If a node is HCN as per Definition 5, then Fig. 23 shows this for the monotone case in the following example (where $a$=0, $b$=2, [N]={3,4}):

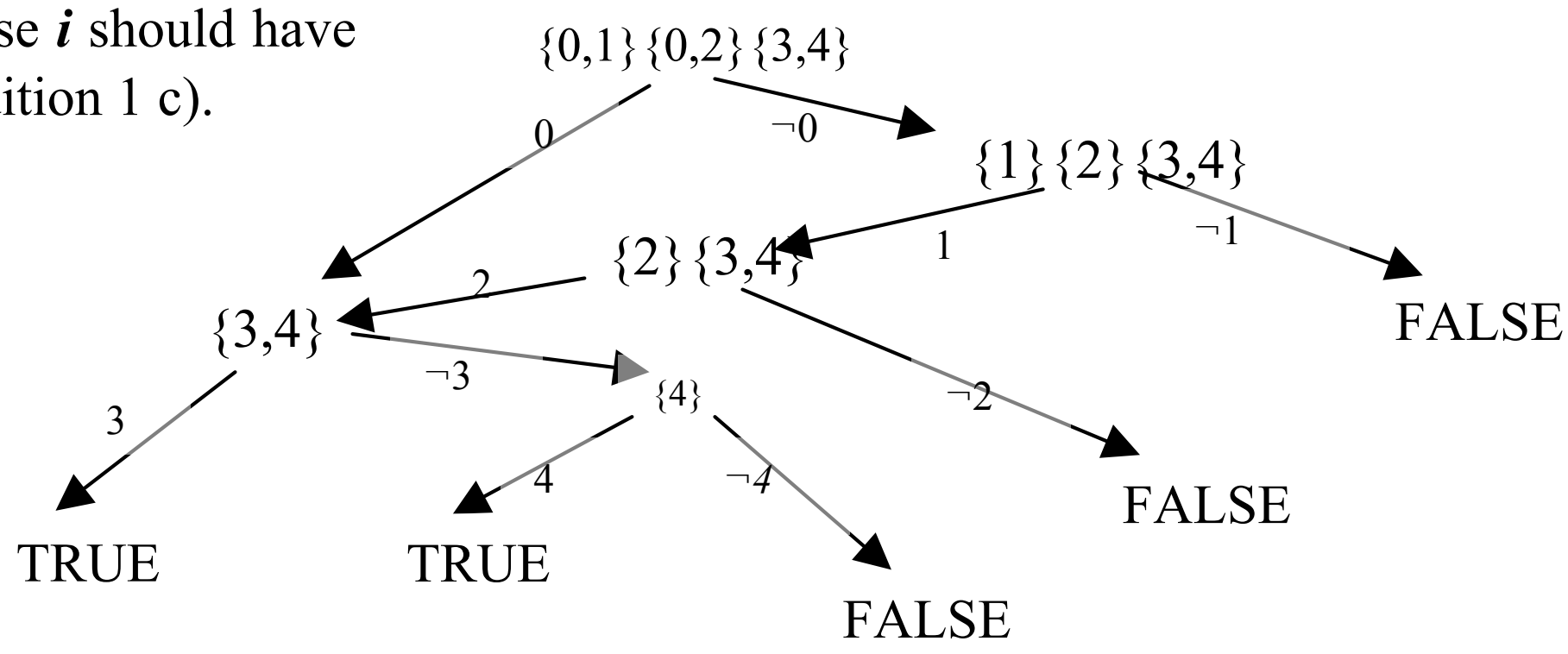

**Figure 23:** HCN-node

The below tree in Fig. 24 demonstrates the existence of TCNs in ordinary 2-SAT-SRTs which are s.o. (lo.o. Sets are similar).

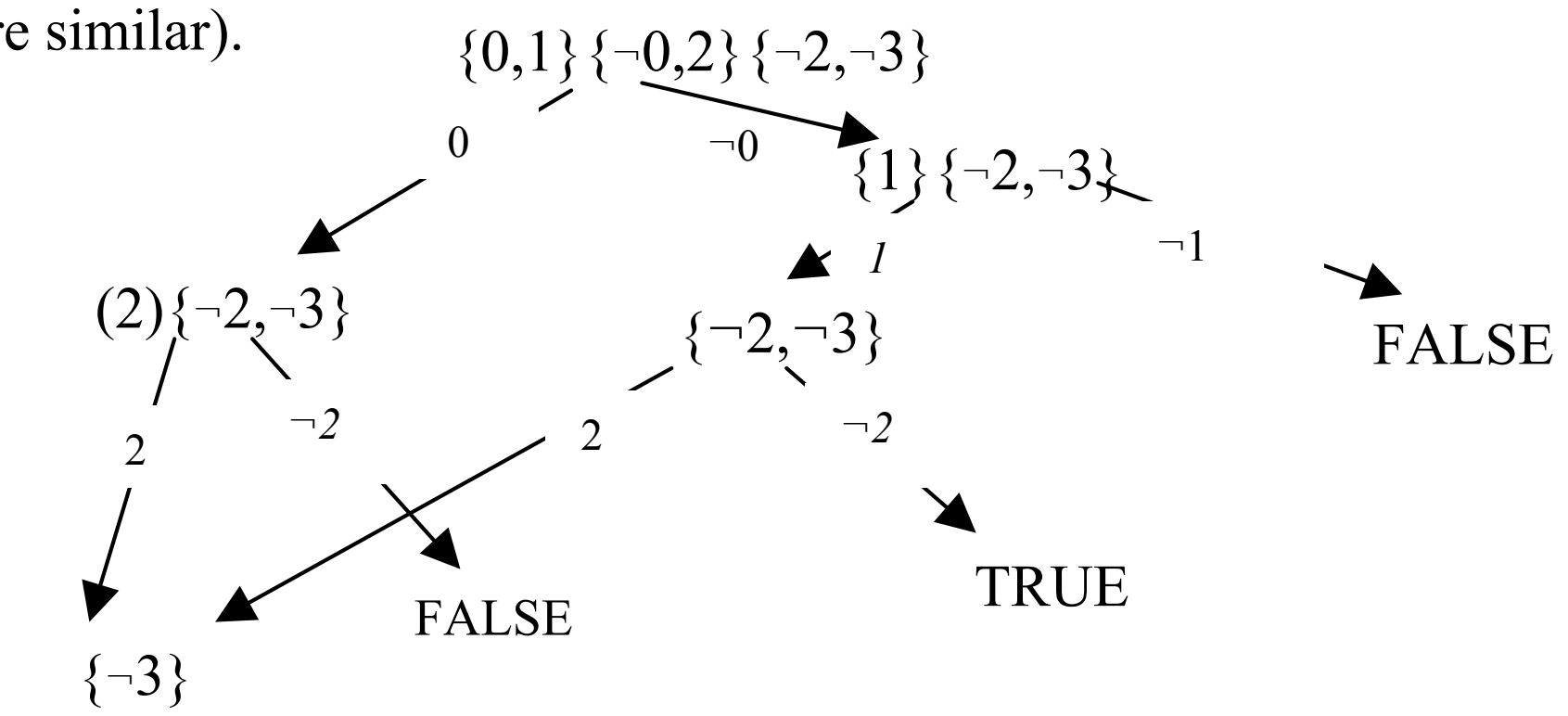

**Figure 24:** TCN-node

d) Suppose S={{$a,\neg x,\neg y$}{$b,\neg x,y$}{$c,x,\neg y$}{$x,y,z$}}, $a<b<c<x<y<z$, which is obviously l.o. The question is: Can we scatter clauses of the block $B_x$={{$\neg x,\neg y$}{$\neg x,y$}{$x,\neg y$}{$x,y,z$}} between mutually exclusive branches of an SRT so that it looks for example like:

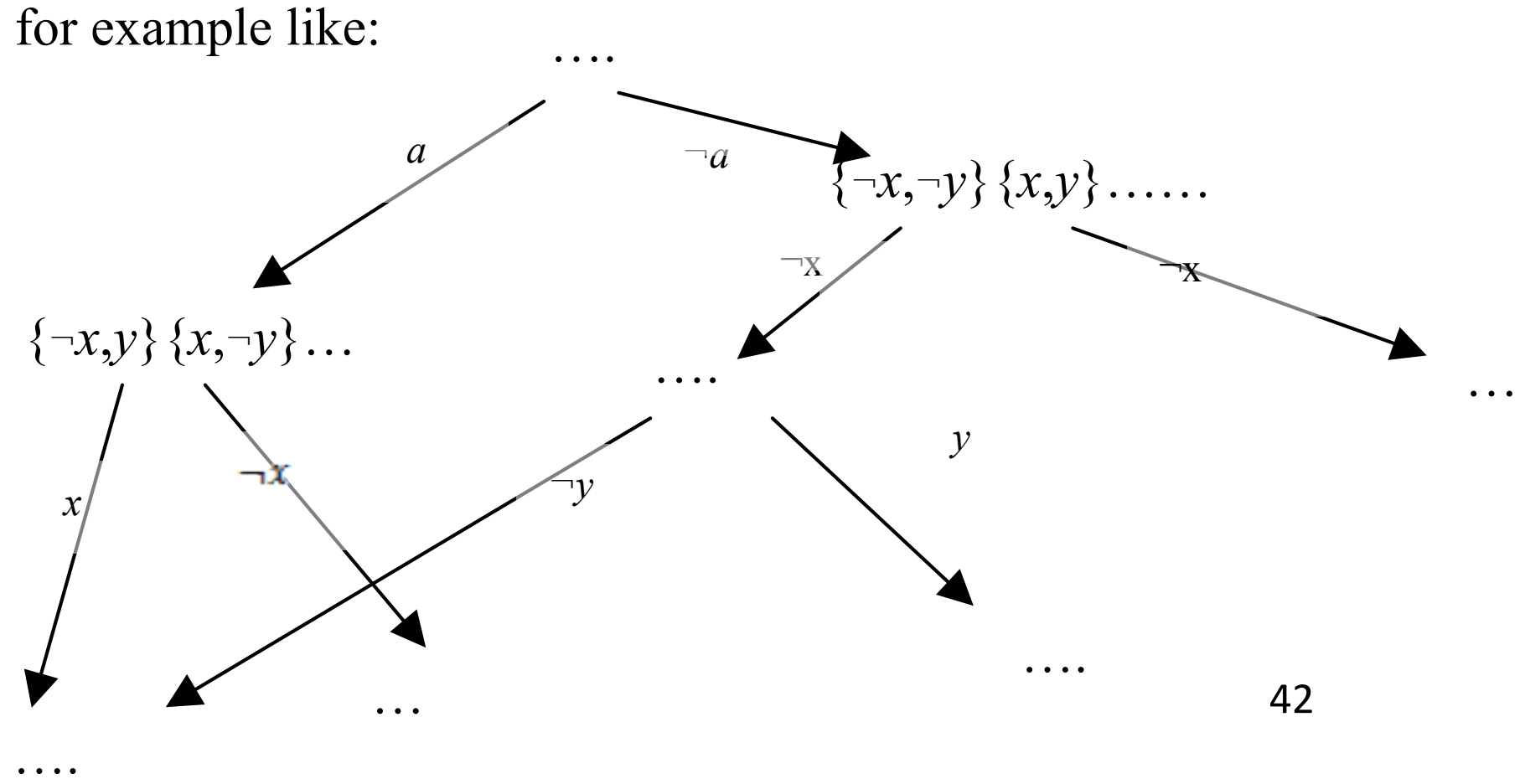

**Figure 25**

where different members of $B_x$ can be found in different nodes mutually excluding each other? The answer is that this is only possible if $a$=$b$=$c$, i.e., if S={{¬$a$,¬$x$,¬$y$}{$a$,¬$x$,$y$}{¬$a$,$x$,¬$y$}{$x$,$y$, $z$}} for example. The obvious reason for that being the fact that exclusivity of branches in any SRT relates to different instantiations of one and the same literal. To be able to disperse at least some members of $B_x$, a common literal in a parent-set needs to be instantiated in two mutually exclusive ways. Note that the clause {$x$, $y$, $z$} is included in all branches, since its leading literal is the block literal of $B_x$ (not $B_a$) and $x$>$a$. $B_x$ of the form seen here is called partially embedded.

e) Suppose a $DSCN_x$ [$q$] which is not an $SSCN_y$ for any $y$ is formed within a block $B_x$ through instantiations of variables such that two branches of the SRT contain edges marked with $x$ (¬$x$ respectively) connect to the CN as in Fig. 26 (left):

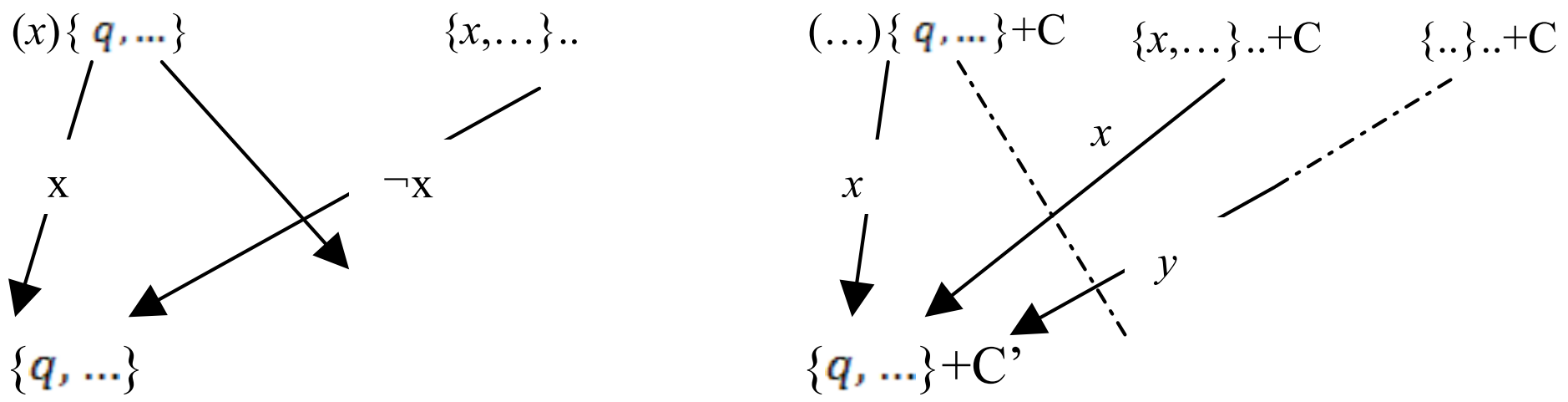

**Figure 26:** $DSCN_x$ [$q$] formed within block $B_x$. SSCN augmenting.

Obviously, any clause C attempting to augment the size of [$q$] cannot use for this the distinguished literal $x$, because otherwise a split would occur as two different derivations of C must result from any instantiation efforts. Thus, the only way to increase the size of [$q$] in that case is by adding clauses from a block $B_y$ different from $B_x$ to the parents of [$q$] before propagating them down to [$q$], $y$>$x$. This is precisely what is done when [$q$] is supported (c.f. Definition 5, Section A).

On the other hand: Suppose [$q$] is an SSCN ($SSCN_x$ and $SSCN_y$ as well Fig. 26 right where $x$<$y$). The question is: Can this node split after its formation and before a clause C attempts to augment it? This is theoretically possible if a clause containing literal y is used. Clauses containing literal $x$ may only cause such a split if $x$ doesn't occur before y in that branch. Now suppose its size is to be augmented to become >1 using clause C, then: Only one derivation of C (in Fig. 26: C') can be propagated down to [$q$] from all possible directions. C' must be the result of instantiating none, one, or more than one literals of C in the same way throughout all branches. Obviously $y$ in Fig. 26 cannot be CNAL, since it is not instantiated in any other branches containing $x$ and can thus cause a split as mentioned before. This leaves $x$ as the only possible CNAL in the constellation illustrated in Fig. 26. An edge marked $x$ must exist in the branch containing $y$ prior to edge $y$ (in the dashed lines region) otherwise C' cannot be the same for all edges. After [$q$] is augmented in size with C' in step $k$ (through CNAL $x$): Any clause CC attempting in a further step to split [$q$]+C' through instantiations of $x$ will have no effect on this CN, since all branches contain an edge marked $x$. Meaning: If CC uses a +*ve* literal $x$, it's value shall become TRUE and neither CC nor any of its derivations will be propagated down to [$q$]+C'. If on the other hand, CC uses a -*ve* literal $x$, one and the same derivation CC' shall be propagated through all branches and the size of [$q$]+C' is increased again. Thus, CC cannot split [$q$]+C' using CNAL $x$ in any step $>k$. What about $y$? Since [$q$] was augmented, C must have been free of $y$. Now if CC contains $y$, this would mean a parent Clause Set of the form {..$y$..}{<no $y$>}{..$y$..} in some node which is not l.o. [76]

Putting all this together: GSPRA's generic way of augmenting the size of a CN [$q$] in step $k$ using a clause C is depicted in following Fig. 27 where $a<b<x$. Assuming [$q$] is augmented in size through CNALs (otherwise it can only be augmented in size when supported as seen above or using distinguished literals before a tCN is created as seen in the next point). This means C belongs to either $B_x$, $B_a$, or $B_b$ (head-literal of C and potential CNAL being: $x$ or $a$ or $b$ respectively). Instantiation of one of those literals must result in one single derivation C'' propagated through all branches increasing the size of [$q$] as before. Obviously, literal $x$ cannot be CNAL as just demonstrated, Literal $b$ cannot be used either since $a$ is an edge-literal of [$q$] and hence $b$ is not being instantiated at all on that branch contradicting what is happening on the branch containing $N_4$+C'', i.e., those two branches cannot produce the same derivation if $a<b$. This leaves literal ***a*** only as CNAL. For ***a*** to be CNAL the newly added clause C must be equal to $\{\neg a\} \cup$C'' where literals $b$, $x$ are not $\in$ C''.[77] [$q$] must be $SSCN_a$ and all dashed lines reaching from the Base-Node to $N_2$, $N_3$,$N_4$ must have edges marked with ***a*** similar to the branch of $N_1$. Otherwise the same derivation C'' cannot be propagated from those directions. Let us now try to split the augmented node {$q$}+C''. Any new clauses from $B_x$ or $B_b$ attempting this will breach the l.o. condition in nodes $N_2$,$N_3$ and $N_4$, since they already contain C'' which is free from both $b$,$x$.

A new clause from $B_a$ cannot do this, since ***a*** is CNAL. Thus [$q$]+C'' cannot be split in any step $>k$.

76 Clause-Sets of the form: {... {$y$,...},{<no literal $y$>},{.,$y$,..} ...} are not l.o. for any $y$, because in l.o. Sets heads of clauses (at least) must be sorted in ascending order. This form means that the head of {<no literal $y$>} can only be $>y$ contradicting the occurrence of $y$ again in the last clause in any position.

77 C cannot be {$a$} $\cup$C'', because C would then get the value TRUE when instantiated through edge a and no derivation would be passed down to [$q$] from that direction. Also: C'' has to be the same derivation propagated from all branches to [$q$]. Therefore, it cannot contain neither $b$ nor $x$ whose existence would cause different derivations through $N_1$,$N_4$ and $N_2$,$N_3$ respectively.

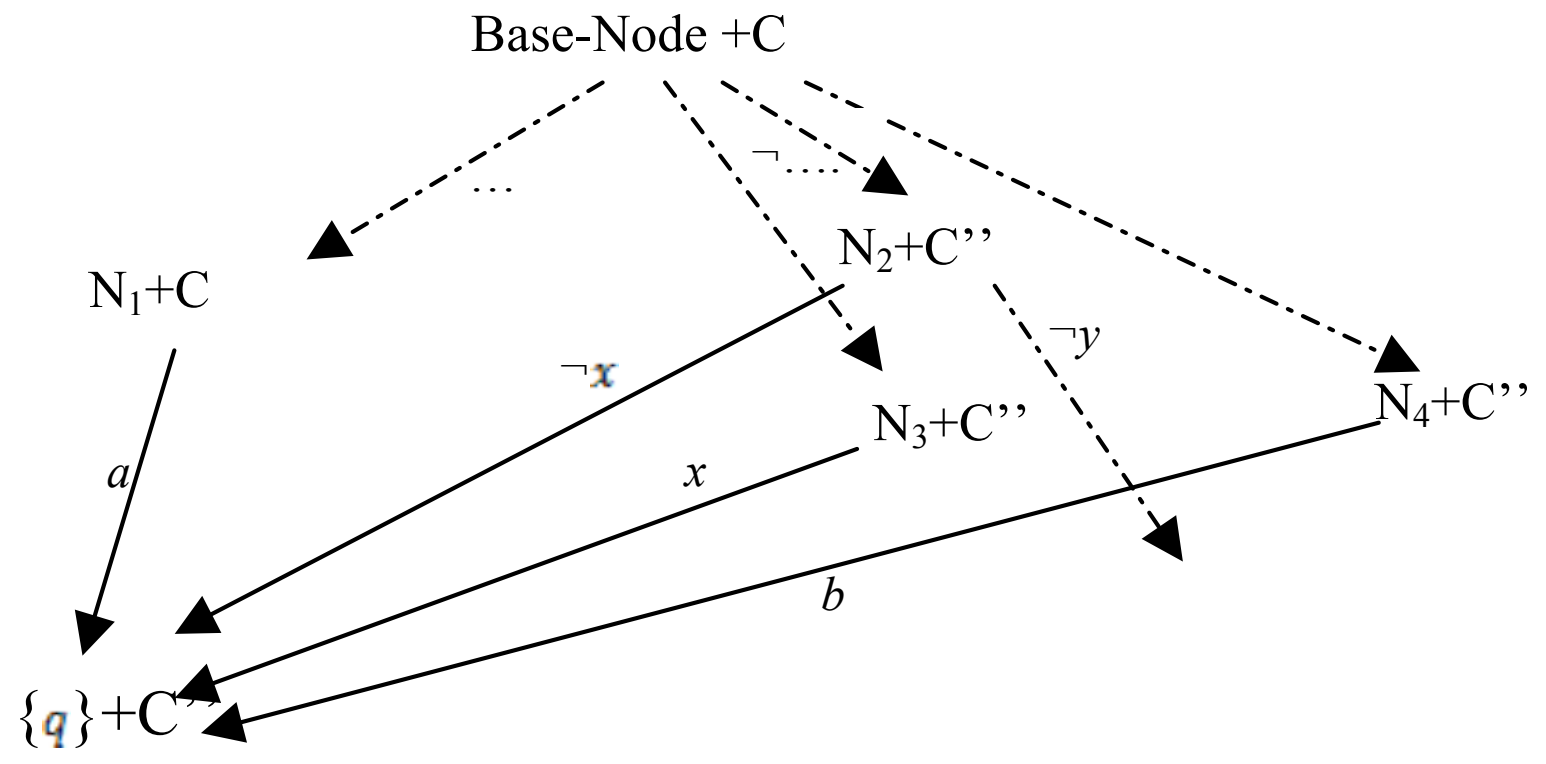

**Figure 27: Generic way of augmenting a CN**

Here is yet another argument showing the same: Suppose a CN [*q*] is augmented using any CNAL ***a*** in step *k*, then, ***a*** cannot be used to split [*q*] in any step >*k* as seen above. It is then sufficient to show that [*q*] cannot be split in steps >*k* neither using any distinguished literal ***x*** nor any non-distinguished literal ***y*** which is not a CNAL. For a distinguished ***x***: [*q*] could have only been augmented in size in *k* through parents who already have clauses containing ***a*** or its negation from a block $B_y$, *y*>*x* otherwise a split would have occurred. Any attempt, then, to split [*q*] in steps >*k* using clauses containing ***x*** assumes therefore parent-nodes containing Clause-Sets of the form: $B_xB_y${..*x*..} which are not l.o.

For a non-distinguished literal ***y*** which is not CNAL: [*q*] could not have been augmented in step *k* with any clause containing ***y*** or its negation (otherwise it would have been a CNAL similar to ***a*** if all branches agree on its instantiation or a split would have occurred if they disagree). If this is the case then any attempt to spilt [*q*] using ***y*** in steps >*k* must assume some parent Clause-Sets of the form: $B_y$<no *y*>{..*y*..} (c.f. footnote 76) which are not l.o. either.

f - For tCNs: A block $B_x$ is called Symmetric Block (SB) if *–ve* and/or *+ve* instantiations of block literal *x* result in the same Clause Set. It is called Dissymmetric Block (DB) if *–ve* and/or *+ve* instantiations of block literal *x* result in Sets $S_1$, $S_2$ respectively and either $S_1 \subseteq S_2$ or $S_2 \subseteq S_1$. A tCN is obviously per Definition (see linear expansion Property 2 above) created in an SB, since two nodes can only be merged into one if their respective Clause Sets are equivalent with respect to a given instantiation. The following example illustrates a case where such a tCN is formed and then split, because $B_a$ became dissymmetrical (Fig. 28).

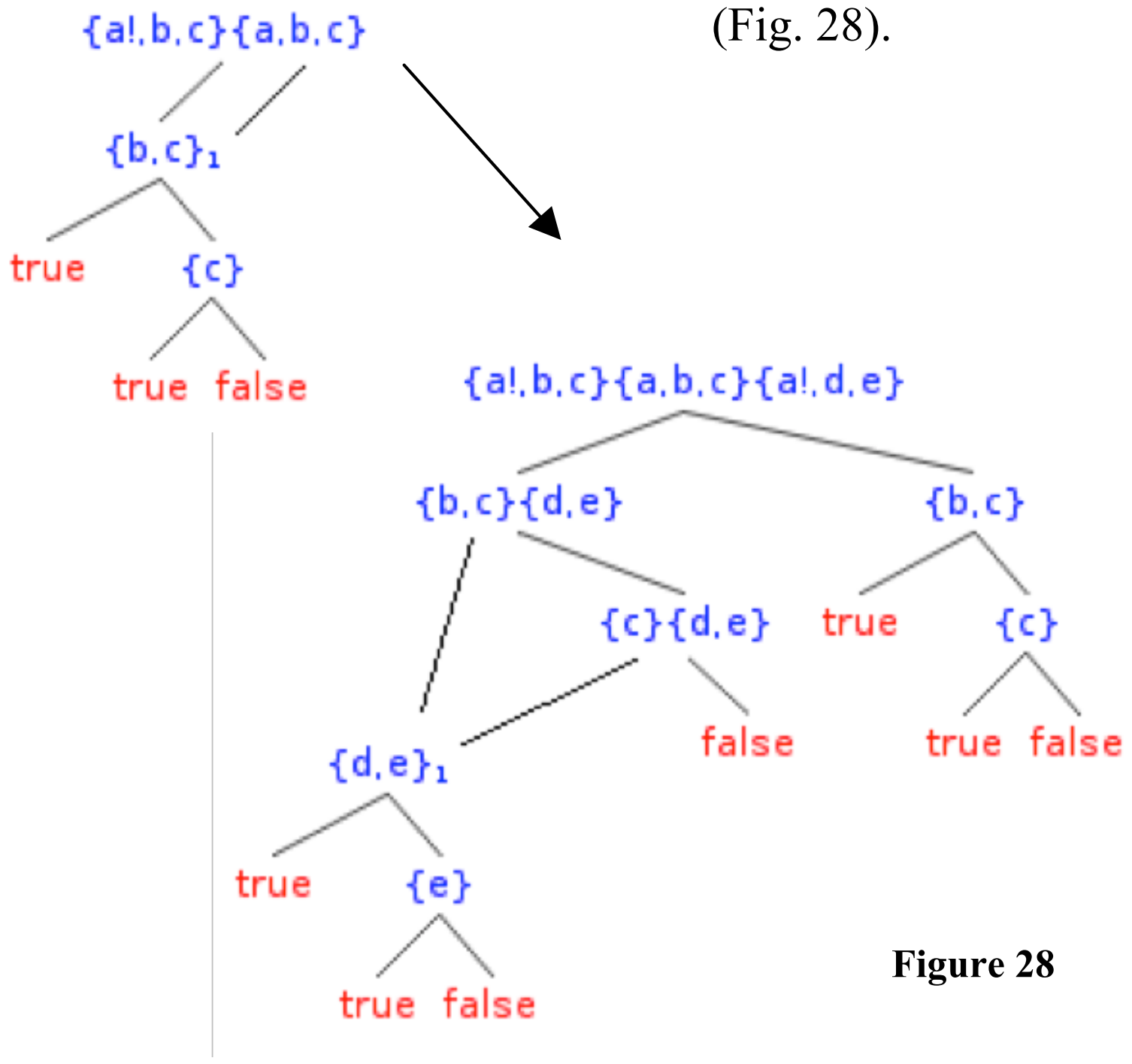

**Figure 28**

Note that this split could only occur because literal *a* was distinguished for tCN {*b*,*c*}. If clauses are sorted such that the Base-Set becomes {¬*a*,*b*,*c*} {¬*a*,*d*,*e*} {*a*,*b*,*c*}, this situation is prevented like in Fig. 29:

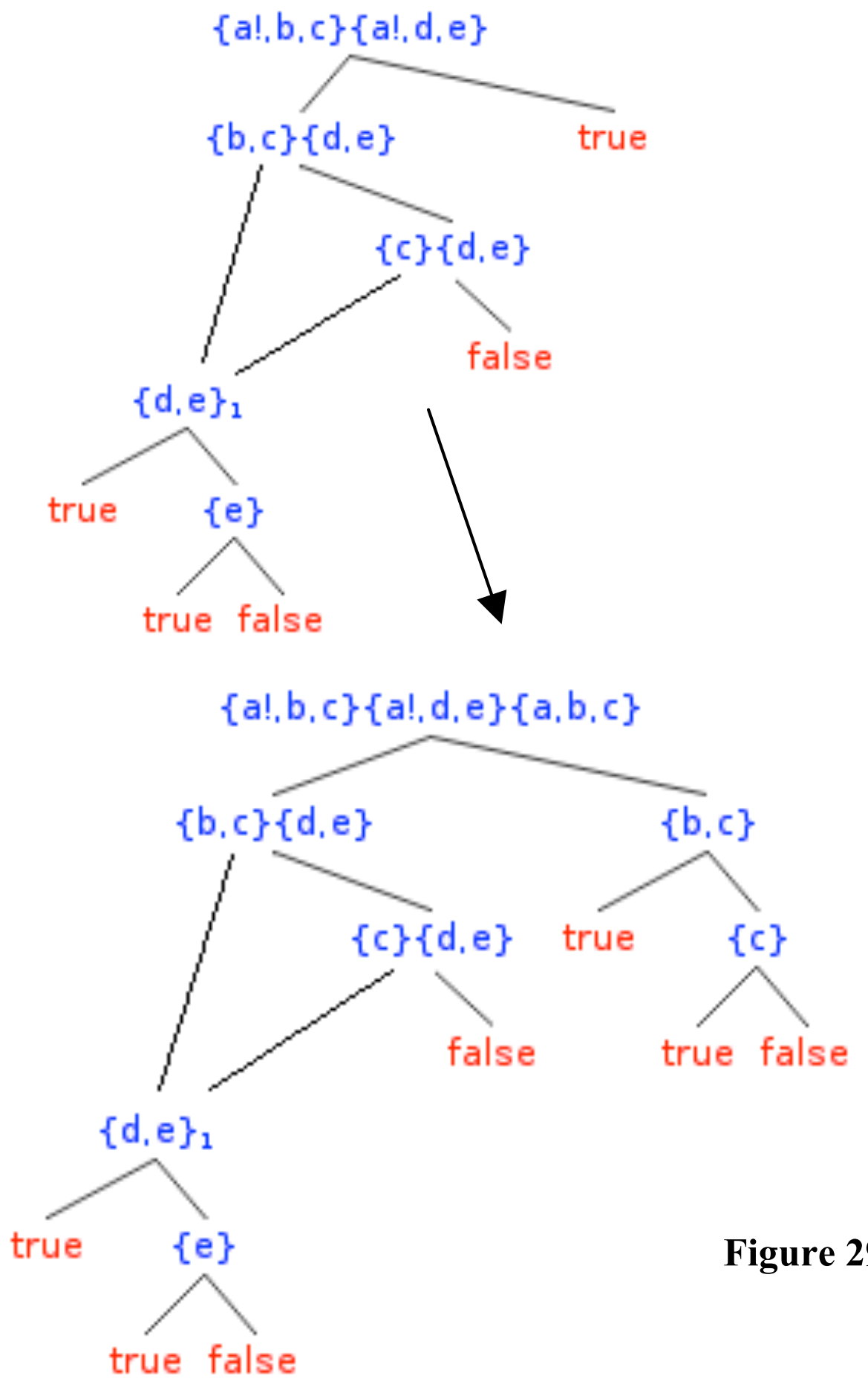

**Figure 29**

Note also that if the Clause Set is sorted the other way round: {{*a*,*b*,*c*}{¬*a*,*b*,*c*}{¬*a*,*d*,*e*}} the split still occurs, i.e. to prevent this type of splits altogether, sorting must take into account the number of clauses containing *–ve* and *+ve* instances of the leading literal and prioritize the instantiations with the most clauses. We call this additional condition on sorting: l.o.s (linearly ordered, but stretched). Is imposing l.o.s. conditions on SBs and DBs really necessary, i.e., how expensive is it to allow this type of trivial splits to occur? Obviously, {*b*,*c*} is copied once. This becomes only expensive if more clauses are added to the tCN before a split occurs. One such situation can happen with the Clause Set: S={{¬*a*,*b*,*c*}{¬*a*,*d*,*e*}{*a*,*b*,*c*}{*a*,*d*,*e*}}. Here the tCN is {{b,c}{d,e}} and of size 2. In that case tCN {*b*,*c*} is augmented in size using the distinguished literal ***a*** before the tCN is formed. Any further clause of the form: {*a*,..} or {¬*a*,..} causes therefore a split of a node of size 2 which, if admitted in its general form, may create a bigger number of new nodes in any step[78]. It seems at first sight that imposing l.o.s. on SBs and/or DBs is a must. Such an undertaking comes with the additional complication that sometimes l.o. and l.o.s. conditions are contradictory, i.e. some Clause Sets may be either l.o. or l.o.s, but not both[79]. Fortunately tCNs are only formed and split in SBs and DBs respectively. This means that applying the least literal rule to the whole block will result in both cases in the same Clause Sets in children nodes whether the block is l.o.s. or l.o., since the difference between those two conditions lies merely in the positions taken by negative and/or positive occurrences of the least literal in respective clauses. This difference has no effect on the overall instantiation effort of the least literal in the base Clause Set. Such a property makes SRTs - resulting from imposing any of those two conditions on the whole block - equivalent [80]. Counting the number of

[78] In this paper the effect of allowing splits of CNs of rank <3 and sizes >1 is not investigated except for the tCN case on hand here.

[79] Such a case is S={{*a*,*b*,*c*}{¬*a*,*b*,*d*}{*a*,*c*,*e*}} as the reader may wish to verify.

[80] May be formally shown using induction on the length of blocks, realizing that top-parts of SRTs applying any one of those conditions are always equivalent. This is not done here to avoid unnecessary length.

unique nodes resulting from imposing l.o. conditions being subject of section E, it remains here to ask: What if SBs or DBs are scattered between different nodes? How can this influence the equivalence between l.o. and l.o.s. conditions for SBs and DBs ? As seen in point d) of this lemma: Blocks can only be scattered when they are at least partially embedded in other blocks. This means that scattered SBs or DBs always have common base nodes like in the below constellation (Fig. 30):

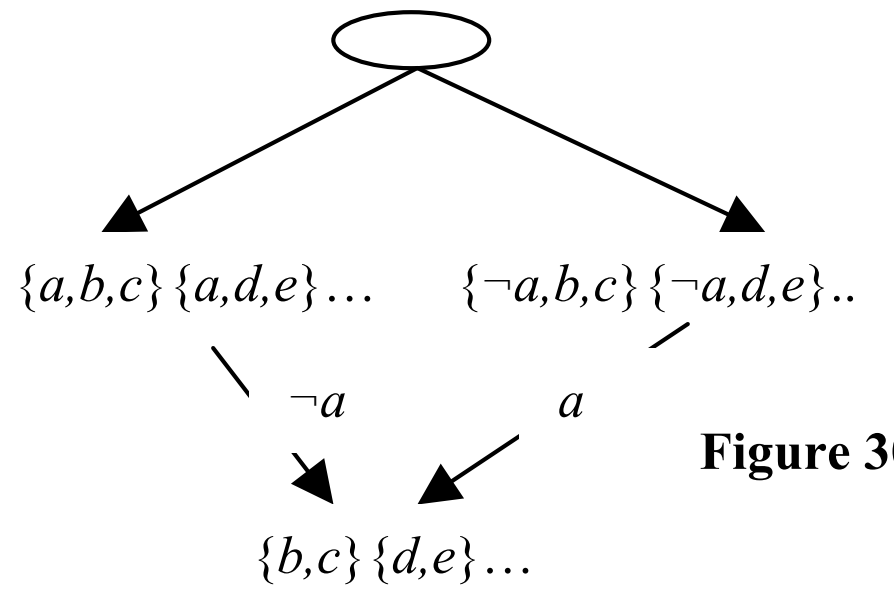

**Figure 30**

As long as all nodes are guaranteed to be l.o. in such an SRT, no node shall be containing clauses of the form: $\{\{x,..\}\{y,..\}\{x,..\}\}$ for some $x$ and $y$. This means that scattered fragments of block $B_a$ have, with respect to the tCN, always priority on other blocks within the same node as seen in the figure. Here again: As the difference between l.o. and l.o.s. conditions within those fragments is related only to positions and priorities of clauses containing the signed/unsigned least literal ***a***, the order of clauses within such fragments and hence within the overall, scattered $B_a$ is irrelevant for any instantiation effort done using the least literal rule (here applied on ***a***) either during the formation of this tCN or during its split. L.o.s and l.o. blocks produce therefore even when they are scattered the same SRTs.
(Q.E.D.)

In furtherance we show detailed sample cases of size augmentation trials within parent-instantiation blocks for the purpose of clarification of findings of Lemma 2:

**Case 1:** All parent-sets belong to the same block (in the example above of Fig. 28). For tCN $\{b,c\}$ try to increase its size by adding clauses to $S=\{\{\neg a,b,c\}\{a,b,c\}\}$. If we add a clause of the same block $B_a$ such as $\{\neg a,\neg b,c\}$,$\{a,\neg b,c\}$,$\{\neg a,d,e\}$ or $\{a,d,e\}$ a situation like the above occurs and $\{b,c\}$ is always split[81]. In all those cases $B_a$ becomes dissymmetrical.
In the previous example, S only had one parent-node containing the block $B_a$ of the tCN $\{b,c\}$. Here, a situation where such a tCN has parents from different nodes in different branches although one and the same block ($B_b$) is being instantiated by direct-parents (Fig. 31). A split is occurring for tCN $\{c\}$.

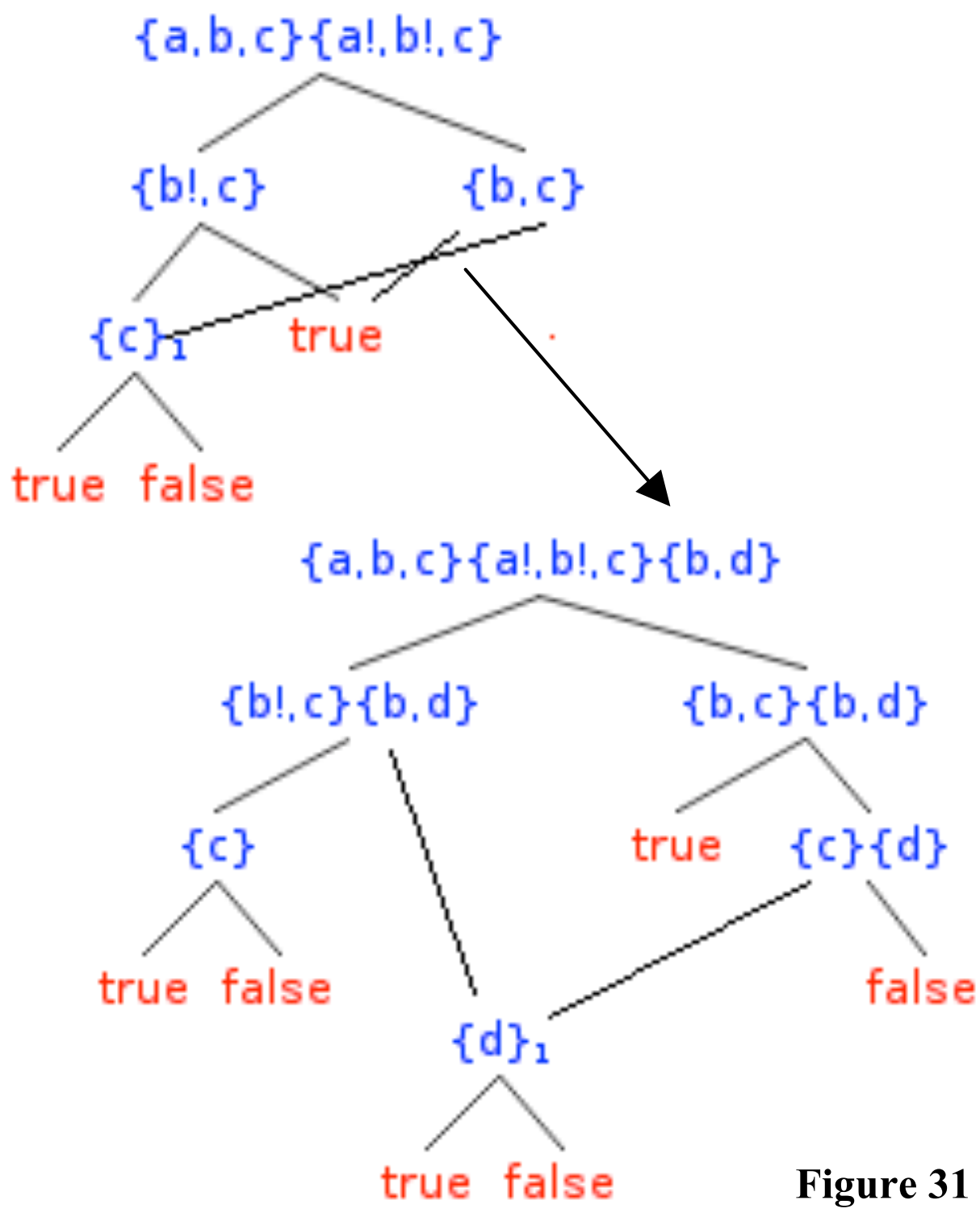

**Figure 31**

[81] The reader is encouraged to check this by him/herself.

Can we make the size of $\{c\}$ bigger than 1 in any step before a split occurs afterwards? The answer is: No! If we use clauses of $B_a$ like $\{a,b,d\}$ or $\{\neg a,c,d\}$ then $\{c\}$ is split, because $B_b$ loses its symmetry. If we take clauses like $\{b,c,d\}$, $\{\neg b,\neg c,d\}$ or $\{\neg b,c,d\}$ from block $B_b$ similarly happens. For all clauses from $B_x$, $x>b$, like $\{c,d,e\}$ or $\{\neg c,d,e\}$, this tCN will be supported and cannot split in any further step.

**Case 2:** Parent-sets belong to different blocks. One interesting constellation is seen in below Fig. 32 where block $B_y$ has only one edge (either +*ve* or -*ve*) going to the tCN[82].

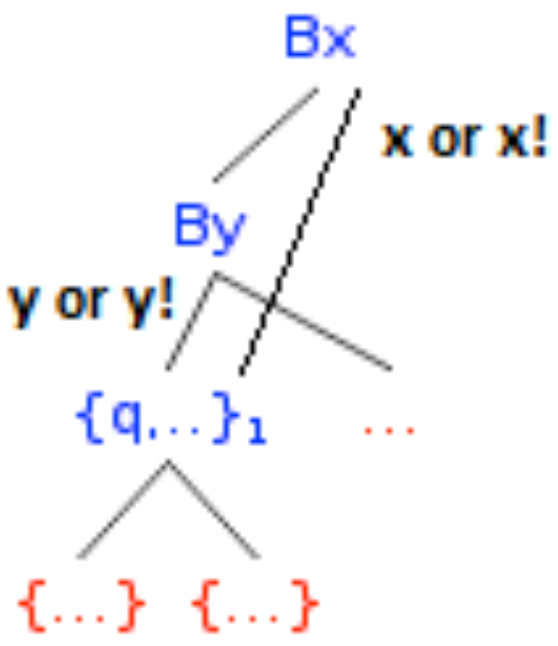

**Figure 32**

Let us try to increase the size of $[q]$ through clauses from block $B_y$, i.e., attempting to use $y$ as CNAL. A clause C attempting this will have two derivations, one in which $y$ is not instantiated and one in which it is. This makes $[q]$ split. Thus, for this case as well, $[q]$ cannot be increased in size before a split occurs. Another practical example : Set S=$\{\{a,\neg b,d\}\{b,c,d\}\}$. The following tree (Fig. 33) contains two tCNs $\{c,d\}$ and $\{d\}$:

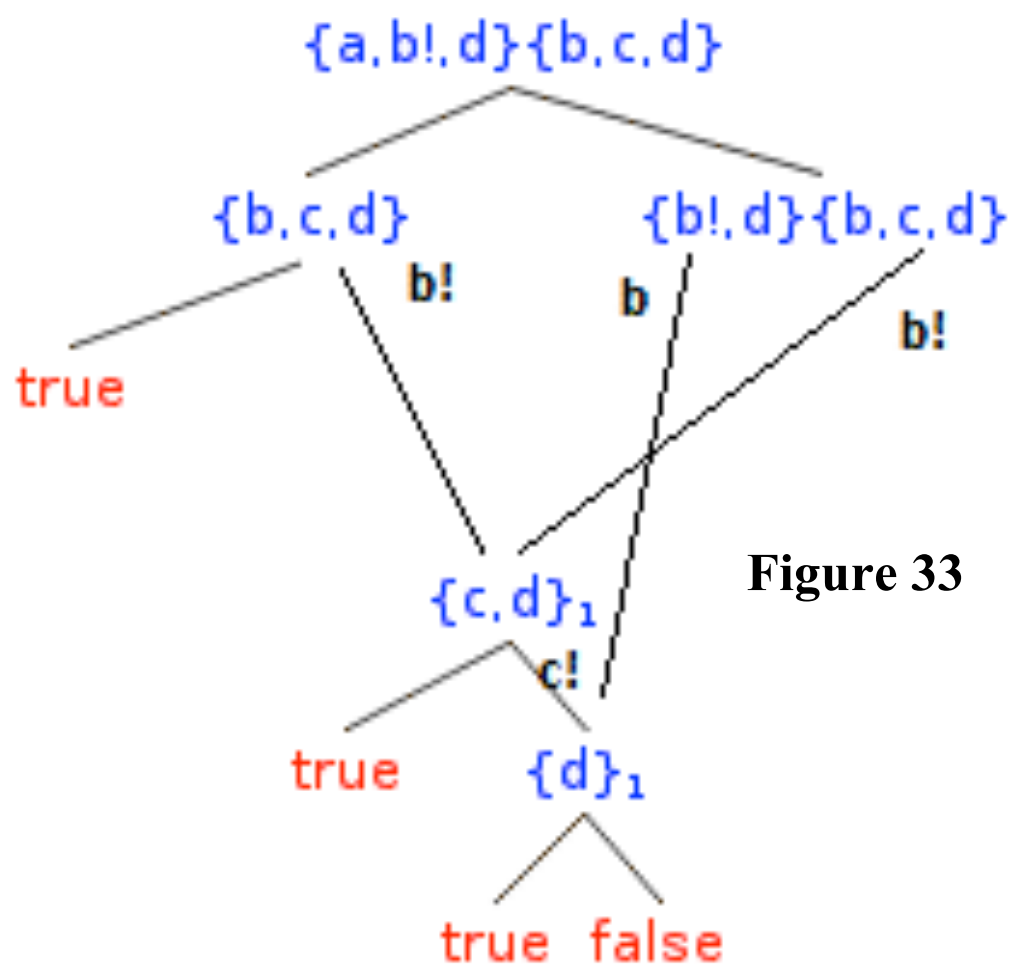

**Figure 33**

Now let's try to increase the size of tCN $\{c,d\}$ with any clause from block $B_b$ or $B_c$ with an example (Fig. 34) for a clause from $B_b$: $\{b,e\}$[83]

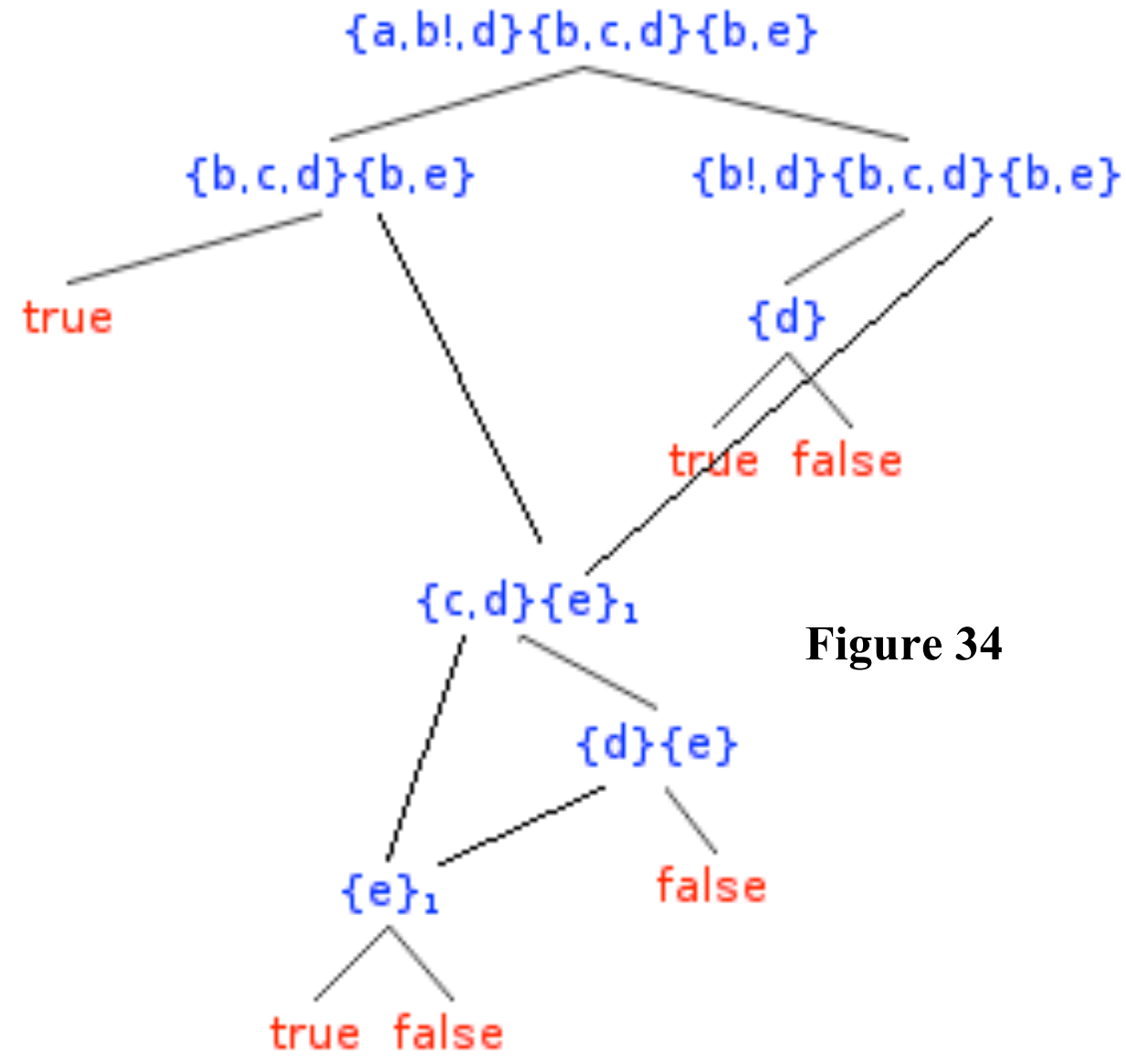

**Figure 34**

Now we try to split $\{\{c,d\}\{e\}\}$. This cannot be done using any $B_a$-clause (because of the l.o. condition) or $B_b$-clause, because $b$ is CNAL. What about the tCN $\{d\}$? If we attempt to augment its size using clauses from $B_b$ like $\{b,e\}$ (the one we used above) it will be split as just seen before. Let us try $\{c,e\}$ from $B_c$. In that case we have the following picture (Fig. 35) where $\{d\}$ is also split and the split results in a breach of the l.o. condition as well[84].

[82] Note that if $B_y$ is allowed to have two edges of opposite signs linked to $[q]$, making $[q]$ a $DSCN_y$, any attempt to augment the size of $[q]$ using $B_y$ shall clearly result in a split as for $B_x$.

[83] The reader is encouraged to try other examples of the mentioned blocks.

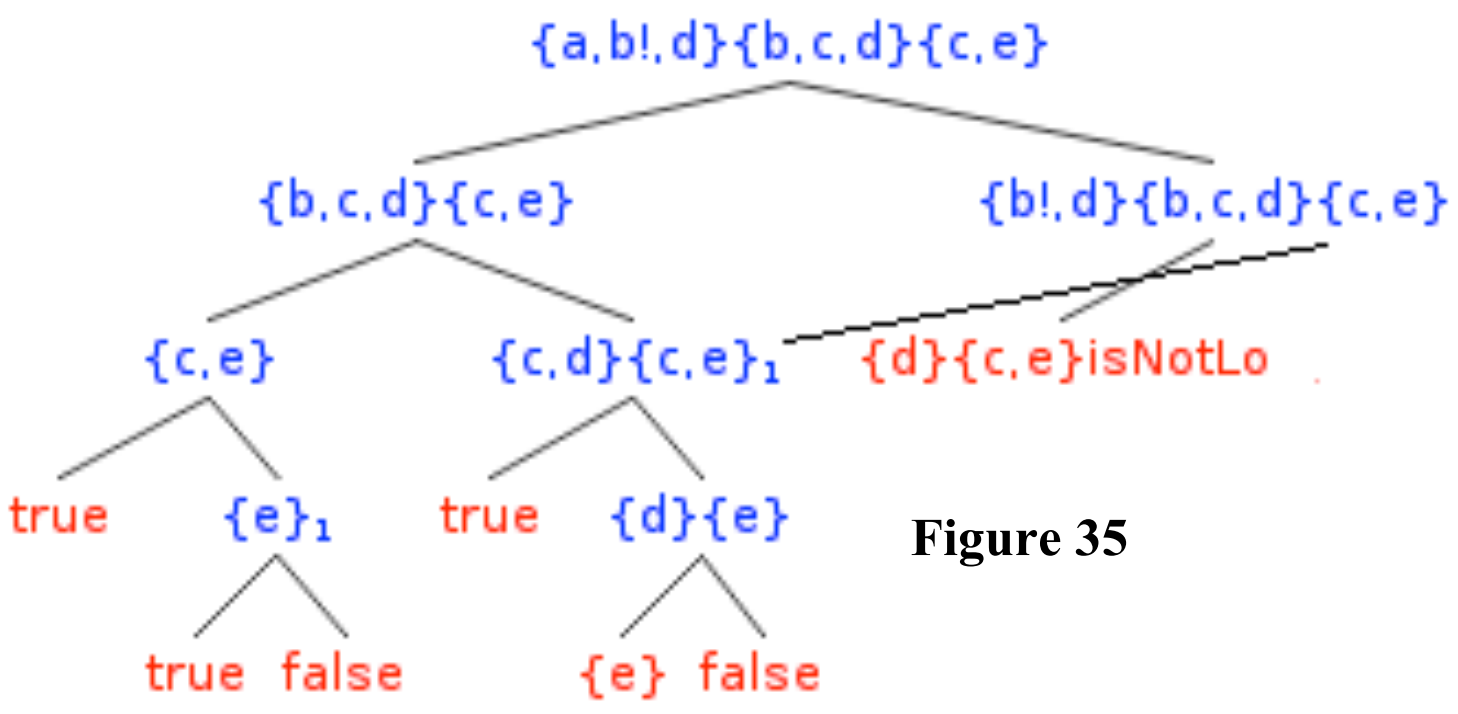

**Figure 35**

**Remark 1:** (TCNs occur in s.o. Sets of monotone, *+ve* 2-SAT when the least-literal/head-clause-rule of GSPRA is dropped).
The above assumes the application of GSPRA with a least-literal/head-clause-rule. Fig. 36 below shows two examples, the resolution-tree for the same Set ({0,1}{1,2}{2,3}), i.e, for the monotone *+ve* 2-SAT case when the least-literal/head-clause-rule is dropped (left) and when it is applied (right).

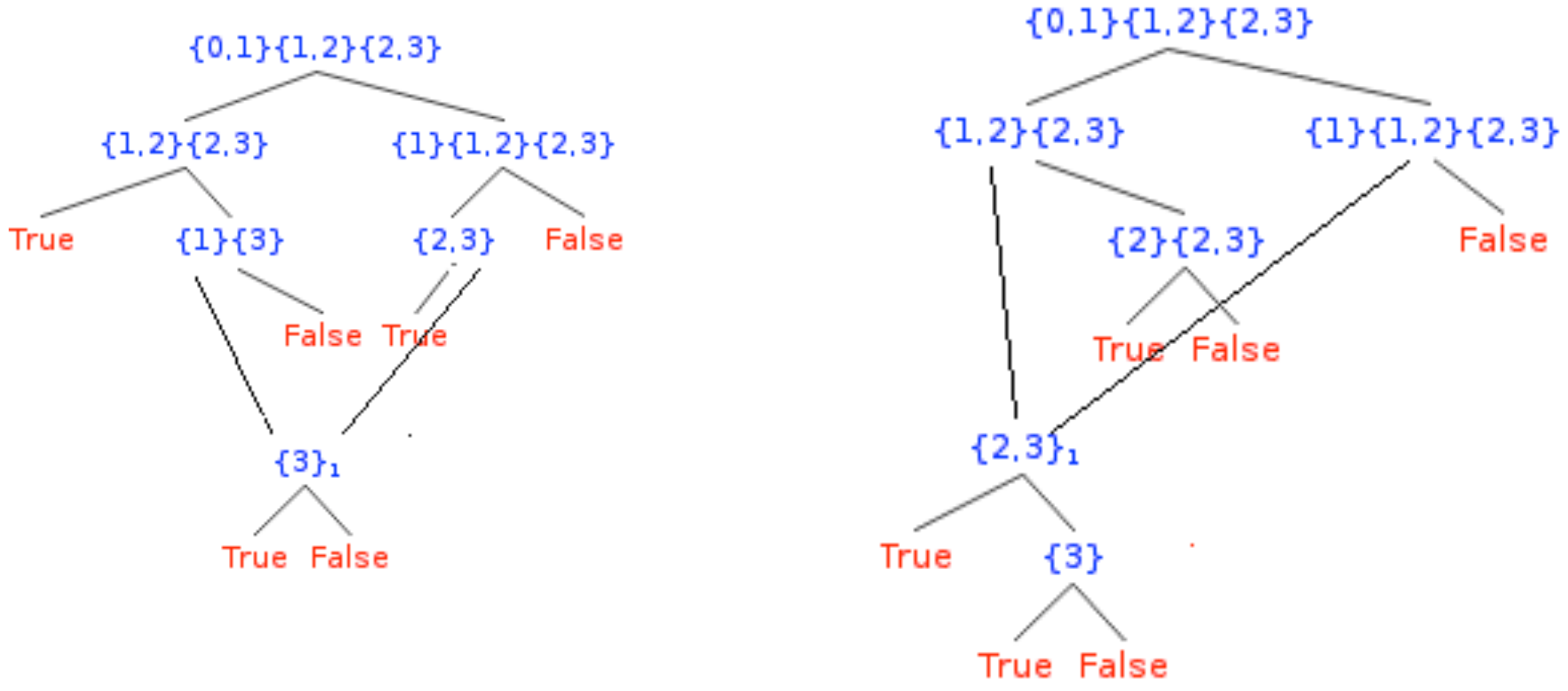

**Figure 36:** The left tree shows a TCN appearing in case the clauses {1,2}{2,3} are instantiated along the literal 2 common to both clauses (instead of 1, the least-literal, as the rule instructs). In the right tree an HCN is formed instead ({2,3}).

[84] The reader is encouraged to try other possibilities from blocks $B_b$ and $B_c$ (like $\{\neg b,e\}$ or $\{\neg c,e\}$ for example).

**Remark 2:** (TCNs in resolution-trees of lo.o. Sets for monotone +*ve* 2-SAT Type 1 TCN)
Suppose S is an lo.o. Set, i.e., at least for some Clause-Set S' in a node of the SRT, S' is l.o.u., then the following form of a TCN exists:

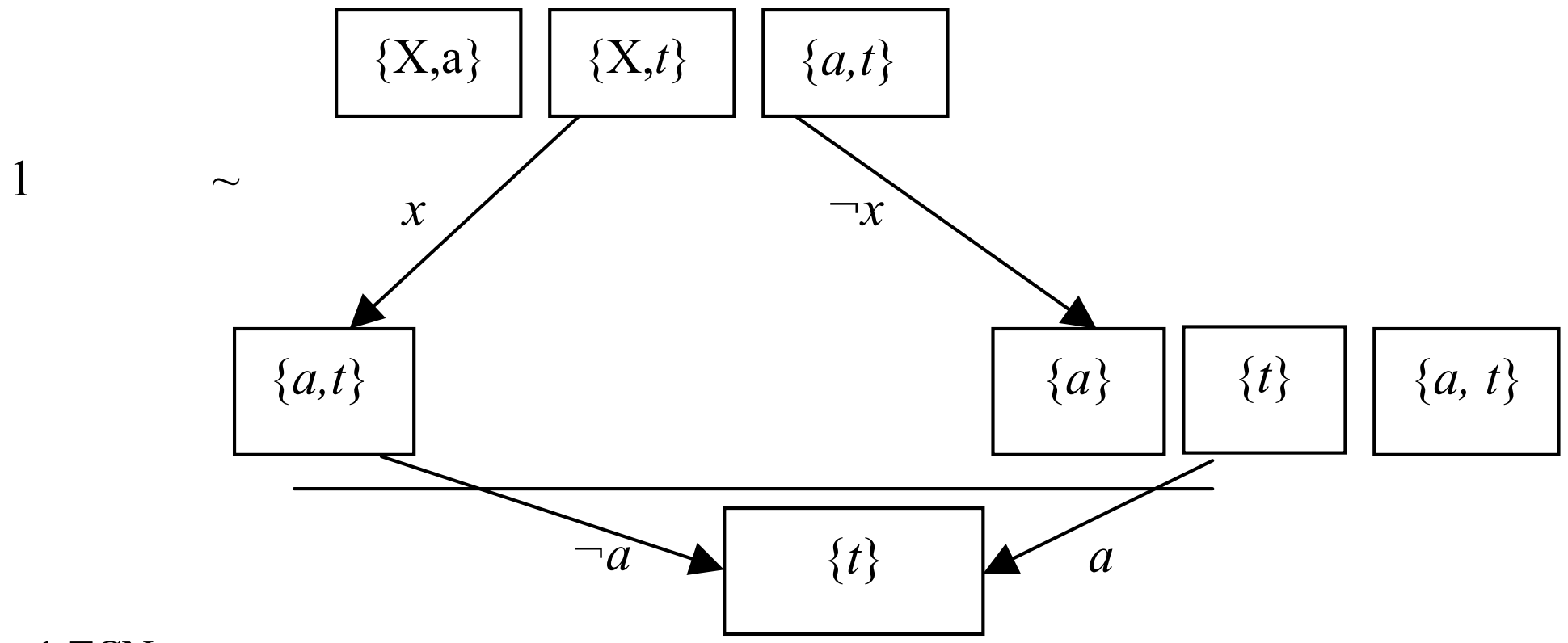

**Figure 37:** Type 1 TCN

**Remark 3:** (TCNs in resolution-trees of lo.o. Sets for monotone +*ve* 2-SAT Type 2 TCN).
In the previous example for a TCN, literals written on **edges leading** to the TCN (called: TCN edge-literals) were identical.

The constellation in Fig. 38 below shows an example where those literals are different. Literals on edges of **branches leading** to a TCN are called branch-literals of the TCN. Every edge-literal is a branch-literal, but not vice versa.

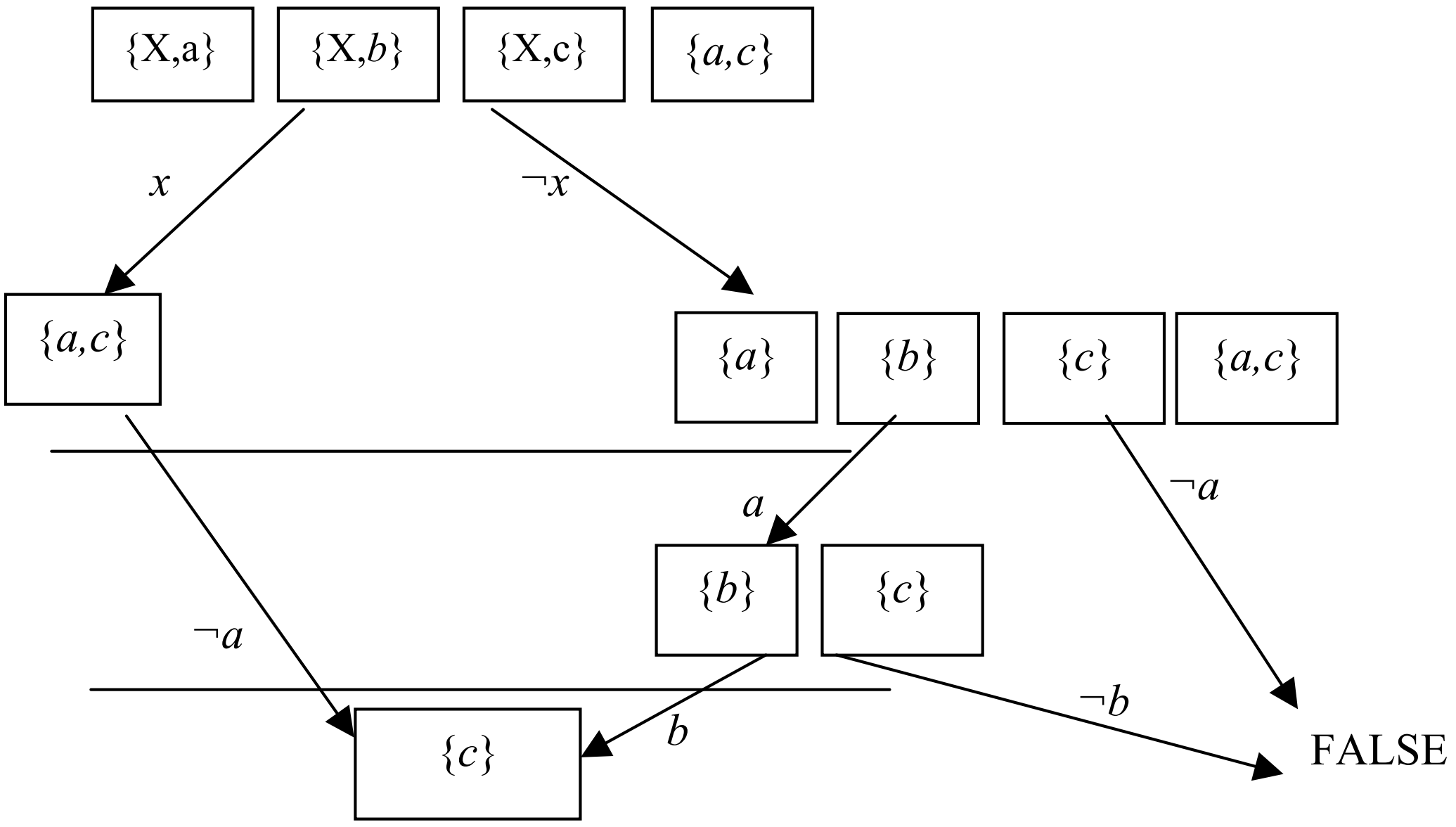

**Figure 38:** Branch-literals of the TCN where ***b*** is an edge-literal for node {*c*} while ***a*** and ¬X are branch-literals

## C) STUDY OF SPLIT CONDITIONS IN SRTs

**Definition 8:** An SRT produced by GSPRA is said to possess a split **if**:
**Either** node ***n*** containing Clause-Set S constructed in step *k* is duplicated one or more times in steps >*k* together with all or parts of its sub-nodes, the cause of this duplication being that S is resolved with a clause whose least-literal is new and has an index < **all** or **any** indices of head-literals in S (called: N-splits).
**Or** a CN [*q*] constructed in step *k* and/or any of its sub-nodes are duplicated with variations[85] one or more times in steps >*k* (called: CN-splits).
As examples of N-splits have been discussed in detail in Section I, we focus on CN-splits in furtherance:

### C-1 Example for a CN-split of a type 1 TCN node:

The reason why different CN-splits occur is generally that different derivations of C get resolved with a CN through different branches of the SRT linked to this CN as mentioned before. New nodes [*q*]'=[*q*]+C' are formed where C' is a possible derivation.
[*q*'] is called: split-node.
This situation is illustrated in the below Fig. 39.
Split-nodes are causes of exponential behavior of GSPRA when it is applied to a.a. or l.o.u. Clause-Sets.

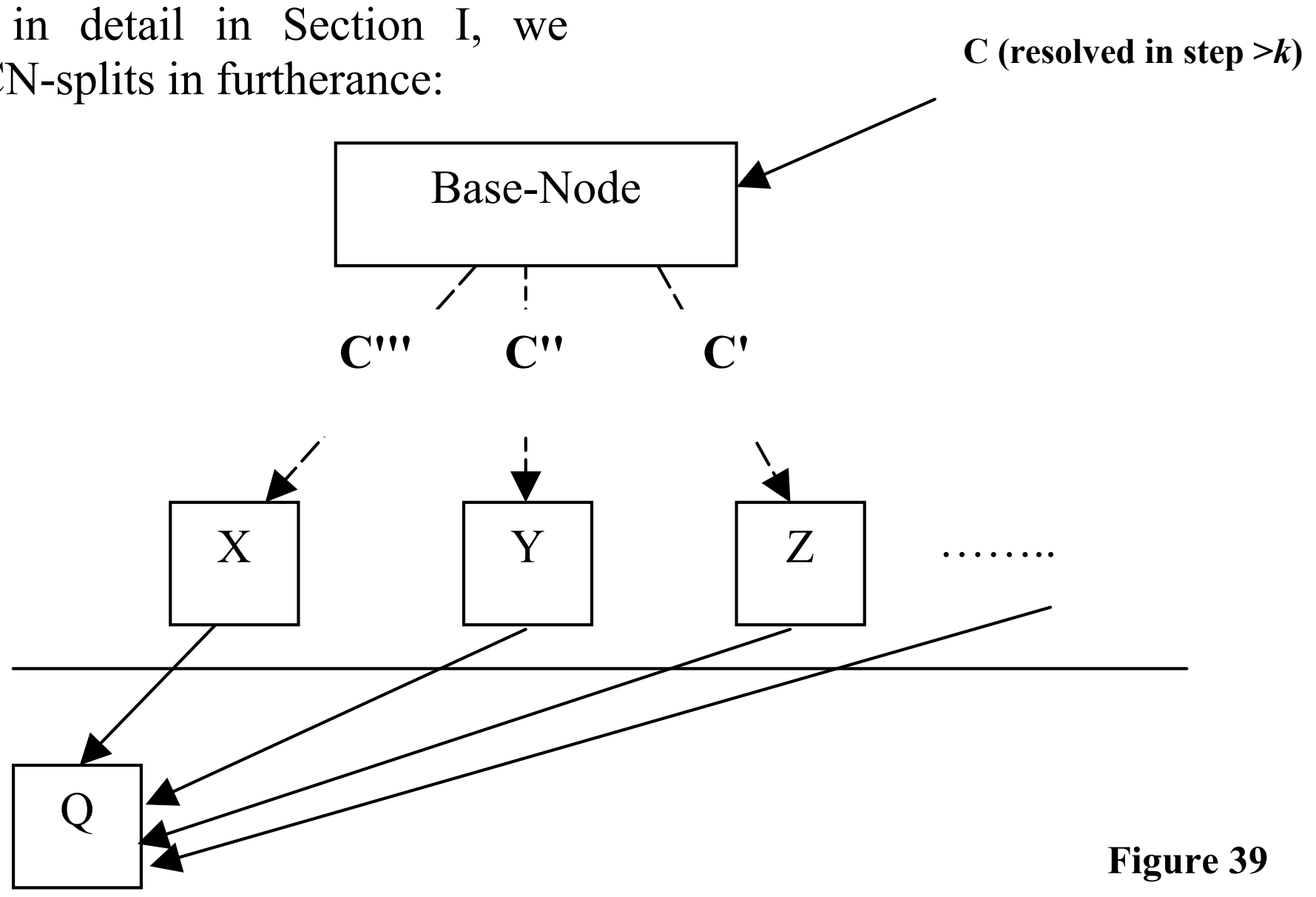

**Figure 39**

[85] Different variations of the duplicated CN correspond to the resolution of different derivations of a newly resolved clause C with the CN.

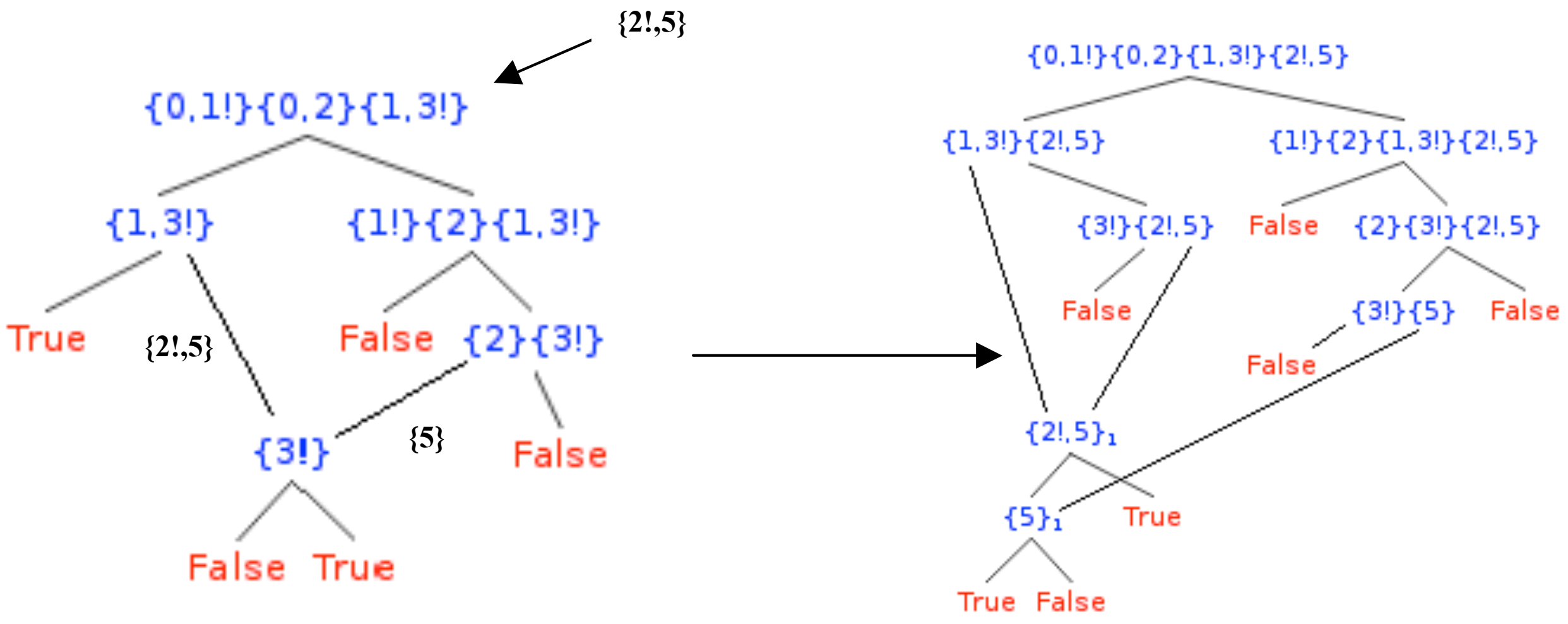

**Figure 40:** A concrete example for the ordinary 2-SAT case.

**C-2: CN-splits in s.o. Sets**

Important properties of CNs in s.o. Sets are shown in the following lemma.

**Lemma 3 (a):** HCNs in SRTs of a 3-SAT-CNF s.o. Clause-Set S generated by GSPRA cannot be split.

**Proof:** Suppose [*q*] is such an HCN formed in step *k*, i.e., [*q*] is head-literal of a clause $C_i$ of S which is the first clause in [*q*], then according to Lemma 2 b), all parent-nodes in the branches for which [*q*] is a sink, [X],[Y],[Z] have NL *q*> NLs *x,y,z* (edge-literals). *q* is also >*s,t,v*.., where *s,t,v*.. <*x,y,z*,.. and *s,t,v*,... are branch-literals which are not edge-literals. To be able to split [*q*] in any step >*k*, a new clause C causing such a split needs to traverse branches leading to [*q*] and contain literals from the Set {*x,y,z*,..., *s,t,v*,...} making the overall value of C =TRUE according to respective signs. This cannot be the case, because $C_i$ ={*q*,..} ∈ S has already been processed in all parent-sets to form the HCN [*q*] and parent-sets are all l.o., i.e., *i,j*, indices of clauses: if *i*<*j* then head-literal of $C_j$ >= head-literal of $C_i$, because of the sorting condition (c.f. Definition 1 b), Section A). Any head-literal of C must therefore be >=*q*.
(Q.E.D.)

**Lemma 3 (b):** CNs of rank 3 in SRTs of 3-SAT-CNF s.o. Clause-Sets cannot be split either.

**Proof:** Suppose [*q*] formed in step *k* is any CN including in its Clause-Set a clause {*x,y,z*}. To be able to split [*q*] in any step >*k*, a new clause C causing such a split needs to traverse branches leading to [*q*] and contain literals from a Set of branch- or edge-literals leading to [*q*] which are all <*x* making the overall value of C =TRUE according to respective signs. This cannot be the case because of the l.o. condition imposed on all parent-nodes as seen in 3(a) which requires the head of C to be >=*x*.
(Q.E.D.)

**Lemma 3 (c):** In any resolution-step: SRTs of 3-SAT-CNF s.o. Clause-Sets possess at most CN-splits of size 1[86] or trivial Block-Splits (**BS**). BS relate to tCNs and can be avoided altogether. The maximum number of size 1 splits possible for each CN is $RCC_{3\text{-SAT}}$.

**Proof:** We have just shown that no splits for rank 3 nodes can exist in s.o. SRTs. For nodes of rank<3: Suppose [*q*] is such an arbitrary CN produced in steps <*k*. According to the least-literal-rule: To split [*q*] in steps >=*k*, the new clause(s) causing the split must possess a head-literal which is equivalent to some least-literal (= edge- or branch-literal) instantiated in a parent-node prior or during the creation of [*q*]. Because of the l.o. condition: This leaves out supporting parents as they already started (per definition) different blocks than the ones which they were instantiating when [*q*] was created. Direct-parents are also excluded, since they cannot admit any new clause with a head-literal <*q*[87]. A split of a rank<3 node in any step >=*k* can thus only occur if the new clause belongs to the same block as the one **some or all** of its parents are still instantiating.

[86] We call splits of size 1 nodes : Size 1 splits.

In such a case : DSCNs and SSCNs may produce size 1 splits before their sizes are augmented as seen in Lemma 2 and its subsequent examples. As per Lemma 2 also: If a new clause C of a parent block succeeds in augmenting the size of [*q*] to become >1 using any non-distinguished literal (=CNAL) in step k, [*q*] will not be able to split in any further steps >k (Lemma 2-e). On the other hand, trying to augment the size of [*q*] using distinguished literals for DSCNs can be done before the formation of [q]. [q] is, then, a tCN formed in step k, but augmented in size in steps <k. Such nodes may - as per Lemma 2-f - be avoided altogether using l.o.s. conditions on SBs and/or DBs.

If size 1 splits exist in any step k, then one copy of a size 1 CN built in steps <*k* is needed in the worst case. What happens if the CN is a sink of many nodes, not just two as in the examples above? Every derivation of a newly resolved clause may cause a different split of the CN. As the number of possible derivations GSPRA produces for newly resolved clauses is always $RCC_{3\text{-SAT}}$[88], only $RCC_{3\text{-SAT}}$ copies of a size 1 CN is produced in any step in the worst case[89].

(Q.E.D.)

**Lemma 3 (d):** All IRTs of a 3-SAT-CNF Clause-Set S are free of non-trivial splits (also called: Big Splits, BigSps)[90] **iff** ∃3-SAT-CNF Clause-Set *S':*S' is s.o., S=S'.

**Proof:** If S=S' is an s.o. 3-SAT-CNF Clause-Set, then all Clause-Sets formed during resolution must be l.o. per Definition 4. N-splits cannot occur in any IRT, because otherwise (per Definition 8) there must exist a clause C resolved with a Clause-Set $S_n$ of some node *n* of an IRT whose least-literal is new and has an index < all or any indices of head-literals of clauses in $S_n$ so that $S_n \cup C$ contradicts the l.o. condition imposed. For rank 3 CN-splits: Lemma 3(a) and (b) above show that any IRT of S is free of them. Lemma 3(c) just showed that no splits of sizes >1 exist as well.

**Other way around:** Consider IRTs of a 3-SAT-CNF Clause-Set S which don't possess BigSps. This means that neither Condition b) nor c) of Definition 1 were breached in course of the resolution of Clause-Sets forming those IRTs (their breach causes N- and rank 3 CN-splits respectively), i.e., all Clause-Sets formed in subsequent IRTs were l.o. which means that (including the final one) all IRTs were s.o. Put S'=Base Clause-Set of the final SRT.

(Q.E.D.)

---

[87] c.f. Definition 5 in Section A

[88] Because the number of possible permutations of a 3-SAT clause is constant ($RCC_{3\text{-SAT}}$).

[89] This observation is important and shall be used in Section E when properties of GSPRA$^{+}$ are discussed and the maximum number of nodes produced in each step is calculated.

[90] BigSps include therefore : Both N- as well as rank 3 CN-splits in addition to any splits of sizes >1

Lemma 3's main statement is that GSPRA - when working with 3-SAT-CNF s.o. Sets - doesn't copy any sub-tree solving a problem of the same order of magnitude as the original one. As seen in above illustrations of CN-splits in a.a. Sets, this was the reason for doubling sizes of IRTs during resolution-runs. It can be shown that for lo.o. Sets such an exponential behavior still exists[91]. Moreover, splits which are still possible in s.o. Sets are of trivial nature, i.e., not costing more than a constant amount of copies of size 1 CNs (in each step) in the worst case.

Fig. 41-a resumes what has been discussed in Sections II-B and II-C for a.a., lo.o. and s.o. Sets while Fig. 41-b shows findings of Lemmas 2 and 3 related to splits of CN types :

[91] Take the case of l.o. Set S={{0,1,2}{0,3,4}} when resolved with {2,3,5} for example.

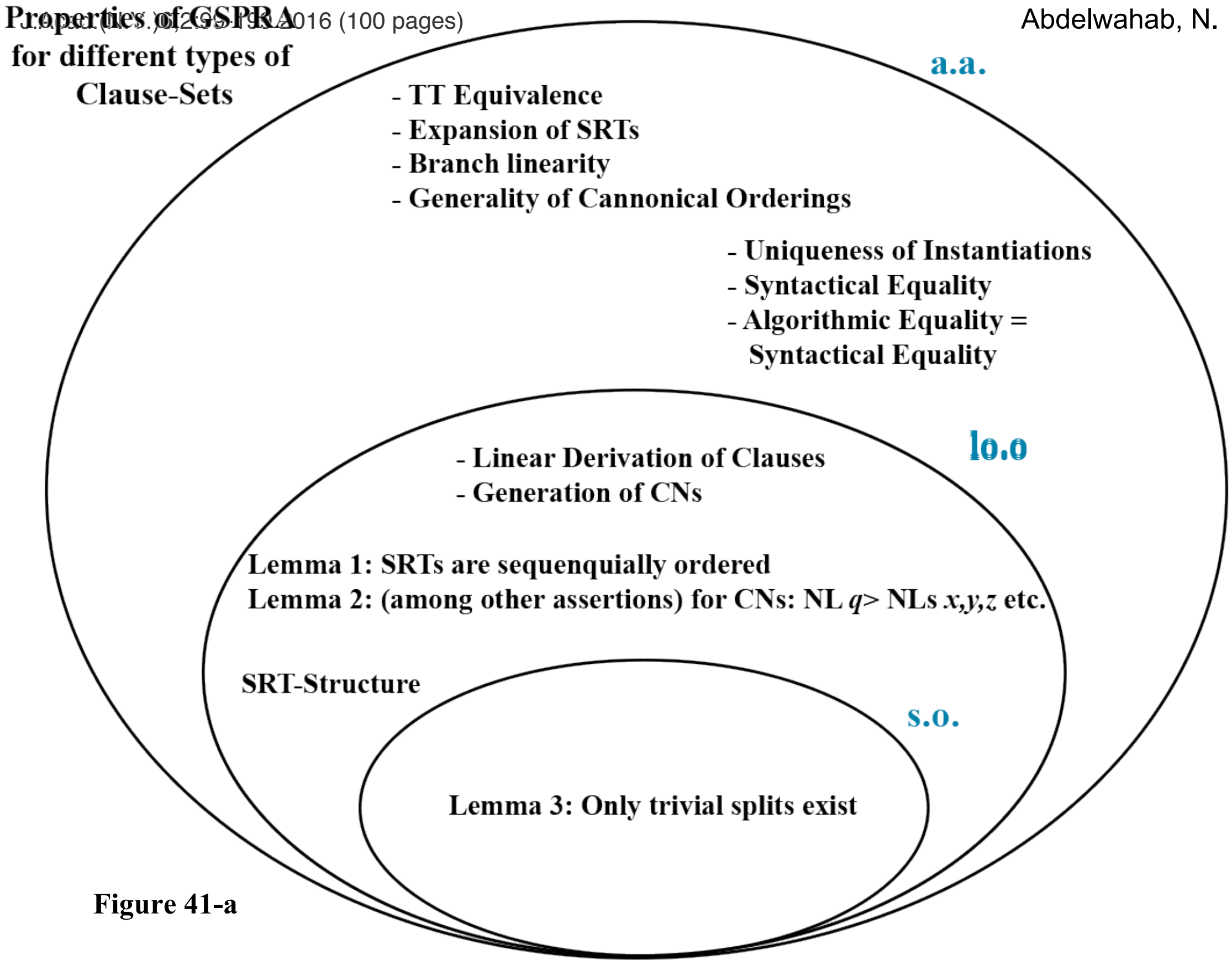

**Figure 41-a**

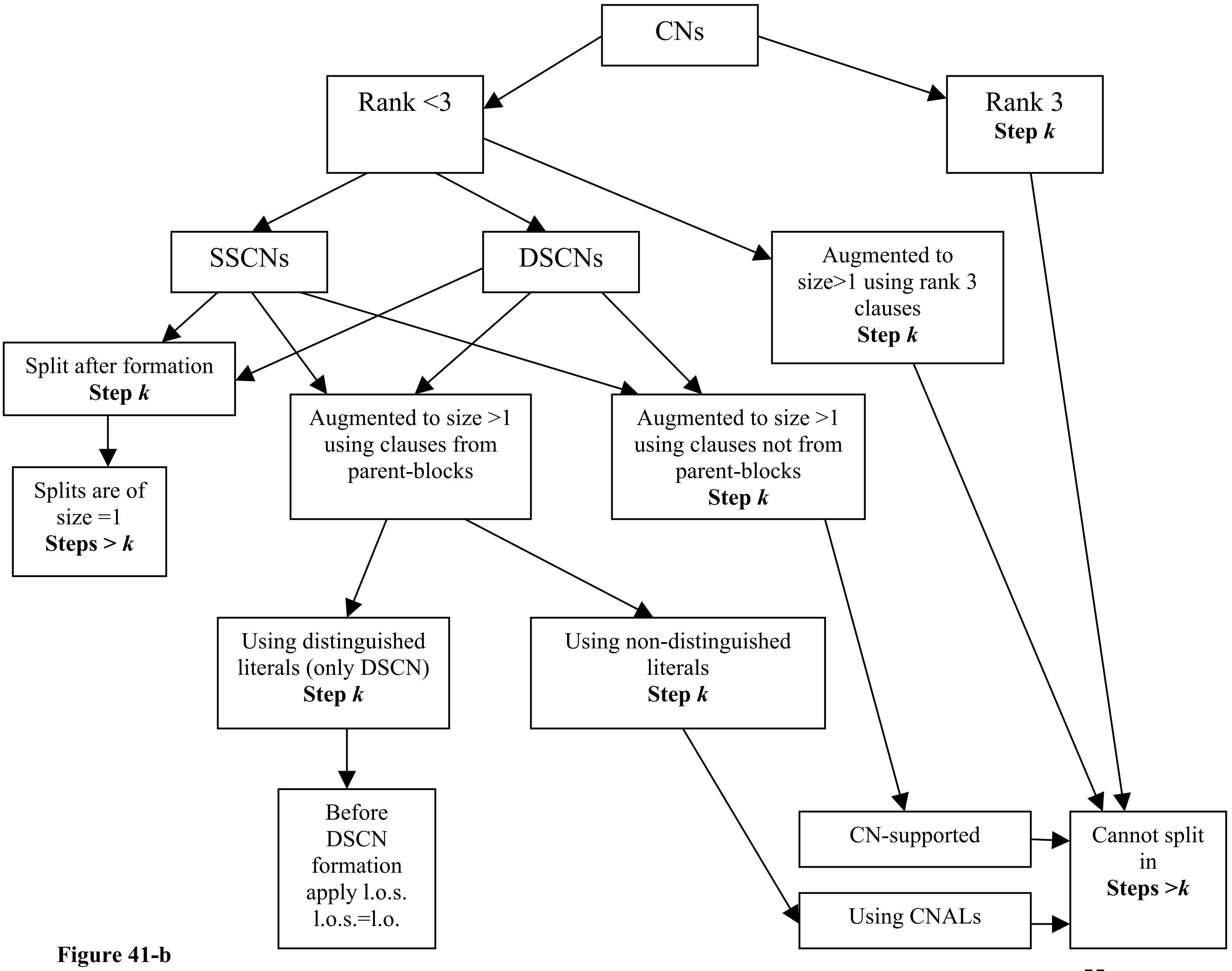

**Figure 41-b**

### D) CONVERTING A.A. 3-SAT-CNF CLAUSE-SETS TO S.O. AND LO.O. SETS

Above lemmas were concerned with s.o. Clause-Sets which are generally not present as such. Can we convert arbitrary Sets to s.o. or lo.o. ones? To answer this question we need to investigate how to convert a.a. Clause-Sets[92] to l.o.u. and l.o. ones (c.f. Definition 1).

**Definition 9:** The Clauses Renaming Algorithm (**CRA**) is a procedure which takes an a.a. Clause-Set S as input, renames its literals yielding a new S' (equivalent to S[93]) as output which is guaranteed to be l.o.u. This procedure consists of the following steps:

1. Index clauses in S (starting with 0) in ascending order.
2. For each clause $C_i$:
   a) Arrange literals in ascending order within $C_i$ so that literals which appear more often in other clauses come before those which appear less often or which only appear in $C_i$. This condition shall hereafter be called: Renaming Precedence Condition (**RPC**).
   b) For all literals, one by one, arranged in step a) do the following: For any literal in the clause not having already a row, create a new row and write column values TRUE or FALSE according to whether the literal appears in the corresponding clause or not. The matrix resulting from this step is called **Connection Matrix of S**. Rows in this matrix represent variable/literal names/indices while columns represent clauses.

Example: If S = {{0,5} {0,2} {1,3} {1,4} {2,3}}, then the Connection Matrix of S is:

| | $C_0$ | $C_1$ | $C_2$ | $C_3$ | $C_4$ |
|---|---|---|---|---|---|
| 0 | **True** | **True** | False | False | False |
| 5 | **True** | False | False | False | False |
| 2 | False | **True** | False | False | **True** |
| 1 | False | False | **True** | **True** | False |
| 3 | False | False | **True** | False | **True** |
| 4 | False | False | False | **True** | False |

3. Rename all variables in the Connection Matrix in ascending order. The matrix in the example thus becomes:

| | $C_0$ | $C_1$ | $C_2$ | $C_3$ | $C_4$ |
|---|---|---|---|---|---|
| **0** | **True** | **True** | False | False | False |
| **1** | **True** | False | False | False | False |
| **2** | False | **True** | False | False | **True** |
| **3** | False | False | **True** | **True** | False |
| **4** | False | False | **True** | False | **True** |
| **5** | False | False | False | **True** | False |

4. Reconstruct the clauses again using the new variable names. This reconstruction may be done by simply substituting each literal in the original Clause-Set with its new literal name/index.

The new clause list for the above reads S: S' = {{0,1}{0,2}{3,4}{3,5}{2,4}}. Note that S' is l.o.u. Note also that if we would want to convert S' to a l.o. Set by sorting clauses via their least-literals (as required by Condition b) in Definition 1) we would get: S'' = {{0,1} {0,2} {2,4} {3,4} {3,5}} which is not fulfilling Condition c) because of literal 3 (i.e., S'' is neither l.o. nor even l.o.u.).

---

[92] Converting an arbitrary Clause-Set to an almost arbitrary one (a.a.) being a trivial exercise needing only sorting literals inside each clause in ascending order and taking care that clauses have unique occurrences.

[93] here: Logical, not Syntactical Equivalence.

To convert an a.a. Clause-Set to a l.o. Clause-Set, an extension to CRA is needed, introduced hereafter with some definitions:

**Definition 10:** Mapping: N => N is a bijective function giving a literal index in a Clause-Set its new name/index after a renaming operation using CRA. If each literal index is given itself, the function is called trivial Mapping (*tMapping*). If a subset of literal indices is mapped to itself, this subset is called a Stable-Set. If all literal indices of a clause are members of a Stable-Set, it is called stable clause. If all clauses of a Clause-Set are stable, the Set itself is called Stable-Clause-Set. The function Mapping#: <Clause-Sets, Mappings> => Clause-Sets applies a certain mapping to a Clause-Set S changing the names/indices of all literals/indices in this Clause-Set to the new names/indices given by the mapping using direct substitution, the new resulting Clause-Set S' is said to be in a Variable Space (VS) different than S.
If S and S' are Clause-Sets of the same node, then this node is called a Mixed-Space Node (MSN) as opposed to Single-Space Nodes (SSN) whose Clause-Sets are not renamed. Trees with MSN nodes are called Mixed-Space Trees (MSTs). Trees with only SSNs are called Single-Space Trees (SSTs).
Example:

For S = {{0,5}{0,2}{1,3}{1,4}{2,3}} and S' = Mapping# (S,M) = {{0,1}{0,2}{3,4}{3,5}{2,4}} in the above example, Mapping M has the following extension: {{0,0}{5,1}{2,2}{1,3}{3,4}{4,5}}, Stable-Set = {0,2}

**Definition 11:** The Clauses Renaming & Ordering Algorithm (**CRA**+) is a procedure which takes an a.a. Clause-Set S as input and applies CRA repetitively generating a new mapping each time. After each step the intermediate Clause-Set is sorted as required by Definition 1b) before iterating back. This is done until renaming literal indices in two consecutive steps yields *tMapping*, i.e, the Stable-Set becomes equivalent with the Set LIT(S)) while the output Clause-Set S' becomes l.o. A recursive pseudo-formal description of this procedure is used in the below proofs:

**CRA+:**
Inputs: a.a. Clause-Set **S**
Output: l.o. Clause-Set **S'**
Steps:

4- set CurrentMapping = null, CurrentSet=S
5- while (CurrentMapping != tMapping)
   i. currentSet=CRA(CurrentSet)
   ii. sort CurrentSet as instructed in Definition 1 b)
   iii. set CurrentMapping=Mapping passed by CRA
6- S'=CurrentSet
7- return S'

Example: Following this procedure for the above Set S = {{0,5}{0,2}{1,3}{1,4}{2,3}} applying CRA to get S' = {{0,1}{0,2}{3,4}{3,5}{2,4}} and a sorting step giving the above S''={{0,1}{0,2}{2,4}{3,4}{3,5}}.
A new CRA-iteration will yield the following Connection Matrix:

| | $C_0$ | $C_1$ | $C_2$ | $C_3$ | $C_4$ |
|---|---|---|---|---|---|
| 0 | **True** | **True** | False | False | False |
| 1 | **True** | False | False | False | False |
| 2 | False | **True** | **True** | False | False |
| 4 | False | False | **True** | **True** | False |
| 3 | False | False | False | **True** | **True** |
| 5 | False | False | False | False | **True** |

It is then transformed to:

| | $C_0$ | $C_1$ | $C_2$ | $C_3$ | $C_4$ |
|---|---|---|---|---|---|
| 0 | **True** | **True** | False | False | False |
| 1 | **True** | False | False | False | False |
| 2 | False | **True** | **True** | False | False |
| 3 | False | False | **True** | **True** | False |
| 4 | False | False | False | **True** | **True** |
| 5 | False | False | False | False | **True** |

Mapping: {{0,0}{1,1}{2,2}{4,3}{3,4}{5,5}}, Stable-Set: {0,1,2,5} yields S'''={{0,1}{0,2}{2,3}{3,4){4,5}} when applied on S'' which is l.o. already and needs no further sorting. Note that in the last matrix all literals are forming an ordered sequence which means that any further renaming would result in *tMapping*. This is the termination condition.

**Lemma 4:** CRA is guaranteed to convert an a.a. Clause-Set S into a l.o.u. Clause-Set. It takes O(N*M) steps[94] to do this for M = number of clauses, N = number of variables.

**Proof:** c.f. the three conditions of Definition 1 for a Clause-Set to be l.o.u.:

a) $\forall a_i, b_{ij} c_{ij} \in C_{i+j}$: $a_i < b_{ij} < c_{ij}$

c) $\forall x \in$ LIT(S), $\forall$C $\in$ S: if $x$ not $\in$ LEFT($x$,C) then $\forall y \in$ LEFT($x$,C): $x > y$

d) Clauses appear only once in S

It is clear that a) and d) are fulfilled by any output of the CRA as they constitute the mere definition of a.a. Sets. For Condition c): Suppose some literal L in a clause $C_i$={... L ...} $\in$ S' (S' = output Set) breached Condition c): This means that L is new in the clause sequence starting with $C_0$ until $C_i$, but there exists L' to its left where L<L'. This cannot be the case, since any such L' would have to appear in a row before L in the connection matrix (step 2-) and thus get a smaller index in the renaming step 3-. For the complexity assertion: The number of cells to be created in a Connection Matrix is always N*M.
(Q.E.D.)

**Lemma 5:** A Set S is l.o. **iff** it reaches a Stable-Set of literals equivalent to LIT(S) through application of $CRA^+$.

**Proof:** Suppose S is l.o. This means that it is fulfilling all Conditions a)-d) of Definition 1. Any attempt to use $CRA^+$, i.e., rename the literals and then sort them, must generate a Stable-Set = LIT(S) after only one CRA- and sorting iteration, since otherwise (i.e., if a literal gets a new name/index after such an iteration) this would mean a breach of one or all of those conditions. Other direction: Suppose S reached such a Stable-Set through application of $CRA^+$, i.e., $CRA^+$ terminated. If S is not l.o., then it must be at least l.o.u. (because of Lemma 4). The only reason for S not to be l.o. would thus be that clauses are not sorted correctly. This is not possible because $CRA^+$ can only become a Stable-Set equivalent to LIT(S) if two consecutive renaming iterations assign literals with the same names/indices, the first of which is followed per Definition by a sorting operation.
(Q.E.D.)

**Lemma 6:** $CRA^+$ takes a number of steps which is $O(M^2(\log M+N))$. More precisely M CRA-iterations and M sorting operations[95] (M = number of clauses in S, an a.a Set).

**Proof:** (by induction on M)

**Base-Case: M=1:** For S={$a,b,c$} $CRA^+$ takes one CRA and one sorting operation to generate *tMapping* per definition.

**Illustration Case: M=2**[96]
Let S={{$a,b,c$},{$d,e,f$}}={$C_0$,$C_1$}

[94] Steps are invocations of primitive operations as usually understood in complexity analysis.

[95] Assuming that a sorting operation takes O(M log M) primitive operations.

[96] Monotone 3-SAT case is used here and in the next lemma (w.l.o.g.), since $CRA^+$'s behavior does not depend neither on literal signs nor on clause breadth.

**Case 1:** No literals in common between $C_0$ and $C_1$: In that case *a<b<c<d<e<f*.
S is l.o. No CRA- or sorting iterations needed.
**Case 2:** Only head-literal in common: S={{*a,b,c*}{*a,e,f*}} for example[97]: Same as case 1, S is also l.o.
No CRA or sorting needed.
**Case 3:** Only middle-literal in common: S={{*a,b,c*}{*b,d,f*}} for example: S' is converted after one CRA-iteration to S={{*a,b,c*}{*a,e,f*}}, because of Definition 9, 2a), Renaming Precedence Condition (RPC).
Thus, no sorting needed.
**Case 4:** Only tail-literal in common: S={{*a,b,c*}{*c,d,e*}} for example: S' is converted after one CRA-iteration to S={{*a,b,c*}{*a,e,f*}} (RPC as well), no sorting needed.
**Case 5:** Two literals in common: All forms are converted to the form S={{*a,b,c*}{*a,b,d*}} (RPC) requiring only one CRA-operation and no sorting.

**Resuming the Base-Cases M=1,2:** Although we may not need CRA or sorting, CRA$^+$ takes at least one iteration (i.e., one CRA- and one sorting operation) to generate *tMapping* and to terminate.

**Induction Hypothesis:** For M clauses M CRA-iterations as well as M sorting operations are needed in the worst case to make S l.o.

**Induction step:** For any additional clause $C_{M+1}$ = {*x,y,z*} we have the following cases (c.f. Definition 11, pseudo formal procedure):

1. ***x,y,z* are new literals not appearing before in any clause $C_i$:** This case is straightforward in that no sorting is needed, i.e., only CRA (renaming) in the worst case.
2. **One or more literals of *x,y,z* appeared in a previous clause:** Suppose S={{0,1} {0,2,3} {0,4,5} {0,6,7} {2,8} {9,10} {11,12}} which is l.o. adding the clause {3,5,15}, then the following steps are required:
   a) S={{0,1}{0,2,3}{0,4,5}{0,6,7}{2,8}{9,10}{11,12}{3,5,15}} input
   b) S={{0,1}{0,2,3}{0,4,5}{0,6,7}{2,8}{3,5,15}{9,10}{11,12}} sort
   c) S={{0,1}{0,2,3}{0,4,5}{0,6,7}{2,8}{3,5,9}{10,11}{12,13}} CRA

S in step c) is already l.o., i.e., for a Clause-Set of size M S={{*a,b*,..} {*b*,…} {*d*,…}…} where as per induction hypothesis it is assumed that it is l.o. and we add a clause containing one or more literals which appeared before (*s,t,u*,…) and one or more literals which are new (*x,y*,…), we note that S is l.o.u. A sorting step is what is required to align the new clause to its right place. If this step is done (c.f. step b)), then another CRA-step (c.f. step c)) guarantees both l.o.u (per Lemma 4) and sorting condition. This means that we need an additional CRA (renaming) as well as a sorting step for this case.

**Resuming the induction step:** One additional CRA- and one additional sorting step is needed in the worst case for M+1
(Q.E.D.)

This section concludes with a lemma showing **that any a.a. Set can be converted to a l.o. Set**, i.e., application of CRA$^+$ on any a.a. Set always terminates yielding the right result.

**Lemma 7:** CRA$^+$ terminates always converting any a.a. 3-SAT-CNF-Set S of size M to a Stable-Clause-Set.

[97] Another example could be: {{*a,b,c*}{*b,e,f*}}.

**Proof:** (by induction on M)

**Base-Case M=1:** For S={*a,b,c*} as seen in the Base-Case of Lemma 6, $CRA^+$ terminates after one iteration yielding the Clause-Set S'={*a*',*b*',*c*'} where *a*',*b*',*c*' are new indices/names for *a,b,c*. S' is stable.

**Illustration Case M=2:** Let S={{*a,b,c*}{*x,y,z*}}. As seen in all Base-Cases for M=2 of Lemma 6: One iteration of CRA and one sorting operation converts S to a l.o. Set. This means any further iteration of $CRA^+$ yields a Stable-Set (per definition of $CRA^+$) letting the algorithm terminate.

**Induction Hypothesis:** Application of $CRA^+$ for a number of iterations *k* on an a.a. 3-SAT-CNF Clause-Set S of size M converts S to a Stable-Clause-Set (i.e., $CRA^+$ produces M stable clauses after *k* iterations).

**Induction Step:** Per induction hypothesis for S = M+1, there are M stable clauses in iteration *k*. Let C={*x,y,z*} be the clause which is not stable. After step *k* C's position cannot be before any other stable clause C'={*i,j,k*}, e.g., as in {*a,b,c*}…{*x,y,z*} {*i,j,k*}…, because this would mean that CRA-operations will have to change indices *i,j,k* to new ones for C' contradicting its stability assumption, i.e., C has to be the last clause in S.
In that case, even if literals in C would not fulfill the l.o. condition for whatever reason other than sorting (because C is already in its place), further CRA-steps in iterations >=*k* guarantee to convert C into a stable clause (per definition of $CRA^+$)[98] causing $CRA^+$ to terminate with a Stable-Clause-Set of size M+1.
(Q.E.D.)

[98] CRA renders {S *U* C} l.o.u., i.e., any new literal *v* of C is >left(*v*) in {S *U* C} after an iteration.

## E) GSPRA$^+$ FOR ARBITRARY 3-SAT-CNF CLAUSE-SETS

It is possible to convert GSPRA to an algorithm using MSTs (called: GSPRA$^+$) which produces for arbitrary Clause-Sets of 3-SAT-CNF-problems SRTs of similar properties as s.o. ones[99]. Consider the following programs written in pseudo code:

**GSPRA$^+$(S):**

**Inputs**: Arbitrary 3-SAT-CNF Clause-Set **S** of size M
**Outputs**: Final SRT
**Data Structure**: list of Tuples: < Clause-Sets, Node index> (called: LCS) initially empty

**Steps**: -

1- convert arbitrary clauses in S to a.a. ones (sorting literals inside each clause).
2- choose a clause $C_0$ which guarantees a minimal top-part of the SRT containing unique-nodes[100]. For that choice apply the below procedure **SelectFirstClause**.
3- convert S to a l.o. Set using CRA$^+$
4- convert $C_0$ of S to an SRT in a way similar to the one described in Definition 2
5- set IRT (Intermediate Resolution Tree) = SRT produced in 4
6- for all the rest Clauses $C_i$ of S
   a. convert $C_i$ to an SRT
   b. IRT=Align(IRT,$C_i$)
7- return IRT

**Align (SRT, C):**

**Inputs**: an SRT with base-node of size M, an a.a. 3-SAT-CNF-clause C
**Outputs**: an SRT for a Base-Set of size M+1
**Steps**:

1. Let *n* be base-node of SRT, S Clause-Set of *n*
2. If M>1
   a. S $\cup$ C is l.o.
      i. set S'= S $\cup$ C, put S' Clause-Set of *n*
      ii. instantiate C to become C' and C'' according to left- and right edges of *n* taking into consideration the least-literal/clause-rule of Definition 2 (respectively)
      iii. if (BaseSet(leftTree(*n*)) $\cup$ C' is found in LCS)
      Then leftTree(*n*)= foundNodeIndex
      else leftTree(*n*)=Align(leftTree(*n*),C'), Store <BaseSet(leftTree(*n*)), NewNodeIndex> in LCS
      iv. if (BaseSet(rightTree(*n*)) $\cup$ C'' is found in LCS)

---

[99] Remember that s.o. Sets were SSTs not MSTs as the ones produced by GSPRA$^+$.

[100] Recall that the top-part of an SRT of a node of size M contains at most *k* unique M-sized (THN) and *k* not necessarily unique <M-sized (TBN) nodes where *k* is the size of the first clause chosen (c.f. Property 8 in Section B). The idea here is to produce top-parts with minimal unique-nodes.

Then rightTree(*n*)=foundNodeIndex
else rightTree(*n*)=Align(rightTree(*n*),C''), Store <BaseSet(rightTree(*n*)),NewNodeIndex> in LCS

v. return SRT of *n*

b. S $\cup$ C is not l.o.

i. set S'= S $\cup$ C
ii. convert S' to l.o. using CRA$^+$
iii. set S' Clause-Set of *n*
iv. separate last clause of S', i.e., S'=S'' $\cup$ A
v. if (C<>A), i.e., the last clause in S' is not C
   1. set SRT'= GSPRA$^+$(S''), i.e., rebuild the SRT for *n* once more
   2. set SRT of *n* = SRT'
vi. instantiate A to become A' and A'' according to left- and right edges of *n* taking into consideration the least-literal/clause-rule (respectively)
vii. if (BaseSet(leftTree(*n*)) $\cup$ A' is found in LCS)
Then leftTree(*n*)= foundNodeIndex
else leftTree(*n*)=Align(leftTree(*n*),A'), Store <BaseSet(leftTree(*n*)),NewNodeIndex> in LCS
viii. if (BaseSet(rightTree(*n*)) $\cup$ A'' is found in LCS)
Then rightTree(*n*)=foundNodeIndex
else rightTree(*n*)=Align(rightTree(*n*),A''), Store <BaseSet(rightTree(*n*)),NewNodeIndex> in LCS
ix. return SRT of *n*

3. If M=1

a. let S'= S $\cup$ C, put S' Clause-Set of *n*
b. if Convert S' to become l.o. using CRA$^+$ if it is not already. C is converted to C$^\#$
c. for all nodes *n*' of SRT , SS Clause-Set of *n*':
   i. propagate C$^\#$ to *n*' instantiating it according to the edges in SRT leading to *n*' so that it becomes CC. Use least-literal/clause-rule
   ii. set SS = SS $\cup$ CC
d. return SRT of *n*

**SelectFirstClause(S):**

**Inputs**: an a.a. 3-SAT-CNF-Clause-Set S
**Outputs**: a clause C from S
**Steps**:

1- For all Clauses C' in S:
   i. choose C' to be first clause
   ii. for all possible literal arrangements in C':
      1. set MinNumber = Number of unique-nodes resulting from instantiation of literals of C' in the top-part of the SRT

2. if the newly calculated number of unique-nodes for C’ < previously stored MinNumber, then set MinNumber=newly calculated one, bestClause=C’ in the current literal arrangement

2- return bestClause.

**E-1: Example for the usage of GSPRA$^+$ on a.a. 3-SAT-CNF Clause-Sets**[101]

Let S = {{0,¬1}{0,2,¬3}{0,4,¬5}{2,¬6}{¬3,4,5}{4,6,7}} be an a.a. 3-SAT-CNF-set. Following are the steps taken as per the above description of GSPRA$^+$:

1- **Step 0**: S is already l.o.

2- **Step 1: End**: The following SRTs (Fig. 42-45) are generated in subsequent steps using GSPRA$^+$

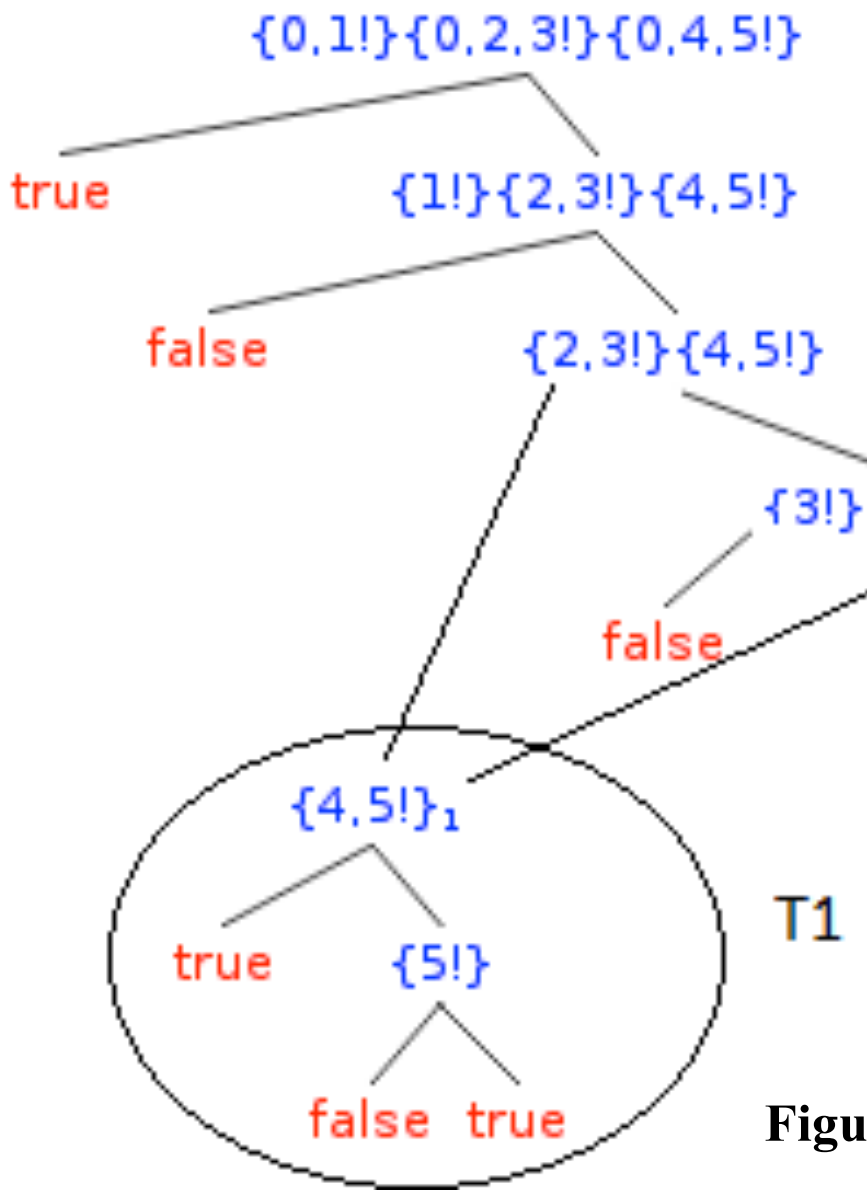

Figure 42: T1

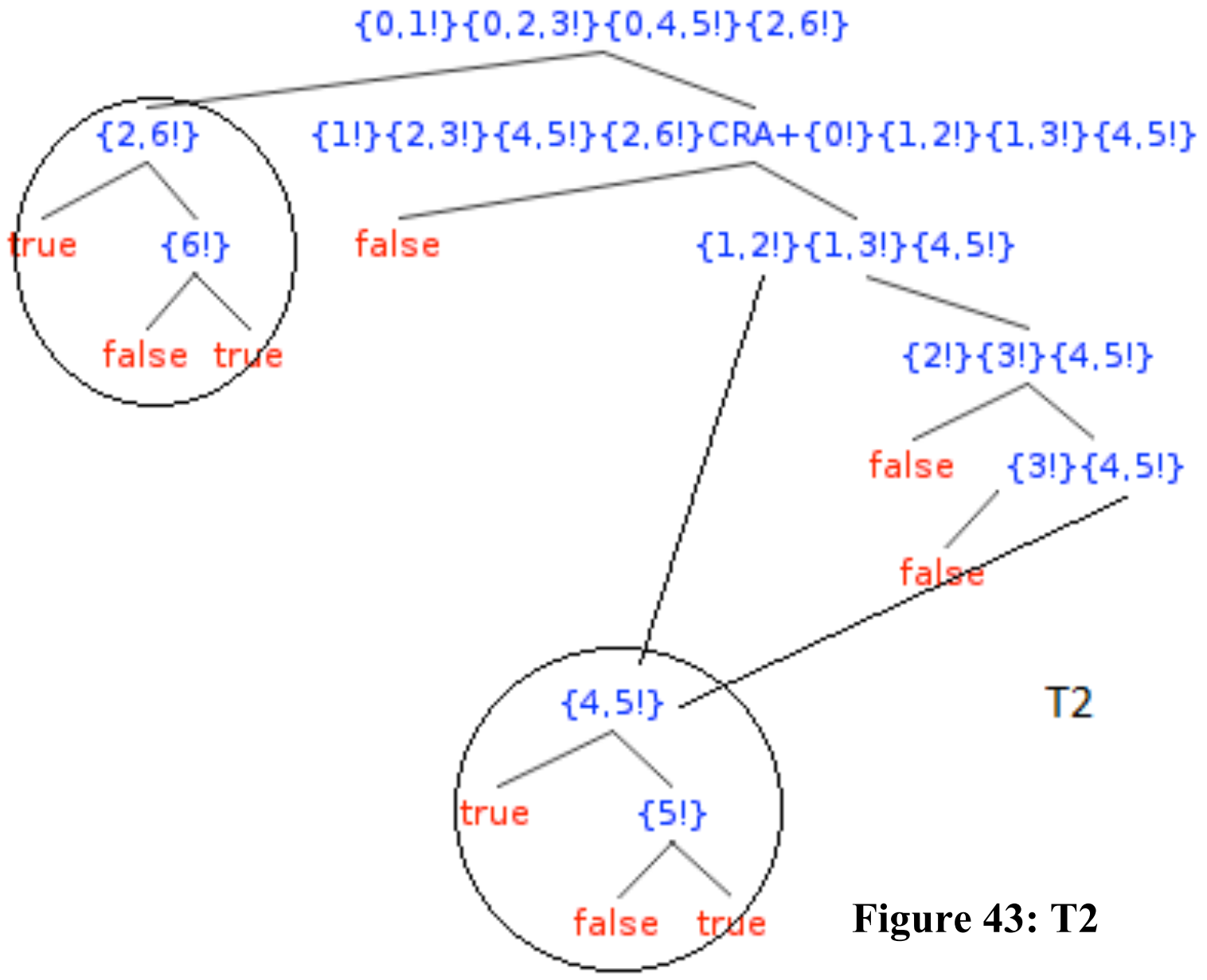

Figure 43: T2

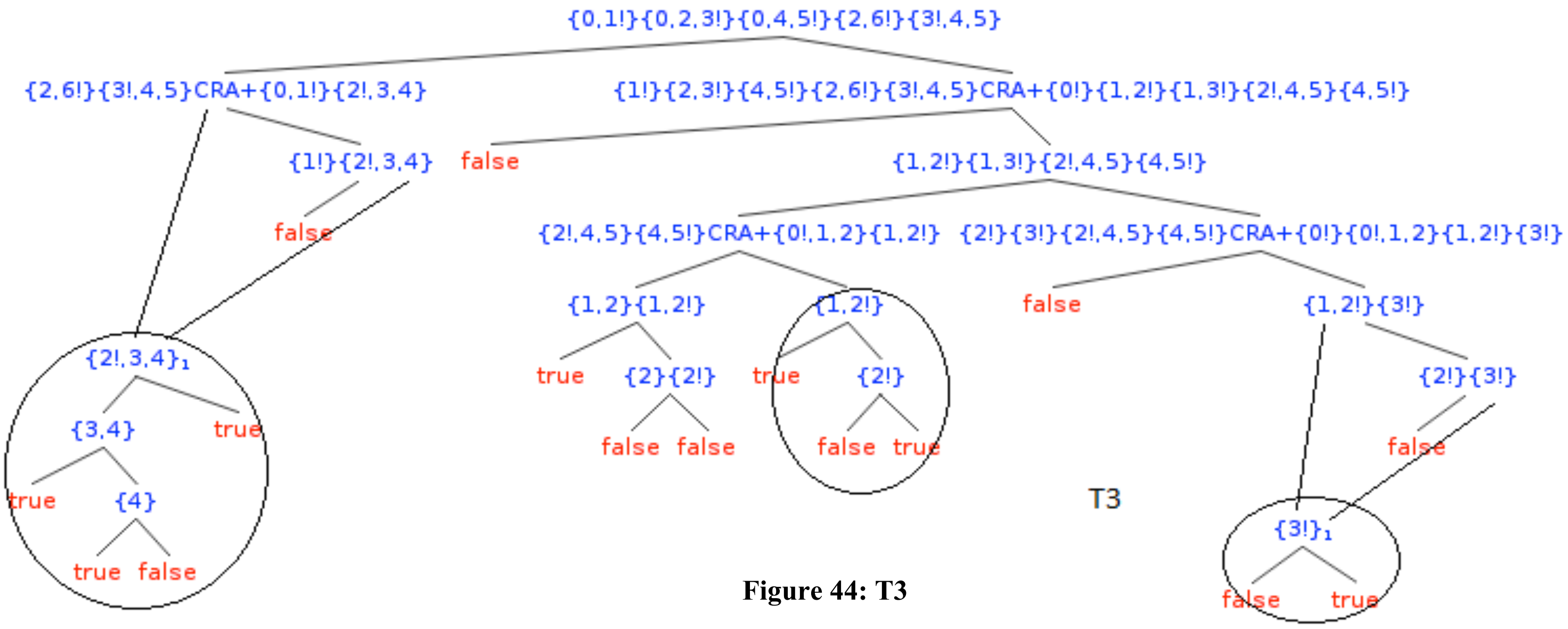

Figure 44: T3

---

[101] For the sake of simplicity in the example it is assumed that SelectFirstClause uses the shortest clause.

**Figure 45: T4**

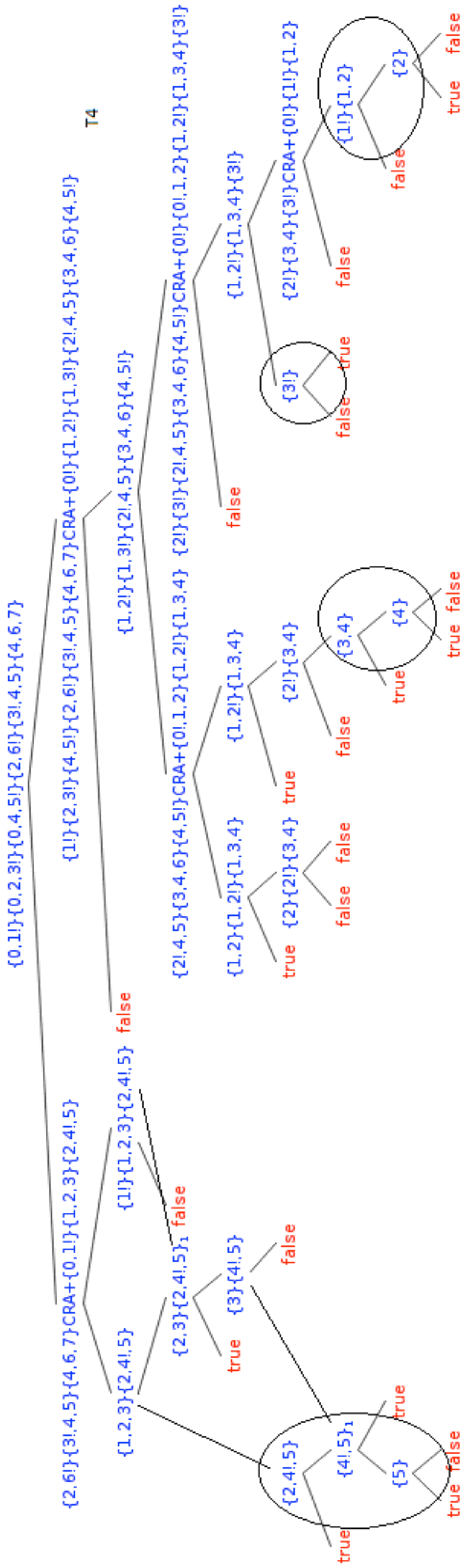

From analyzing the above trees we realize the following:

1. The number of unique-nodes which are not leafs increases in a non-exponential way from T1 to T4: 6,9,17,24.
2. Trees T2,T3,T4 generated by GSPRA+ are MSTs (c.f. Definition 10).
3. All trees end with special sub-trees which have one distinguished clause tailing all their Clause-Sets. Those clauses are called alignment-clauses. They are circled in the foregoing trees (Fig. 42-45). Trees possessing alignment-clauses are called aligned and are further defined below. They shall play an important role in understanding the way GSPRA+ works.
4. As CRA+ is used for converting clauses and Clause-Sets to l.o., it might be the case that the original order of literals in clauses changes even deliberately in the new variable space (as allowed in step 1.ii in SelectFirstClause procedure). Consider for example the Set S={{¬0}{1,¬2} {1,¬5} {¬2,3,4} {3,¬4}{3,5,6}}[102]. CRA(S)={{¬0} {1,¬2} {1,¬3} {¬2,4,5} {4,¬5} {3,4,6}} where M(5)=3, M(3)=4, M(6)=6, for the last clause {3,5,6} thus changing places of literals in that clause. GSPRA+ has therefore no property similar to the linear derivation property (Property 3) of GSPRA.
5. Choice possibility of both clauses and literal arrangements in the SelectFirstClause procedure enable GSPRA+ to implement any strategy producing an SRT whose Clause-Sets are all l.o. provided that this strategy minimizes the number of unique-nodes in its top-part. This is an important property to be used later.

[102] This is an intermediate Set produced by CRA+ while converting *{¬1}* {2,*¬3*} {4,*¬5*} {2,*¬6*} {*¬3,4,5*} {4,6,7} to {*¬0*} {1,*¬2*} {1,*¬3*} {*¬2,4,5*} {3,4,6} {4,*¬5*} in T4.

**E-2: Node Equivalence vs. Clause-Set Equivalence revisited**

The notion of equivalence between Clause-Sets used in Section B (Properties: 5,6,10) can be extended through mapping functions as follows:

**Definition 12:** Two Clause-Sets $S_1$, $S_2$ are said to be equivalent via mapping (written $S_1 \approx_m S_2$) **iff** $S_1=S_2$[103] **or** there exists a Mapping# function M and a mapping *m* such that: $M(S_1,m)=S_2$ or $M(S_2,m)=S_1$.

As GSPRA+ uses an algorithm to determine whether a node/SRT has been already created or not to avoid redundancies and minimize nodes (c.f. steps: 2-a-iii & iv, 2-b-vi & vii in the definition of GSPRA+), this effort can be restricted to comparing already resolved Clause-Sets with newly created ones in each step. Because of Property 10 (Algorithmic Equivalence = Syntactical Equivalence), this implies the necessity of investigating whether the above equivalence-via-mapping concept disturbs that property or not. It could disturb the property only in case "Syntactical Equivalence" and "equivalence-via-mapping" were not one and the same property as formalized in the following lemma.

**Lemma 8:** Assuming $S_1 \approx_m S_2$, then there always exist Clause-Sets $S_1$' and $S_2$' fulfilling $S_1 \approx_m S_2$ **iff** $S_1$'=$S_2$'

**Proof:** Per Definition 12 either $S_1=S_2$ and in that case this assertion is trivially valid or there exists a Mapping# function M and a mapping $m$ such that: $M(S_1,m)=S_2$ or $M(S_2,m)=S_1$.
Suppose $M(S_1,m)=S_2$, i.e., literals of $S_1$ can be substituted by others in a way prescribed by $m$ to yield $S_2$. What happens if we apply $CRA^+$ to both, $S_1$ and $S_2$?
Since structures of those Clause-Sets are the same (else substitution of different literal names wouldn't have rendered them syntactically equivalent), connection matrices constructed by CRA will also be the same getting $S_1'=S_2'=CRA^+(S_1)=CRA^+(S_2)$.
(Q.E.D.)

[103] Syntactical Equivalence implies the existence of a trivial mapping between $S_1$, $S_2$.

Lemma 8 asserts that equivalent Clause-Sets used under $GSPRA^+$ can have one common syntactical form, namely the one produced by $CRA^+$. This amounts to instructing $GSPRA^+$ to only store Clause-Sets in its LCS data structure after converting them to a canonical form using $CRA^+$ (which we call further down: CRA-form).
Having established that checking Syntactical Equivalence of Clause-Sets in SRTs produced by $GSPRA^+$ is sufficient to identify redundant nodes/SRTs, a straightforward algorithm will demonstrate how to do that:

**CompareSets: -**
  **Inputs**: two a.a. 3-SAT-CNF Clause-Sets S1,S2 of sizes <=M
  **Outputs**: TRUE/FALSE
  **Steps**: - for all clauses C in S1: Set C'=next clause in S2
    if (CompareClauses(C,C')=FALSE) return FALSE
   return TRUE

**CompareClauses: -**
  **Inputs:** two a.a. 3-SAT-CNF-clauses C1,C2
  **Outputs**: TRUE/FALSE
  **Steps**: - for all literals lit in C1: Set lit'=next literal in C2
    if (lit<>lit') return FALSE
   return TRUE

Note that CompareSets has O(M) primitive operations, since CompareClauses is in O($c$), $c$ constant.

**E-3: Splits in MSTs**
Although MSTs produced by $GSPRA^+$ are not s.o., a lemma about splits (similar to Lemma 3) can be formulated after two definitions:

**Definition 13:** A MST whose Clause-Sets are all l.o. is called: Sequentially-Ordered, Multi-spaced SRT (notation: **$MSRT_{s.o.}$**). A block $B_x$ whose Clause-Sets or derivations thereof (all or part of them) belong to more than one Variable Space is called a Multiple Space Block, **MSB** (notation: $B_x^{S1,S2,..}$,$S_1$,$S_2$,.. Variable Spaces). A node in a space $S_T$ (called: Target Space, TS) which is common between two or more Variable Spaces is called Multiple Space Common-node, MSCN (written: $[q]_{ST}^{S1,S2,..}$,$S_1$,$S_2$,..,$S_T$ Variable Spaces). More formally[104]: A node is called MSCN **if** in step $k$ of the resolution it becomes common child to two or more nodes of different spaces ($[x]^{S1}$, $[y]^{S2}$, $[z]^{S3}$, … in Fig. 46)[105] generated in steps <=$k$. This happens when there exist mappings $M_1$,$M_2$,$M_3$…, such that: $x=M_1(x')$,$y=M_2(y')$,$z=M_3(z')$,…, where $x$, $y$, $z$ are literals in $S_T$, and $x'$, $y'$, $z'$ are

104 c.f. with Definition 5, Section A)
105 The notation $[x]^{S1}$ is read: Node $[x]$ in Variable Space $S_1$.

literals replaced by TRUE or FALSE in their respective Clause-Sets and respective Spaces. The common-node $[q]_{ST}^{S1,S2,..}$ contains the first appearance of its name literal (NL) $q$ in all branches of the $MSRT_{s.o}$ containing $[x']^{S1}$, $[y']^{S2}$, $[z']^{S3}$, … etc. and there exist literals $q'$, $q''$, $q'''$, etc. in Spaces $S_1,S_2,S_3,\dots$ such that: $q=M_1(q')=M_2(q'')=M_3(q''')=\dots$ etc.

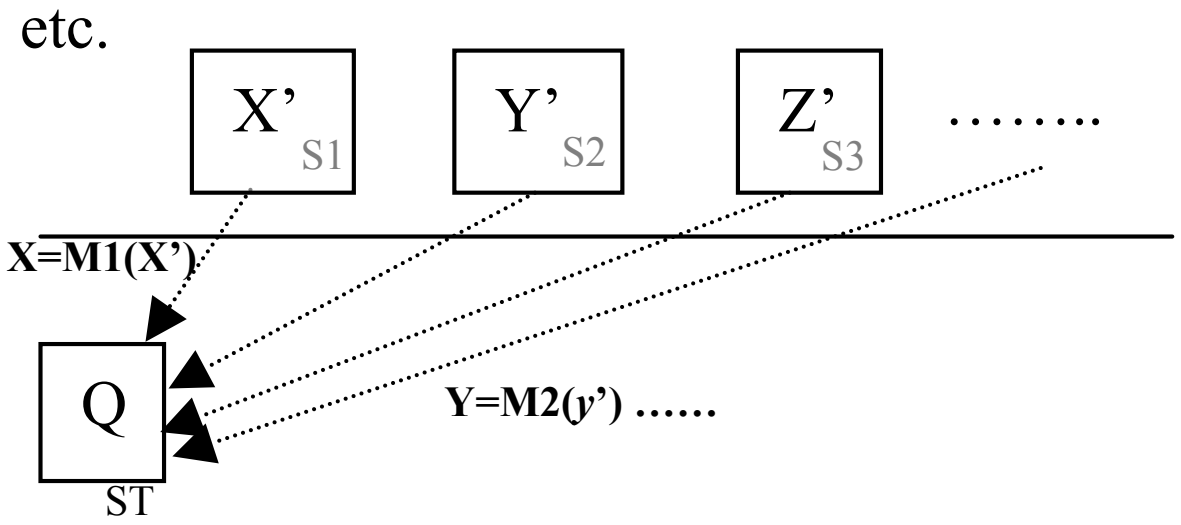

**Figure 46:** Multiple Space Common-Node (MSCN)

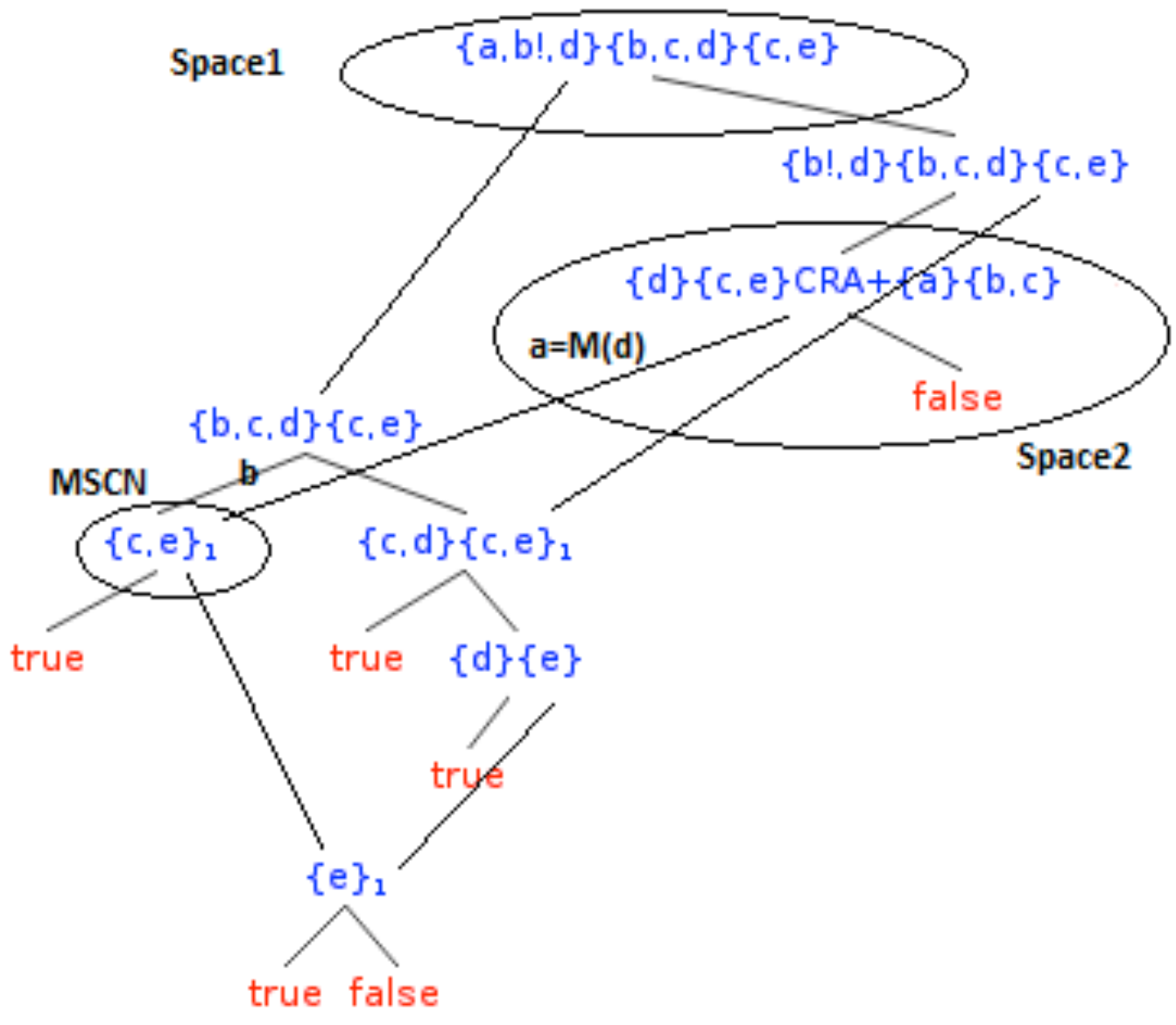

**Figure 47:** Illustration of Definition 13 where $S_T$=Space$_1$, $[c]_{ST}^{Space2}=\{\{c,e\}\}$ is an MSCN, $B_c^{ST,Space2}=\{\{c,d\}\{c,e\}\}$, $M=\{(d>a), (c>b), (e>c)\}$. $[c]_{ST}^{Space2}$ is obviously child to both, $[b]^{ST}=\{\{b,c,d\}\{c,e\}\}$ and $[a]^{Space2}=\{\{a\}\{b,c\}\}$ with edge-literals $b$ and $a=M(d)$.

**Definition 14:** An MSCN is called DS-MSCN$_z$ (Double-Sided MSCN with respect to literal $z$)[106] **if** there exist at least two edge- or branch-literals $x$, $y$ from Spaces $S_1$, $S_2$ respectively and a literal $z$ from the target space $S_T$ such that: $x=M_1(z)$, $y=M_2(z)$, $y=\neg x$.

Literals $x$ and $y$ are also called distinguished (c.f. Lemma 2 and 3).

If an edge- or branch-literal $\boldsymbol{a}$ from Spaces (for which an MSCN is a sink) is not distinguished then the MSCN is called SS-MSCN$_a$ (Single-Sided MSCN with respect to $\boldsymbol{a}$). An MSCN is called trivial MSCN, tMSCN, if it is formed through a newly resolved clause in step k, who belongs to an MSB to which one or more of its parents belonged in steps <k.

$[c]_{ST}^{Space2}=\{\{c,e\}\}$ in the example above of Figure 47 is thus a SS-MSCN$_b$. We are now ready for the following important lemma which is basically a generalization of Lemma 3 of Section C:

**Lemma 9:** For any trees produced while resolving an a.a. 3-SAT-CNF Clause-Set S the following is valid:

a) MSCNs generated by $GSPRA^+$ in step $k$ and containing rank 3 clauses cannot be split in any step $>k$.

b) The only splits possible in IRTs generated by $GSPRA^+$ for S are trivial ones related to size 1 MSCNs. Their maximum number is $RCC_{3\text{-}SAT}$ for any MSCN in any step.

c) For SPR-like- or unlike resolution-procedures: If all IRTs produced during resolution of S are free of BigSps then the final SRT must be an $MSRT_{s.o.}$

**Proof:** We recall the generic form of an MSCN $[q]_{ST}^{sp1,sp2,sp3,..}$ whose edge- or branch-literals can be either distinguished or not:

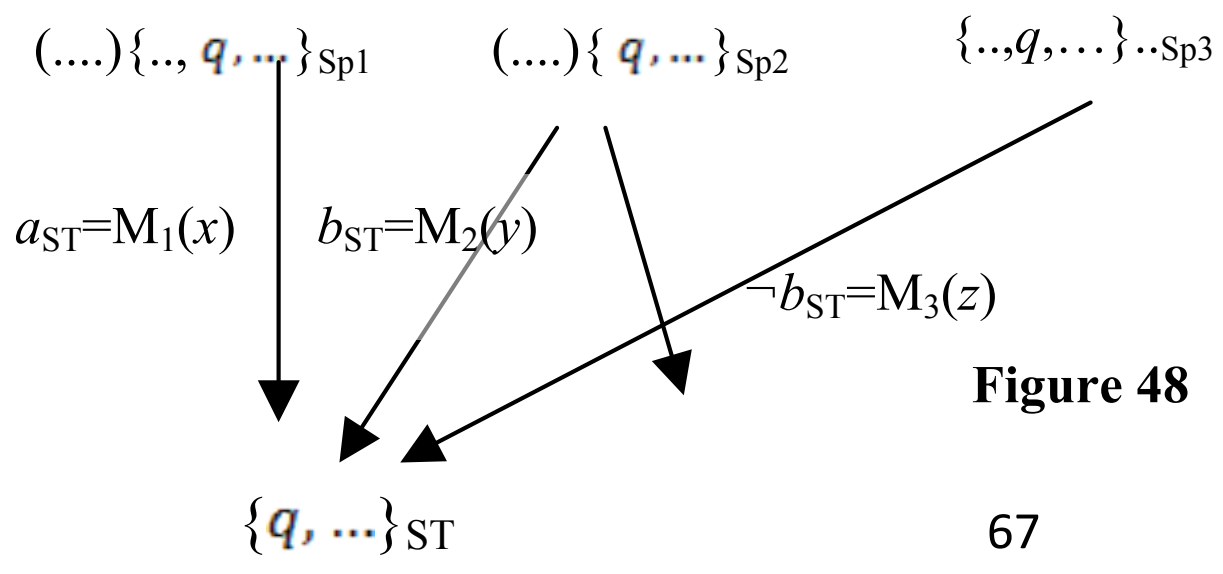

**Figure 48**

106 c.f. definitions provided here with definitions of DSCNs and SSCNs in Lemma 2.

a- If $[q]_{ST}^{sp1,sp2,sp3,..}$ has a rank 3 clause C'={$a$',$b$',$c$'}$_{ST}$ among its clauses, then obviously, when C' was added, equivalent images $C_i$ of a certain clause C from S (the Base-Set) traversed all branches leading to the MSCN in respective spaces $Sp_i$ to augment $[q]_{ST}^{sp1,sp2,sp3,..}$ with C'[107]. All literals of those images $C_i$ were >= literals occurring to their left in their respective Clause-Sets in all branches and all spaces, because of the l.o. condition. Thus:

$C_1$={$M^{-1}_1(a')$,$M^{-1}_1(b')$,$M^{-1}_1(c')$}$_{Sp1}$, $C_2$={$M^{-1}_2(a')$,$M^{-1}_2(b')$,$M^{-1}_2(c')$}$_{Sp2}$, $C_3$={$M^{-1}_3(a')$,$M^{-1}_3(b')$,$M^{-1}_3(c')$}$_{Sp3}$, etc…

were clauses added to parent Clause-Sets of $[q]_{ST}^{sp1,sp2,sp3,..}$ in all respective branches. On the other hand: Any clause attempting to split $[q]_{ST}^{sp1,sp2,sp3,..}$ after it is created through any Space $Sp_i$ must traverse those branches again and contain literals <$M^{-1}_i(a')$ in $Sp_i$. This contradicts the l.o. assumption imposed on all Clause-Sets and all spaces.

b- Similar to the argument used in the proof of Lemma 3(c): Supported- and direct-parent MSCNs[108] cannot cause a split after the size of an MSCN is increased anyway, because they already started an MSB in their respective spaces different than the one they were instantiating when the MSCN was created. Suppose the size of $[q]_{ST}^{sp1,sp2,sp3,..}$ is augmented through a clause of the same MSB, one or more of [$q$]'s parent(s) are instantiating to become >1. This obviously cannot be done – after the MSCN formation - through distinguished edge- or branch-literals or their images, because such an attempt would result in a split. Suppose the MSCN is augmented by the non-distinguished literal $a_{ST}$ and its images (which becomes CNAL in that case). It means that equivalent images of the same derivation of a clause of S, say C'', are passed through all branches and edges of different spaces to the Clause-Set of $[q]_{ST}^{sp1,sp2,sp3,...}$. This can only happen if all other edge- or branch-literals of this MSCN (and their images in all involved spaces representing literals other than $a_{ST}$) were not present in C'' and/or its images. Otherwise more than one derivation would have resulted and would have caused a split. An attempt to split this MSCN in a further step using literals other than $a_{ST}$ or its images is therefore only fruitful if Clause-Sets of the form:

**{{*b*,..}…{<no *b*>}…{…*b*…}..}$_{sp\boldsymbol{i}}$**

for arbitrary literals $b$ are allowed in an $Sp_i$. This is not the case since all Clause-Sets in all spaces must be l.o. On the other hand: Literal $a_{ST}$ itself or any of its images cannot be used for splitting the node because all branches in all spaces must have

107 All images $C_i$ traversing branches leading to the MSCN need to be equivalent (via mapping) otherwise a split would occur. Moreover, they need to be images of C (vs. a derivation of it) because C' is of rank 3.

108 A formal definition of supported- and direct-parent MSCNs is omitted here to avoid unnecessary length. Their definition is similar to the one given for the single-spaced case (c.f. Definition 5, Section A).

agreed upon its instantiation when the node was increased in size. This means that an MSCN whose size becomes >1 in a step cannot be split further in any subsequent step. What about tMSCNs which relate to symmetric and dissymmetric MSBs and which may be augmented to sizes >1 before formation?[109] Recall that per Lemma 3: L.o.s. conditions imposed on SBs and MBs were sufficient to avoid the formation of similar single spaced tCNs. Is this also possible for tMSCNs? To see that it is indeed the case, it is sufficient to remember that, if not scattered, MSBs are always represented in one CRA-form. This means that even though they span different spaces, a target space can always be found in which those blocks are expressed. In such a case all results of Lemma 3 are applicable, i.e. tMSCNs can be avoided, because l.o. and l.o.s. conditions produce equivalent SRTs. If MSBs are scattered : Fragments in different spaces are, again, represented in CRA-form allowing all clauses headed by an MSB-head-literal to be sorted in a way preserving the property that l.o.s. and l.o. conditions produce the same SRTs. The only difference between those two conditions being related to positions of –ve and +ve occurrences of the MSB-head-literal, it is irrelevant which occurrences come first, since only the instantiation of this literal in the whole scattered block is producing the tMSCN (compare with the proof of Lemma 3-c). In Summary: When MSCNs are increased to sizes >1 they cannot split while tMSCNs can be avoided altogether whether MSBs are scattered or not. This means that there are at most only size 1 splits.

As seen before in Lemma 3(c), the number of possible derivations of a new to-be-resolved clause in $GSPRA^+$ is $RCC_{3\text{-}SAT}$. Therefore, only $RCC_{3\text{-}SAT}$ splits of any size 1 MSCN are possible per step in the worst case.

c- If all IRTs produced during resolution are free of BigSps: This means that neither Condition b) nor c) of Definition 1 were breached in the course of resolution in all Clause-Sets of all IRTs (their breach causes N- and rank 3 CN-splits respectively), i.e., all Clause-Sets in all subsequent IRTs (including the final one) were l.o.. Hence, the final one is an $MSRT_{s.o.}$ as well. Note that this assertion is weaker than the one made in Lemma 3(d), but generic enough to allow a broader understanding of properties not only of $GSPRA^+$ and SPR-like procedures, but also of SPR-unlike ones (for reasons stated in Section E-5).

(Q.E.D.)

**E-4: Way of Work of $GSPRA^+$**

As in the case of s.o. SRTs, Lemma 9 guarantees that using $GSPRA^+$ produces only trivial splits which is a significant indication concerning its efficiency. It is still necessary to understand what

[109] Formal Definitions of symmetric and dissymmetric MSBs are not given here to avoid unnecessary length. The reader is referred to Lemma 2 to recall those notions.

$GSPRA^+$ really does when it imposes the l.o. condition on clauses.

**Definition 15:** An $MSRT_{s.o}$ produced by $GSPRA^+$ for an arbitrary 3-SAT-CNF Clause-Set S is said to be aligned **if** ∃C ∈S:∀*n* node not leaf ∈$MSRT_{s.o.}$,

∀S', X where: S' Clause-Set of *n*,

X ∈ S' the following is true :

a) SortOrder(C **or** its derivations C',S') >SortOrder(X,S')
b) S' is l.o.

In other words: Either C or one of its derivations C' are the last clauses in any Clause-Set of the $MSRT_{s.o.}$. C is called alignment-clause.

**Definition 16:** A node *n* of size M is said to be *aligned* **if**:

a) For M<=2: *n* possesses a Clause-Set with an aligned $MSRT_{s.o}$
b) For M>2:
   (i) All sub-nodes of size M are l.o.
   (ii) All sub-nodes of size <M are aligned

The Set of all unique clauses and their derivations used for the alignment of all nodes of an $MSRT_{s.o}$ of an arbitrary 3-SAT-CNF Clause-Set S is called Alignment Clause-Set of (ACS). Obviously, ACS cannot have more than $RCC_{3\text{-}SAT}$*M elements/clauses containing all possible permutations of literals in linear- or non-linear sequences.

**Lemma 10:** All size 1,2 nodes of any $MSRT_{s.o}$ of a 3-SAT-CNF Clause-Set S produced by $GSPRA^+$ are aligned.

**Proof:** For size 1 nodes it is clear that the $MSRT_{s.o}$ representing any single clause is aligned per definition with the single clause itself being the alignment-clause. For size 2 nodes of the form S={{*a,b,c*}{*x,y,z*}}let's recall that $GSPRA^+$ converts any such Clause-Set to a l.o. Clause-Set using $CRA^+$ (step 3 in the $GSPRA^+$ algorithm description). This leads to the following cases:

**Case 1 (Fig. 49)**: No literals are common between the two clauses. The clause {*x,y,z*} is then the alignment-clause (for this type of situation, c.f. for example the circled sub-tree {2!,3,4} of T3 in Fig. 44). Here one of the general forms:

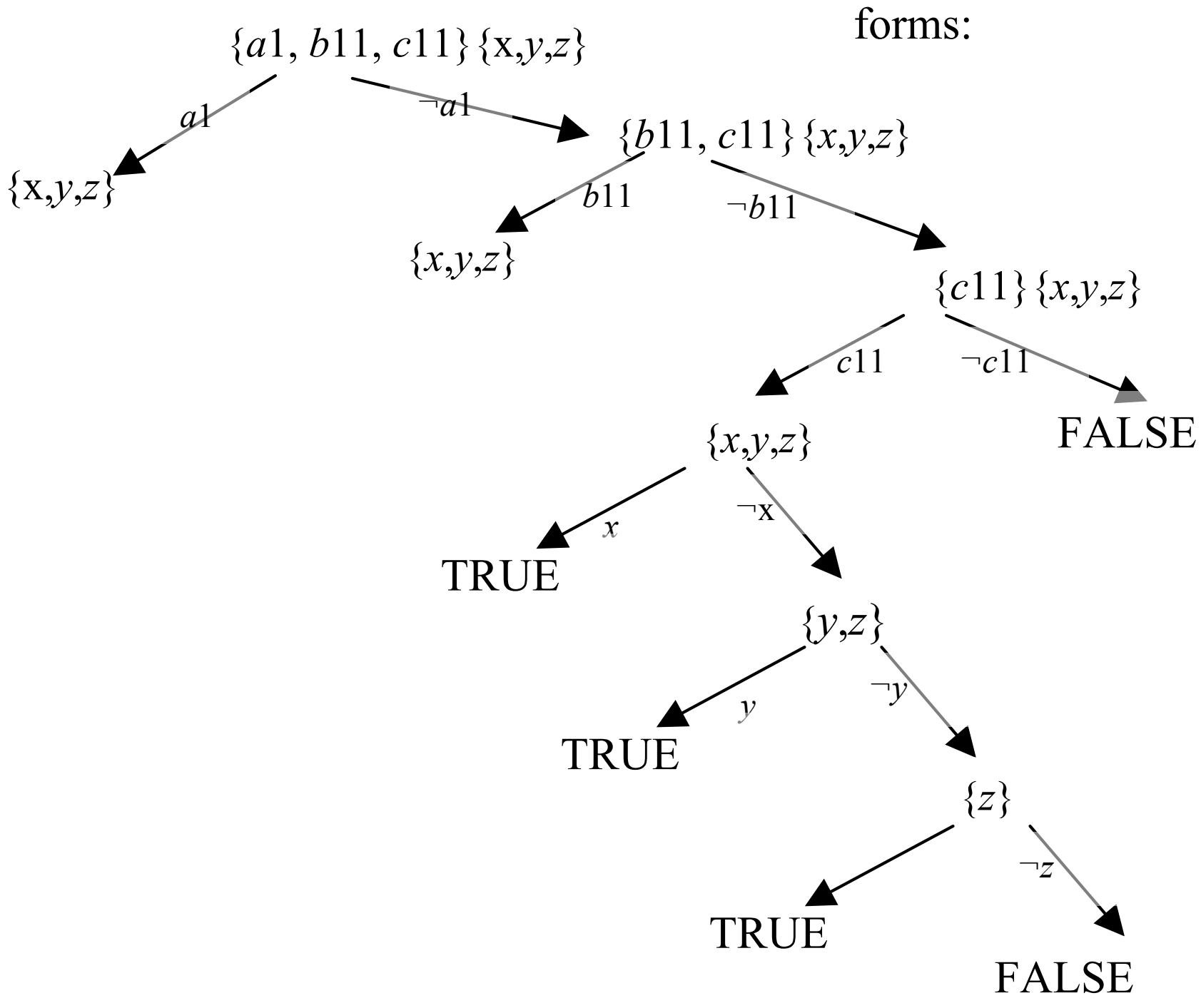

**Figure 49**

**Case 2 (Fig. 50)**: There is one literal in common independent of the specific place of this literal. Because of the renaming precedence condition of CRA, all Clause-Sets will be converted via CRA$^+$ to the form *{a,b,c}{a,y,z}* which has *{a,y,z}* as alignment-clause[110]:

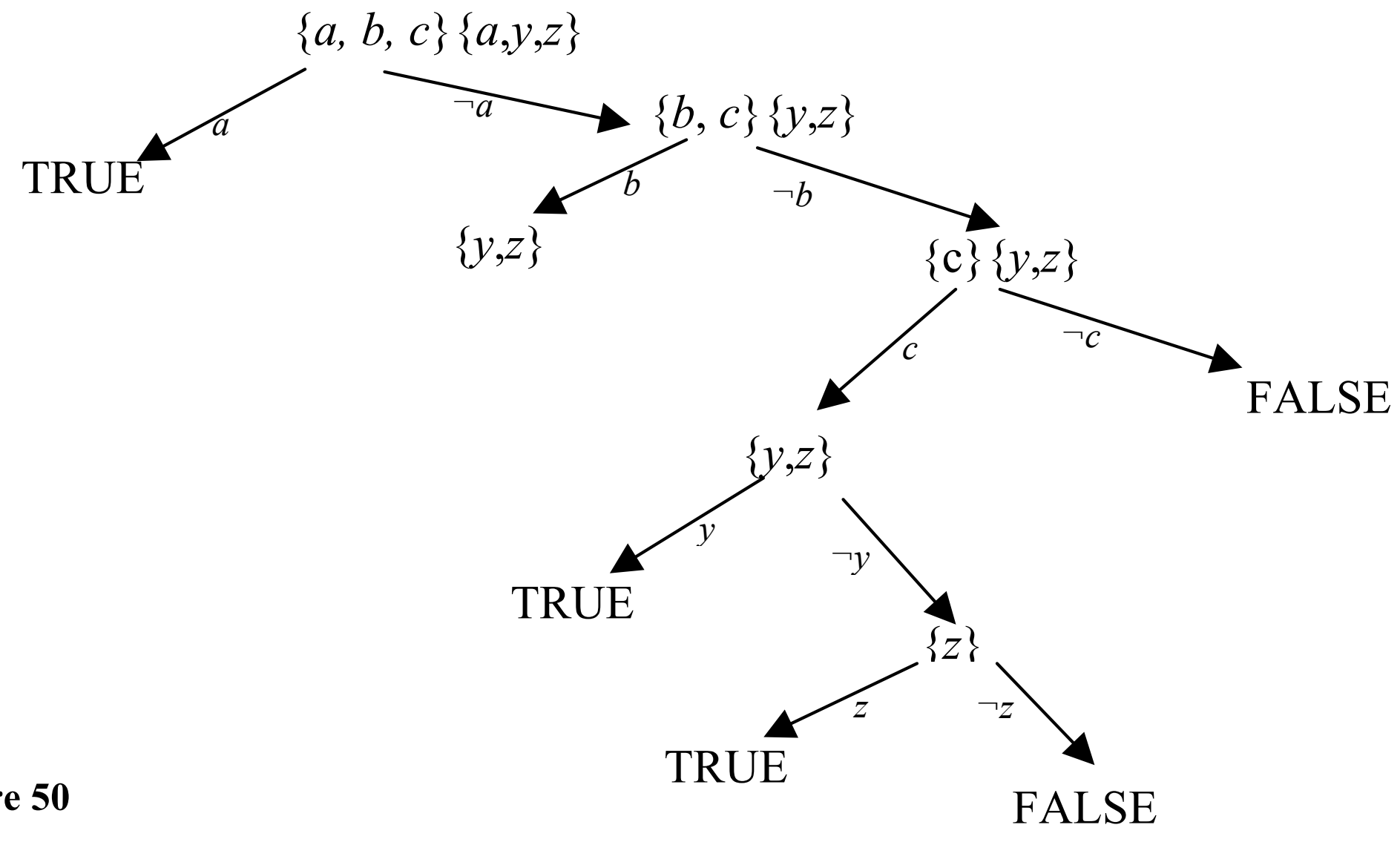

**Figure 50**

[110] Only clauses with *+ve* literals are used in the illustration (w.l.o.g.), since *-ve* signs are irrelevant in the current context.

**Case 3**: There is more than one literal in common. Because of the renaming precedence condition all Clause-Sets will be converted via CRA$^+$ to the form $\{\{a,b,c\}\{a,b,d\}\}$. This form has $\{a,b,d\}$ as alignment-clause and has a general form similar to the above tree (Fig. 50). (Q.E.D.)

**Lemma 11:** GSPRA$^+$ produces aligned MSRT$_{s.o}$s and possesses all properties of GSPRA except the linear derivation property (Property 3).

**Proof:**

1. **Aligned MSRT$_{s.o}$s:** M=3-sized nodes are aligned because their M=2-sized sub-nodes produced by GSPRA$^+$ in a resulting MSRT$_{s.o}$ are all aligned (Lemma 10) and (as per Definition 16) their M=3-sized sub-nodes are l.o. The fact that all size M=3 sub-nodes are aligned makes in the same way all size M=4 nodes aligned and so forth. In general: All M-sized nodes are aligned because all their M-1-sized sub-nodes are aligned and their M-sized sub-nodes are l.o. This implies that the final MSRT$_{s.o}$ is aligned.
2. **Properties** (c.f. Section II B):
   a) **Property 1 (Completeness, truth table equivalence)**: It is sufficient to see that GSPRA$^+$ only renames variables/literals in Clause-Sets whereby per definition of a truth table, this operation doesn't affect any argument related to truth table equivalence.
   b) **Property 2** (**Expansion of MSRT$_{s.o}$s**): GSPRA$^+$ uses the least-literal/clause-rule (c.f. steps 2 a) and b) in the Align Algorithm). This rule is applicable within a space as well as between spaces in the same way described in the proof of this property in Section B. Thus: If nodes $n_1$ and $n_2$ belong to different spaces and were not directly connected in step $k$, they cannot be directly connected in steps $>k$ because newly resolved clauses don't affect old results of an application of the least-literal-rule (except for tMSCN cases like the one shown in the proof of Section B). Also: Nodes of sizes $j$<=M generated in step $k$ are at most as many as nodes of sizes $j$-1 existent in step $k$-1 (not counting nodes generated through splits in level $j$), since newly resolved clauses have to traverse all branches of IRTs, if necessary. Assuming only unique-nodes are permitted, then we have for the worst case, i.e., the case where all $j$-sized nodes in step $k$-1 become $j$+1-sized in step $k$, the following important consequence to be used in Lemma 13:
   There are as many newly generated nodes at any size-level $j$<=M in step $k$ as there are newly generated nodes at size-level $j$-1 in step $k$-1, not counting splits produced in $k$ at size-level $j$[111].
   c) **Property 4** (**Generation of non-trivial CNs**): Has to be slightly changed to the following property:
   **Property 4' (Generation of non-trivial CNs/MSCNs in GSPRA$^+$)**: The only non-trivial CNs/MSCNs generated in any step $n$ by GSPRA$^+$ while resolving any clause C of a Set of 3-SAT-CNF-clauses are identical with either C or derivations (**not**

[111] Illustration of the worst case: Suppose we have $x$ size-1 nodes in step 0 (and no other sizes in the tree), then additional $x$‘ size-1 nodes are generated in step 1, another $x$‘‘ size-1 in step 2 etc. This means that size-2 nodes in step 1 will be $x$ and become $x$+$x$‘ in step 2, $x$+$x$‘+$x$‘‘ in step 3 etc., not counting splits in size-2 level.

**necessarily linear**) of C which are $\in$ ACS.

To see that this property holds suppose a non-trivial CN/MSCN is generated in step *k* (either between nodes of the same- or different spaces) which is neither C nor a derivation of it. This implies that a "legacy" node constructed in steps <*k* became non-trivial CN/MSCN in step *k*. This can only mean that at least one new connection has been established in step *k* between two nodes which were previously not connected. This is only possible in the trivial case as per the Linear Expansion Property 2. Moreover, as renaming actions can alter the sequence of literals in any clause, this means that derivations of C forming CNs/MSCNs could also be non-linear. C and all its derivations are $\in$ ACS.

d) **Property 5 (Uniqueness of instantiation results**) is a general property valid for a.a. Clause-Sets and SPR-like algorithms using the least-literal-rule and thus valid for all l.o. Clause-Sets in $MSRT_{s.o}$s of $GSPRA^+$ as well.

e) **Property 6 (Syntactical Equivalence**): Identical with Property 5.

f) **Property 7 (FBDD Equivalence**, **Branch-Linearity**): The argument in Section B uses the least-literal-rule also used by $GSPRA^+$ which basically makes variables/literals disappear from all child-nodes of a Clause-Set once they are instantiated explaining why a variable/literal can appear in a branch only once. The difference between GSPRA and $GSPRA^+$ in this respect is that $GSPRA^+$ changes names of variables/literals using $CRA^+$. This doesn't affect this property. Part b) is also still valid for $GSPRA^+$

g) **Property 8:** Has to be slightly changed to the following property:

**Property 8' (Minimal top-part):** $MSRT_{s.o}$s produced by $GSPRA^+$ for a.a. 3-SAT-CNF Clause-Sets S of size M have a minimal unique-node-count in their top-part compared to any SPR-like- or unlike procedure.

This property holds per definition of $GSPRA^+$ for SPR-like procedures (c.f. step 2, SelectFirstClause procedure). Note that the difference between GSPRA and $GSPRA^+$ related to the top-part of an SRT/$MSRT_{s.o}$ is that $GSPRA^+$ minimizes the number of unique-nodes and uses $CRA^+$ to convert Clause-Sets to l.o. Clause-Sets. $GSPRA^+$ avoids redundant <M-sized nodes, since it searches LCS for any already created similar Clause-Sets/nodes and links them to the currently processed one if found (c.f. steps 2-a-iii,iv and 2-b-vi,vii). To see why $GSPRA^+$'s unique-node-count in the top-part of any $MSRT_{s.o}$ is minimal for SPR-unlike procedures as well, recall that such a procedure PR must use more than one clause for instantiation in any top-part of generated trees for Clause-Sets of size M. All literals of one of those clauses have to be instantiated to complete such a top-part (per definition of Top-parts in Property 8, Section B). Suppose the best count of unique-nodes reachable by PR in such a top-part is $\boldsymbol{v}$. Suppose now that instantiating all literals of the shortest clause C of S generates $\boldsymbol{w}$ unique-nodes, then obviously ***v*>=*w***, because PR could pick a clause which is not shortest to fully instantiate. Moreover, even if it picks the shortest, it must still instantiate at least one literal from another clause which is an operation generating at least one additional unique-node of size M and not necessarily unique node of size <M in such a top-part. As GSPRA$^+$, uses the SelectFirstClause procedure it checks all clauses, including the shortest. For the least unique-node-count, it will produce ***w*** or less (c.f. Argument in Property 8 Section B for reference and comparison). Here is another argument using refutation : Suppose PR produces the best count, say **v**, of unique nodes in the top-part of a resolution tree. This must have been produced through instantiation of literals of at least two clauses of S : $C_1$, $C_2$, one of which, say $C_2$, until it is fully instantiated. Let us say that the full instantiation of $C_2$ alone produces **w** unique Clause Sets. Obviously **v>=w**, since any literals in $C_1$ which are not present in $C_2$ must generate at least one additional unique node of size M and not necessarily unique node of size <M. If this is the case, then GSPRA$^+$ will definitely choose an initial Clause producing **<=w** unique node count, because it must check $C_2$ in the course of its action, contradicting thus the assumption.

h) **Property 9** (**generality of canonical orderings**) is a general property valid for a.a. Clause-Sets independent of the difference(s) between GSPRA and GSPRA$^+$.

i) **Property 10** (**Algorithmic Equivalence = Syntactical Equivalence**) is a general property valid for a.a. Clause-Sets and dependent on the validity of Properties 5,6. Lemma 8 enhanced Syntactical Equivalence with the idea of CRA-form as seen above.

(Q.E.D.)

It is about time to find an upper bound for the total number of nodes in $MSRT_{s.o}$ produced by GSPRA$^+$. The following lemma is straightforward:

**Lemma 12:** $MSRT_{s.o}$s produced by GSPRA$^+$ for 3-SAT-CNF Clause-Sets of Sizes M=1, M=2 have at most 3*M, O(M) non-leaf-nodes.

**Proof:** As seen in the proof of Lemma 10 above: $MSRT_{s.o}$s of M=1 have at most 3 non-leaf-nodes per definition. The alignment procedure for M=2 Clause-Sets produces 3 more non-leaf-nodes at most (in the case when there are no literals in common between both clauses).

(Q.E.D.)

The following lemma is an important assertion related to the upper bound of unique-nodes produced by GSPRA$^+$.

**Lemma 13:** In any step *i* of GSPRA$^+$ resolving an a.a. Base-Clause-Set S of size M[112] with clause $C_{i-1}$: Newly added size 1 nodes used to align any sub-trees of Clause-Sets S' of size <M produced in steps <*i* can **only** be $\in$ ACS[113]. The total number of unique-nodes produced

112 When *i*=1, M=2 with clauses $C_0$, $C_1$

113 The realization that aligning a clause to an $MSRT_{s.o}$ is actually the process of rearranging resolution priorities of clauses in all sub-nodes/Clause-Sets of this $MSRT_{s.o}$ (making the base problem equivalent to sub-problems using different clauses of the same S for alignment with sub-trees) is central for this lemma.

by GSPRA$^+$ for S in the final $MSRT_{s.o}$, including those generated by splits, is hence bounded above by:

$$3+ c*RCC_{3\text{-}SAT}^2 *M^4 + RCC_{3\text{-}SAT} *M^3,\ c<=3, \text{ i.e., } \mathbf{O(M^4)}$$

**Proof:** (by induction on M)

**Base-Case: M=1:** As in Lemma 10: For size 1 nodes the $MSRT_{s.o}$ representing a single clause which is aligned per definition, the single clause itself being the alignment-clause. For M=1 we have:

$$i=0: 3 <3+ 3*(15)^2 *(1)^4$$

**Illustration Case: M=2**: The alignment of clause $C_1$ to $C_0$ in step $i$=1 of the resolution adds 3 to the nodes of the $MSRT_{s.o}$ of clause $C_0$ which are also 3 at most (Lemma 12). Thus, for step M=2 we have:

$$i=1: 3+3 <3+ 3*(15)^2 *(2)^4$$

The practically used ACS-portion is comprised of clause $C_1$ and/or its derivations as seen in Lemma 10.

**Induction Hypothesis (size M):**
If an IRT with a base-node of size M in the form of Fig. 51 is produced by adding in each step only elements of the ACS to the size 1 nodes levels (while aligning clauses to the intermediate IRTs of previous steps), the total number of unique-nodes, including those resulting from splits, not exceeding $3+c*RCC_{3\text{-}SAT}^2 *M^4 + RCC_{3\text{-}SAT}*M^3$ in this IRT, then:

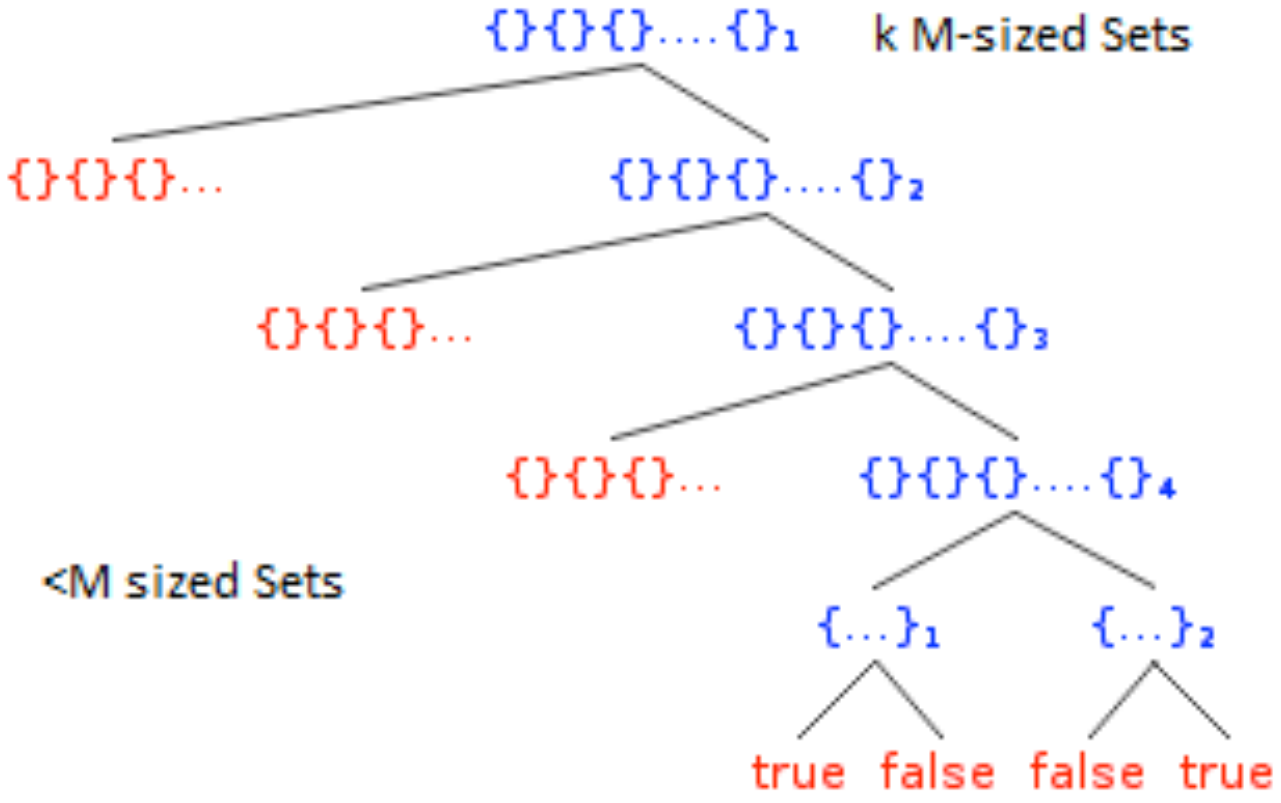

**Figure 51: IRT with** base-node **M**

**Induction Step (size M+1):**
When IRT is resolved in step $i$ via GSPRA$^+$ with a clause $C_{i-1}$:

1. $k$ M-sized nodes shall become $k$ M+1-sized nodes and l.o. as well (per definition of GSPRA$^+$ and the fact that the base Clause-Set is l.o.). The breadth $k$ of the first clause $C_0$ in S is not altered. No other M+1-sized nodes can be formed.
2. For <M-sized nodes (when they are resolved with $C_{i-1}$ forming nodes of Sizes <= M) the induction hypothesis applies, i.e., step $i$ produces for each one of them at most $|ACS|=RCC_{3\text{-}SAT}*M$ new nodes of size 1 in their respective sub-trees (not counting trivial size 1 splits). Suppose that a clause $C=C_i$ is aligned to such a node $n$ (Fig. 52) needing for the alignment of sub-trees of $n$ (not necessarily in the same space) some other clauses C', C'' ϵ ACS.

   If two or more $MSRT_{s.o}$s of node $n$ or any other node are aligned with the same clause C, C' or C'', then a CN/MSCN possessing one unique CRA-form (Lemma 8) will be built within a space or between different spaces for each one of C, C' or C''. In addition: All such non-trivial CNs/MSCNs[114] can only represent members of ACS as per Property 4'. Thus, the total number of newly formed unique size 1 nodes for all nodes and sub-nodes in this step (which may or may not become non-trivial CNs/MSCNs) cannot exceed |ACS| in the worst case, i.e., $RCC_{3\text{-}SAT} *M$.

114 Trivial CNs/MSCNs are not accounted for, because they can be avoided w.l.o.g. as per Lemma 9.

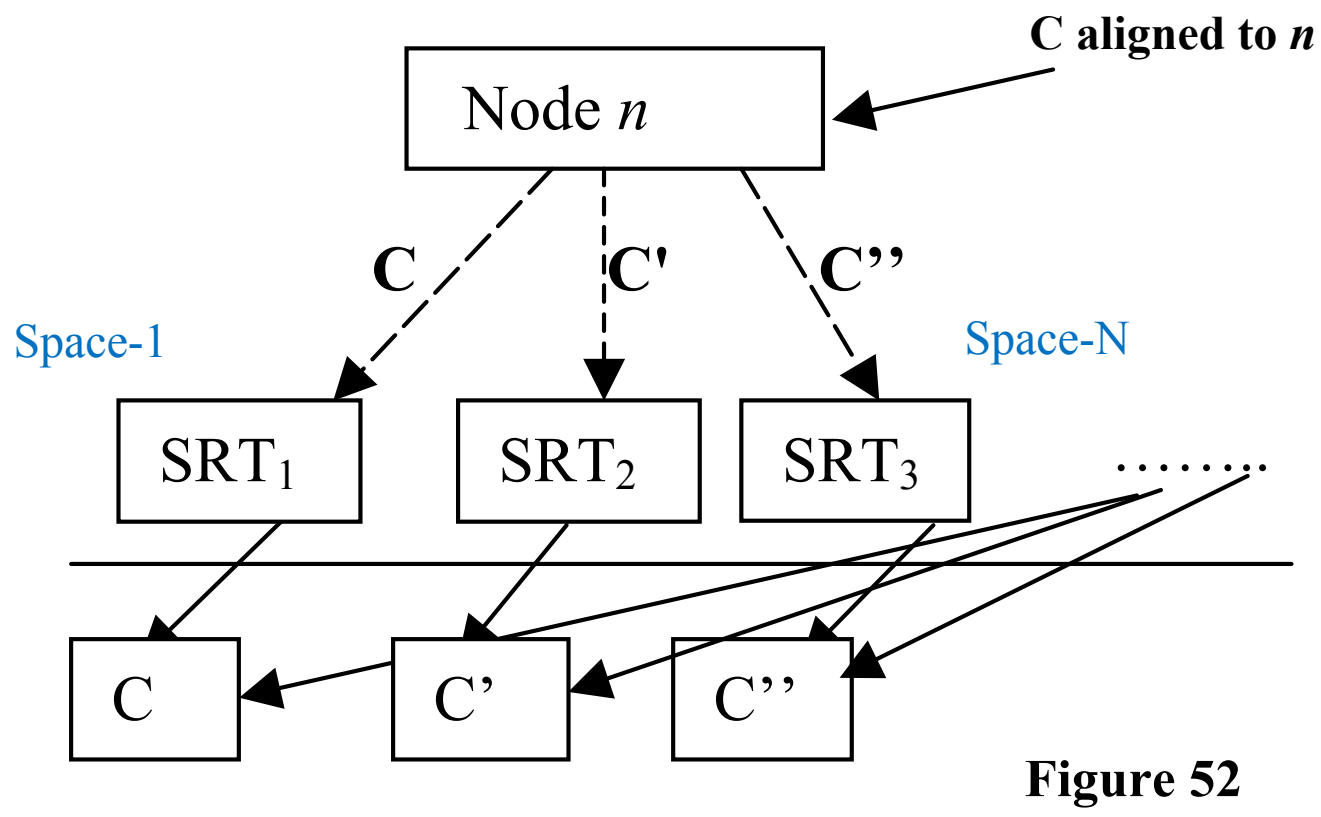

**Figure 52**

3. As per 2., the total number of generated non-trivial CNs/MSCNs cannot exceed $RCC_{3\text{-SAT}} *M^2$ in any step. Assuming for the worst case that each one of those nodes is split by the newly resolved clause in step *i*: Lemma 9(b) states that there are $RCC_{3\text{-SAT}}$ ways to do this for any CN/MSCN and that this split is for size 1 nodes only, i.e., it produces only a constant amount *c*(=<3) of new nodes for each split[115].
   This takes the maximum amount of newly added size 1 nodes (including ACS-elements of point 2) in step *i* to:

   $$c*RCC_{3\text{-SAT}}^2 *M^2+ RCC_{3\text{-SAT}} *M$$

   What about newly added nodes of sizes >1? Lemma 9 assures that there are no splits of nodes of sizes>1. Moreover, Property 2, the Expansion Property of $MSRT_{s.o}$s (Lemma 11) asserts that for the worst case, i.e., when all size-*j*-level nodes are assumed to become size-*j*+1-level during resolution, the number of new size *j* (<=M) nodes in any step cannot exceed the number of new size *j*-1 nodes created in a previous step (if splits at level *j* are not counted as is the case here). This indicates for step *i* that the previously created

   $$c*RCC_{3\text{-SAT}}^2 *M^2 + RCC_{3\text{-SAT}} *M$$

   nodes of size 1 (created in step *i*-1) may in the worst case all be propagated up the hierarchy of sizes to form for each size-*j*-level of nodes

   $$c*RCC_{3\text{-SAT}}^2 *M^2 + RCC_{3\text{-SAT}} *M$$

   new ones in that level in the worst case[116]. This means that

   $$c*RCC_{3\text{-SAT}}^2 *M^3 + RCC_{3\text{-SAT}} *M^2$$

   new nodes are added in step *i* in all levels at most, thus confirming the given $O(M^4)$ bound.

(Q.E.D.)

**E-5: Minimal $MSRT_{s.o}$s**

In this section we prove that an $MSRT_{s.o.}$ produced by $GSPRA^+$ is minimal compared to outputs of SPR-like- or unlike resolution-procedures[117] using canonical orderings and working with non-l.o. 3-SAT-CNF Clause-Sets S causing BigSps. Property 9 (Generality of Canonical Orderings) enables us to conclude then that it is also minimal for similar procedures which use any arbitrary type of orderings. In a further step we show that any $MSRT_{s.o.}$ produced by $GSPRA^+$ is near-to-optimal in terms of the number of nodes. This finding prepares for the next section

115 We are assuming hence that each newly resolved clause in each step *i* comes with a least-literal equivalent to previously instantiated block literals of parent-nodes of every non-trivial CN/MSCN created before in every space and splits this non-trivial CN/MSCN in all possible ways without breaching any l.o. condition, a clear exaggeration.

116 In other words: In any step *i*, nodes of all sizes may be propagated up the size-hierarchy if they were not propagated before, but only non-trivial CN/MSCNs are split at size-level 1. This makes the max number of new size-1 nodes generated in such a step *i* : $O(M^2)$ at all times as seen. Per Lemma 9: No other splits exist in any upper level of sizes, therefore, only those $O(M^2)$ are counted as new additions in every size-level of the hierarchy making the total number of newly added nodes in all levels in step *i*: $O(M^3)$.

117 Remember: SPR-unlike procedures require the use of more than one clause for instantiation in a top-part of a resolution-tree allowing such a top-part not to be minimal.

where FGPRA$^{+}$ (the parallel counterpart of GSPRA$^{+}$) is shown to be a polynomial-time 2-approximation algorithm for the problem of finding the minimal FBDD solving S (MinFBDD Problem).

**Lemma 14:** Suppose S is a 3-SAT-CNF a.a. Clause-Set of size M and PR is any SPR-like- or unlike resolution-procedure producing BigSps using a canonical ordering to resolve S, $C_{PR}$ is the total number of unique-nodes produced by PR for S, $C_{GSPRA+}$ is the total number of unique-nodes produced by GSPRA$^{+}$ for the same S, then the rate of expansion of trees[118] produced by PR in any step (α) is greater than or equal to the rate of expansion of trees produced by GSPRA$^{+}$ in the same step (β),

i.e., ∀*i*<=M:

$$C_{PR}{}^{i}=\alpha * C_{PR}{}^{i-1},$$
$$C_{GSPRA+}{}^{i}= \beta * C_{GSPRA+}{}^{i-1}$$

Where α>=β, and hence:

$$C_{PR} >= C_{GSPRA+}$$ [119]

**Proof:** (Induction on M, the size of S)

**Base-Case M=1:** In this trivial case both, GSPRA$^{+}$ and PR produce an SRT/MSRT$_{s.o.}$ for the only clause of S which per definition have the same number of nodes.

**Illustration Case M=2:** A characteristic Set S fulfilling a.a. conditions can have a literal ***a*** breaching Condition c) in Definition 1 in any node of its tree (like the one chosen in the below example permutations of clauses)[120]. Below are resolution-trees containing BigSps produced by PR compared to MSRT$_{s.o.}$s of GSPRA$^{+}$ for the following cases:

[118] A factor expressing an upper bound of the number of new nodes possibly created in any step

[119] $C_{PR}{}^{i}$ reads: Number of unique-nodes produced by PR for Clause-Sets of size *i*. Similarly for $C_{GSPRA+}{}^{i}$.

[120] Monotone 3-SAT-cases are used (w.l.o.g.), since negation signs are irrelevant for the discussion here. Also: Not all possible permutations of literals are demonstrated for the described cases to avoid unnecessary length.

**Case 1 (Fig. 53):** S={{*b*,*c*,*d*}{*a*,*e*,*f*}} no literals in common between the two Clause-Sets. Alpha=3, Beta=2, $\alpha>\beta$

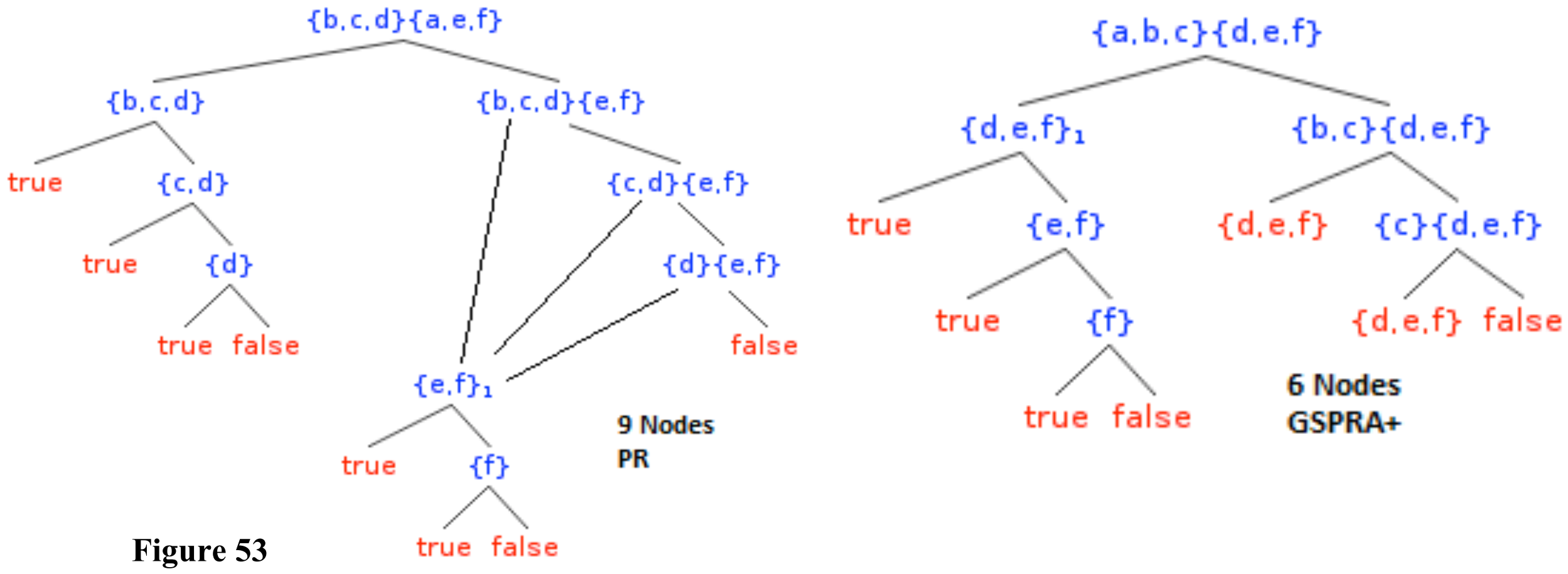

Figure 53

**Case 2 (Fig 54):** S={{*b*,*c*,*d*}{*a*,*c*,*f*}}one literal in common. Alpha=3, Beta=5/3, $\alpha>\beta$

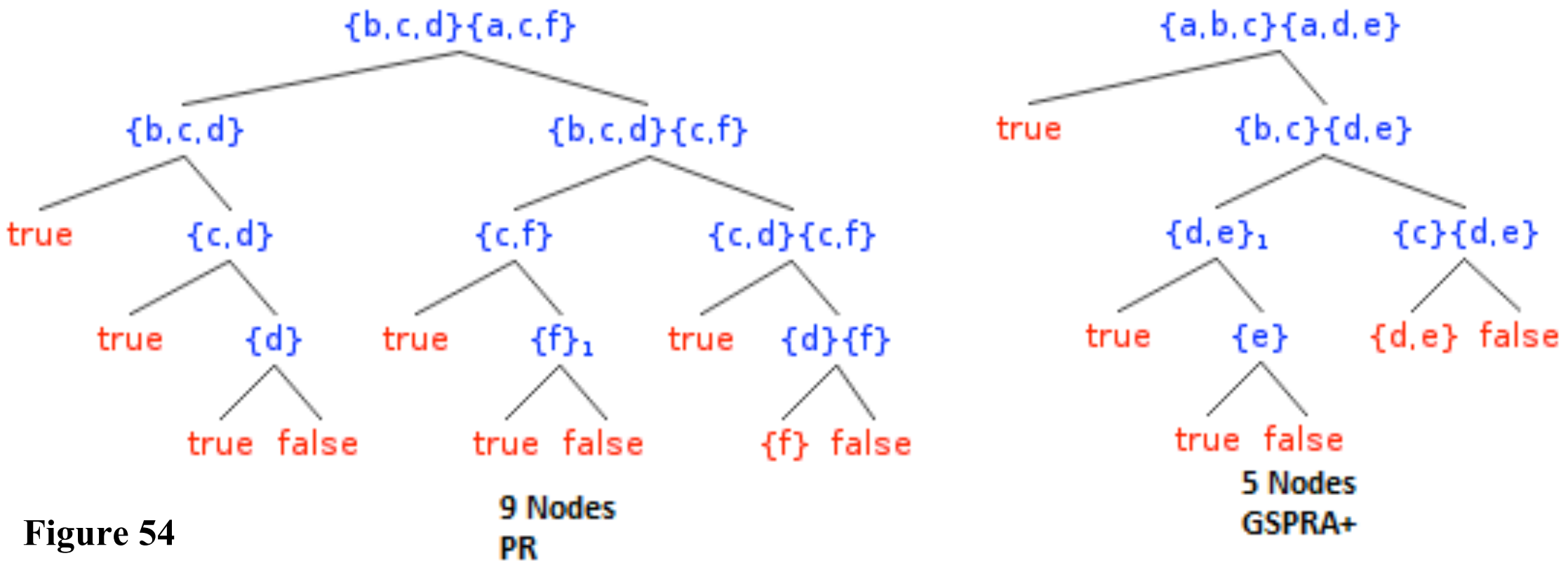

Figure 54

**Case 3 (Fig. 55):** S={{*b*,*c*,*d*}{*a*,*c*,*d*}} two literals in common. Alpha=5/3, Beta=4/3, α>β

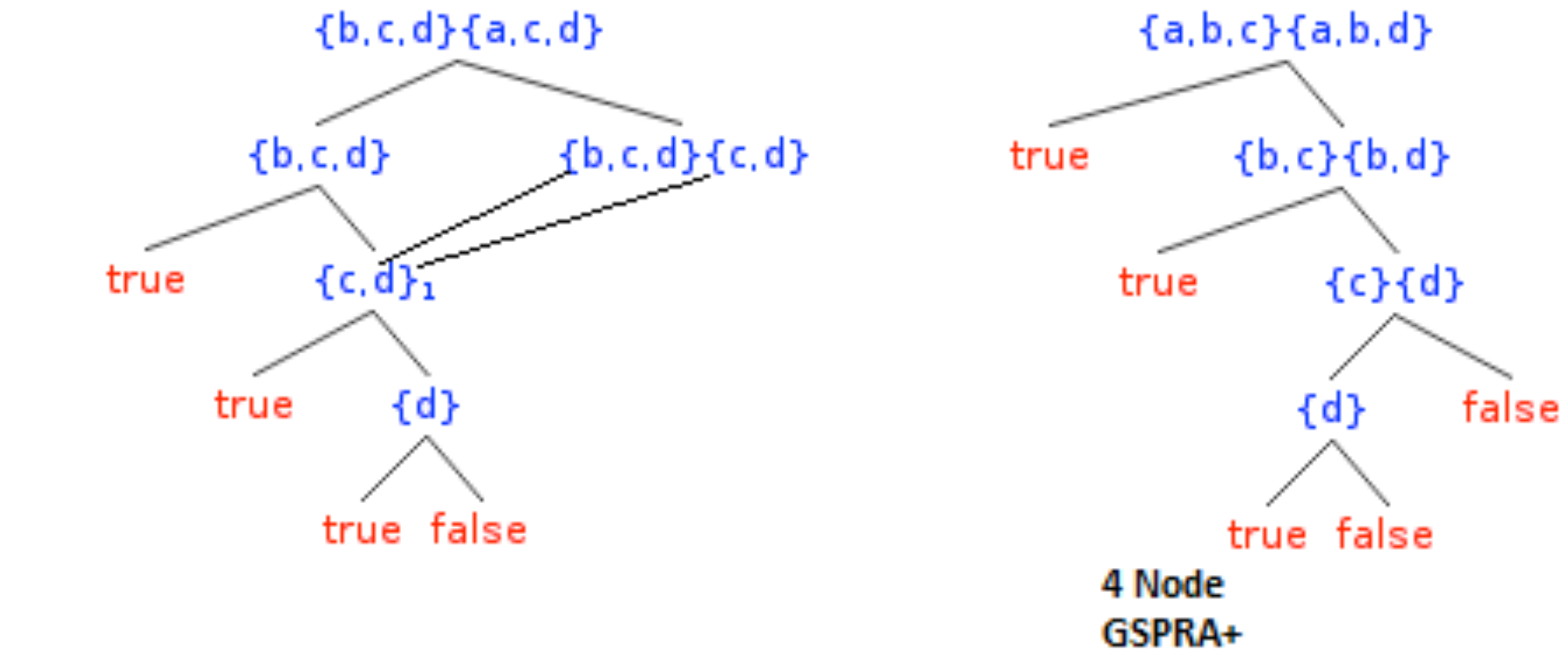

Figure 55

BigSps are clearly causing the larger expansion rates. $MSRT_{s.o.}$s of $GSPRA^{+}$ are minimal in all the above cases.

**Induction Hypothesis:** For S of Size M: $C_{PR}{}^{M} >= C_{GSPRA+}{}^{M}$, $\forall i<=M$:

- $C_{PR}{}^{i}=\alpha * C_{PR}{}^{i-1}$
- $C_{GSPRA+}{}^{i}=\beta * C_{GSPRA+}{}^{i-1}$
- $\alpha >= \beta$

**Induction Step:** If S has size M+1, then per definition of α and the induction hypothesis we know that:

$$C_{PR}{}^{M+1} = \alpha * (C_{PR}{}^{M}) >= \alpha * (C_{GSPRA+}{}^{M})$$

In the same time :

$$C_{GSPRA+}{}^{M+1} = \beta * (C_{GSPRA+}{}^{M})$$

Therefore:

$$C_{GSPRA+}{}^{M} = C_{GSPRA+}{}^{M+1} / \beta$$

Substitution yields:

$$C_{PR}{}^{M+1} >= \alpha * (C_{GSPRA+}{}^{M+1} / \beta),$$

where α>=β which means that:

$$C_{PR}{}^{M+1} >= C_{GSPRA+}{}^{M+1}$$

(Q.E.D.)

A similar lemma for l.o.u Clause-Sets is the following:

**Lemma 15:** Suppose S is a 3-SAT-CNF l.o.u. Clause-Set of size M and PR is any SPR-like- or unlike resolution-procedure producing BigSps using a canonical ordering to resolve S, $C_{PR}$ is the total number of unique-nodes produced by PR for S, where $C_{GSPRA+}$ is the total number of unique-nodes produced by $GSPRA^{+}$ for the same S, then the rate of expansion of trees produced by PR in any step is greater than or equal to the rate of expansion of trees produced by $GSPRA^{+}$ in the same step, i.e., $\forall i<=M$:

$$C_{PR}{}^{i}=\alpha * C_{PR}{}^{i-1},$$
$$C_{GSPRA+}{}^{i}= \beta * C_{GSPRA+}{}^{i-1}$$

where α>=β, hence:

$$C_{PR}{}^{M} >= C_{GSPRA+}{}^{M}$$

**Proof:** (Induction on M, the size of S)

**Base-Case M=1:** As before: $GSPRA^{+}$ and PR produce an $SRT/MSRT_{s.o.}$ for the only clause of S which per definition have the same number of nodes.

**Illustration Case M=3:** A generic Set S fulfilling l.o.u. conditions can have a clause $C_2$ breaching Condition b) of Definition 1 like the one chosen in the below example permutations of clauses. As there are for M=3 much more permutations than can be included here, a representative case is chosen. This is the case where $C_0$ and $C_2$ have only one common literal ***a*** causing the breach and no other literals in common between any two or more clauses[121]. Fig. 56a and 56b illustrate resolution-trees produced by PR compared to $MSRT_{s.o.}$ of $GSPRA^{+}$:

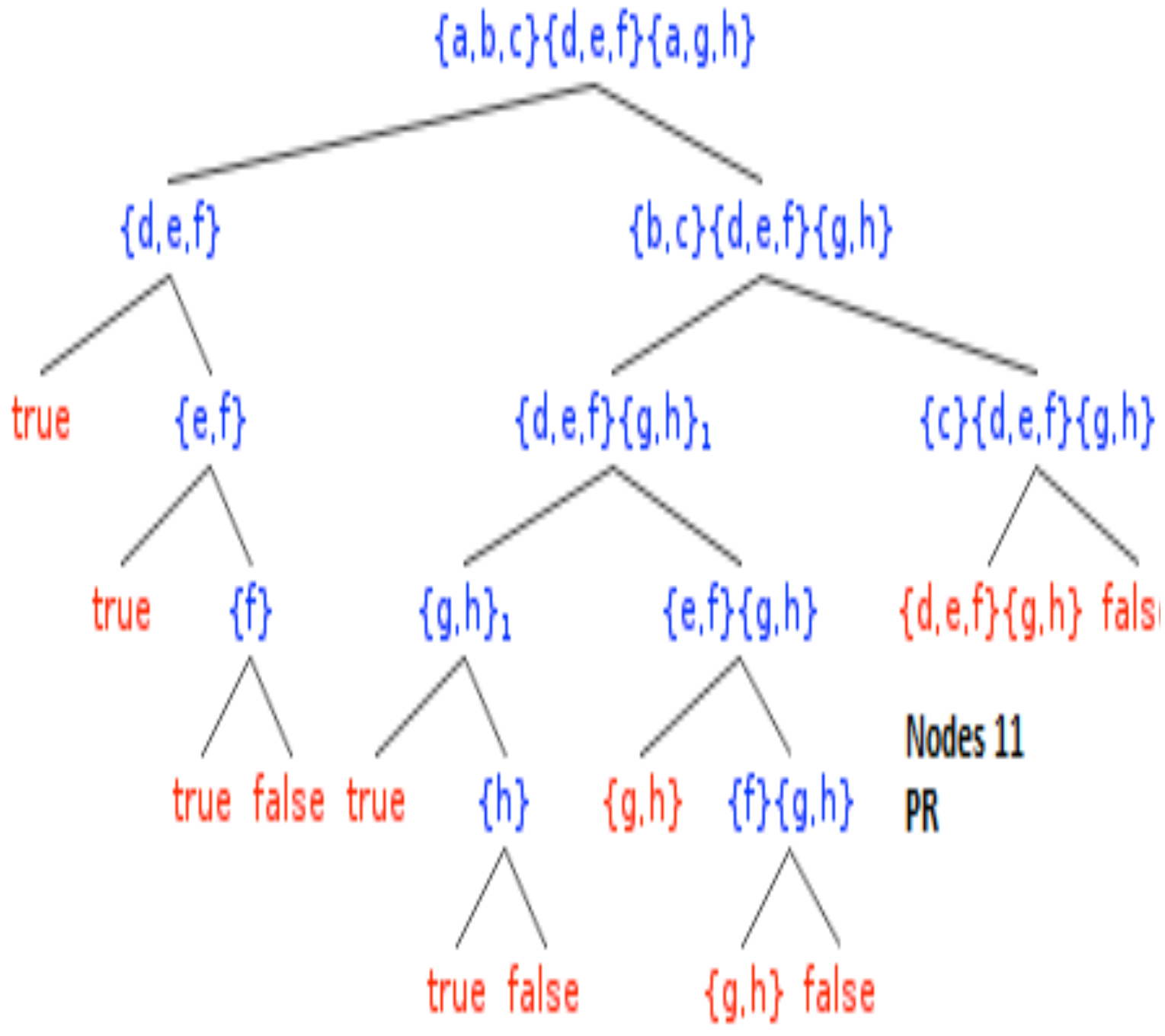

**Figure 56a**

121 The reader is encouraged to use shown trees to check other permutations as well (for example: *g*=*f*=*c*).

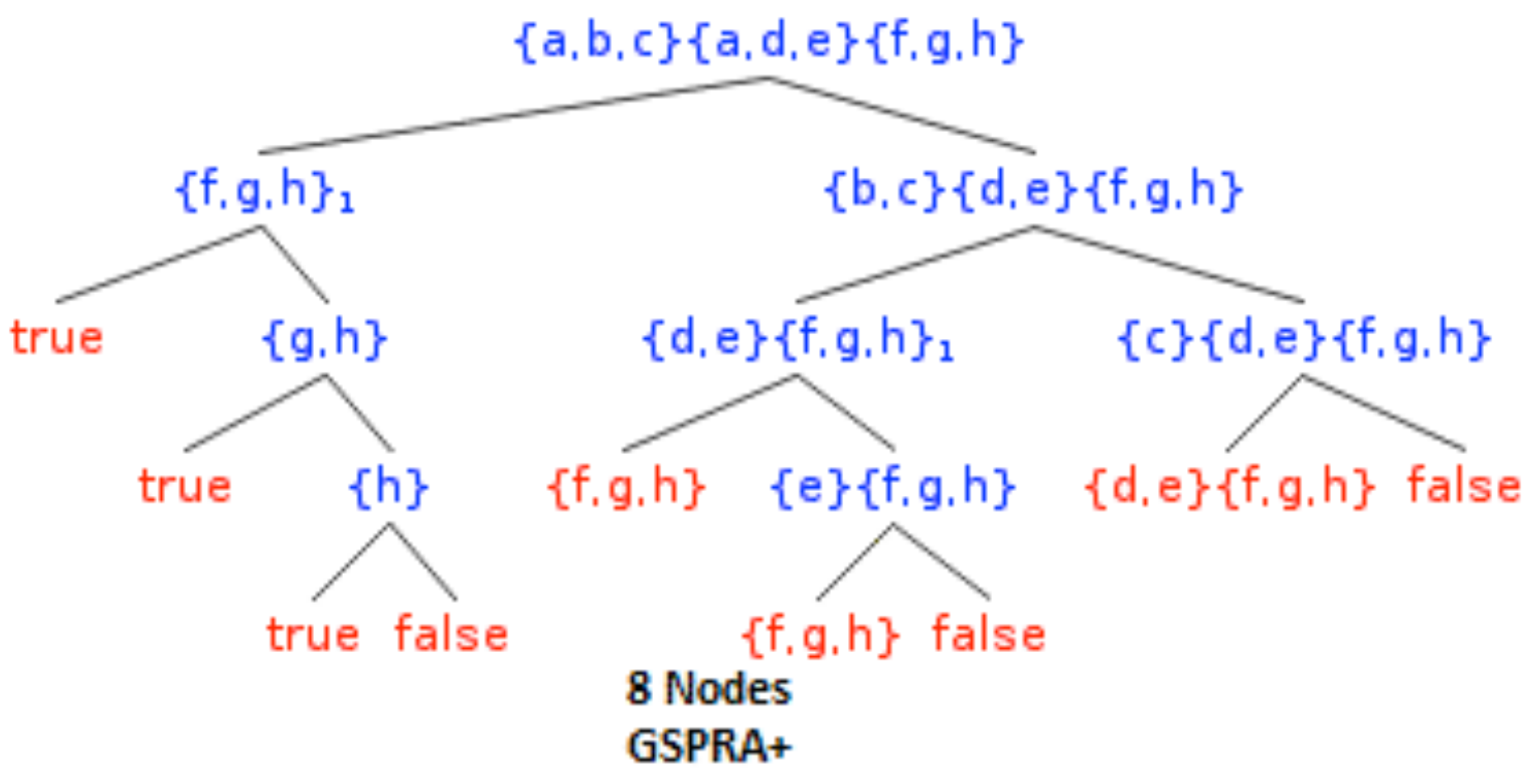

**Figure 56b**

Note that for S'={{*a,b,c*}{*d,e,f*}} in above upper tree (Fig. 56a) the number of nodes of a tree produced by PR is 6 making Alpha = 11/6 while S''={{*a,b,c*}{*a,d,e*}} in the above lower tree (Fig. 56b) as per $GSPRA^+$ has 5 nodes making Beta = 8/5, i.e., $\alpha>\beta$. If we set $g=c$ Clause-Sets become {{*a,b,c*}{*d,e,f*}{*a,c,h*}} for PR and {{*a,b,c*} {*a,b,d*} {*e,f,g*}} for $GSPRA^+$ (after $CRA^+$ is used) giving us the below two trees (Fig. 57a and 57b) verifying the claim as well, since Alpha=11/6 and Beta=7/4. Induction Hypothesis & Induction Step are both as in Lemma 14.

(Q.E.D.)

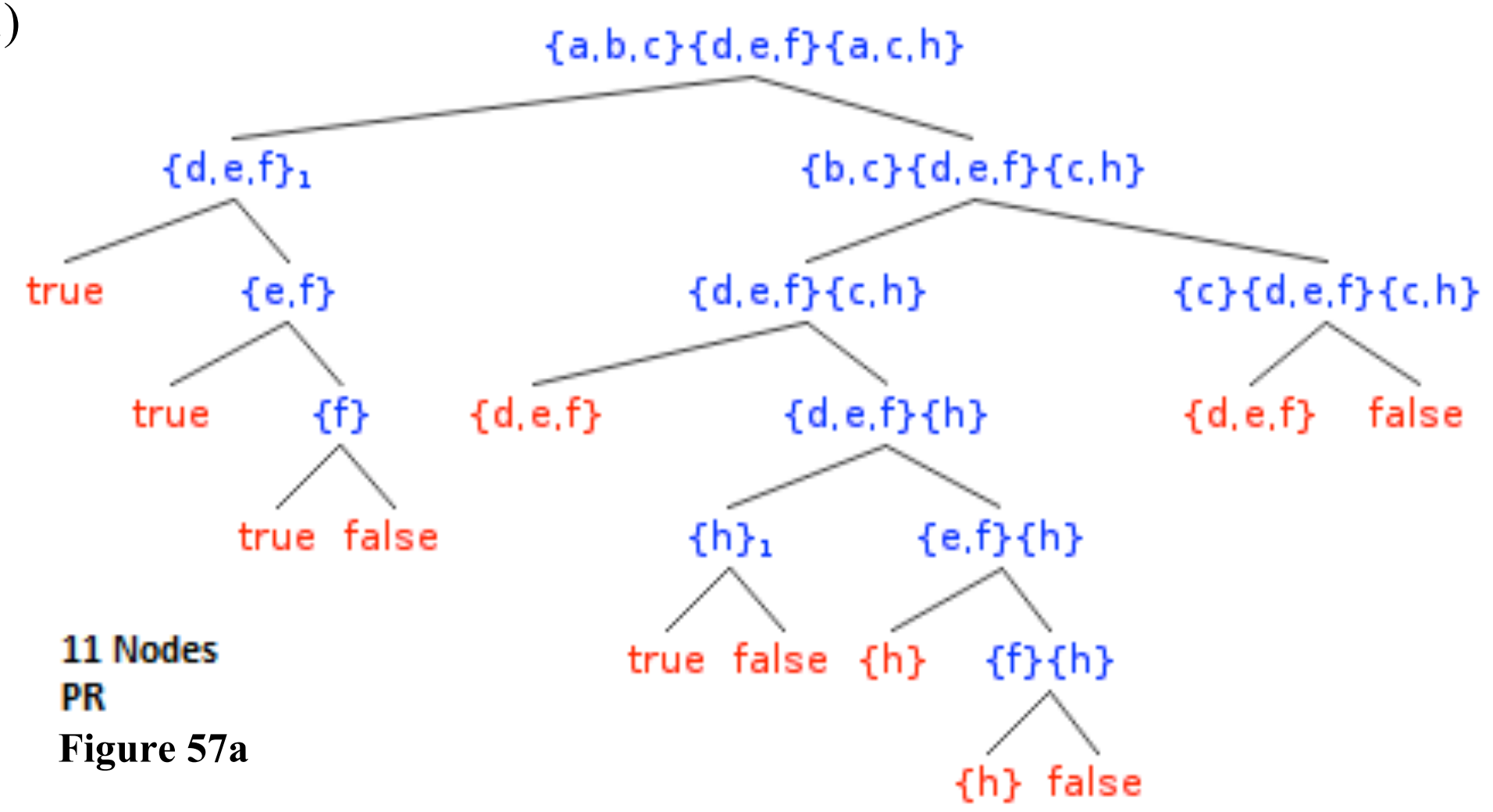

**Figure 57a**

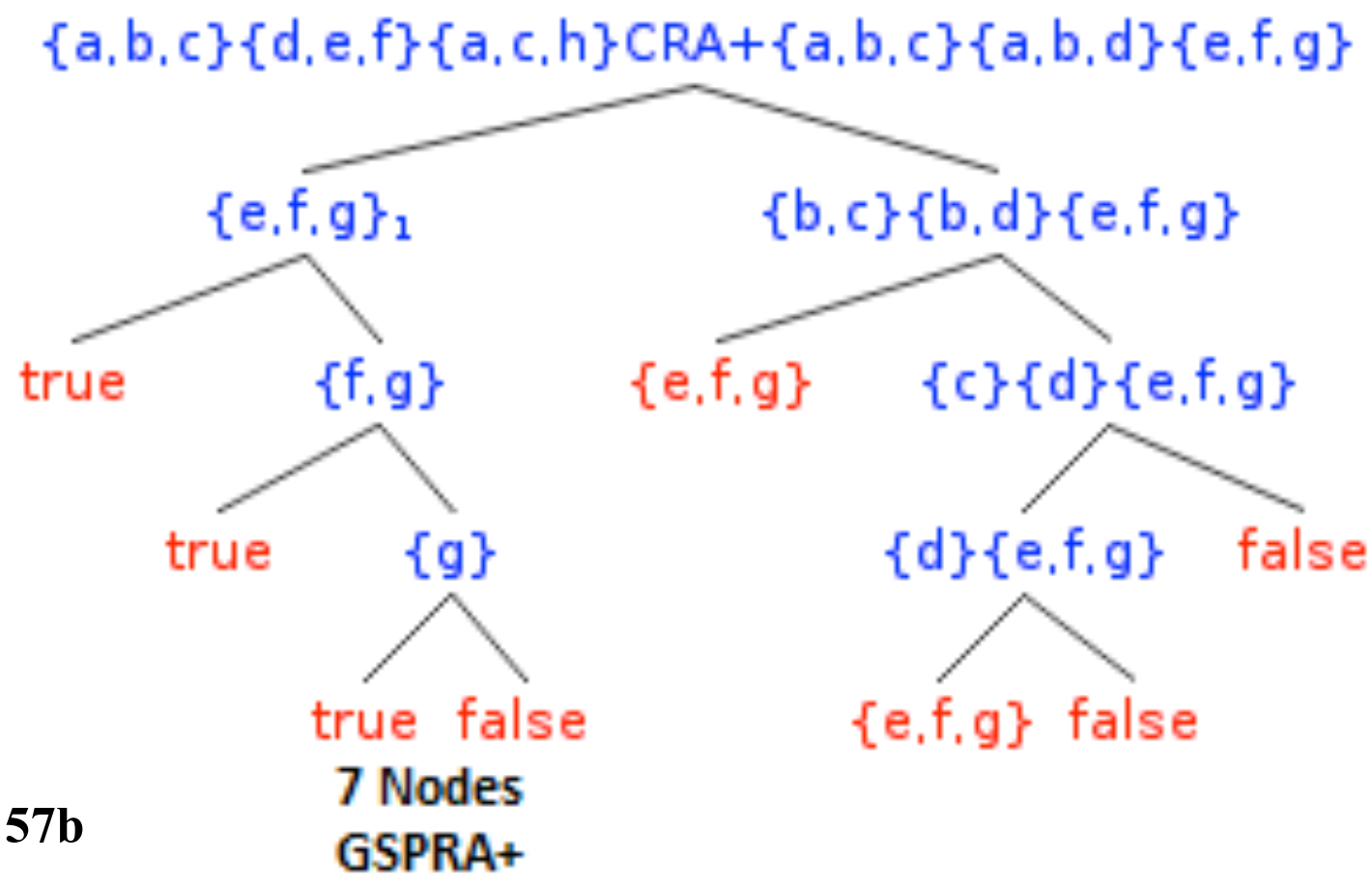

**Figure 57b**

**Lemma 16:** Suppose S is a 3-SAT-CNF Clause-Set which is a.a. or l.o.u., then the $MSRT_{s.o.}$ produced for S by $GSPRA^+$ is both, free of redundancy and minimal in the number of unique-nodes generated while resolving S compared to any other SPR-like or unlike resolution-procedure producing BigSps.
**Proof:** Let PR be an arbitrary SPR-like- or unlike resolution-procedure producing BigSps using an arbitrary ordering to resolve an S which is either a.a. or l.o.u. As per Property 9: If an $MSRT_{s.o.}$ produced by $GSPRA^+$ for S has a minimal number of unique-nodes with respect to all possible canonical orderings used by procedures, then it is minimal for all non-canonical orderings used by those procedures as well. Lemma 14, 15 state that the $MSRT_{s.o.}$ produced by $GSPRA^+$ for an a.a. or l.o.u. S has a number of unique-nodes which is minimal compared to any SPR-like- or unlike procedure using canonical orderings for S and producing BigSps. Hence, an $MSRT_{s.o.}$ produced by $GSPRA^+$ is minimal with respect to the number of nodes produced by PR as well. Moreover, $GSPRA^+$ is redundancy free because of Property 10 (Algorithmic Equivalence = Syntactical Equivalence) as well as Lemma 8 and steps 2-a-iii and 2-b-vi which implement both by guaranteeing that no new node is created if a node with the same Clause-Set (in CRA-form) already exists.
(Q.E.D.)

The above three lemmas of this section were concerned with the difference in node-counts between resolution-trees possessing BigSps and $MSRT_{s.o.}$s which are free of them. **The main statement of those lemmas being that any FBDD-minimizing SPR-like- or unlike procedure cannot produce BigSps while resolving an a.a. or l.o.u. Clause-Set, otherwise $GSPRA^+$ yields better results.** This means also as per Lemma 9(c) that optimal minimization procedures must produce $MSRT_{s.o}$s. It is still necessary to investigate the difference in size between $MSRT_{s.o.}$s, which may have a variety of node-counts. $GSPRA^+$ allows many strategies of selection of the first clause $C_0$ as long as they all result in a minimal top-part of the final tree. Also: SPR-unlike resolution-procedures may generate $MSRT_{s.o.}$s of competitive sizes although they don't guarantee minimal top-parts. In what follows we will show that any strategy used for generating a minimal top-part of the $MSRT_{s.o.}$s in $GSPRA^+$ produces a tree near-to-optimal with respect to the number of unique-nodes. This is sufficient for the lemmas and theorems of the next section.

**Lemma 17:** Suppose $A_1$ is an SPR-like- or unlike algorithm which always generates an SRT with minimal number of unique-nodes for an a.a. 3-SAT-CNF Clause-Set S, say $C_{Min}$ in its resolution-tree. $A_2$ is an algorithm using $GSPRA^+$, then there exists an integer $z$ such that the total number of nodes generated by $A_2$, say C', is bounded above by $C_{Min}*(1+1/z)$, $z>1$.

**Proof:** (Induction on M, the size of S)

**Base-Case M=1:** Both $A_1$ and $A_2$ produce the same number of nodes for the single clause, which is $k$[122], its breadth, $k<=3$, $C'=C_{Min}$.

**Induction Hypothesis:**
S is an a.a. 3-SAT-CNF Clause-Set of size M, T1 is a tree generated by $A_1$ with Base-Set $S_1$ and unique-node-count $C_{Min}$, T2 is a tree generated by $A_2$ with Base-Set $S_2$ and unique-node-count C', where $S=S_1=S_2$ via mapping and there exists an integer $z>1$ such that: $C'=C_{Min}*(1+1/z)$

122 Excluding TRUE and FALSE nodes.

**Induction Step: For Clause-Sets S of size M+1**:

Suppose $T=T_2$ is resolved in the induction step with a clause $C_M$ to give T', then there are two cases:

**Case 1:** $C_M$ doesn't affect the top-part of T, i.e., no re-arrangement of clauses of the Base-Set $S_2$ is needed and $C_M$ is just appended to the rear of all Clause-Sets of size M of the top-part of T to form the top-part of T', then:

The top-part of T' is still minimal if constructed by $A_2$ per definition of $GSPRA^+$. It is thus sufficient to investigate what happens in the bottom part. There are two possibilities depending on which algorithm is used:

$$C_{Min}^b=[a+b+c]-1/x*[a+b+c]=[a+b+c]*(1-1/x) \quad (1)$$

where $C_{Min}^b$ is the total number of unique nodes generated in the bottom part, *a,b,c* are the total number of nodes generated by $A_1$ for each respective TBN of T when it is resolved with $C_M$ (there are always maximum three of those TBNs), $1/x$ is the portion of nodes counted more than once ($C_{Min}^b$ should be redundancy free).

Or:

$$C^b= [a'+b'+c']-1/y * [a'+b'+c'] \quad (2)$$

where $a',b',c'$ and $y$ have similar meanings for $A_2$. Obviously if we choose $z = \mathrm{Max}(x,y)$ we have per minimalism of $A_1$:

$$C_{Min}^b<=C^b<=[a'+b'+c'](1-1/z) \quad (3)$$

Using induction hypothesis: There exists $z'>1$ such that:

$$C^b<=[a'+b'+c'](1-1/z)=(1+1/z')[a+b+c](1-1/z) \quad (4)$$

where $(1/z')[a+b+c]$ are surplus-nodes generated by $A_2$[123], then:

$$C^b<=(1+1/z')[a+b+c](1-1/z)$$
$$<=[a+b+c](1-1/z+1/z'-1/(z*z'))$$
$$<=[a+b+c]\ (1+1/z')$$

substituting for $[a+b+c]$ using (1)

$$<=C_{Min}^b * x/(x-1) * (1+1/z')$$
$$<= C_{Min}^b * [x/(x-1) + x/((x-1)*z')]$$
$$<= C_{Min}^b * [xz'+x/(xz'-z')]$$
$$<= C_{Min}^b * [(1+1/z')/(1-1/x)]$$

we want thus to select a $z''$ such that:

$$(1+1/z')/(1-1/x)=1+1/z''$$
$$1+1/z' =(1+1/z'')*(1-1/x)$$
$$= 1-1/x+1/z''-1/(x*z'')$$
$$<= 1+1/z''$$

i.e., $1/z'<=1/z''$ and $z'>=z''$

It is therefore sufficient to choose $z''=z'$ to see that

$$C^b <= C_{Min}^b * (1+1/z') \quad (5)$$

i.e., that there exists an integer $z'>1$ which, if assumed to count the surplus portion of nodes generated by $A_2$ in step *i*, is also counting the surplus-nodes generated by the same algorithm in step *i*+1.

**Case 2:** $C_M$ does affect the top-part of T, i.e., re-arrangement of clauses of the Base-Set $S_2$ is needed, then $GSPRA^+$ will convert $S_2$ to an l.o Clause-Set $S_2'=S_2$ of size M+1 (using $CRA^+$) whose last clause is $C_M'$ and reconstruct the tree for the first M clauses (as per step 2-b.v. of $GSPRA^+$'s definition). Set T equal to this newly constructed tree, then resolve $C_M'$ with T. The rest of the argument is similar to Case 1 since $C_M'$ will not affect the top-part of T per definition of l.o. Clause-Sets.

(Q.E.D.)

---

123 Recall that $A_1$ and $A_2$ are both applied on the same TBNs of T.

## F) FGPRA+ FOR ARBITRARY, 3-SAT-CNF CLAUSE-SETS

GSPRA+ is a constructive algorithm which served well to give us an upper bound for the number of nodes resulting from pattern-oriented resolution-methods, but it is hardly practical, because of the fact that it needs to rebuild trees of nodes again and again in each sequential step if they are found to be not l.o. (c.f. step 2.b.iv in the above description of Align). To overcome this drawback, the following algorithm (called: FGPRA+) works directly with the whole Set of 3-SAT-CNF-clauses and resolves them in parallel.

**FGPRA+:**

**Prerequisites**:

- convert arbitrary clauses in S to a.a. ones (sorting literals inside each clause)
- convert S to a l.o. Set S' using **CRA+**
- make sure to sort S so that a clause which produces the least top-part is selected first. Use for this **SelectFirstClause** procedure which was used for GSPRA+
- create a base-node having S' as Clause-Set and pass it to FGPRA+

**Inputs**: base-node

**Outputs**: an $MSRT_{s.o}$

**Data Structures used:** list of Tuples: < Clause-Sets, Node index> (called: LCS, initially empty)

**Steps**:

For current node *n*, Clause-Set S':

if size of *n* > 1

1. use least-literal/clause-rule to instantiate all clauses in S' accordingly
   a. **form** leftInstantiatedClauseSetOfS' and use CRA+ to make it l.o. if it is not
   b. form rightInstantiatedClauseSetOfS'and use CRA+ to make it l.o. if it is not
   c. make sure to sort both right- and left Clause-Sets of S' so that the clause which produces the least top-part is selected first. Use for this SelectFirstClause procedure which was used for GSPRA+
2. **search** for leftInstantiatedClauseSetOfS' in LCS
   a. if found: Set leftNodeIndex of *n* = node index found
   b. if not found:
      i. **create a new node** for leftInstantiatedClauseSetOfS'
      ii. give it an index *I* and **Store** the tuple <leftInstantiatedClauseSetOfS', *I* > in LCS
      iii. set leftNodeIndex of *n* = *I*
3. search for rightInstantiatedClauseSetOfS' in LCS
   a. if found: Set rightNodeIndex of *n* = node index found
   b. if not found:
      i. create a new node for rightInstantiatedClauseSetOfS'
      ii. give it an index *J* and store the tuple <rightInstantiatedClauseSetOfS', *J* > in LCS
      iii. set rightNodeIndex of *n* =*J*
4. set current node = leftNodeIndex of *n*

5. call yourself recursively with FGPRA$^+$(current node) if leftNode was newly formed and not TRUE or FALSE
6. set current node = rightNodeIndex of *n*
7. call yourself recursively with FGPRA$^+$(current node) if rightNode was newly formed and not TRUE or FALSE
8. return $MSRT_{s.o}$

If size of *n* =1

- **create $MSRT_{s.o}$ for this single clause** as described in Definition 2 step 1 if it is not already done
- return $MSRT_{s.o.}$

Now we need to show the equivalence between FGPRA$^+$ and GSPRA$^+$, i.e., that the same $MSRT_{s.o}$ is produced in both cases.

**Lemma 18:** For any arbitrary 3-SAT-CNF Clause-Set S: FGPRA$^+$ (S) =GSPRA$^+$ (S).

**Proof:** (induction on M, the size of S) Remembering that both algorithms use CRA$^+$ in their preparation phases on the same S.

**Base-Case:** M=1: They convert $C_0 \in S$ into the same $MSRT_{s.o}$ as prescribed in Definition 2. Thus, FGPRA$^+$ (S) =GSPRA$^+$ (S).

**Induction Hypothesis:**
For all 3-SAT-CNF Clause-Sets S of size M: FGPRA$^+$ (S) =GSPRA$^+$ (S).

**Induction Step:** If S is of size M+1, then it is sufficient to show that top-parts of both $MSRT_{s.o}$s constructed by GSPRA$^+$ and FGPRA$^+$ are equivalent, then use the induction hypothesis for the remainder. First note that Property 5 (Uniqueness of Instantiation Results) is valid for FGPRA$^+$ as it is for GSPRA$^+$ because they both are SPR-like and use the same least-literal/head-clause-instantiation rule. This property states that if S is a.a.[124] ($S_1$, $S_2$ = any children of S produced through instantiations of literals *i,j*, respectively), then $S_1$=$S_2$ **iff** *i=j*. As both algorithms instantiate the same $C_0$ using the same rule, all literals chosen for instantiation in the top-parts of both final $MSRT_{s.o}$s must be equal, thus Clause-Sets resulting from those instantiations (to be resolved further in the TBN bottom-parts) must per Property 5 also be equal. We use the induction hypothesis for all TBN M-sized nodes and conclude that FGPRA$^+$(S) =GSPRA$^+$ (S).
(Q.E.D.)

The following **main lemma** of this paper paves the way to a **new Solver** algorithm and Theorem 1.

**Lemma 19:** For the following Assistance Operations[125] used by FGPRA$^+$ on 3-SAT-CNF Clause-Set S of size M: Node creation, $MSRT_{s.o}$ creation for a single clause, CRA$^+$, Selecting a FirstClause, Forming new Clause-Sets using least-literal-rule (instantiation), Storing (nodes), Searching Clause-Sets in LCS: The total, worst case number of Primitive Operations[126] performed by any single one of them during a run of FGPRA$^+$ is: **$O(M^9)$** which therefore also represents the complexity of FGPRA$^+$.

[124] i.e., also valid for l.o. Clause-Sets.

[125] By *Assistance Operations* we mean modules and/or sub-functions used in the pseudo-code of FGPRA$^+$.

[126] Primitive Operations take a constant amount of time in the RAM computing model.

**Proof:** Because of Lemma 13 and 18 we know that the total number of unique-nodes in the final $MSRT_{s.o}$ produced by $FGPRA^{+}$ cannot exceed

$$3+ c*RCC_{3\text{-}SAT}^{2} *M^{4} + RCC_{3\text{-}SAT} *M^{3}.$$

The following are then upper bounds of the total number of invocations of Primitive Operations for all Assistance Operations listed above:

1. $3+ c*RCC_{3\text{-}SAT}^{2} *M^{4} + RCC_{3\text{-}SAT} *M^{3}$ times $CRA^{+}$ (each node needs renaming in the worst case).
   Through Lemma 6 it is known that $CRA^{+}$ takes $O(M^{2}(\log M+N))$. This makes the total worst case number of Primitive Operations of this category: $O(M^{6}(\log M+N)$.
2. $2*(3+ c*RCC_{3\text{-}SAT}^{2} *M^{4} + RCC_{3\text{-}SAT} *M^{3})$ times instantiation (two new Clause-Sets are formed for each node in the worst case). Instantiating a Clause-Set by substituting values TRUE or FALSE for a certain literal in all M clauses is an operation in O(M). This makes the total number of Primitive Operations for instantiation: $O(M^{5})$.
3. $3+ c*RCC_{3\text{-}SAT}^{2} *M^{4} + RCC_{3\text{-}SAT} *M^{3}$ times selecting the first clause to minimize top-parts using a procedure which tries all permutations of literals in a clause ($RCC_{3\text{-}SAT}$ in number) for each clause, i.e., O(M) taking the total number of Primitive Operations to $O(M^{5})$.
4. $3+ c*RCC_{3\text{-}SAT}^{2} *M^{4} + RCC_{3\text{-}SAT} *M^{3}$ times node creation assuming that it is in $O(c)$, i.e., $O(M^{4})$.
5. $3+ c*RCC_{3\text{-}SAT}^{2} *M^{4} + RCC_{3\text{-}SAT} *M^{3}$ times Storing/Appending in/to LCS assuming that it is in $O(c)$, i.e., $O(M^{4})$.
6. $MSRT_{s.o}$ creation for a single clause: $O(c)$, since independent of M any clause can have at most 3 literals where 3 nodes are created for each one of them.
7. $3+ c*RCC_{3\text{-}SAT}^{2} *M^{4} + RCC_{3\text{-}SAT} *M^{3}$ times Searching Tuples in LCS. This search operation can be accomplished in the least efficient way[127] by comparing the sought Clause-Set with all Clause-Sets stored in the LCS using the CompareSets algorithm of Section E-2 which is O(M). In the worst case there are $3+ c*RCC_{3\text{-}SAT}^{2} *M^{4} + RCC_{3\text{-}SAT} *M^{3}$ Clause-Sets in LCS, i.e., $O(M^{8})$ comparisons are needed. This makes the total number of Primitive Operations for Searching $O(M^{9})$.

(Q.E.D.)

[127] The least efficient way is chosen to avoid any assumptions regarding sort- and search orders of Clause-Sets in LCS.

The following algorithm (hereafter referred to as: Solver) uses $FGPRA^+$ to produce the answers TRUE or FALSE for any Set of arbitrary 3-SAT-CNF-clauses.

**Solver:**

**Prerequisites:**

- set $MSRT_{s.o} = FGPRA^+(S)$
- set current node = base-node of $MSRT_{s.o}$
- call Solver with the current node

**Inputs**: node of an $MSRT_{s.o}$
**Outputs**: **TRUE** or **FALSE**

**Steps** for current node N in $MSRT_{s.o}$:

- if N is leaf
  - if *-ve* then
    - Return FALSE
  - if *+ve* then
    - Stop giving **TRUE**
- if N not leaf
  - set current node = leftnode of N in $MSRT_{s.o}$
  - call yourself recursively: Solver(current node)
  - set current node = rightnode of N in $MSRT_{s.o}$
  - call yourself recursively: Solver(current node)

Step 4: stop giving **FALSE**

**Theorem 1:** For any Set of arbitrary 3-SAT-CNF-clauses of size M Solver takes a polynomial number of steps bounded by $O(M^9)$. Moreover: P=NP.

**Proof:** In the preparation phase, Solver has to use $FGPRA^+$ and (in the worst case) has then to traverse all nodes of the final $MSRT_{s.o}$ produced by it (which is an operation in $O(M^4)$). The result follows immediately from Lemma 19. Since 3-SAT is a NP-complete problem, P=NP.
(Q.E.D.)

**Thus, the formulated objective in Section I has been attained.**

In the last part of this section we show that $FGPRA^+$ is a polynomial 2-Approximation algorithm for the MinFBDD problem[128]. We conclude by affirming for any Boolean function the existence of minimal FBDDs, which are polynomial in M, the number of clauses.

**Definition 17 (Approximation Algorithms)**:
Let X be a minimization problem. Let $\varepsilon > 0$, and set $\rho = 1+\varepsilon$. An algorithm A is called a $\rho$-approximation algorithm for problem X, if for all instances I of X it

[128] In [Seshia et al. 2000] the problem of minimizing level *i* of a FBDD where the number of nodes is less than $2^{i-1}$ is shown to be not approximable to within a factor of $2^{\log^{(\varepsilon-1)} n}$ for any $\varepsilon>0$, unless the class NP is contained in RQP (the class of all problems solvable in random quasi-polynomial time). This is the closest *-ve* result related to minimizing FBDDs known to us. A 2-approximation algorithm like the one shown here is not conclusive enough to deduce the inclusion NP $\subseteq$ RQP using this result because it might still be the case that whatever is lost in terms of minimization at level *i* is compensated in other levels of the FBDD to produce an overall 2-approximation.

delivers a feasible solution with an objective value A(I) such that:

$$|A(I) - Opt(I)| \leq \varepsilon * Opt(I)$$

In this case, the value ρ is called the performance guarantee or the worst-case ratio of the approximation algorithm A.

**Theorem 2:** Let ***f*** be a Boolean function of N variables described by an a.a. 3-SAT-CNF Clause-Set S of size M. **MinFBDD** is the problem of computing a FBDD for *f* which has minimal size. Then:

1- $FGPRA^+$ is a 2-Approximation Algorithm for **MinFBDD**.
2- *f* has a minimal FBDD with polynomial number of nodes in M.

**Proof:**
**First Claim:** Suppose *f* has some minimal $FBDD_1$, then per Property 7(b) there is a procedure PR (SPR-like- or unlike) which uses variable orderings to generate $FBDD_1$. On the other hand, $FGPRA^+$ is per Lemma 19 a polynomial time algorithm and produces per Lemma 17 for S a $FBDD_2$ which has $C <= C_{Min*}(1+1/z)$ number of nodes for some $z>1$, where $C_{Min}$ is the optimal node-count produced by PR in $FBDD_1$.
In general we can say that for any instance I of MinFBDD (representing any function *f* expressible in a.a. 3-SAT-CNF form):

$$|FGPRA^+(I) - Opt(I)| \leq \varepsilon * Opt(I),$$
$$Opt(I)=C_{Min}, \varepsilon=1/z, z>1, \text{ making}$$
$$\rho=1+1/z<=2.$$

Therefore, $FGPRA^+$ is a polynomial time 2-approximation algorithm for MinFBDD.

**Second claim:** Since C is in $O(M^4)$ per Lemma 13 and $C_{Min}<=C$ we know that both $FBDD_1$ and $FBDD_2$ must have a polynomial count of unique nodes. As for the time being we don't have a clue on how PR is calculating its optimal $FBDD_1$, we cannot conclude through this direction that the MinFBDD problem has an exact solution produced by a polynomial algorithm[129].
(Q.E.D.)

**The conjecture formulated in Section I is therefore proven.**

## III) Application: Solving Blocking Sets Problems of Projective Plains using $FGPRA^+$

As early as [Bryant 1986] it is known that practically important Boolean functions such as *Integer Multiplication* which are in P can also possess exponentially sized OBDDs and FBDDs.
Since $FGPRA^+$ produces FBDDs and in light of our new results, we are also concerned with discussing exponential lower bounds related to FBDDs (c.f. [Ponzio 1998] for example).
The way of work of $FGPRA^+$ imposes yet another consideration, namely CNF-compactness, i.e.: In case an input-function to be resolved is already expressed in exponentially many clauses M (with respect to the number of input variables N), $FGPRA^+$ can only produce, relative to N, exponentially big FBDDs[130]. For example : Known CNF-representations of Integer Multiplication are exponentially long (including those assumed in Bryant's proof). This leaves such a function out of discussion here.

129 Although Theorem 1 is essentially an indication that PR can be efficient.

130 It is important to remember that the complexity of an algorithm in P is bounded by a polynomial function in the size of the input (here represented by M, the number of clauses), i.e., if any given input is already exponentially big with respect to certain reference variables, this doesn't affect the asymptotic behavior of the algorithm which will still produce a polynomial function of that exponential size.

There has been some work comparing BDDs to CNF-representations.
However, they aren't really comparable in that there are functions with small BDDs but exponential CNF-representations (e.g., the odd parity function, XOR) and vice-versa (c.f. [Devadas 1993]).
The latter type of functions represents a challenge to the ideas presented in this paper. As per Lemma 13 and Theorems 1,2 it should be always the case that any compact CNF-representation of a Boolean function yields polynomial-sized FBDDs. For Devadas' function, the exponential-size result has been proven for OBDDs and there is indeed a polynomial-sized FBDD solving that function. Are there problems for which a compact CNF-representation has a proven exponential lower bound for FBDDs?

*Yes.*

In [Gal 1996], the function defined by blocking sets of a finite projective plane

> **Definition 18:** Let ∏ = (P, L) be a projective plane of order *q*. *(P* is the set of points and L is the set of lines, viewed as subsets of P). Let $n=q^2+q+1$ and $m=q+1$. So Abs(P) = Abs(L) = *n,* each line has *m* points, and each point is incident with *m* lines.

is shown to have FBDDs of size $2^\Omega(\sqrt{n})$. This result practically says that FBDDs constructed for blocking Sets of PPs have their **first *q* levels equivalent to a complete binary tree**.
How does $FGPRA^+$ perform in the blocking-sets-of-finite-projective-planes case? Appendix A and B show runs of $FGPRA^+$ for planes of order *q*=2 (Fano) and *q*=3 respectively (only first part of the final FBDD is presented), whose original problems are expressed in 3-SAT, 4-SAT Clause-Sets as follows:

| $PG_2(2)$ | $PG_2(3)$ |
|---|---|
| 0 1 2 | 0 1 2 3 |
| 0 3 4 | 0 4 5 6 |
| 0 5 6 | 0 8 9 12 |
| 1 3 5 | 0 7 10 11 |
| 2 4 5 | 1 4 7 8 |
| 2 3 6 | 1 6 9 11 |
| | 1 5 10 12 |
| | 3 4 9 10 |
| | 2 4 11 12 |
| | 2 5 7 9 |
| | 3 6 7 12 |
| | 3 5 8 11 |
| | 2 6 8 10 |

Note the following observations:
In the k-SAT-representation of this problem N (number of variables/literals) and M (number of clauses) represent number of points and lines (denoted in Definition 18 as P and L), M=N=*n*. The breadth of clauses *k*=*m* is the number of points per line as well as intersecting lines per point.
$FGPRA^+$ utilizes a 3-SAT representation of k-SAT clauses and uses $CRA^+$ to produce l.o. Base-Sets (c.f. Appendix B) before resolving.
For all projective planes with *q*>=3 the FBDDs generated by $FGPRA^+$ ***do not show a complete binary tree in the first q levels*** (c.f. circled parts of the trees in appendix B). This is mainly caused by Property 8' of $FGPRA^+$. In addition to that: Practical implementations of $FGPRA^+$ realized for *q*=3, M=51 clauses show a total number of nodes of 176,839 with polynomial growth ($O(M^4)$ as predicted)[131] illustrated in following Fig. 58:

[131] There are of course more efficient, not necessarily equisatisfiable ways of translating 4SAT Clause Sets into 3Sat ones. For example the transformation : $\{x_1,x_2,x_3,x_4\}$ => $\{x_1,x_2,z_1\}\{\neg z_1,x_3,x_4\}$ yields only 26 Clauses and 2832 nodes for q=3

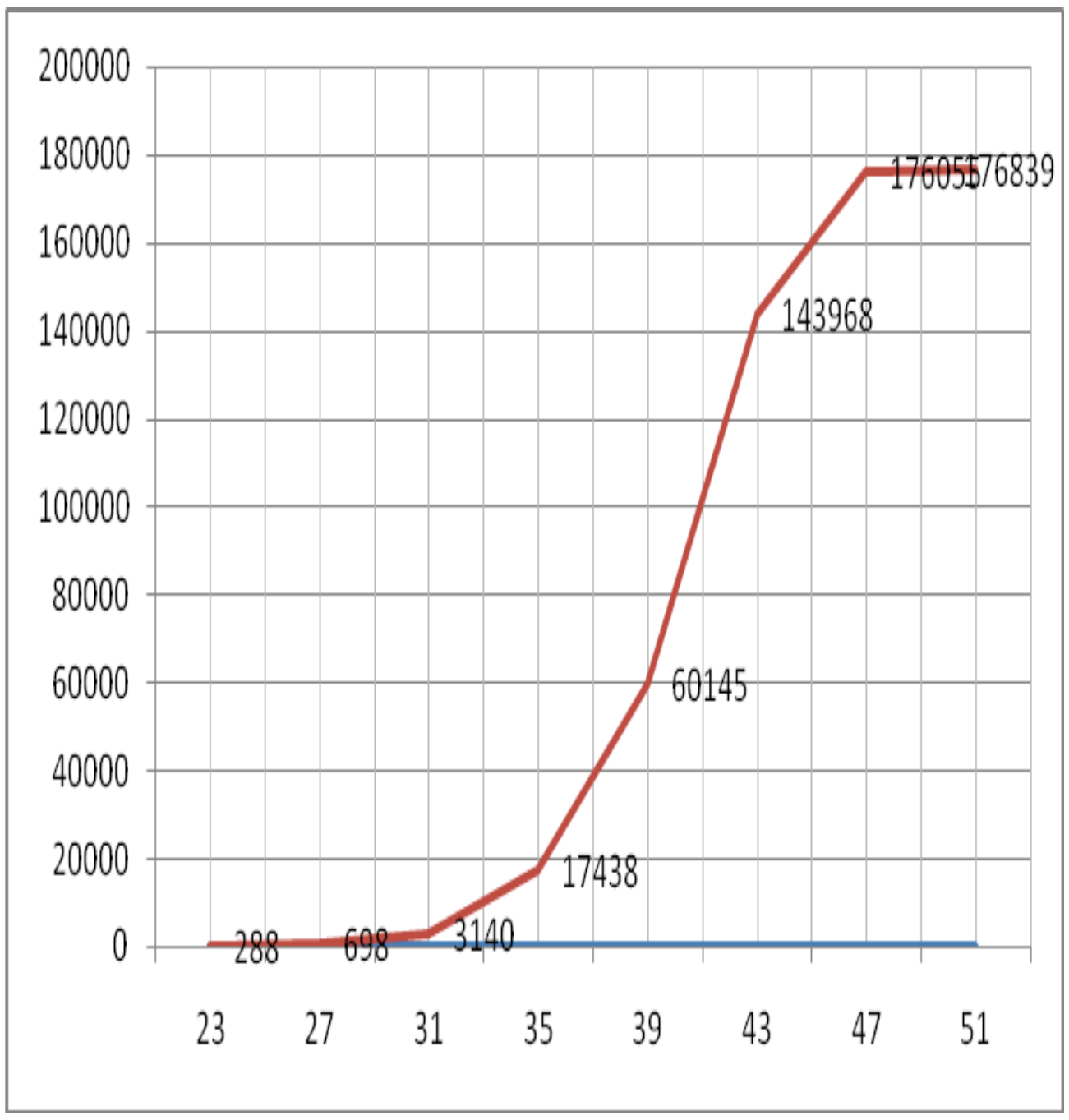

**Figure 58:** The x-axis represents the number of clauses where the y-axis represents the number of nodes generated in the FBDD.

The final theorems of this paper use the following practical definition of compact CNF-representations.

**Definition 19:** A Boolean function *f*(N) (for N = input variables) is said to possess a k-SAT-CNF-compact-form if the number of clauses M expressing *f* in k-CNF is polynomial in N.

**Theorem 3:** The function defined by blocking sets of a finite projective plane (Definition 18) possesses a FBDD of polynomial size. The number of nodes of this FBDD is in $O((q+1)^4M^4)$ with M= number of points/lines, q order of the plane.

**Proof:** Note first that the proof of the exponential lower bound given in [Gal 1996] only holds under the assumption that clauses represent complete lines:

"**Proof of the theorem.** We show that for every $q$-element subset $A$ of the variables, $N(f_{\Pi}, A) = 2^q$ holds, i.e., each truth assignment to the variables in $A$ yields a different sub-function on the remaining variables. Since each line defines a clause $\vee_{i\in\lambda} x_i$ of the function $f_{\Pi}$, it follows from the Fact[132] that for an arbitrary $q$-element subset $A$ of the variables there exist $q$ clauses such that each variable from $A$ appears in exactly one of them, and each variable appears in a different clause."[133]

When FGPRA+ converts the k-SAT description to 3-SAT, this problem structure dissolves and original lines/clauses are split into smaller ones not fulfilling the condition that "each line defines a clause of the function" described in the citation and crucial for establishing the lower bound result. Moreover, converting clauses to 3-SAT in an equisatisfiable way produces for each original line/clause $m=q+1$ additional "portions" at most (connected using newly introduced variables/literals) thus making the total amount of clauses given to FGPRA+ $(q+1)$*M. Eventually, this fact along with Lemma 13 explain $O((q+1)^4M^4)$ as the worst case for the number of nodes in the FBDD constructed for this problem by FGPRA+.
(Q.E.D)

132 *Fact* is a combinatorial property of projective planes.
133 [Gal 1996] p.15

It is highly probable that other exponential lower bounds for functions possessing a compact CNF which use the following common observation for Boolean functions in their proofs have to be revised in the light of our results in a way similar to Theorem 3, i.e., taking into consideration that the 3-SAT formulation may destroy the assumed clause-structure of *f* in the given problem:

> "Lemma: Let *f* be a Boolean function of *n* variables. Assume that *m* is an integer, $1 < m < n$, if for *m* any *m*-element subset Y of the variables $N(f, Y) = 2^m$ holds[134], then the size of any read-once branching program computing *f* is at least $2^{m-1}$."[135]

Independent of that we can establish:

**Theorem 4:** Suppose a Boolean function *f* possesses a compact CNF-form, then it possesses FBDDs of polynomial size in N as well as M (for N = number of variables and M = number of clauses).

**Proof:** Follows immediately from Definition 19, Lemma 13 as well as from Theorem 2.
(Q.E.D)

134 N(*f*,Y) denoting the number of different sub-functions obtained under all possible assignments to Y.
135 c.f. proof of this lemma in [Gal 1996]

## VI) RESULTS AND DISCUSSION

Efforts to solve the NP-Problem have been going on for decades. In the here presented work, the inability to solve it for so much time has been taking on heuristically by considering the contemporary computer scientific- and mathematical paradigm as inappropriate.[136]

*A new paradigm is required.*

In the context of the 3-SAT-problem, we have introduced a constructive view which identifies a *variable* with an *intrinsic logical truth pattern* holding information about its semantics and interacting with similar ones to facilitate calculating the desired overall truth value of a formula in an efficient way.
We have shown that this new perception divides 3-SAT-CNF-formulas into three main types:

1. almost arbitrary (a.a.)
2. linearly ordered but unsorted (l.o.u.)
3. linearly ordered (l.o.).

New *pattern-oriented algorithms* (AP), which are shown to be efficient, make use of this distinction and of the canonical ordering(s) induced by renaming operations in respective sub-problems. In addition, FBDDs constructed by AP are *near-to-minimal* with respect to all other possible variable orderings. *The first result answers the P=NP question positively* while the second answers the open question whether there are constant approximation algorithms for the MinFBDD problem *positive* as well.
Finally, the positive answer of this second problem entails a practical result showing that it is always possible for compactly expressed Boolean functions to construct *near-to-minimal* FBDDs with polynomial number of nodes.
As per the introduction, the work presented in this paper is inspired by ideas originating from ancient Muslim scientists who effectively built their theoretical and practical insights to serve humanity.

They must have truly believed in the epistemological statement:

"[…] of knowledge it is only a little that is communicated to you […]" [137]

It motivated ancient Muslim scientists to keep looking for practical solutions even for seemingly hard problems. Ultimately, human knowledge is not a reflection of ontological *reality*, but a matter of subjective perception.[138]

136 c.f. [Daghbouche 2012 (1)]

137 Quran: Al-Israa, 85

138 c.f. [Daghbouche 2012 (2)]

## ACKNOWLEDGEMENT

I would like to thank *Menassie Ephrem* for his valuable feedback, revisions of definitions, and proofs in the Sections II, A and B.

My deepest thanks go to my family, to my mother *Nadra Shannan* and father *Mahmoud Abdelwahab* and to my brother and sister for their continuous moral support.

This paper would not have been possible without the continuous motivation and the dedicated effort provided by my friend and brother *Karim Daghbouche* whose sound epistemological framework and fruitful discussions were of mission-critical value in pursuing and précising the approach to finally reflect the theoretical and practical results.

This work is dedicated to my beautiful wife and children wishing they could forgive me my way too long, unexcused absence.

## APPENDIX

A) *Fano Plane*, $PG_2(2)$,S={0,1,2}{0,3,4}{0,5,6}{1,3,5}{1,4,6}{2,3,6}{2,4,5},S is l.o.

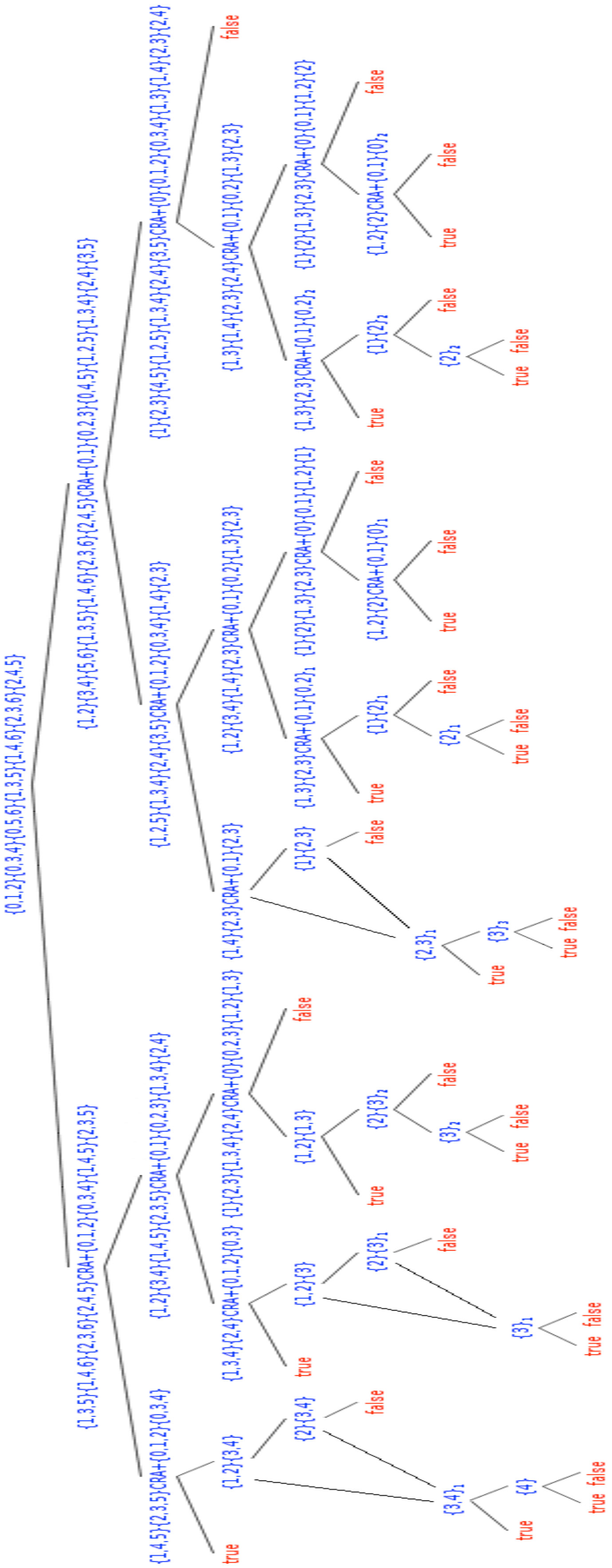

B) $PG_2(3)$

After conversion to 3-SAT: N=63 variables, M=51 clauses:
S={{0,1,2!}{0,3,4!}{0,5,6!}{0,7,8!}{1,3,9!}{1,10,11!}{1,12,13!}{2,14,15!}{3,16,17!}{3,18,19!}{4,20,21!}{5,7,22!}{5,23,24!}{5,25,26!}{6,27,28!}{7,29,30!}{7,31,32!}{8,33,34!}
{9,22,35!}{10,12,20!}{10,16,36!}{10,18,37!}{11,38,39!}{12,16,40!}{12,18,41!}{13,42,43!}
{14!,16,18}{15,21,44!}{17,45,46!}{19,47,48!}{23,25,33!}{23,29,47!}{23,31,38!}{24,40,49!}
{25,29,42!}{25,31,45!}{26,37,50!}{27!,29,31}{28,34,51!}{30,36,52!}{32,41,53!}{35,43,54!}
{39,48,55!}{44,51,56!}{46,52,57!}{49,58,59!}{50,53,60!}{54,44,61!}{56,61,62!}{57,58!,60}
{59,62}}. S is l.o.
The conversion is done in the following way. The Clause
(A[1] or A[2] or A[3] or A[4] or A[5] or A[6] or A[7])
yields the following set of clauses.
(A[1] or A[2] or ~X[1])
(A[3] or A[4] or ~X[2])
(A[5] or A[6] or ~X[3])
(X[1] or X[2] or X[3] or A[7])
The last clause is not 3-sat so the algorithm is re-run on this last clause, yielding the following new clauses:
(X[1] or X[2] or ~Y[1])
(X[3] or A[7] or ~Y[2])
(Y[1] or Y[2])

(developed for illustration stepwise from M=4 to until M=8 only)

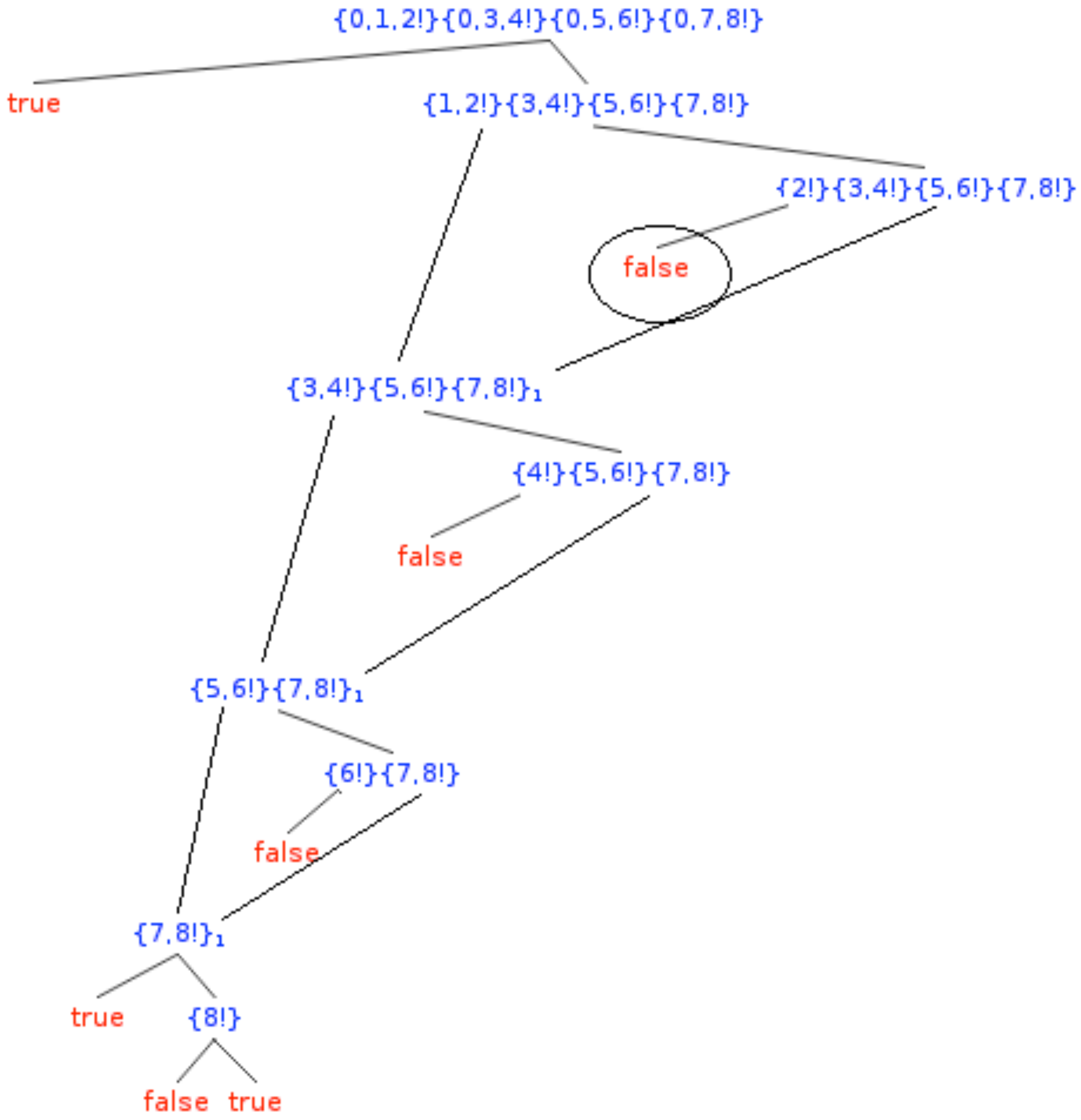

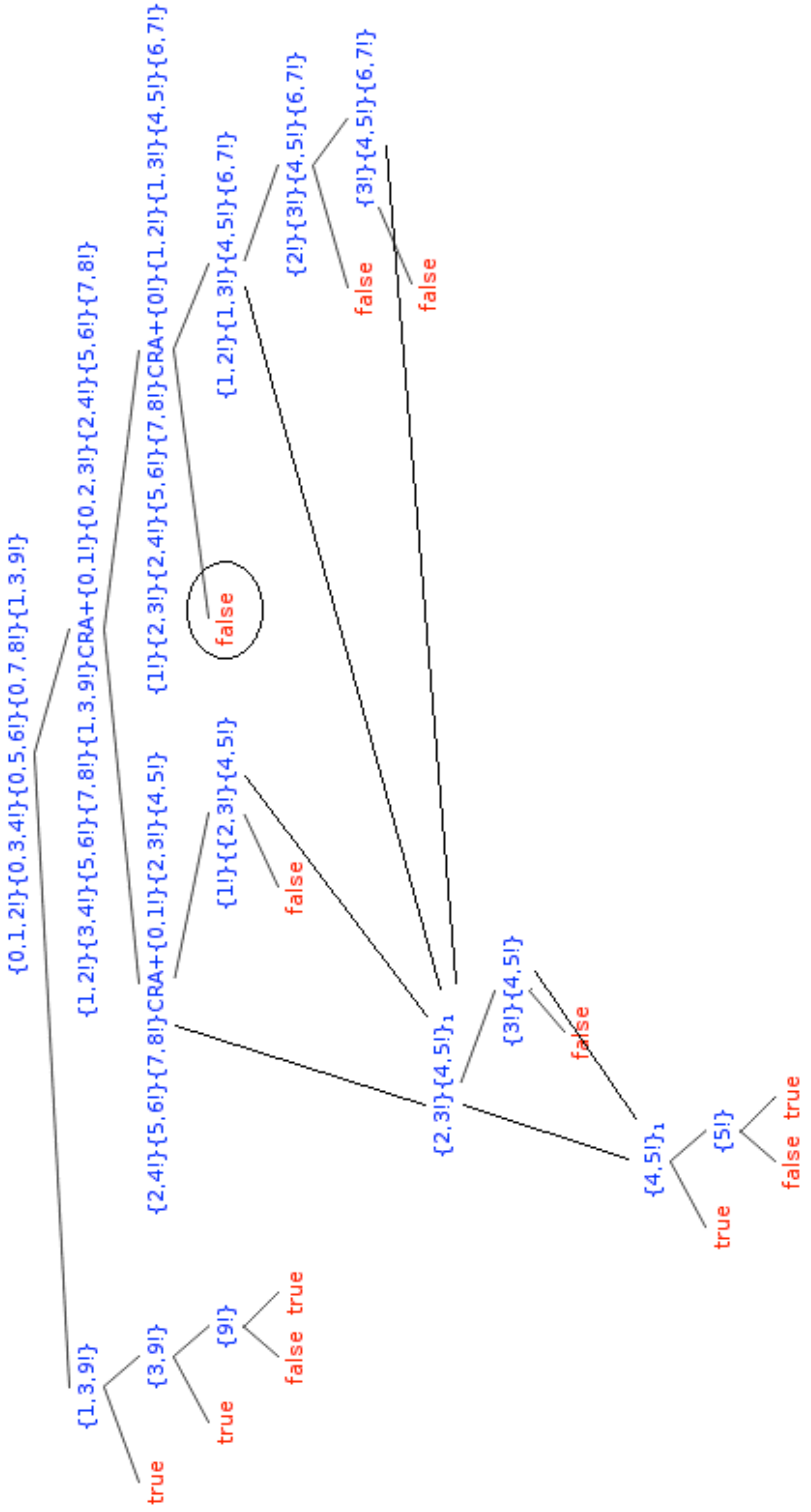
{0,1,2!}{0,3,4!}{0,5,6!}{0,7,8!}{1,3,9!}
{1,2!}{3,4!}{5,6!}{7,8!}{1,3,9!}CRA+{0,1!}{0,2,3!}{2,4!}{5,6!}{7,8!}
{2,4!}{5,6!}{7,8!}CRA+{0,1!}{2,3!}{4,5!}
{1!}{2,3!}{2,4!}{5,6!}{7,8!}CRA+{0!}{1,2!}{1,3!}{4,5!}{6,7!}
false
{1,2!}{1,3!}{4,5!}{6,7!}
{2!}{3!}{4,5!}{6,7!}
false
{3!}{4,5!}{6,7!}
false
{1!}{2,3!}{4,5!}
false
{2,3!}{4,5!}1
{3!}{4,5!}
false
{4,5!}1
true
{5!}
false
true
{1,3,9!}
true
{3,9!}
true
{9!}
false
true

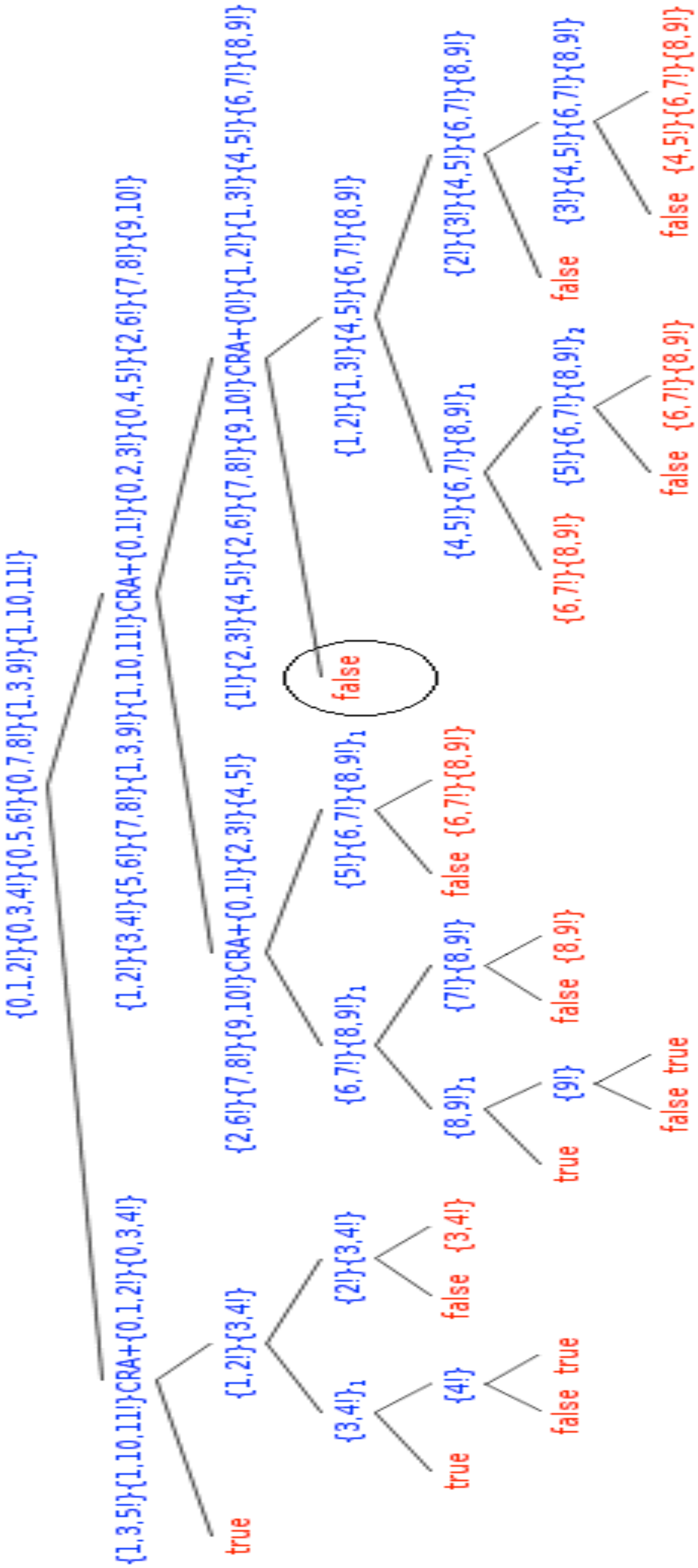
{0,1,2!}{0,3,4!}{0,5,6!}{0,7,8!}{1,3,9!}{1,10,11!}
{1,3,5!}{1,10,11!}CRA+{0,1,2!}{0,3,4!}
{1,2!}{3,4!}{5,6!}{7,8!}{1,3,9!}{1,10,11!}CRA+{0,1!}{0,2,3!}{0,4,5!}{2,6!}{7,8!}{9,10!}
true
{1,2!}{3,4!}
{2,6!}{7,8!}{9,10!}CRA+{0,1!}{2,3!}{4,5!}
{1!}{2,3!}{4,5!}{2,6!}{7,8!}{9,10!}CRA+{0!}{1,2!}{1,3!}{4,5!}{6,7!}{8,9!}
{3,4!}$_1$
{2!}{3,4!}
{6,7!}{8,9!}$_1$
{5!}{6,7!}{8,9!}$_1$
false
{1,2!}{1,3!}{4,5!}{6,7!}{8,9!}
true
false
{3,4!}
{8,9!}$_1$
{7!}{8,9!}
false
{6,7!}{8,9!}
{4,5!}{6,7!}{8,9!}$_1$
{2!}{3!}{4,5!}{6,7!}{8,9!}
{4!}
false
true
true
{9!}
false
{8,9!}
{6,7!}{8,9!}
{5!}{6,7!}{8,9!}$_2$
false
{3!}{4,5!}{6,7!}{8,9!}
false
true
false
{6,7!}{8,9!}
false
{4,5!}{6,7!}{8,9!}

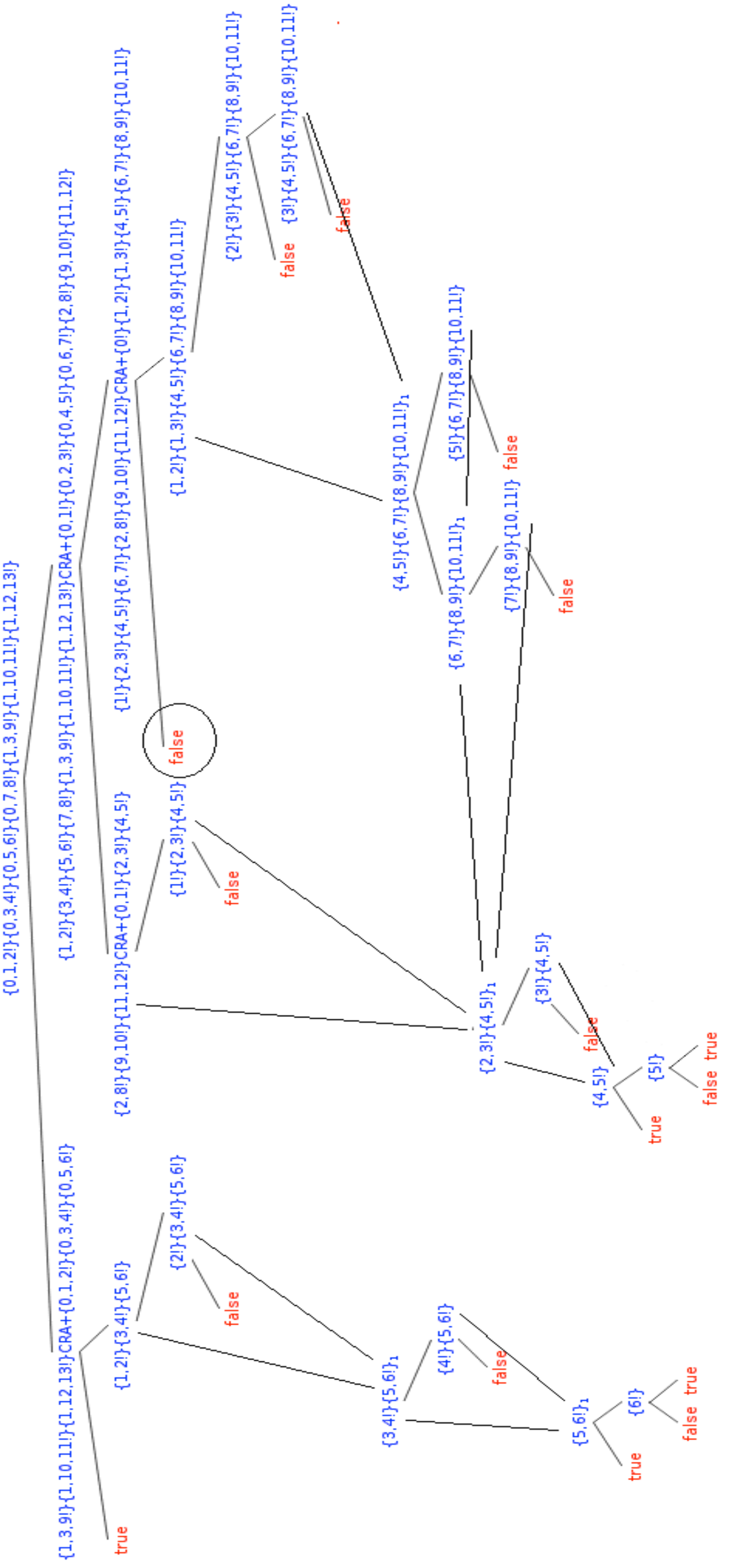
{0,1,2!}{0,3,4!}{0,5,6!}{0,7,8!}{1,3,9!}{1,10,11!}{1,12,13!}
{1,2!}{3,4!}{5,6!}{7,8!}{1,3,9!}{1,10,11!}{1,12,13!}CRA+{0,1!}{0,2,3!}{0,4,5!}{0,6,7!}{2,8!}{9,10!}{11,12!}
{1,3,9!}{1,10,11!}{1,12,13!}CRA+{0,1,2!}{0,3,4!}{0,5,6!}
true
{1!}{2,3!}{4,5!}{6,7!}{2,8!}{9,10!}{11,12!}CRA+{0!}{1,2!}{1,3!}{4,5!}{6,7!}{8,9!}{10,11!}
{2,8!}{9,10!}{11,12!}CRA+{0,1!}{2,3!}{4,5!}
{1,2!}{3,4!}{5,6!}
{1,2!}{1,3!}{4,5!}{6,7!}{8,9!}{10,11!}
false
{1!}{2,3!}{4,5!}
false
{2!}{3,4!}{5,6!}
false
{2!}{3!}{4,5!}{6,7!}{8,9!}{10,11!}
false
{3!}{4,5!}{6,7!}{8,9!}{10,11!}
false
{4,5!}{6,7!}{8,9!}{10,11!}₁
{5!}{6,7!}{8,9!}{10,11!}
false
{6,7!}{8,9!}{10,11!}₁
{7!}{8,9!}{10,11!}
false
{2,3!}{4,5!}₁
{3!}{4,5!}
false
{4,5!}
true
{5!}
false
true
{3,4!}{5,6!}₁
{4!}{5,6!}
false
{5,6!}₁
true
{6!}
false
true

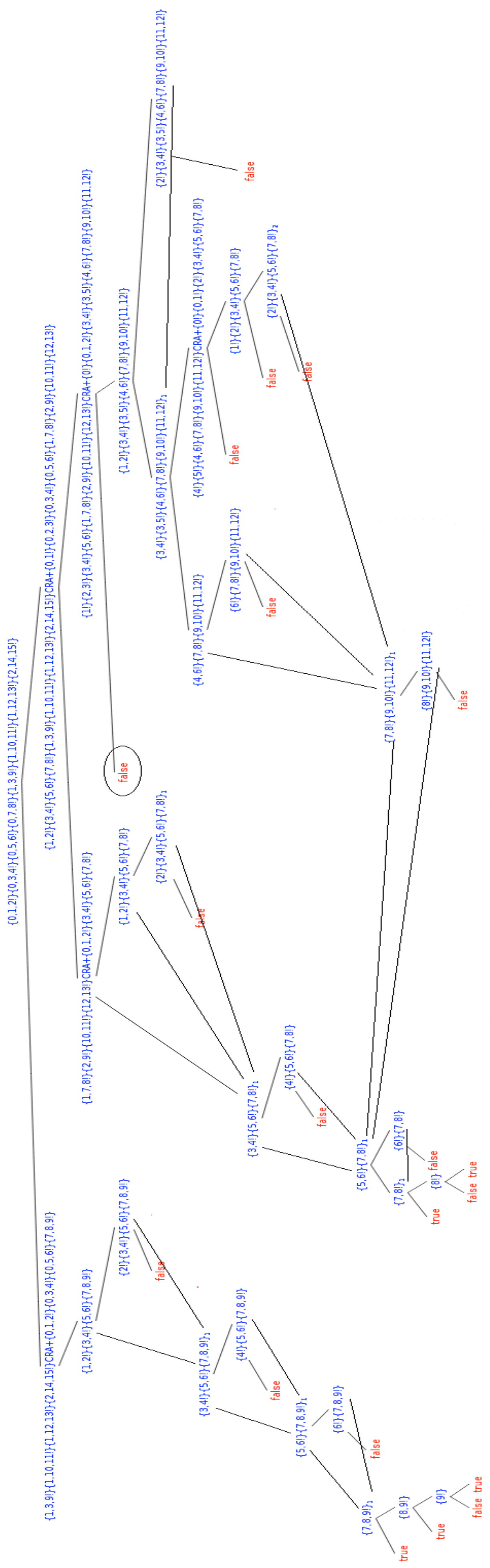